\documentclass[a4paper,12pt]{book}
\usepackage[dutch,english]{babel}
\usepackage{latexsym}
\usepackage{amsbsy}
\usepackage{fancyhdr}
\usepackage{sectsty}
\usepackage[dvips]{graphicx}
\usepackage{epsf}
\usepackage{epsfig, color}
\usepackage{a4}
\usepackage{amsmath}
\usepackage{amsfonts}
\usepackage{mathrsfs}
\DeclareMathSymbol{\mg}{\mathrel}{symbols}{"1D}

\newcommand{\ga}{\alpha}
\newcommand{\gb}{\beta}
\renewcommand{\gg}{\gamma}
\newcommand{\gd}{\delta}
\renewcommand{\ge}{\epsilon}
\newcommand{\gve}{\varepsilon}
\newcommand{\gf}{\phi}
\newcommand{\gvf}{\varphi}
\newcommand{\gc}{\chi}
\newcommand{\gx}{\xi}
\newcommand{\gm}{\mu}
\newcommand{\gn}{\nu}

\newcommand{\gk}{\kappa}
\newcommand{\gl}{\lambda}
\newcommand{\gr}{\rho}
\newcommand{\gth}{\theta}
\newcommand{\gs}{\sigma}
\newcommand{\gt}{\tau}
\newcommand{\go}{\omega}
\newcommand{\gz}{\zeta}
\newcommand{\gp}{\pi}
\newcommand{\gps}{\psi}
\newcommand{\get}{\eta}

\newcommand{\gG}{\Gamma}
\newcommand{\gD}{\Delta}
\newcommand{\gF}{\Phi}

\newcommand{\gX}{\Xi}
\newcommand{\gL}{\Lambda}

\newcommand{\gTh}{\Theta}
\newcommand{\gO}{\Omega}

\newcommand{\gPs}{\Psi}

\newcommand{\BgG}{{\boldsymbol \Gamma}}

\newcommand{\cA}{{\cal A}}
\newcommand{\cB}{{\cal B}}
\newcommand{\cC}{{\cal C}}
\newcommand{\cD}{{\cal D}}
\newcommand{\cE}{{\cal E}}
\newcommand{\cF}{{\cal F}}
\newcommand{\cG}{{\cal G}}
\newcommand{\cH}{{\cal H}}

\newcommand{\cK}{{\cal K}}
\newcommand{\cL}{{\cal L}}
\newcommand{\cM}{{\cal M}}
\newcommand{\cN}{{\cal N}}
\newcommand{\cO}{{\cal O}}

\newcommand{\cR}{{\cal R}}
\newcommand{\cS}{{\cal S}}

\newcommand{\cU}{{\cal U}}

\newcommand{\cW}{{\cal W}}

\newcommand{\cZ}{{\cal Z}}
\newcommand{\wK}{\mathscr{K}}

\newcommand{\wZ}{\mathscr{Z}}
\newcommand{\fb}{{\mathfrak b}}

\newcommand{\fs}{{\mathfrak s}}

\newcommand{\fu}{{\mathfrak u}}

\newcommand{\tg}{{\tilde g}}

\newcommand{\tA}{{\tilde A}}

\newcommand{\tgx}{{\tilde\xi}}

\newcommand{\tgo}{{\tilde\omega}}

\newcommand{\tgF}{{\tilde\Phi}}

\newcommand{\ba}{{\bar a}}
\newcommand{\bb}{{\bar b}}
\newcommand{\bc}{{\bar c}}

\newcommand{\bff}{{\bar f}}

\newcommand{\bi}{{\bar \imath}}
\newcommand{\bj}{{\bar \jmath}}
\newcommand{\bk}{{\bar k}}
\newcommand{\bl}{{\bar l}}
\newcommand{\bm}{{\bar m}}
\newcommand{\bn}{{\bar n}}

\newcommand{\bs}{{\bar s}}

\newcommand{\bx}{{\bar x}}

\newcommand{\bz}{{\bar z}}
\newcommand{\bA}{{\bar A}}
\newcommand{\bB}{{\bar B}}
\newcommand{\bC}{{\bar C}}
\newcommand{\bD}{{\bar D}}
\newcommand{\bE}{{\bar E}}
\newcommand{\bF}{{\bar F}}

\newcommand{\bM}{{\bar M}}

\newcommand{\bS}{{\bar S}}

\newcommand{\bV}{{\bar V}}
\newcommand{\bW}{{\bar W}}
\newcommand{\bX}{{\bar X}}

\newcommand{\bZ}{{\bar Z}}
\newcommand{\bgg}{{\bar\gamma}}

\newcommand{\bge}{{\bar\epsilon}}

\newcommand{\bgc}{{\bar\chi}}
\newcommand{\bgx}{{\bar\xi}}

\newcommand{\bgn}{{\bar\nu}}

\newcommand{\bgl}{{\bar\lambda}}

\newcommand{\bgth}{{\bar\theta}}
\newcommand{\bgs}{{\bar\sigma}}

\newcommand{\bgo}{{\bar\omega}}
\newcommand{\bgz}{{\bar\zeta}}

\newcommand{\bgps}{{\bar\psi}}
\newcommand{\bget}{{\bar\eta}}

\newcommand{\bgTh}{{\bar\Theta}}
\newcommand{\bgO}{{\bar\Omega}}

\newcommand{\bcF}{{\bar{\cal F}}}
\newcommand{\bcG}{{\bar{\cal G}}}

\newcommand{\Tr}{\mbox{Tr}}

\newcommand{\Id}{\text{\small 1}\hspace{-3.5pt}\text{1}}
\newcommand{\slashed}{\hspace{-1.1ex}/}

\renewcommand{\Re}{\text{Re}}
\renewcommand{\Im}{\text{Im}}

\newcommand{\der}{\partial}
\newcommand{\bder}{\bar\partial}

\newcommand{\sder}{\der\slashed}

\newcommand{\half}{\frac 12 }

\newcommand{\Kh}{K\"{a}hler }

\newcommand{\beq}{\begin{gather}}
\newcommand{\eeeeeq}{\end{gather}}
\newcommand{\barr}{\begin{array}}
\newcommand{\earr}{\end{array}}
\newcommand{\equ}[1]{\begin{gather} #1 \end{gather}}

\newcounter{oldcounter}

\newcommand{\be}{\begin{equation}}
\newcommand{\ee}{\end{equation}}
\newcommand{\bea}{\begin{eqnarray}}
\newcommand{\eea}{\end{eqnarray}}

\renewcommand{\d}{\mbox{d}}

\newcommand{\commentary}[1]{}

\newcommand{\mZ}{\mathbb{Z}}
\newcommand{\mC}{\mathbb{C}}
\newcommand{\mR}{\mathbb{R}}
\newcommand{\mH}{\mathbb{H}}
\newcommand{\hH}{{\hat{H}}}
\newcommand{\hN}{{\hat{N}}}
\newcommand{\hA}{{\hat{A}}}
\newcommand{\hF}{{\hat{F}}}
\newcommand{\hC}{{\hat{C}}}

\newcommand{\hB}{{\hat{B}}}
\newcommand{\hD}{{\hat{D}}}
\newcommand{\hg}{{\hat{g}}}
\newcommand{\hR}{{\hat{R}}}
\newcommand{\hQ}{{\hat{Q}}}

\newcommand{\nn}{\nonumber}
\newcommand{\R}[1]{(\ref{eq:#1})} 
\newcommand{\pp}[2]{\frac{\partial #1}{\partial #2}}

\newcommand{\p}[1]{\partial_{#1}}
\newcommand{\pu}[1]{\partial^{#1}}
\newcommand{\bp}[1]{\bar{\partial}_{#1}}

\newcommand{\3}{\frac{1}{3}}
\newcommand{\4}{\frac{1}{4}}

\newcommand{\2}{\frac{1}{2}}

\newcommand{\I}{\mathrm{i}}
\newcommand{\iz}{{i_0}}
\newcommand{\jz}{{j_0}}
\newcommand{\kz}{{k_0}}
\newcommand{\iq}{{i_q}}
\newcommand{\jq}{{j_q}}

\def\dga{\dot{\ga}}

\def\pt{\;\;.}
\def\km{\;\;,}

\def\susy{supersymmetry }
\def\sugra{supergravity } 
\def\quat{quaternionic } 
\def\5b{$5$-brane }
\def\fb{$5$-brane }
\def\mem{membrane }

\newcommand{\quart}{\tfrac{1}{4}}

\newcommand{\tab}{\quad\,}
\newcommand{\lp}{\begin{pmatrix}}
\newcommand{\rp}{\end{pmatrix}}

\newcommand{\bbW}[2]{\bW_{\!#1}^{#2}}
\newcommand{\W}[2]{W_{\!#1}^{#2}}
\newcommand{\G}[3]{\Gamma_{\!#1}{}^{#2}{}_{#3}}
\newcommand{\OO}[3]{\gO_{\!#1}{}^{#2}{}_{#3}}

\providecommand{\slsh}[1]{#1\!\!\!\!/\,}

\newcommand{\cl}{\mathrm{cl}}
\newcommand{\inst}{\mathrm{inst}}
\newcommand{\eff}{\mathrm{eff}}
\newcommand{\class}{\mathrm{class}}

\DeclareSymbolFont{AMSb}{U}{msb}{m}{n}
\DeclareMathSymbol{\fieldC}{\mathalpha}{AMSb}{"43}
\DeclareMathSymbol{\fieldR}{\mathalpha}{AMSb}{"52}
\DeclareMathSymbol{\fieldZ}{\mathalpha}{AMSb}{"5A} 
\usepackage{type1cm}

\subsubsectionfont{\large}
\subsectionfont{\large}
   
\renewcommand{\chaptermark}[1]{\markboth{ #1}{}}

\renewcommand{\headrulewidth}{0pt}
\renewcommand{\footrulewidth}{0pt}
\begin{document}
\frontmatter
\thispagestyle{empty}
\vspace*{-2.2cm}

\begin{flushright} \small
 ITP--UU--06/10 \\ SPIN--06/08  \\ hep-th/0603073
\end{flushright}
\smallskip

\begin{center}
 {\LARGE\bfseries Membrane and fivebrane instantons \\[3mm]
  and  quaternionic geometry}
\\[10mm]
Marijn Davidse \\[5mm]
 {\small\slshape
 Institute for Theoretical Physics \emph{and} Spinoza Institute \\
 Utrecht University, 3508 TD Utrecht, The Netherlands \\[2mm]
 {\upshape\ttfamily M.Davidse@phys.uu.nl} \\[3mm]
 }
\vspace{12mm}

\begin{minipage}[]{11.5cm}
\begin{center}{\bfseries Abstract}\end{center}
{\small
We investigate membrane and fivebrane instanton effects in type IIA string theory
compactified on rigid Calabi-Yau manifolds. These effects contribute
to the low-energy effective action of the universal hypermultiplet, in four dimensional spacetime.
To compute the nonperturbative effects due to the fivebrane instanton to the universal hypermultiplet, an instanton calculation is performed.\\
In the absence of fivebrane instantons, the quaternionic
geometry of the hypermultiplet is determined by solutions of the
three-dimensional Toda equation. We construct solutions describing
membrane instantons and find perfect agreement with string
theory predictions. In the context of flux compactifications we 
discuss how membrane instantons contribute to the scalar potential
and the stabilization of moduli. Finally, we
demonstrate the existence of meta-stable de~Sitter vacua.
}
\end{minipage}\\
\end{center} 
\newpage
\thispagestyle{empty}
\clearpage{\pagestyle{empty}\cleardoublepage} 
\makeatletter
\def\thickhrulefill{\leavevmode \leaders \hrule height 1ex \hfill \kern \z@}
\def\@makechapterhead#1{%
  \vspace*{10\p@}%
  {\parindent \z@ \centering \reset@font
        \Huge \bfseries #1\par\nobreak
        \par
        \vspace*{10\p@}%
    \vskip 60\p@
  }}
\def\@makeschapterhead#1{%
  \vspace*{10\p@}%
  {\parindent \z@ \centering \reset@font
        \Huge \bfseries #1\par\nobreak
        \par
        \vspace*{10\p@}%
    \vskip 60\p@
  }}

\tableofcontents
\thispagestyle{plain}
{\fancyhead{}
\fancyhead[LE, RO]{\thepage}
\fancyhead[CO]{\slshape\leftmark}
\fancyhead[CE]{\slshape\leftmark}

\chapter{Preface}
The following text is based on my Ph.D. thesis titled \emph{Nonperturbative effects in supergravity} which was defended on the thirtieth of  January $2006$. The research on which this thesis is based was performed at the Spinoza Institute at the University of Utrecht.  Naturally, I would not have succeeded in writing my thesis alone.
I am primarily indebted to Stefan Vandoren for being my supervisor and for  his invaluable help during my years as a Ph.D. student. Furthermore I am grateful to Ulrich Theis and Frank Saueressig for our years of collaboration. For carefully reading this thesis I also would also like to thank my promotor Bernard de Wit and Ulrich. \\

\begin{flushright}
Marijn Davidse, March $2006$
\end{flushright}

 \clearpage{\pagestyle{empty}\cleardoublepage} 
\chapter{Introduction}
\section*{String theory}
In recent decennia string theory has become a leading candidate for a unified theory of particle physics, gravity and quantum mechanics, see \cite{Green:1987sp, Green:1987mn, Polchinski:1998rq, Polchinski:1998rr}. As the name suggests, the starting point of string theory is to consider strings instead of point particles. One can consider closed or open strings, which both sweep out two-dimensional  surfaces as they move through spacetime. Such a surface is called the world-sheet.\\
String theory has one dimensionfull coupling constant called $\ga'$ with dimension of length squared. The vibrations of the string correspond to particles in spacetime. For example, in closed string theory there is a massless rank two traceless symmetric tensor which we must identify with the graviton if we view string theory as a unified theory. There is also a massless rank two antisymmetric tensor. Furthermore, there is a whole tower of massive states, with masses proportional to $\sqrt{1/\ga'}$. \\
When considering string theory as a unified theory, as we are, $\ga'$ is of the order of the natural scale determined by the fundamental constants of gravity and quantum mechanics. This scale is given by $M_P^{-2}$ where $M_P\equiv \sqrt{\hbar c/G_N}$ is the Planck mass, roughly $1.2 \times 10^{19}$ GeV. Put differently, $\sqrt{\ga'}$ which sets the scale of the string length is of the order of the Planck length $l_P\equiv \sqrt{\hbar G_N/c^3}$, which is about $1.6\times 10^{-33}$cm. \\
Furthermore, there is a dimensionless parameter called the string coupling constant $g_s$. This constant organizes the perturbative expansion of string theory. Just as in normal field theory we can set up a perturbation theory with the difference that in- and outgoing lines are now tubes, in the case of the closed string. A one-loop diagram, for a four-point amplitude for example,  then takes the form of a donut with four tubes attached. This diagram comes with a higher power of $g_s$, compared with the tree level diagram (a sphere with four tubes attached). Higher order loop diagrams come with higher powers of $g_s$, this parameter therefore counts the order of the diagram. In general the perturbative expansion of string theory is a genus expansion in the world-sheet where each genus comes with a certain power of $g_s$. The result is a power series in $g_s$: $\sum_n g_s^{2n} A^{(n)}$. Here $A^{(n)}$ denotes the diagram, that is, $A^{(0)}$ the tree level diagram, $A^{(1)}$ the one-loop diagram (the donut) and so on. We see that such an expansion makes sense as long as $g_s \ll 1$.\\
The parameter $g_s$ is not an arbitrary parameter. It is related to the vacuum expectation value of the dilaton, a scalar field which is one of the massless modes. However, in general it is not clear what the vacuum expectation value of the massless scalar fields should be. 
In chapter \ref{ch4} we will  consider a scenario in which we can dynamically provide a vacuum expectation value for the dilaton. We will come back to this issue below on page \pageref{moduli1}.\\

If one considers only the coordinates which describe the embedding of the string in spacetime, one has bosonic string theory. A next step is to add fermions on the world-sheet. This gives rise to the so-called superstring theories, which contain additional massless states besides the massless states of the bosonic string. There are actually five different superstring theories, which seems to be a bit of a setback if one  would like string theory to be a unique and unifying theory. About ten years ago though, it became clear that these five theories are actually different phases of a more general theory dubbed M theory, \cite{Witten:1995ex, Hull:1994ys}. However, not only is it as yet  unclear what this theory looks like, it is also unclear what the `M' stands for: dial M for mystery!\\
Without going into the specifics of these five string theories, we remark that we will be concerned only with one of them in this thesis, namely with the `type IIA' theory. For reasons of consistency, the superstring theories have to live in $10$ spacetime dimensions and the bosonic string in $26$. Naturally this seems excessive and has to be remedied. We will come back to this issue in a moment. \\
The name `string' theory has become a misnomer since the discovery \cite{Polchinski:1995mt} of `D-branes'. These are extended objects on which strings can end. For example, in type IIA theory there are membranes and four-branes which sweep out three- and five-dimensional world-volumes respectively.\\
Such D-branes are nonperturbative objects, like solitons in field theory,  because these object are heavy when the string coupling constant is small, but become lighter when it increases.  The reason is that their tension is inversely proportional to $g_s$, so when the perturbative expansion breaks down (large $g_s$) these branes are light. \\
Moreover, string theory contains also a NS five-brane. This object is dual to the fundamental string in the same sense that a magnetic monopole is dual to an electric charge by means of electric-magnetic duality. The tension of the NS five-brane is proportional to $1/g_s^2$, this NS five-brane will be very important to us and we will discuss it more extensively in chapter \ref{ch2}. 
For some general references on D-branes see \cite{Polchinski:1998rq, Polchinski:1998rr, Johnson:2003gi, Forste:2001ah}.\\
These branes are solitonic objects, but below we will see that there are situations in which the soliton will appear in a lower dimensional spacetime as an instanton. \\  
This reminds us of nonperturbative effects in Yang-Mills theories. Apart from the trivial vacuum, (Euclidean) Yang-Mills theory has a vacuum which consists of an infinite number of topologically distinct vacua. The instanton solution represents a transition from one vacuum class to another.  The tunnelling amplitude for such a transition is proportional to $\exp(-S_E)$, with $S_E$ the Euclidean action. The action for a Yang-Mills instanton is proportional to $1/g_{\textrm{ym}}^2$, with $g_{\textrm{ym}}$ the Yang-Mills coupling constant. 
So when one computes amplitudes in the presence of such instanton configurations in doing a semiclassical approximation, one finds an additional weight factor of the form $\exp(-1/g_\textrm{ym}^2)$. We therefore see that instanton contributions to physical processes are heavily suppressed for small $g_{\textrm{ym}}$.
However, when $g_{\textrm{ym}}$ becomes large this factor becomes more important. Thus starting from the perturbative description of Yang-Mills theory, one can construct solutions which give insight into nonperturbative effects. Instantons are, for instance important in the context of CP violating processes, see \cite{Forkel:2000sq} for a review and references. For a very nice introduction to instantons in Yang-Mills theory see \cite{Coleman:1978ae, Coleman}, for a clear account of the topological details of instantons see \cite{Nash:1983cq}.\\
In this thesis we will approach the NS five-brane in a similar fashion.
That is, in chapter \ref{ch3} we will use the semiclassical approximation to compute its contribution to correlation functions. We should stress that we will use an effective description of the NS five-brane. We will not make use of a worldvolume description of the brane. How this brane can be effectively described is discussed below.
For more information on the semiclassical approach see for instance  \cite{Coleman}.

\section*{Effective theories}
As remarked above, the massive states of the string theories have masses proportional to the Planck mass. This means that only when considering processes with extremely high energies the massive states start to play a role. Such energies are far beyond the reach of any man-made accelerator.
Therefore one can restrict to the massless modes only and describe them by an effective theory.  As long as one considers processes with energies much lower than the Planck mass, this is a good approximation. The low energy effective theories of the superstring theories take the form of supergravity theories. The low energy theory of the type IIA superstring, for instance, is type IIA supergravity which is a theory with two local supersymmetry generators. 
The universal (since it occurs in other superstring theories as well) part of its bosonic sector is given by 
\equ{
\int \!d^{10}x\sqrt{-g}\left(R-\2 (\p{}\gf)^2 -\frac{1}{12}e^{-\gf}H^2\right) \nn \pt
}
$H$ is the field strength of the massless rank two antisymmetric tensor and $\gf$  is the dilaton. The Einstein-Hilbert term describes the graviton. This action captures the massless states already present in the bosonic string theory. In type IIA superstring theory there are in addition massless tensor- and fermionic fields.
The nonperturbative objects from string theory, i.e., the D-branes are then described by so-called $p$-branes which are solitonic solutions of the supergravity theory. For references on branes from a supergravity point of view see \cite{Duff:1994an, Stelle:1998xg}.\\

We can illustrate this by the well-known example in the context of type IIB string theory and its low energy effective theory, type IIB supergravity. This is the example of the D-instanton. It is a D-brane with no spatial extension and which is moreover located in time, hence the name instanton. In the context of type IIB it is a BPS solution which preserves half of the spacetime supersymmetries. In  \cite{Gibbons:1995vg} the supergravity solution corresponding to the D-instanton was found. Consider the following bosonic subsector of the Euclidean type IIB action
\equ{
 \int \!d^{10}x \sqrt{g}\left(R-\2 (\p{}\gf)^2 + \2 e^{2\gf}(\p{}a)^2\right) \nn \pt
}
In flat Euclidean space we can solve the equation of motion for $\gf$ by
\equ{
e^\gf=e^{\gf_\infty}+\frac{c}{r^8} \nn \km
}
with $e^{\gf_\infty}$ and $c$ integration constants. The axion $a$ is determined in terms of $\gf$. The above solution for the dilaton solves the field equation $\p{}^2 e^\gf=0$ except at the origin where  a delta function source term is needed. Consequently we can add the source term $\gd^{(10)}(x)\,e^{-\gf}$ to the action which cancels the singularity in the field equation. This source term tells us that we have a new object in the theory at $x=0$, namely the D-instanton. So we see that we can use supergravity to describe the objects in string theory.\\
The action of the D-instanton is given by
\equ{
S_\textrm{D-inst}=\frac{|Q|}{g_s} \nn \km
}
where we have defined $g_s\equiv e^{\gf_\infty}$, note the $1/g_s$ behaviour. This action derives from the fact that one has to consider an additional boundary term to the action. The charge $Q$ is due to the existence of a conserved current and is related to the integration constant $c$, see also \cite{Belitsky:2000ws}.
By studying certain amplitudes in the background of the D-instanton one can compute the effective vertices due to this instanton, \cite{Green:1997tv}. For a review of D-instantons and more references see \cite{Gutperle:1997iy}. 

We will  make a similar effort in this thesis for the NS five-brane and two-brane. However, we will not be working in ten dimensions but in four dimensions, as we will discuss first.  

\section*{Compactification}
The superstring theories, and  therefore their low energy effective theories, live in ten spacetime dimensions. Of course we are interested in the case of four spacetime dimensions. One way to go from ten to four spacetime dimensions, is to compactify six space dimensions on some internal manifold. If these six dimensions are taken to be very small, they are effectively not seen anymore in experiment. Consequently, the fields have to be expanded in the compact directions. This leads to massless states and a tower of massive states, determined by the manifold the theory is compactified on. The mass is inversely proportional to the characteristic radius $R_c$ of the compactified dimensions. This defines a compactification scale $m_c=1/R_c$, the characteristic mass of states with momentum in the compactified directions. 
When $R_c$ is sufficiently small, $m_c$ is high enough to enable us to work with the massless modes only. This means that a four-dimensional theory can be constructed: compactify the low energy effective action of string theory on some (small) internal manifold. This then gives a theory in four spacetime dimensions. If $m_c$ is high enough this theory can be approximated by an effective theory which only describes the massless modes. This is done by integrating out the heavy degrees of freedom (which can give corrections to the massless degrees of freedom). The idea of compactifying higher dimensional theories to four dimensions was initiated by Kaluza and Klein. See \cite{O'Raifeartaigh:1998pk} for a historical overview.\\
The effective theory in four dimensions can be (depending on the internal manifold) rather complicated. 
An interesting choice for manifolds to compactify on are the so-called Calabi-Yau manifolds, which we will describe in chapter \ref{ch1}. 
In the eighties it was found that compactifying the heterotic string theory on a Calabi-Yau manifold gives rise to phenomenologically interesting models, see   \cite{Candelas:1985en}.
Furthermore, these manifolds preserve (a quarter of the) supersymmetry. For example, compactifying type IIA supergravity in ten dimensions on a six-dimensional Calabi-Yau manifold gives an $N=2$ supergravity theory with vector and hypermultiplets \cite{Bodner:1990zm}. This theory describes the massless degrees of freedom. The fact that it has (local) supersymmetry means that the theory is more constrained, thus making it easier to work with. In chapter \ref{ch2} we will review the process of compactifying type IIA supergravity on a Calabi-Yau manifold of six  dimensions.\\
To summarize: the reason why we study type IIA supergravity on a Calabi-Yau manifold is that this allows us to work with a theory which has just enough supersymmetry. In contrast, $N=4$ or $N=8$ (which does not even allow for matter multiplets) \susy is too constraining and $N=1$ \susy is not constraining enough, see also \cite{Aspinwall:2000fd}.\\

The aim of this thesis can now be formulated as follows: \emph{to study the nonperturbative effects of NS five-branes and membranes on the underlying $N=2$ supergravity theory in four dimensions.} \\

Just as one can describe these branes in the supergravity approximation in ten dimensions, so can one construct corresponding solutions which describe them in four dimensions. Intuitively this can be understood as follows. Consider the five-brane with its six dimensional worldvolume\footnote{Assume for the moment that we have performed a Wick rotation, such that we are wrapping an Euclidean brane around the Calabi-Yau manifold.}, which we can embed entirely in the six-dimensional internal Calabi-Yau manifold. This means that the five-brane is then completely localized in four-dimensional spacetime. 
 From a four-dimensional point of view it will then appear as an instanton. This will also be described  in chapter \ref{ch2}. In chapter \ref{ch3} we will perform a semiclassical computation to compute the instanton effects. Contrary to the case of Yang-Mills instantons, we will not consider instantons as giving tunnelling amplitudes between different vacua, because such an interpretation is not  clear in our case. Instead instantons will be regarded as solutions to the equations of motion which have finite (Euclidean) action. These solutions furthermore saturate lower bounds of the action and will be derived using Bogomol'nyi equations in chapter \ref{ch2b}. \\
Similarly to the five-brane, the three-dimensional worldvolume of the membrane can be wrapped around a three-dimensional submanifold of the Calabi-Yau. This will also correspond to an instanton solution in four dimensions which will be constructed in chapter \ref{ch2b} as well.
In chapter \ref{ch4} we will then consider the effect of these membranes on the four-dimensional supergravity theory.

\section*{Geometry}
As said above, we will be working with the supergravity theory describing the massless sector of the theory which arises when compactifying type IIA supergravity on a Calabi-Yau manifold. This is an $N=2$ supergravity theory coupled to hypermultiplets and vector multiplets. \\
Such a theory can be thought of in geometrical terms. We can consider the various scalars, appearing in these multiplets, as maps from spacetime to some target space manifold. The scalars are then coordinates on this manifold. The kinetic terms of the scalars function as the metric on such a target space. Supersymmetry relates the various terms in the action and puts constraints on them. This means that the `metrics' are also constrained to be of a certain type, thus determining the geometry of the target space.\\
This has a (relatively) long history. For instance, $N=1$ matter-coupled supergravity in four dimensions has a number of complex scalar fields. These fields are governed by \Kh geometry: they parametrize a \Kh manifold, as discovered by Zumino and others \cite{Zumino:1979et, Cremmer:1982en}. \\
The geometric interpretation of hypermultiplets coupled to $N=2$ supergravity in four dimensions was emphasized by Bagger and Witten \cite{Bagger:1983tt}. For more information on the interplay between supergravity and geometry see for instance \cite{Fre:1995bc} and its references. \\
This geometrical formulation of the constraints obeyed by the scalars of the hypermultiplets coupled to $N=2$ supergravity will be very important to us, because this is the sector of the theory we will be working with. The target space of the hypermultiplet sector is a so-called quaternionic manifold, which we will review in chapter \ref{ch1}.  The crucial point is that the conditions for unbroken supersymmetry force the hypermultiplet sector of the supergravity theory to have such a target space. So when we compute our instanton corrections to this theory, the corrections will in general perturb the target space but in such a way that it is still quaternionic. In chapter \ref{ch3} we will establish that after computing the five-brane instanton effects on the supergravity theory, and thus on the target space, this target space is still of the quaternionic type. \\
We will be focussing on the hypermultiplet sector of the theory because this is where the instanton effects appear, as we will discuss. Furthermore, since the vector multiplet sector decouples, we will not consider it in our calculations.\\
In chapter \ref{ch1} we will present some background material on the quaternionic geometry. Furthermore, we will (briefly) consider the geometry corresponding to the vector multiplets, for the sake of completeness. These geometries will reappear in chapter \ref{ch2} where we sketch how the $N=2$ supergravity theory emerges when compactifying type IIA supergravity theory on a Calabi-Yau manifold.\\ 
$\,$\\
We can now reformulate the aim of this thesis: \emph{to compute the nonperturbative corrections induced in the quaternionic target space of the hypermultiplet sector of the four-dimensional $N=2$ supergravity theory.}

\section*{Moduli}
\label{moduli1}
The massless scalar fields resulting from a compactification are sometimes referred to as `moduli'. The reason is that they are closely related to the parameters labelling the geometry of the internal manifold, which are also called moduli\footnote{In chapter \ref{ch2} we will closely investigate this relation for Calabi-Yau manifolds}. The point is that the various couplings in four dimensions are determined by the internal geometry and typically depend on the vacuum expectation values of the scalar fields in four dimensions.  As long as these vacuum expectation values are undetermined, the couplings are free as well and the four-dimensional theory has no predictive power. In other words, every vacuum expectation value corresponds to a different groundstate of the four-dimensional theory. One would like to find a mechanism that (spontaneously) fixes these moduli, this is called moduli stabilization. One such mechanism is that of `compactification with fluxes'. When compactifying to four dimensions one can switch on fluxes in the internal manifold. This alters the four-dimensional theory and in particular gives rise to a potential for the scalar fields, i.e., the moduli.\\
Alternatively one can describe such effects by  staying in four dimensions and gauging certain isometries in the supergravity theory. This also gives rise to a potential which can in principle stabilize (some of) the moduli, we will come back to this in chapter \ref{ch4}.
 This subject has attracted considerable and renewed attention in recent years. For some recent reviews see \cite{Grana:2005jc, Louis:2005te}. \\
In chapter \ref{ch4} we will gauge an isometry in the hypermultiplet\footnote{We will consider only one hypermultiplet.This is related to the fact that we choose a simple type of Calabi-Yau compactification, as explained in chapter \ref{ch2}.} which will generate a potential.   The membrane instanton corrections to the potential make it possible to stabilize all the fields in the hypermultiplet. The membrane instanton correction furthermore makes it possible for the potential to have a positive minimum.
This value appears in the action as a cosmological constant term. This is very interesting because a positive cosmological constant corresponds to a de Sitter universe and at the moment our own universe seems to be of that type, see for instance the review \cite{Trodden:2004st} and references therein.

\section*{Outline of this manuscript}
After having introduced the main topics of this thesis, let us for clarity's sake summarize them and specify where they will be discussed. This will hopefully make the relations between the various chapters and sections clearer.
\begin{itemize}
\item \textbf{Chapter \ref{ch1}}\\
Because quaternionic geometry will be  important in this thesis, we devote the first half of this chapter to reviewing the  geometry relevant to quaternionic manifolds and Calabi-Yau compactifications in general. The second part will be a short review of Calabi-Yau manifolds and some differential geometry, which will be especially relevant for chapters \ref{ch2} and \ref{ch4}. 

\item \textbf{Chapter \ref{ch2}}\\
We discuss the process of compactifying type IIA supergravity on a Calabi-Yau manifold. We shall derive the bosonic field content of the resulting four-dimensional supergravity theory. We will concentrate in particular on the massless fields obtained by compactifying on a simple type of Calabi-Yau manifold.

\item \textbf{Chapter \ref{ch2b}}\\
By making use of Bogomol'nyi bounds, we construct solutions to the equations of motion that provide  local minima to the action and as such can be interpreted as instantons. This will be done for a certain description of a hypermultiplet that is introduced in chapter \ref{ch2}. We construct two types of instanton solutions. The first corresponds to the NS five-brane and the second to the membrane. Their actions, which agree with results from string theory, and theta angles will be given.

\item \textbf{Chapter \ref{ch3}}\\
This is the chapter where we will perform a detailed instanton calculation for the NS five-brane in the context of the four-dimensional effective supergravity theory. This is a fairly traditional field theoretic semiclassical approximation, although it does contain (apart from the final results) some novel features. \\
The result is the effective $N=2$ supergravity theory. Effective now means that the theory can reproduce at tree level the corrections to the action due to the five-brane instanton. We will examine the consequences for the quaternionic target space, in particular the consequences for its isometries. 

\item \textbf{Chapter \ref{ch4}}\\
A chapter in which the membrane instanton corrections to the four-dimen\-sional action are derived. The material from chapter \ref{ch1} is used and elaborated on.  Where we can compare our results with string theory we find beautiful agreement. Furthermore, the effect of these membrane instanton corrections to the potential obtained by gauging a certain isometry is studied. This is the mechanism referred to  above, it will be explained in this chapter. The result can lead to a de Sitter vacuum in which all the moduli of the universal hypermultiplet sector are stabilized.  The effect of the membrane instanton is to make it possible for the minimum to be positive. 

\item \textbf{Appendices}\\
Because there are a number of technical details we have collected them in a few appendices. These contain further information on calculations or extra background material. They will be referred to in the text as needed. 
\end{itemize}

We must stress again that, apart from a section in chapter \ref{ch4}, we will always perform our calculations within the supergravity approximation. This means that when we speak of branes we think of them in a supergravity sense. Moreover, they will always be treated from a four-dimensional point of view, where they appear as instanton field configurations.\\
Furthermore, this thesis is not about supergravity as such. We shall use it as a tool to consider nonperturbative effects. Consequently we will treat it as such. We will neither explain nor construct the $N=2$ supergravity theory we will be working with from first principles, as this would warrant another thesis. Instead we will present the necessary information such as supersymmetry transformation rules and such when needed, especially in chapter \ref{ch3}. We will mainly focus on the geometry of the target space manifold of the theory. 

Lastly let us indicate which chapters contain new material. Chapter \ref{ch2b} is largely based on \cite{Davidse:2003ww}, chapter \ref{ch3} on \cite{Davidse:2004gg} and chapter \ref{ch4} on \cite{Davidse:2005ef}.

 \clearpage{\pagestyle{empty}\cleardoublepage}

}
 
\makeatletter
\def\thickhrulefill{\leavevmode \leaders \hrule height 1ex \hfill \kern \z@}
\def\@makechapterhead#1{%
  \vspace*{10\p@}%
  {\parindent \z@ \centering \reset@font
        {\fontsize{35}{15.6pt}\selectfont \bfseries \thechapter \fontsize{13}{15.6pt}\selectfont}
        \par\nobreak
        \vspace*{15\p@}%
        \interlinepenalty\@M
        \vspace*{10\p@}%
        \Huge \bfseries #1\par\nobreak
        \par
        \vspace*{10\p@}%
    \vskip 60\p@
  }}
\def\@makeschapterhead#1{%
  \vspace*{10\p@}%
  {\parindent \z@ \centering \reset@font
        {\fontsize{35}{15.6pt}\selectfont \bfseries \vphantom{\thechapter} \fontsize{13}{15.6pt}\selectfont}
        \par\nobreak
        \vspace*{15\p@}%
        \interlinepenalty\@M
        \vspace*{10\p@}%
        \Huge \bfseries #1\par\nobreak
        \par
        \vspace*{10\p@}%
    \vskip 60\p@
  }}

\renewcommand{\chaptermark}[1]{\markboth{\thechapter\ #1}{}}
\fancyhf{}
\fancyhead[LE, RO]{\thepage}
\fancyhead[CO]{\slshape\rightmark}
\fancyhead[CE]{\slshape\leftmark}
\renewcommand{\headrulewidth}{0pt}
\renewcommand{\footrulewidth}{0pt}
\addtolength{\headheight}{0pt}
\mainmatter
\chapter{Some geometrical concepts}
\label{ch1}
In a non-linear sigma model the scalars can be viewed as maps from spacetime to a `target-space'. The scalars are then coordinates on this target-space. The kinetic terms of the scalars function as a metric on the target space. If the theory is supersymmetric, this imposes conditions on the fields. In particular, the conditions on the metric of the target-space determine its geometry. In this thesis we will mainly be concerned with $N=2$ supergravity in four dimensions coupled to vector multiplets and hypermultiplets. The scalars of the hypermultiplet sector parametrize a `quaternionic' target-space. The main purpose of this chapter is to explain the geometrical concepts involved in such spaces. Furthermore, we present some material on Calabi-Yau manifolds. For more information on the geometry discussed in this chapter see \cite{Nash:1983cq, Nakahara:1990th, CY, Greene:1996cy, Candelas:1990pi, Joyce}. For more specific texts on supergravity and geometry see \cite{Fre:1995bc, Fre:1995dw, Andrianopoli:1996cm, Andrianopoli:1996vr, Aspinwall:2000fd}.

\section{\Kh manifolds}
\label{kahlergeo}
Let $\cM$ be a manifold of real dimension $2n$ with a complex structure $J$ which maps the tangent space into itself, i.e. $J: T\cM \to T\cM$ with  $J^2=-\Id$. A metric $g$ on $\cM$ is called Hermitian with respect to $J$ if
\equ{
g\left(J \vec u, J \vec w\right) =g\left(\vec u,  \vec w\right) \label{eq:Hermco1} \pt
}
We can always choose a local frame such that 
\equ{
J^\gb_{\,\,\,\ga} = \left(%
\begin{array}{cc}
  0 & \Id \\
  -\Id & 0 \\
\end{array}%
\right) \nn \pt
}
In this frame we can define complex coordinates such that 
\equ{
J \pp{}{z^i}=+\I \pp{}{z^i} \quad \textrm{and} \quad J \pp{}{\bz^\bi}=-\I \pp{}{\bz^\bi} \label{eq:compls} \km
}
furthermore we can write
\equ{
g(u,w)=g_{ij} u^i w^j + g_{\bi\bj}u^\bi w^\bj + g_{i\bj} u^i w^\bj + g_{\bi j} u^\bi w^j \nn \pt
}
Reality and symmetry of $g(u,w)$ dictate
\bea
&& g_{ij}=\left(g_{\bi\bj}\right)^\star  \qquad g_{ij}=g_{ji}  \nn \\
&& g_{\bi j}=\left(g_{i\bj}\right)^\star \qquad  g_{\bj i}=g_{i\bj} \nn 
\eea
and the Hermiticity condition \R{Hermco1} gives 
\equ{
g_{ij}=g_{\bi\bj}=0 \label{eq:Hermiticity} \pt
}
Given this metric, one can define a two-form $K$: 
\equ{
K\left(\vec u,  \vec w\right)= g\left(J \vec u,  \vec w\right) \label{eq:khform1} \km
}
or in components, 
\equ{
K_{\ga\gb}= g_{\gb\gg}J^\gg_{\,\,\,\ga} \nn \pt
}
This $2$-form is called the \emph{\Kh form}.
Note that this implies that $g$ is Hermitian if and only if $K$ is antisymmetric.
In the complex basis the $2$-form is written as
\equ{
K=\I g_{i \bj} \,dz^i \wedge d\bz^\bj \nn \pt
}
An important theorem states that a  Hermitian metric $g$ on a complex manifold is called \Kh if one of the following three equivalent conditions is met: 
\equ{
d K=0 \qquad \nabla J=0 \qquad \nabla K=0 \label{eq:Kcond1}
}
and consequently the manifold is called a \Kh manifold, see \cite{CY}. The $\nabla$ denotes the covariant derivative containing the Levi-Civita connection of $g$. The symbol $d$ denotes the exterior derivative. For a short review on holonomy, differential forms and cohomology, both real and complex, see page \pageref{dvg1}-\pageref{holonomy}.\\
The holonomy of a \Kh manifold is contained in $U(n)$: $Hol(g)\subseteq U(n)$. This means that a metric on the $2n$-dimensional manifold $\cM$ is \Kh with respect to an (integrable\footnote{Alternatively we could have considered an almost complex structure instead of the complex structure $J$, but the $U(n)$ holonomy ensures the vanishing of the Nijenhuis tensor and thus the integrability of the structure.}) complex structure if and only if $Hol(g)\subseteq U(n)$. \\
This is often used as the definition of a \Kh manifold: if the holonomy group of a Riemannian manifold is contained in $U(n)$, then it is a \Kh manifold and therefore a complex manifold.

As a $2$-form, the \Kh form lies in a cohomology class: $K\in H^2(\cM)$, the second De Rahm cohomology class. In the complex basis the \Kh form is a $(1,1)$-form and therefore lies in the Dolbeault cohomology class $H^{(1,1)}(\cM)$. This cohomology class is referred to as the \Kh class of the metric.
In any local chart we can solve $d K=0$ by
\equ{
g_{i\bj} =\p{i}\p{\bj}\wK \label{eq:ch1Kpot} \km
}
where $\wK=\wK^\star=\wK(z,\bz)$ is a real function of the complex coordinates $z^i,\bz^\bi$. This function is called the \emph{\Kh potential}, defined up to so-called \Kh transformations:
\equ{
\wK\to \wK + f(z) + \bff (\bz) \label{eq:kahlertr1} \km
}
since this does not change the \Kh metric. \\
Note that $K$ is closed but not exact in general. Locally it can be defined as  $K=\I \p{}\bp{} \wK,$ a closed real $(1,1)$-form. If two \Kh metrics $g, \tg$ on $\cM$ belong to the same \Kh class, they differ by a \Kh transformation. 

\section{Hodge-\Kh manifolds}
A next step is to consider a line-bundle (a complex $1$-dimensional vector bundle) over the \Kh manifold: $\cL \to \cM$. There is only one Chern class for such a bundle, namely the first: 
\equ{
c_1(\cL)=\I \bder \left(h^{-1} \der h\right)=\I\bder \der \ln (h) \label{eq:ch1a} \km
}
where $h(z,\bz)$ is some Hermitian fibre metric on $\cL$. If we define
\bea
&&\gth\equiv h^{-1} \der h =\frac{1}{h}\p{i}h dz^i \nn \\
&&\bgth \equiv h^{-1} \bder h =\frac{1}{h}\p{\bi}h d\bz^\bi \nn \km
\eea
then $\gth$ and $\bgth$ are the canonical Hermitian connections on $\cL$. The connection $Q$ on the associated principal $U(1)$-bundle is given by
\equ{
Q= \Im\, \gth =-\frac{\I}{2}\left(\gth -\bgth\right) \nn \pt
}

If we introduce a holomorphic section $s(z)$ of $\cL$ then we can re-express \R{ch1a} as
\equ{
c_1(\cL)=\I  \bder \der \ln |\!|s(z)|\!|^2 \nn \km
}
where the norm of $s(z)$, i.e. $|\!|s(z)|\!|^2$ is given by $|\!|s(z)|\!|^2=h(z,\bz) s(z) \bs (\bz)$.\\
This leads us to the definition of a \emph{Hodge-K\"ahler} manifold. \\
A \Kh manifold is a Hodge-\Kh manifold \emph{if and only if} there exists a line bundle $\cL\to\cM$ such that $c_1(\cL)=[K]$ where $[K]$ denotes the cohomology class of the \Kh form. Locally this implies that there exists a holomorphic section $W(z)$ such that 
\equ{
K=\I g_{i\bj}\, d z^i \wedge d\bz^\bj=\I \bder \der \ln |\!|W(z)|\!|^2 \pt \nn
}
 So if $\cM$ is a Hodge manifold, comparing with \R{ch1Kpot} tells us that $h(z,\bz)=e^{\wK(z,\bz)}$ and the $U(1)$ connection is also determined in terms of $\wK$: 
\equ{
Q=-\frac{\I}{2}\left(\p{i}\wK dz^i -\p{\bi}\wK d\bz^\bi\right) \nn \pt
}
 The Hodge condition has profound topological implications in the case of compact \Kh manifolds. It can be shown that the first Chern class of any $U(1)$-bundle $\cL\to \cM$ belongs to the integral cohomology class, i.e. $H^2(\cM, \mZ)$. The Hodge condition in turn implies that the same property holds for the \Kh form: $K\in H^2(\cM, \mZ)$. \\
There is much more to be said about Hodge-\Kh manifolds, especially in relation with \mbox{$N=1$} supergravity. However, they will only be needed here as an ingredient of \emph{special \Kh geometry}.\\
We must distinguish between two kinds of special \Kh geometry: the one of the local type and the one of the rigid type. The first type describes the geometry of the scalar fields in the vector multiplet sector of \mbox{$N=2$} \sugra and the second type describes the same sector in \mbox{$N=2$} super Yang-Mills theories with rigid supersymmetry. The second type will not play a role in the rest of this thesis and will therefore be ignored.\\
One way to define a special \Kh manifold in the local case is to consider again the line bundle $\cL$ whose first Chern class equals the \Kh form $K$ of an $n$-dimensional Hodge-\Kh manifold $\cM$. 

\section{Special \Kh manifolds}
\label{spkg}
For more information and references on special \Kh manifolds see \cite{Strominger:1990pd, Craps:1997gp}.
Let $\cH$ be a flat projective holomorphic $Sp(2n+2,\mR)$ vector bundle over $\cM$ and $-\I \langle \,|\, \rangle$ the symplectic  Hermitian metric on $\cH$. Then $\cM$ is a special \Kh manifold if there is a holomorphic section $\gO$ of $\cH \otimes \cL$ with the property
\equ{
K=\I\bder\der \ln\left(\I \langle\gO|\bgO\rangle\right) \label{eq:ch1K1}\pt
}
We define the compatible metric  as
\equ{
\I \langle\gO|\bgO\rangle =-\I \gO^\dagger \left(%
\begin{array}{cc}
  0 & \Id \\
  -\Id  & 0 \\
\end{array}%
\right) \gO \nn \pt
}
Such a section $\gO$ has the following structure: 
\equ{
\gO=\left(%
\begin{array}{c}
  X^A \\
  F_B \\
\end{array}%
\right) \qquad A,B=0,1,\ldots,n \label{eq:Osection1}
}
and is related on two different patches $\cU_i, \cU_j \subset \cM$ by 
\equ{
\left(%
\begin{array}{c}
  X \\
  F \\
\end{array}%
\right)_i =e^{f_{ij}} M_{ij} \left(%
\begin{array}{c}
  X \\
  F \\
\end{array}%
\right)_j \km \nn
}
where $f_{ij}$ are holomorphic maps $\cU_i \cap \cU_j \to \mC$ and $M_{ij}$ is a constant $Sp(2n+2, \mR)$ matrix. Equation \R{ch1K1} gives us an expression for the \Kh potential: 
\equ{
\wK=-\ln |\!| \gO |\!|=-\ln \left(\I \langle\gO|\bgO\rangle\right)=-\ln \left(\I \bX^A F_A -\I \bF_B X^B\right) \label{eq:kahlerpot1} \pt
}
Note that the $X^A$ and $F_B$ are projective coordinates, i.e., they are defined up to multiplications with a complex number\footnote{Because we are considering the bundle $\cH\otimes\cL$  one has to multiply the bundle $\cH$ and its section with a (local) complex number.} which give rise to an equivalence class of \Kh forms and potentials. Indeed, if one multiplies $\gO$ with a complex number (function) the effect on \R{kahlerpot1} is precisely that of a \Kh transformation, see \R{kahlertr1}.\\
Historically, this `special geometry' was introduced in so-called \emph{special coordinates} which are defined by $z^i=\frac{X^i}{X^0} \quad i=1, \ldots,n$, see \cite{deWit:1984pk, Cremmer:1984hj}. If we then define $\cF(z)=(X^0)^{-2} \cF(X)$, where $\cF(X)$ is homogeneous of degree two and identify $F_A=\p{A} \cF$, the \Kh potential becomes
\equ{
\wK(z,\bz)=-\ln \I \left[2(\cF-\bcF) -(\p{i}\cF + \p{\bi}\bcF)(z^i-\bz^\bi)\right] \nn \pt
}

\section{Hypergeometry}
\label{hypergeo}
So far we have discussed geometry relevant for the vector multiplets of an \mbox{$N=2$} \sugra theory in four spacetime dimensions. Next we turn to the geometry relevant for the hypermultiplets of such a theory. We again focus on the geometry itself. The supergravity and its relation with this geometry will be discussed later on, see also \cite{Breitenlohner:1981sm, deWit:1984px, Bagger:1983tt}. More details on the group theory and quaternionic geometry have been collected in appendices \ref{quatapp},  \ref{appsusy1} and \ref{geopt}. The presentation of the group theory aspects follows \cite{Andrianopoli:1996cm}, we use the notation of \cite{Davidse:2005ef}.\\
In the hypermultiplet sector of an \mbox{$N=2$} \sugra theory in four spacetime dimensions, there are four real scalars  per hypermultiplet whose target space is a \emph{quaternion-K\"ahler} manifold\footnote{Contrary to what the name suggests, quaternion-K\"ahler manifolds are not necessarily K\"ahler. In the following we will simply use the term `quaternionic' to denote these manifolds, for more precise definitions see for instance \cite{Bergshoeff:2004nf, Bergshoeff:2002qk}.} in the case of local \susy and a \emph{hyperk\"ahler} manifold in the case of rigid supersymmetry. 
In both cases \susy requires the existence of a principal $SU(2)$-bundle $\cS$ which plays a similar role for hypermultiplets as the line bundle $\cL \to \cM$ in section \ref{spkg}. The difference between hyperk\"ahler and quaternionic geometry is that in the first case the $SU(2)$-curvature is zero, whereas in the second case the $SU(2)$-curvature is proportional to the \Kh forms.\\
Consider as before a product bundle $T\cM$ such that its structure group is of the form
\equ{
SU(2) \otimes Sp(2n, \mR) \nn \pt
}
$T\cM$ (the tangent bundle on $\cM$) is a real $4n$-dimensional manifold. The connection on the tangent bundle is just the sum of the connection on the $SU(2)$-bundle and on the $Sp(2n, \mR)$-bundle.\\
Both the hyperk\"ahler and the quaternionic manifold are real $4n$-dimensional manifolds with a metric $G$:
\equ{
ds^2=G_{AB}\, dq^A \otimes dq^B \qquad A,B=1,\ldots,4n \nn
}
and three \emph{almost} complex structures
\equ{
J^r: T\cM \to T\cM \qquad r=1,2,3 \nn \km
}
which satisfy a quaternionic algebra
\equ{
J^s J^t=-\gd^{st} \Id - \gve^{str}J^r \label{eq:quatal1} \pt
}
The metric is Hermitian with respect to each of these complex structures and instead of just one \Kh form we can introduce three: 
\bea
&& K^r=\2 K^r_{AB}\, dq^A \wedge dq^B \nn \\
&& K^r_{AB}=G_{AC} (J^r)^C_{\,\,B} \label{eq:quat2form1} \km
\eea
$K^r$ is now called the hyperk\"ahler form. This is a generalization of the concept \Kh form. The presence of the hyperk\"ahler form on a quaternionic manifold does not mean that the manifold is a \Kh manifold as defined in section \ref{kahlergeo} (we will come back to this below).\\

The $SU(2)$-bundle $\cS \to \cM$ over the $4n$-dimensional real manifold $\cM$ has  a \label{su2conn1} connection $\go^r$. To obtain a hyperk\"ahler or \quat manifold we must impose\footnote{For any $p$-form $\gPs$ we have $d\gPs=\nabla \gPs$ where $\nabla$ is the covariant derivative with respect to the Levi-Civita connection.}
\equ{
D K^r\equiv d K^r -\gve^{rst} \go^s \wedge K^t =0 \label{eq:condqh1} \pt
}
The difference between the two manifolds is that in the hyperk\"ahler case $\cS$ is flat and in the \quat case it is proportional to the hyperk\"ahler form. Specifically, if the $SU(2)$ curvature is defined by
\equ{
\gO^r\equiv d\go^r - \2 \gve^{rst}\go^s \wedge \go^t \label{eq:SU2curv1} \km
}
 then 
\bea
&& \gO^r=0 \quad \,\,\qquad \textrm{in the hyperk\"ahler case} \nn \\
&& \gO^r=\lambda K^r \qquad \textrm{in the \quat case,}\label{eq:hyperkrel1}
\eea 
where $\gl$ is a real nonvanishing number, related to the scale of the manifold. In the context of the effective \sugra theory which will be introduced in a later chapter, it will turn out that appropriate normalizations of the kinetic terms determine $\gl$ to be $-\gk^2$, where $\gk$ is the Newton constant. We will come back to this in chapter \ref{ch4}.\\
For practical purposes we  introduce a set of quaternionic $1$-form vielbeins $V_i^{\,a}\equiv V_{i\,\,A}^{\,a} dq^A$, where we have the flat indices $i=\{1,2\}$ and $a=\{1,\ldots,2n\}$ that run over the fundamental representations of $SU(2)$ and $Sp(2n,\mR)$ respectively. In terms of these vielbeins we can write the metric of the quaternionic manifold as
\equ{
G_{AB}= V_{i\,\,A}^{\,a} V_{j\,\,B}^{\,b}  \gve^{ij} \gve_{ab} \km \label{eq:metrict1}
}
where $\gve_{ab}=-\gve_{ba}$ and $\gve_{ij}=-\gve_{ji}$ are the  $Sp(2n,\mR)$ and $SU(2)$ invariant metrics\footnote{Note that $SU(2)$ is isomorphic to $Sp(1)$ and $U\!Sp(2)$. Furthermore, $Sp(2n,\mR)$ is isomorphic to $Sp(n)$.}. 
A different expression for the metric (line element) is given by
\equ{
ds^2=G_{\ba b}V_i^{\,b} \otimes \bV^{i\ba} \label{eq:quatmet1} \pt
}
$G_{\ba b}$ is the tangent space metric appearing in front of the kinetic terms of the fermions in the \sugra action. $\bV^{i \ba }$ is the complex conjugate of the vielbein, i.e.
\equ{
\bV^{i \ba } \equiv \left(V_i^{\,a}\right)^\star \nn 
}
and $G_{ \ba b}$ is given in terms of the \emph{inverse} vielbeins $V^{i\,\,A}_{\,b}$ satisfying 
\equ{
V^{i\,\,A}_{\,b} V_{i\,\,B}^{\,b}=\gd^A_{\,B} \nn \km
}
as follows:
\equ{
G_{a\bb }=\2 G_{AB} V^{i\,\,A}_{\,a} \bV_{i\bb }^{\,\,B} \label{eq:metrict2} \pt
}
An important result is that the Riemann tensor $R_{ABCD}$,  is decomposed into two parts, one containing the field strength of the $Sp(2n,\mR)$ connection and the other the $SU(2)$ curvature $\gO^r$, see \cite{Alekseevski}. This means that the Levi-Civita connection associated to the metric $G_{AB}$ has a holonomy group contained in \mbox{$SU(2)\otimes Sp(2n,\mR)$}. In the  hyperk\"ahler case the $SU(2)$-bundle is flat, so we have the following identification:
\bea
&& H\!ol(\cM) = SU(2) \otimes \cH \qquad \textrm{quaternionic manifold} \label{eq:quat2} \\
&& H\!ol(\cM) = \Id \otimes \cH \qquad \, \qquad \textrm{hyperk\"ahler manifold} \nn \\
&& \cH \subseteq Sp(2n,\mR) \nn \pt
\eea 
Equation \R{quat2} is often used as the definition of a quaternionic manifold, i.e., if we have a manifold whose holonomy is of this form, then it is a quaternionic manifold. 
A consequence of the above definition of quaternionic manifolds is that for $n>1$ the Ricci tensor is proportional to the metric:
\equ{
R_{AB}=\gl(2+n)G_{AB} \label{eq:Einstein1} \pt
}
In other words, quaternionic manifolds are Einstein. Contracting with the metric then gives the Ricci scalar curvature
\equ{
R=\gl 4n(2+n)=-\gk^2 4n(2+n) \nn \km \label{eq:einstein1}
}
which is negative. However, for $n=1$,  a manifold of real dimension four, the holonomy group is $SU(2)\otimes Sp(2, \mR) \cong O(4)$ which is the maximal holonomy group of a manifold anyhow, so the definition is of no use. Consequently the Einstein property\footnote{The metric still has this property for $n=1$ if certain conditions on the Riemann curvature tensor, which are automatically satisfied for $n>1$, are imposed as part of the $n=1$ definition. For a complete discussion see \cite{Bergshoeff:2002qk} and references therein.} is no longer guaranteed. Instead we will define a quaternionic manifold of four real dimensions ($n=1$) to be a manifold such that it is Einstein, as in \R{Einstein1},  and has a (anti-)selfdual Weyl curvature
\equ{
W=\pm ^\star W \qquad \textrm{that is } \qquad W_{ABCD}=\pm \2 \gve_{AB}^{\quad EF} W_{EFCD} \pt\nn
}
It is not very  surprising that the holonomy group factorizes as in \R{quat2} because we started with a product bundle with exactly such a structure group. However, starting from the scalars in the \sugra theory and investigating their \susy transformation rules precisely determines the bundle $T\cM$. Supersymmetry requires that the $SU(2)$ curvature is non-zero, giving rise to the geometry described above.\\ 
A note on the the introduced terminology. When the $SU(2)$-bundle is trivial, i.e., when the curvature \R{SU2curv1} is zero, we can take its connection $\go^r$ to vanish. If its connection vanishes we can write for \R{condqh1} $dK^r=0$. From \R{Kcond1} we see that this means that the metric is \Kh with respect to these three \Kh forms and their combinations\footnote{If $J^1, J^2, J^3$ are the three complex structures then there exists in fact a whole family of complex structures parametrized by $J=a J^1 + bJ^2 + cJ^3$, where $a^2+b^2+c^2=1$.}. This is why the manifold with such a metric is called a \emph{hyper}k\"ahler manifold.

\section{Calabi-Yau manifolds}
\label{CY}
If we compactify type IIA/B supergravity on a Calabi-Yau manifold, the effective supergravity theory in four dimensions has $N=2$ supersymmetry. In chapter \ref{ch2} the $4$-dimensional bosonic field content obtained after compactifying on a Calabi-Yau manifold will be derived. For the geometry of Calabi-Yau manifolds see for instance \cite{Candelas:1990pi, Strominger:1990pd}.\\
By a Calabi-Yau manifold of complex dimension $n$ (CY $n$-fold, often denoted by $Y_n$),  we shall mean a compact \Kh manifold of vanishing first Chern class. As a differential form, the first Chern class is the cohomology class of the curvature two-form: 
\equ{
c_1= \left[ \cR\right] \qquad \cR\equiv \I R_{i\bj} \, \,dz^i \wedge d\bz^\bj \nn \km
}
where $R_{i\bj}$ is the Ricci tensor and $\cR$ the Ricci $2$-form. Therefore the CY condition, $c_1=0$ implies that the Ricci form is exact. In complex coordinates that means,
\equ{
R_{i\bj} \, dz^i\wedge d\bz^\bj = \left(\p{i}A_\bj -\p{\bj}A_i\right) dz^i \wedge d\bz^\bj \km \label{eq:een}
}
where $\mathbf{A}=A_i dz^i + A_\bi d\bz^\bi$ is a globally defined $1$-form.\\
A famous theorem of Yau states that given a CY manifold $Y_n$ with associated \Kh form $K_0$, there exists a unique Ricci-flat metric for $Y_n$ whose associated \Kh form $K$ is in the same cohomology class as $K_0$.\\
This basically means that the parameter space of CY manifolds is the parameter space of Ricci-flat \Kh metrics. In other words: for a CY  manifold we can find  a representative  metric in each \Kh class that is Ricci-flat\footnote{It is often  difficult to construct this metric explicitly.}.\\
One of the characteristics of a CY $n$-fold is that the Ricci-flatness implies $SU(n)$ holonomy and vice versa. We will treat this in some detail. \\

On a complex $n$-fold with a Hermitian metric, the Levi-Civita connection $\BgG^i_{\,j}\equiv\gG^i_{\,jk} dz^k$ is $U(n)$ Lie-algebra valued. This is nothing else than the fact that a connection on some vector (tangent in this case) bundle, takes values in the structure group, which is $U(n)$ for a complex manifold\footnote{In general one writes for the connection $1$-form $A^i_{\,j}=A^A_{\,\gm}dx^\gm \left(T_A\right)^i_{\,j}$ where the $\left(T_A\right)^i_{\,j}$ are a set of generators of the structural group Lie algebra in the representation carried by the fibre.}. Note that in the case of a \Kh manifold the only nonvanishing components of the connection and the Riemann curvature tensor are respectively $\gG^i_{\, jk}$ and $R^i_{j\bk l}$ (and their complex conjugates).
Next we introduce a vielbein form, adapted to the complex structure: 
\equ{
e^a=e^a_{\,i} dz^i \qquad e^\ba_{\, \bi} d\bz^\bi =\left(e^a\right)^\star \qquad (a=1, \ldots, n) \nn
}
where $a$ and $\ba$ are the flat indices, taking values in $U(n)$. The metric is related to the vielbeins by $g_{i\bj}=e^a_{\, i}e^\bb_{\, \bj} \,\get_{a\bb}\,$, where $\get_{a\bb}=$diag$(+, \ldots, +)$ is the flat Hermitian metric left invariant by $U(n)$ transformations.\\
The $U(n)$ Lie algebra valued curvature
\equ{
R^a_{\, b}\equiv d \go^a_{\, b} -\go^a_{\, c}\wedge \go^c_{\, b} \nn
}
is related to the ordinary Riemann tensor as
\equ{
R^a_{\, b} =R^i_{\, m \bn j} dz^m\wedge d\bz^\bn e^a_{\, i} e^j_{\, b} \nn \km
}
where $\go^a_{\, b}=-\bgo^\bb_{\, \ba}$ is a $U(n)$ valued spin connection\footnote{The spin connection is related to the affine connection in the usual way: 
\equ{
\p{i} e^a_{\, j} -\gG^k_{\, ij} e_k^{\,a} -\go_{i\,\, b}^{\,a} e^b_{\, j} =0 \qquad \p{[\bi} e^a_{\, j]} -\go_{[\bi \,\, b}^{\,a} e^b_{\, j]}=0 \nn \pt }
The second equation states that the torsion vanishes, which means that the connection is in fact the Levi-Civita connection.}.
Since $e^a_{\, i} e^j_{\, a}=\gd^i_{\, j}$, the Ricci $2$-form is equal to the $U(1)$ part of the $U(n)$ curvature\footnote{This can easily be seen by evaluating $R^a_{\, a}$}. Specifically: $U(n)$ is not semi-simple and can be decomposed as $SU(n)\otimes U(1)$, or on the level of the algebra as $\fs\fu(n) \oplus \fu(1)$. For the spin connection this means that we can write
\equ{
\go^a_{\, b}=\tgo^a_{\,b} + \frac{1}{n} \gd^a_{\, b} \cA_{U(1)} \nn \km
}
where $\tgo^a_{\,b}$ is the traceless part and $\cA$ is the $U(1)$ part. This gives us
\equ{
\cR=R^a_{\, a}=\I R_{i\bj} \, dz^i\wedge d\bz^\bj =\cF_{U(1)}\equiv d \cA_{U(1)} \label{eq:twee} \pt
}
When the first Chern class vanishes, the curvature is an exact $(1,1)$-form. The gauge field \R{twee} is precisely the globally defined one from equation \R{een}. For a Ricci-flat metric this means that \R{twee} vanishes, so the $U(1)$ form $\cA$ is globally defined and closed.  The holonomy group is not $U(n)$ but $SU(n)$, since the $U(1)$ part is trivial. \\

We have shown that for a CY $n$-fold the vanishing of the first Chern class implies that it has $SU(n)$ holonomy. We will now discuss how this leads to the existence of a covariantly constant spinor. The existence of such a spinor in turn implies that the Calabi-Yau is Ricci flat, as we shall demonstrate. 
Let us consider for simplicity the case of a CY $3$-fold which has $SU(3)$ holonomy, we can make a similar argument for CY $n$-folds.\\
The covering group of $SO(6)$ is $SU(4)$, so  spinors of $SO(6)$ are in the fundamental representation  (or its conjugate) of $SU(4)$. Consider such a spinor $\gz$. By an $SU(4)$ transformation $\gz$ can always be put in the form $\gz=(0,0,0,\gz_0)$. This spinor is left invariant by $SU(3)$ rotations, i.e., it is left invariant by the holonomy group. This implies that it is covariantly constant.\\
We now show that a consequence of the existence of a covariantly constant spinor is Ricci flatness. We will give this argument for a CY $n$-fold for which there exists a spinor $\gz$ which obeys $D_A \gz =0$, where $(A,B,C, \ldots)$ denote $2n$-dimensional indices. Multiplying with another covariant derivative and antisymmetrizing yields
\equ{
R^{\quad AB}_{CD} \gG_{AB} \gz =0 \nn \km
}
where $\gG_{AB}$ is the (antisymmetrized) product of the gamma matrices. 
Multiplying by yet another gamma matrix gives
\equ{
R^{\quad CB}_{CD}\gG_B \gz=R_{DB}\gG^B \gz=0 \label{eq:rflat} \km
}
where  we used the gamma matrix identity $\gG^{A} \gG^{BC}=\gG^{ABC} + g^{AB}\gG^C - g^{AC}\gG^B$ and  the curvature identity $R_{ABCD}+R_{ACDB}+R_{ADBC}=0$. This implies that the Ricci tensor vanishes. \\
An important property of Calabi-Yau $n$-folds is given by the following theorem, see for instance \cite{Fre:1995bc}. 
A compact \Kh manifold of complex dimension $n$ has a vanishing first Chern class (in which case it is a Calabi-Yau manifold) if and only if it admits a  unique non-vanishing holomorphic $n$-form, \equ{
\gO=\frac{1}{n!} \gO_{i_1 \ldots i_n}(z) \,\,dz^{i_1}\wedge \ldots\wedge dz^{i_n} \nn \pt
}
This form has the following properties:

\begin{enumerate}
\item
$\gO$ is harmonic
\item
the components of $\gO$ are covariantly constant in the Ricci-flat metric.
\end{enumerate}

\label{dvg1}
This $n$-form will play an important role in section \ref{stcal1}. \\
We can characterize CY-folds by their Hodge numbers. To this end we shall first review some material on cohomology.

\subsubsection{Differential geometry and (co)homology}
We will start by considering a real manifold $M$ of dimension $m$. A differential $r$-form $\go_r$ on $M$ is defined as 
\equ{
\frac{1}{r!}\go_{\gm_1\ldots\gm_r} dx^{\gm_1}\wedge \ldots \wedge dx^{\gm_r} \nn \km
}
where $\go_{\gm_1\ldots\gm_r}$ is a totally antisymmetric tensor and the differentials $dx^\gm$ are antisymmetrized, denoted by the \emph{wedge} products $\wedge$. The Hodge $\star$ operation is defined as
\equ{
\,^\star \go=\frac{\sqrt{|g|}}{r!(m-r)!} \go_{\gm_1\ldots\gm_r}\gve^{\gm_1\ldots \gm_r}_{\qquad \gn_{r+1} \ldots \gn_m}dx^{\gn_{r+1}}\wedge \ldots \wedge dx^{\gn_m} \nn \km
}
where $g$ is the determinant of the metric. The Levi-Civita symbol $\gve$ is defined as
\begin{displaymath}
\gve_{\gm_1\ldots\gm_m}=\left\{ 
\begin{array}{ll}
  +1 & \textrm{for $(\gm_1\ldots\gm_m)$ an even permutation of $(1\ldots m)$} \\ \nn
  -1&  \textrm{for $(\gm_1\ldots\gm_m)$ an uneven permutation of $(1\ldots m)$} \\
  \,\,0& \textrm{otherwise}\km \\
\end{array}
\right.
\end{displaymath}
furthermore $\gve^{\gm_1\ldots \gm_m}=g^{\gm_1\gn_1}\ldots g^{\gm_m\gn_m}\gve_{\gn_1\ldots\gn_m}=g^{-1}\gve_{\gm_1\ldots\gm_m}$. So we see that the Hodge operation is a linear map from the space of $r$-forms $\gO^r(M)$ to the space of $(m-r)$-forms $\gO^{m-r}(M)$. Applying the operation twice yields
\equ{
\,^{\star\star}\go =(-1)^{r(m-r)}\go \nn 
}
if $M$ if Riemannian and
\equ{
\,^{\star\star}\go =(-1)^{r(m-r)+1}\go \nn 
}
if $M$ is Lorentzian. The \emph{exterior derivative} $d$ on an $r$-form $\go$ is defined as
\equ{
d\go=\frac{1}{r!}\left(\pp{}{x^\gn}\go_{\gm_1\ldots\gm_r}\right)dx^{\gn}\wedge dx^{\gm_1}\wedge \ldots \wedge dx^{\gm_r} \nn \km
}
it is a map $\gO^r(M)\to\gO^{r+1}(M)$. If we act twice with $d$ on an arbitrary $r$-form the result is zero:
\equ{
d^2=0 \nn \km
}
thus this operator is \emph{nilpotent}. This allows us to define a \emph{cohomology}. An $r$-form $\go_r$ is \emph{closed} if $d\go_r =0$ and \emph{exact} if $\go_r=d\gg_{r-1}$ with $\gg_{r-1}$ some $(r-1)$-form. Although a closed $r$-form can locally always be written as an exact form, in general this is not true globally. \\
This leads to the definition of the $r$-th \emph{de Rahm cohomology group} of a manifold $M$:
\equ{
H^r(M)=\frac{\textrm{closed $r$-forms on $M$}}{\textrm{exact $r$-forms on $M$}} \nn \pt
}
The dimension of $H^r(M)$ is called the \emph{Betti number} $b^r$. They depend only on the topology of $M$. The operator
\equ{
\gD=\,^\star d^\star d + d^\star d^\star= (d+\,^\star d^\star)^2 \label{eq:Laplacian1} \km
}
is a second order differential operator which reduces to the Laplacian in flat space. An $r$-form is \emph{harmonic} if it obeys $\gD \go_r=0$. It can be shown that each equivalence class in $H^r(M)$ has a unique harmonic representative. Therefore, the space of harmonic $r$-forms $\textrm{Harm}^r(M)$ is isomorphic to $H^r(M)$. The Hodge map turns harmonic $r$-forms into harmonic $(m-r)$-forms, which implies $b^r=b^{m-r}$. \\

Similar (in construction) to cohomology one can define \emph{homology} which is based on the boundary operator $\gd$. This boundary operator acts on submanifolds of $M$ and gives their boundary. If the submanifold has no boundary, acting with $\gd$ produces zero. A  boundary has no boundary, so acting with $\gd$ on a submanifold which is a boundary of a higher dimensional submanifold necessarily produces zero. The real\footnote{This defines \emph{real} homology. One can also consider \emph{integer}, indicated with a $\mZ$,  combinations which defines integer homology.} linear combinations of the submanifolds are called chains. A chain that is closed with respect to $\gd$ is called a \emph{cycle} and \emph{exact} if it is obtained by acting with $\gd$ on another submanifold. This leads us to the definition of the $r$-th \emph{homology group} for $r$-dimensional submanifolds ($r$-chains):
\equ{
H_r(M)=\frac{\textrm{closed $r$-chains on $M$}}{\textrm{exact $r$-chains on $M$}} \nn \pt
}
$H_r$ consists of closed submanifolds that are not themselves boundaries. There is a one-to-one correspondence between the homology and cohomology classes. A theorem by de Rham states that if $M$ is a compact manifold, $H^r(M)$ and $H_r(M)$ are finite-dimensional. Moreover, the map
\equ{
\gL: H_r(M)\times H^r(M) \to \mR \nn 
}
is bilinear and non-degenerate. This map is defined by
\equ{
\gL(c,\go)=\int_c \go \nn \km
}
where $c\in H_r(M)$ and $\go\in H^r(M)$. Thus $H^r(M)$ is the dual vector space of $H_r(M)$.\\

On a complex manifold $N$ of real dimension $d=2n$ we can define very similar structures. To start with, we can define $(p,q)$-forms as having $p$ antisymmetric holomorphic indices and $q$ antisymmetric antiholomorphic indices
\equ{
\go^{(p,q)}=\frac{1}{p!q!} \go_{\gm_1\ldots \gm_p \bgn_1\ldots \bgn_q}\, dz^{\gm_1}\wedge\ldots\wedge dz^{\gm_p} \wedge d\bz^{\bgn_1}\wedge\ldots\wedge\d\bz^{\bgn_q} \pt \nn
}
Furthermore any $\go_r$ form can be uniquely decomposed as
\equ{
\go_r=\sum_{p+q=r}\go^{(p,q)} \nn \pt
}
The exterior derivative  splits into a holomorphic and an anti-holomorphic part 
\equ{
d=\p{}+\bp{} \label{eq:extd2} \km
}
 where $\p{}=dz^i\p{i}$ and $\bp{}=d\bz^\bi \p{\bi}$. Then $\p{}$ and $\bp{}$ take $(p,q)$-forms into $(p+1,q)$- and $(p,q+1)$-forms respectively. Each is nilpotent separately: $\p{}^2=\bp{}^2=0$. This means that we can define the complex version of the de Rham cohomology  called the \emph{Dolbeault} cohomology: 
\equ{
H_{\bp{}}^{(p,q)}(N)=\frac{\bp{}\textrm{-closed $(p,q)$-forms in $N$}}{\bp{}\textrm{-exact$(p,q)$-forms in $N$}} \nn \pt
}
The dimension of $H_{\bp{}}^{p,q}(N)$ is called the \emph{Hodge number} $h^{(p,q)}$. One can construct a similar cohomology for $\p{}$, but the information they contain is the same. Because $d$ splits up as in \R{extd2} we can define the two `Laplacians'
\equ{
\gD_{\p{}}\equiv (\p{}\p{}^\dagger + \p{}^\dagger \p{}) \qquad \gD_{\bp{}}\equiv (\bp{}\bp{}^\dagger + \bp{}^\dagger \bp{}) \nn \km
}
where $\p{}^\dagger=\!\,^\star \p{}^\star$ and $\bp{}^\dagger=\!\,^\star \bp{}^\star$.
Then the $\gD_{\bp{}}$-harmonic $(p,q)$-forms are in one-to-one correspondence with $H^{p,q}_{\bp{}}(N)$, similarly for the unbarred version:
\equ{
\textrm{Harm}_{\p{}}^{(p,q)} \cong H_{\p{}}^{(p,q)} \label{eq:A} \pt
}
The de Rham cohomology groups can be written as
\equ{
H^r(N)=\bigoplus_{p+q=r}H^{(p,q)}(N) \label{eq:Dc1} \pt
}
If the manifold is also K\"ahler (see for instance \cite{LS1}) we have the additional property that
\equ{
\gD=2\gD_{\p{}}=2\gD_{\bp{}}\label{eq:B} \km
}
which means that the Hodge numbers obey $h^{(p,q)}=h^{(q,p)}$.
The Hodge numbers furthermore satisfy $h^{(p,q)}=h^{(n-p,n-q)}$, $h^{(p,0)}=h^{(0,n-p)}$ and $h^{(n,0)}=1$. On a $3$-fold of non-vanishing Euler character\footnote{Its definition will be provided shortly.} (we are specifying to CY $3$-folds now, since they are physically the most interesting) $h^{(1,0)}=0$, $h^{(1,2)}$ and $h^{(1,1)}$ are left as free `parameters' (see also \cite{Fre:1995bc}). \\
The Euler character is defined as
\equ{
\gc\equiv \sum_{r=0}^6 (-1)^r b^{r}=\sum_{r=0}^3 (-1)^r \left(\sum_{k=0}^r h^{(r-k,k)}\right) =2\left(h^{(1,1)}-h^{(1,2)}\right) \label{eq:euler1} \km
}
where $b^{r}$ are the $r$th Betti numbers, giving the dimension of the real $r$th cohomology class.  The second equality reflects property \R{Dc1}.
Note that for a so-called \emph{rigid} CY-fold, which has by definition $h^{(1,2)}=0$, the Euler characteristic is given by $2 h^{(1,1)}$ and is therefore manifestly positive. \\
Summarizing,  on a CY $3$-fold the only non-vanishing Hodge numbers are
\bea  
&& h^{(0,0)}=h^{(3,3)}=1 \nn \\
&& h^{(3,0)}=h^{(0,3)}=1 \nn \\
&& h^{(2,1)}=h^{(1,2)} \nn \\
&& h^{(1,1)}=h^{(2,2)} \label{eq:C} \km
\eea
the $(0,0)$-form equals a constant.

\subsubsection{Holonomy}
\label{holonomy}
Consider again an $m$-dimensional Riemannian manifold $M$. Take a vector from the tangent space at point $p$ and parallel transport it around a closed loop back to $p$. The vector will (in general) be transformed, the set of these transformations is called the holonomy group. Explicitly, if we take a vector $X^\gm \in T_p M$, the parallel transported vector is given by $X^{'\gm}=h^\gm_{\,\,\gn}X^\gn$. When we consider an infinitesimal parallelogram, we have the result
\equ{
X^{'\gm}=X^\gm + X^\gn R^\gm_{\,\,\gn\gk\gl} \gd_1^\gk \gd_2^\gl \nn \km
}
where $\gd_1, \gd_2$ are the edges of the parallelogram. We see that (infinitesimally) $h^\gm_{\,\,\gn}=\gd^\gm_{\,\,\gn} + R^\gm_{\,\,\gn\gk\gl}\gd_1^\gk \gd_2^\gl$, so the Riemann curvature tensor determines the holonomy group.\\

\subsubsection{The moduli space of Calabi-Yau manifolds}
As discussed above, Yau's theorem states that, given a \Kh manifold $Y_3$ with associated \Kh form $K_0$, there exists a unique Ricci-flat \Kh metric for $Y_3$ whose \Kh form is in the same cohomology class as $K_0$. We can therefore consider the parameter space of CY-folds to be the parameter space of Ricci-flat \Kh metrics.
Suppose that we have two Ricci-flat metrics for $Y_3$, $g_{\gm\gn}$ and \mbox{$g_{\gm\gn} + \gd g_{\gm\gn}$}:
\equ{
R_{\gm\gn}(g_{\gm\gn})=0 \qquad R_{\gm\gn}(g_{\gm\gn}+\gd g_{\gm\gn})=0 \nn \km
}
with $\gm, \gn$ real indices.
Together with the coordinate condition $\nabla^\gn \gd g_{\gm\gn}=0$ we have that $\gd g_{\gm\gn}$ satisfies the Lichnerowicz equation: 
\equ{
\nabla^\gr \nabla_\gr \gd g_{\gm\gn} + 2 R_{\gm\,\, \gn}^{\,\gr \,\, \gs} \gd g_{\gr\gs} =0 \nn \km
}
$\nabla^\gr$ is the covariant derivative in real coordinates. Using the properties of \Kh manifolds discussed above, we find that  (in complex coordinates) $\gd g_{i\bj}$, $\gd g_{ij}$ and $\gd g_{\bi \bj}$ satisfy this equation separately. This means that there are two distinct deformations of the metric. To the first, $\gd g_{i\bj}$, we can associate the real harmonic $(1,1)$-form 
\equ{
\I \gd g_{i\bj} \,dz^i\wedge d\bz^\bj \label{eq:mixed} \pt 
}
With the second, the pure type variations, we can form
\equ{
\gO_{ij}^{\,\,\bk} \,\gd g_{\bl \bk} \,dz^i \wedge dz^j \wedge d\bz^\bl \label{eq:pure} \km
}
a complex $(2,1)$-form and similarly for its complex conjugate. This means that the zero modes of this Lichnerowicz equation are in one-to-one correspondence with the elements of $H^{(1,1)}(Y_3)$ and $H^{(2,1)}(Y_3)$ . Clearly the mixed type modes, i.e. \R{mixed}, are variations of the \Kh class, giving rise to $h^{(1,1)}$ real parameters.\\
The variations of the pure type give rise to variations of the complex structure. This can be seen as follows: $g_{\gm\gn}+\gd g_{\gm\gn}$ is a \Kh metric on a manifold close to the original one, therefore a coordinate system must exist in which the pure parts of the metric vanish, see \R{Hermiticity}.  Consider e.g. the holomorphic part. Under a change of coordinates $z^i \to z^i + f^i$ and $\bz^\bi \to \bz^\bi + \bff^\bi$ the (holomorphic part of the) metric varies as 
\equ{
\gd g_{ij}\to \gd g_{ij} -\pp{\bff^\bl}{z^i} \, g_{\bl j} -\pp{\bff^\bl}{z^j} \,g_{i\bl} \nn \pt
}
If this is to be zero then $\bff^\bj$ is to be a function of the $z^i$'s and vice versa for the antiholomorphic part. This means that the coordinate transformations are not holomorphic, in other words, the complex structure changes. \\
The conclusion of the above is that the moduli space $\cM_{mod}$ of CY-folds is the product of the space spanned by \Kh class deformations $\cM_{1,1}$ and the space spanned by the complex structure deformations $\cM_{1,2}$, 
\equ{
\cM_{mod}=\cM_{1,1}\times \cM_{1,2} \nn \pt
}
It turns out that both $\cM_{1,1}$ and $\cM_{1,2}$ are special \Kh manifolds. \\
Let us for the moment restrict our attention to the complex structure deformations which parametrize $H^{(1,2)}(Y_3)$ and include $H^{(3,0)}(Y_3)$ and $H^{(0,3)}(Y_3)$ (with elements $\gO$ and $\bgO$)  which together form $H^3(Y_3)=H^{(3,0)}\oplus H^{(0,3)}\oplus H^{(2,1)}\oplus H^{(1,2)}$. Choose a real basis of harmonic $3$-forms $(\ga_I, \gb^J)$ on $H^3(Y_3,\mZ)$, where $I,J=0, \ldots, h^{(1,2)}$, obeying 
\bea
&& \int_{Y_3} \ga_I \wedge \gb^J =-\int_{Y_3} \gb^J \wedge \ga_I = \gd_I^J \nn \\ [1.5mm]
&& \int_{Y_3} \ga_I\wedge \ga_J=\int_{Y_3} \gb^I\wedge \gb^J =0 \label{eq:basissym1} \pt
\eea
Note that these relations are invariant under integer valued symplectic reparame\-tri\-zations
\equ{
\left(
\begin{array}{c}
  \gb \\
  \ga \\
\end{array}
\right)
\to
 \underbrace{\left(
\begin{array}{cc}
  U & Z\\
  W & V \\
\end{array}
\right)}_{A}
\left(
\begin{array}{c}
  \gb \\
  \ga \\
\end{array}
\right)
\qquad A \in Sp(2h^{(1,2)}+2) \label{eq:symplectictransf} \km
}
with the submatrices obeying 
\bea
&& U^T V -W^T Z =V^T U -Z^T W=\Id \nn \\[1mm]
&& U^T W=W^T U \qquad Z^T V=V^T Z \nn \pt 
\eea
When the \emph{periods} of $\gO$ are defined as 
\equ{
Z^I =\int_{Y_3} \gO \wedge \gb^I \qquad \cG_{I}=\int_{Y_3} \gO\wedge \ga_I \nn
}
$\gO$ can be expressed as $\gO=Z^I \ga_I -\cG_I \gb^I$, which is invariant under \R{symplectictransf}. This implies that $(Z^I, \cG_I)$ transforms as a symplectic vector. The $\cG_I$ are in fact functions of the $Z^I$ and can be determined in terms of $\cG(Z)$, a homogeneous function of degree \emph{two} as
\equ{
\cG_I =\pp{\cG}{Z^I}\equiv \p{I}\cG \nn \pt
}
\label{spk1}On the other hand, $\gO$ is of degree \emph{one} in $Z$, i.e. $\gO=Z^I\p{I} \gO$ with $\p{I}\gO=\ga_I -\cG_{IJ} \gb^J$ and $\cG_{IJ}\equiv \p{I}\p{J} \cG$.\\
As a result, the $Z^I$ are projective coordinates, allowing us to choose the patch $Z^I=(1,z^a)$, where $z^a=\frac{Z^a}{Z^0}\, ,  \quad a=1,\ldots,h^{(1,2)}$: precisely the complex structure deformation parameters. The metric on $\cM_{1,2}$ is K\"ahler: $g_{a\bb}=\p{a}\bp{\bb}\wK$ with $\wK$ given by
\equ{
\wK=-\ln \I \int_{Y_3} \gO\wedge \bgO =-\ln \I \left(\bZ^I \cG_I - \bcG^I Z_I\right) \label{eq:kahmod1} \pt
}
Comparing with \R{kahlerpot1} (where $Z^I$ is called $X^I$ and $\cG_{(I)}$ $F_{(I)}$) tells us that $\cM_{1,2}$ is a special \Kh manifold. 

\clearpage{\pagestyle{empty}\cleardoublepage} 
\chapter{Calabi-Yau compactifications}
\label{ch2}
In the first part of this chapter we will schematically review the compactification of $11$-dimensional supergravity to $4$ dimensions, using a CY $3$-fold. Specific attention shall be given to the M$5$-brane and the M$2$-brane.\\
We start by compactifying the bosonic fields of $11$-dimensional supergravity on a circle, which gives the fields of type IIA supergravity. 
We then outline the compactification of type IIA on a general Calabi-Yau manifold. This gives rise to a certain set of fields in four dimensions: the bosonic field content of $N=2$ supergravity coupled to vector multiplets and hypermultiplets.
We focus on the universal sector which arises as a tensor multiplet that does not depend on the details of the Calabi-Yau manifold. Next we will investigate this tensor multiplet by considering its alternative formulations: the universal hypermultiplet and the double-tensor multiplet. We will focus on this sector because we will restrict ourselves for simplicity to compactifications on rigid Calabi-Yau manifolds. 

A point worth emphasizing is that although we speak about `branes' frequently, no worldvolume actions or boundary states will be used. They will always be dealt with from the point of supergravity. The branes will always appear as solitonic objects and consequently will be treated in a `macroscopic' manner.

\section{$11$ dimensions}
\label{11d}
In this section we follow the conventions of \cite{Becker:1999pb}.\\
The bosonic part of the $11$-dimensional \sugra action \cite{Cremmer:1978km}, is given by
\equ{
S_{11}=\2 \int d^{11}x \sqrt{-\hg} \hR -\4 \int \left(\hF_4 \wedge ^\star \hF_4 -\3 \hA_3 \wedge \hF_4\wedge\hF_4\right) \label{eq:11dsugra1} \pt
}
We have set the $11$-dimensional gravitational coupling constant $\gk_{11}$ to $1$. 
The \sugra action contains a $3$-form potential $\hA_3$ with field strength $\hF_4=d \hA_3$ which obeys the Bianchi identity 
\equ{
d \hF_4=0 \label{eq:bianchi1}
}
and the field equation
\equ{
d \,^\star \hF_4 + \2 \hF_4^2=d\left( ^\star \hF_4 +\2 \hA_3 \wedge \hF_4\right)=0 \label{eq:eom1} \pt
}
This equation gives rise to the conservation of an `electric' type charge\footnote{Actually, these charges are quantized, as we will see later explicitly, in a slightly different setting in section \ref{stcal1}.}
\equ{
q_e = \int_{\p{}\cM_8} \left( ^\star \hF_4 +\2 \hA_3 \wedge \hF_4\right) \label{eq:electric1} \km
}
the integral is over the boundary at infinity of some spacelike $8$-dimensional subspace of $11$-dimensional spacetime, \cite{Page:1984qv, Duff:1990xz, Gueven:1992hh}. On the other hand, \R{bianchi1} gives a `magnetic' conserved charge
\equ{
q_m=\int_{\p{}\cN_5} \hF_4 \label{eq:magnetic1} \km
}
where we now integrate over the boundary (at infinity) of a spacelike $5$-dimen\-sional subspace. \\
Just as the Maxwell $1$-form couples naturally to the worldvolume (a worldline in fact) of a charged particle, $\hA_3$ couples naturally to the worldvolume of a $2$-dimensional charged extended object. The charge is given by \R{electric1} and represents the charge of a \emph{membrane} with a $3$-dimensional worldvolume. The integration surface $\p{} \cM_8$ is a surface surrounding the membrane.\\
The second way to come up with an extended object is to think about the dualized field strength $^\star \hF_4$, a $7$-form. Suppose one would find an underlying potential for $^\star \hF_4$, then this potential would couple to a $6$-dimensional worldvolume. This `solitonic' or `magnetic' object is called the \emph{fivebrane}\footnote{M$5$-brane really: the $5$-brane from M-theory, of which the low energy effective theory is presumably given by $11$-dimensional supergravity.}, which means that $\p{}\cN_5$ in \R{magnetic1} actually surrounds this $5$-brane. \\
One procedure to go from $11$ to $10$ dimensions is to compactify on a circle, i.e., $x_{11}\sim x_{11}+2\gp$. By compactifying $11$-dimensional supergravity on a circle one obtains type IIA supergravity in $10$ dimensions\footnote{For more information on this procedure see \cite{Polchinski:1998rr}.}. The  Kaluza-Klein ansatz for the metric is
\equ{
ds_{11}^2=e^{-2\gf/3} g_{AB} dx^A dx^B + e^{4\gf/3} \left(dx^{11} -A_A dx^A\right)^2 \label{eq:stf1} \km
}
where $A_A$ is a vector field, $A=0,\ldots, 9$, $\gf$ is the $10$-dimensional dilaton. The $11$-dimensional $3$-form $\hA_3$ gives the $10$-dimensional objects
\equ{
 A_{ABC}=\hA_{ABC} \qquad  B_{AB}=\hA_{AB 11} \label{eq:hgh1} \km
}
with $10$-dimensional field strengths $H_3 = dB_2$, $F_2=dA$  and $F_4 = dA_3$. \\
Using the same reasoning as before, the $5$-brane charge in $10$ dimensions is associated to 
\equ{
\int_{\p{} \cN_4} H_3\label{eq:magnetic2}
}
and the membrane charge to 
\equ{
\int_{\p{}\cM_7} \left(e^{\gf/2}  ~^\star F_4 + A_3 \wedge H_3\right) \label{eq:electric2} \pt
}
Let us take a closer look at these objects. Consider, for instance, the $5$-brane, the magnetic dual of the fundamental string (in $10$ dimensions). This object was originally discovered as a consistent background for supergravity, see \cite{Duff:1990wv, Callan:1991ky, Stelle:1998xg}. In the case of type IIA \sugra it is characterized by the following equations:
\equ{
e^{2\gf}=e^{2\gf_\infty} + \frac{Q}{r^2} \qquad  H_{\gm\gn\gr}=\gve_{\gm\gn\gr}^{\,\,\,\,\,\,\,\,\,\gs} \p{\gs} \gf \label{eq:nsbrane2} \km
} 
where  $Q$ is the charge of the $5$-brane and $e^{2\gf_\infty}$ is an integration constant. The (conformally) flat $4$-dimensional space transversal to the $5$-brane is parametrized by $x^\gm$. The transversal distance $r=|x|$ is measured from a point $x$ in transversal space to the brane which is located for simplicity at the origin. The longitudinal space can  taken to be flat. Actually,  since the geometry of this solution is simply a direct product, we can insert more complicated geometries in the direction of the $5$-brane, as long as they satisfy the Einstein equations. The transversal geometry will remain the same. For example, one can take the geometry of a CY $3$-fold.
However, note that we  must perform a Wick rotation such that the brane becomes a Euclidean brane before wrapping it around the Calabi-Yau. The result is then a $5$-brane which is wrapped around the CY, thus appearing in the transversal space  as an instanton located at the origin. 
We can construct a similar setup for the membrane, but in this case the membrane cannot wrap around the whole of the CY, but it has to wrap a $3$-cycle.

\section{$4$ dimensions} 
\label{fourd}
One can compactify type IIA supergravity to four dimensions on a $4$-dimensional Minkowski space times an internal Calabi-Yau $3$-fold, i.e., $\cM^{1,3} \times Y_3$. The result is an $N=2$ supergravity theory coupled to vector multiplets and hypermultiplets. It is instructive to consider the amount of supersymmetry preserved by such a compactification first.\\ 
A compactification specifies a background for the $10$-dimensional \sugra theory.  This background preserves a certain amount of supersymmetry. The precise amount preserved by this groundstate is determined by the number of independent supersymmetry parameters for which the fermionic \susy variations vanish.
If one specifies the background by setting the background values for the fermionic fields in the theory to zero, the only \susy variation which is not trivially zero is the variation of the gravitino:
\equ{
\gd \gps_A^i=\nabla_A \ge^i \nn \km
}
with $A$ a 10-dimensional index.
In type IIA \sugra in $10$ dimensions, there are two independent \susy parameters $\ge^i\,,\, i=1,2$ and two gravitino's. We see that the variation only vanishes if there are covariantly constant spinors in our Kaluza-Klein background. \\
For the compactification ansatz of $4$-dimensional Minkowski space times a Calabi $3$-fold, the spinors $\ge^i$ decompose as 
\equ{
\ge^i \propto \bget^i\otimes \gz(y) + \get^i \otimes \gz^\dagger(y) \nn \km
}
where $\gz(y)$ is a covariantly constant spinor on the CY $3$-fold, $\gz^\dagger(y)$ its complex conjugate and the $\get^i$ are complex $4$-dimensional Weyl spinors.\\
A CY $n$-fold admits one covariantly constant spinor, see the discussion in section \ref{CY}. This means that in $4$ spacetime dimensions we are left with the two \susy parameters $\get^1$ and $\get^2$, together providing $8$ supercharges, i.e., $N=2$ supergravity. For more information about the details and consistency of this Kaluza-Klein program see \cite{Lust:1997kx, Duff:1986hr, Pope:1987ad, Duff:1989cr}.\\
In chapter \ref{ch3} we will elaborate on the notion of Killing spinors and perform a detailed analysis on the amount of supersymmetry preserved when we choose a non-trivial bosonic background in $4$ dimensions, namely the instanton background. \\

Next we illustrate how the scalar fields of the hypermultiplets and the gravitational and vector multiplets of the $N=2$ supergravity theory in $4$ dimensions arise.\\
We follow the Kaluza-Klein program and expand the $10$-dimensional fields in harmonics on the internal CY $3$-fold $Y_3$. If we collectively denote the fields by $\gF^{(I)}(x,y)$, where $x^\gm$ are the Minkowski spacetime coordinates and $y^m$ (or $(z^i, \bz^\bi)$) are the internal (complex) coordinates, we can write
\equ{
\gF^{(I)}(x,y)=\sum_\ga \gF^{(I)}_\ga(x) Y^\ga_{(I)}(y) \nn \km
}
with $Y^\ga_{(I)}(y) $ the appropriate harmonics.\\
The light fields appearing in $4$ dimensions are those coefficient functions $\gF^{(I)}_0(x)$ whose harmonics $Y^0_{(I)}(y)$ are zero modes of the internal Laplacian $\gD_{(I)}$ which acts on them in the field equations. \\
For instance, consider a $2$-form field $\hB_2$ in $10$ dimensions with action
\equ{
\int d\hB \wedge ^\star d \hB \nn \pt 
}
Decomposing the $10$-dimensional space  as $\cM^{1,3} \times Y_3$, the $10$-dimensional Laplacian decomposes as $\gD_{10} =\gD_4 + \gD_6$. The equation of motion $d ^\star d \hB_2=0$ together with the gauge condition $d ^\star \hB_2=0$ can be written as $\gD_{10} \hB_2=0$, see \R{Laplacian1}. Upon decomposing this becomes
\equ{
\left( \gD_4 + \gD_6\right)\hB_2 =0 \nn \pt
}
A massless $B_2$ field in $4$ dimensions should obey $\gD_6 \hB_2=0$.
This means that we must expand $\hB_2$ in harmonics of $\gD_6$.
We know, see \R{A}, \R{B} and \R{C}, that such harmonics correspond to the various cohomology groups on $Y_3$. Therefore the following  reductions of $\hB_2$ can be associated to the cohomology groups, in complex coordinates
\bea
&& B_{\gm \gn} \leftrightarrow H^{(0,0)}(Y_3) \nn \\
&& B_{\gm i} \,\leftrightarrow  H^{(1,0)}(Y_3) = \emptyset \label{eq:Bred1} \\
&& B_{i j} \,\leftrightarrow  H^{(2,0)}(Y_3) =\emptyset \nn \\
&& B_{i\bj} \,\leftrightarrow  H^{(1,1)}(Y_3) \nn \pt
\eea
Explicitly this means that we can expand $\hB_2$ as 
\equ{
\hB_2=B_2 + b^i \go_i \label{eq:Bexp1} \km
}
where the $\go_i$ are $(1,1)$-forms, $i=1, \ldots, h^{(1,1)}$. $B_2$ is a $2$-form in $4$ dimensions and 
 the $b^i$ are real scalar fields in $4$ dimensions: $b^i(x)$.\\
The other fields are treated in a similar manner. Reducing to $10$ dimensions gave us the following NS-NS fields: 
\equ{
B_2 \, ,\quad g_{AB}\, , \quad A_A \quad \textrm{and} \quad \gf \nn
}
and the R-R field $A_3$, $B_2$ has just been discussed. The metric field $g_{AB}(x,y)$ can be decomposed as $g_{\gm\gn}(x)$ (the $4$-dimensional graviton) $g_{\gm m}$ and $g_{mn}$, or $g_{m i}$,  $g_{ij}$, $g_{i\bj}$ and their complex conjugates.  In section \ref{CY} we have discussed  the zero modes of $g_{ij}$ and $g_{i\bj}$ and we have seen that they correspond to deformations of the \Kh class and the complex structure, see \R{mixed} and \R{pure}, which means that they can be expanded as 
\equ{
\I \gd g_{i\bj} =\sum_{l=1}^{h^{(1,1)}} v^l(x) \go^l \label{eq:mixed2} 
}
and 
\equ{
\gd g_{ij}=\sum_{a=1}^{h^{(1,2)}} z^a(x) \bb_{a ij} \label{eq:pure2} \pt
}
The $\bb_{a ij}$ are related to the $(1,2)$-forms $\gg_a \,\,\, a=1, \ldots , h^{(1,2)}$ (not in a real basis this time) by 
\equ{
\bb_{aij}=\frac{\I}{|\!|\gO|\!|^2} \gO_i^{\, \bl \bk} \bgg_{a; \bl \bk j} \label{eq:12form} \km
}
where $|\!|\gO|\!|^2 \equiv \frac{1}{3!} \gO_{ijk} \bgO^{ijk}$. Furthermore $g_{\gm i}=0$ because $h^{(1,0)}=0$, which we used in \R{Bred1} as well. Similarly, the $1$-form $A$ only gives a $1$-form in $4$ dimensions: the graviphoton $A^0$. The $10$ dimensional dilaton gives a $4$-dimensional dilaton, $\gf(x)$. Lastly, the R-R $3$-form $A_3$ is expanded as
\equ{
A_3=C_3 + A^i \wedge \go_i + \gx^A \ga_A + \tgx_B \gb^B  \label{eq:3form1e} \km
}
where $C_3(x)$ is a $3$-form on $\cM^{1,3}$, the $h^{(1,1)}$ $A^i(x)$'s are $1$-forms. As before, see \R{basissym1}, the $(\ga_A, \gb^A)$ form a real basis of $H^3(Y_3)$, consequently $A_3$ gives us $2(h^{(1,2)}+1)$ real scalars. \\
These fields organize themselves into the following multiplets:  the gravitational multiplet
\equ{
\left(A^0, g_{\gm\gn}\right) \label{eq:gravm1a}
}
and the vector multiplets
\equ{
\left(A^i, v^i, b^i\right) \nn \km
}
 containing $h^{(1,1)}$ vectors and twice as many real \label{spk2} scalars\footnote{These scalars are often treated in one go by \emph{complexifying the \Kh cone}: write $B_2 + \I J=(b^i + \I v^i) \go_i \equiv t^i \go_i$, where we now have $h^{(1,1)}$ complex scalars $t^i$.}. The hypermultiplets
\equ{
\left(z^a, \gx^a, \tgx_a\right) \label{eq:hyperm1} \km
}
contain the $h^{(1,2)}$ complex scalar fields $z^a$ from \R{pure2} and in total $4h^{(1,2)}$ real scalar fields. \\
Finally there is the \emph{tensor}  multiplet, TM for short:
\equ{
\left(\gx^0, \tgx_0, \gf, B_2\right) \label{eq:tenm1a} \km
}
a multiplet which is always present and does not depend on the details of the CY $3$-fold since $B_2$ and $\gf$ are related to the $0$-form of the $CY$ and $(\gx^0, \tgx_0)$ are the expansion coefficients of $A_3$ w.r.t. the unique $\gO$. This means that if we were to consider a \emph{rigid} ($h^{(1,2)}=0$) CY this multiplet would still be present.\\
Often the tensor multiplet is dualized to an additional hypermultiplet since a $2$-form field in $4$ dimensions is dual to a (pseudo) scalar field, generally called the axion. The resulting multiplet is called the \emph{universal hypermultiplet} (UHM) because it must be present in any CY-compactification. \\
Due to supersymmetry \cite{deWit:1984px} (see also \cite{Aspinwall:2000fd} for a geometrical argument) the total target manifold parametrized by the various scalars factorizes as a product of vector and hypermultiplet manifolds: 
\equ{
\cM_{\textrm{scalar}}=\cM_V\otimes \cM_H \label{eq:scm1} \pt
}
$\cM_V$ is a special \Kh manifold and $\cM_H$ is a quaternionic manifold.\\
This is a very important result, since it means  that string corrections, whether perturbative or nonperturbative, only affect the hypermultiplets, since the dilaton belongs to a hypermultiplet and $g_s\equiv e^{-\gf/2}$. The vector multiplet geometry remains unaffected\footnote{Just as the hypermultiplet metric receives no $\ga'$ corrections, in type IIA that is.  For a general discussion on the corrections to the moduli spaces of both type IIA and type IIB, see \cite{Aspinwall:2000fd}.}. \\

\label{hereI}
The $3$-form $C_3$ is dual in $4$ dimensions to a constant $e_0$. It turns out that taking this space-filling $C_3$ along in the compactification (or conversely $e_0$) has the effect of gauging the axion, see \cite{Louis:2002ny}. The graviphoton $A^0$ acts as the gauge field and the gauging gives rise to a potential, this will be discussed at a later stage. 

In the above we have schematically derived the bosonic field content of ($N=2$) supergravity in $4$ dimensions. Similarly one can reduce the whole $10$-dimensional action to obtain the $4$-dimensional one and relate in this way all the terms appearing in the $4$-dimensional action to the original fields and the geometric data of the CY $3$-fold.  This is done in the seminal paper \cite{Bodner:1990zm}, see also the useful appendices in \cite{Louis:2002ny}.

Conversely, knowing that the reduction gives one \sugra multiplet, $4(h^{(1,2)} +1)$ hypermultiplets and $h^{(1,1)}$ vector multiplets, one could construct such an action from first principles. One could also choose to leave out the vector multiplets, since the moduli space factorizes anyway. Such an action has been constructed in  \cite{Bagger:1983tt}, in which the most general action for hypermultiplets coupled to $N=2$ supergravity is constructed. This action is a non-linear sigma model\footnote{See \cite{deWit:2002vz} and the references therein for non-linear sigma models in supergravity.}  where the Lagrangian contains all kinds of complicated non-linear terms, specifically  the kinetic terms which provide a metric for the target space. Supersymmetry constrains these `complicated terms' such that the hypermultiplet sector of the theory parametrizes a quaternionic manifold. This means that if we assume supersymmetry is unbroken in the $4$-dimensional supergravity theory, quantum corrections cannot give new terms in the effective action.
The various quantities, such as the metrics for the kinetic terms, may change, but only in such a way that the supersymmetry relations are still obeyed. Geometrically speaking, the scalars keep parametrizing a quaternionic manifold, although  perhaps a different one. \\
This is a very powerful  and useful concept which will be put to use in chapter \ref{ch4}. In that chapter we will construct perturbations of the classical hypermultiplet space (corresponding to the classical UHM action). These perturbations respect the quaternionic geometry of the target-space and describe, as it will turn out,  effects that must be attributed to membrane instantons. \\
The conclusion  of the above discussion is that the possible effects the $5$-brane and the membrane can have, must show up in the (universal) hypermultiplet(s) sector of the effective action, which is why we will be neglecting the vector multiplets from now on.

The aim of the next sections is to investigate the $4$-dimensional theory for which we will compute instanton corrections in chapters \ref{ch3} and \ref{ch4}. 

\section{The universal hypermultiplet}
\label{instsol}
The easiest situation to consider is  compactifying  type IIA supergravity on a rigid CY $3$-fold, because we then only have to deal with the universal tensor multiplet. The vector multiplets can be consistently truncated to zero for the purposes of this chapter. The resulting four-dimensional low energy bosonic effective Lagrangian is given by
\bea
&& e^{-1} \cL_T =-R -\2 \pu{\gm}\gf \p{\gm}\gf + \2 e^{2\gf} H^\gm H_\gm -\4 F^{\gm \gn} F_{\gm\gn}  \!\!\quad \textrm{(the NS-NS sector)} \nn \\
&&\,\,\, -\2 e^{-\gf}\left(\pu{\gm}\gc \p{\gm}\gc + \pu{\gm}\gvf \p{\gm}\gvf\right) -\2 H^\gm \left(\gc \p{\gm} \gvf -\gvf \p{\gm}\gc\right) \quad \textrm{(the R-R sector)} \nn
\eea
The R-R scalars $\gvf$ and $\gc$ were formerly known as $\gx^0$ and $\tgx_0$, compare with \R{tenm1a}. Notice that both $\gc$ and $\gvf$ have constant shift symmetries. $H^\gm=\frac{1}{2}\gve^{\gm\gn\gr\gs}\p{\gn}B_{\gr\gs}$ is the dual NS $2$-form fields strength (i.e. $H_3=d B_2$). The tensor $B_2$ also has a shift symmetry and is dual to the axion.
$F^{\gm\gn}$ is the field strength associated to the graviphoton $A_\gm$, previously denoted by $A^0$ (see \R{gravm1a}). As before the dilaton is denoted by $\gf$.\\

We can transform this Lagrangian into the Lagrangian of the UHM by dualizing\footnote{This will be explained later after giving some information about the isometries of the UHM.} $B_2$ to the axion $\gs$ giving
\bea
 e^{-1} \cL_{UH} &=&-R -\2 \pu{\gm}\gf \p{\gm}\gf  -\4 F^{\gm \gn} F_{\gm\gn}  \quad \label{eq:UHM1} \\
&&\,\,\, -\2 e^{-\gf}\left(\pu{\gm}\gc \p{\gm}\gc + \pu{\gm}\gvf \p{\gm}\gvf\right) -\2 e^{-2\gf}\left(\p{\gm} \gs + \gc \p{\gm}\gvf\right)^2 \nn \km
\eea 
which is the classical UHM at string tree-level. It is a non-linear sigma model \cite{Cecotti:1988qn, Ferrara:1989ik}  with a quaternionic target space corresponding to the coset space
\equ{
\frac{SU(1,2)}{U(2)} \label{eq:uhmgeo1} \km
}
where $SU(1,2)$ is the isometry group\footnote{See for instance \cite{Ketov:2001gq, Besse} for some alternative  parametrizations of this  metric.} of the Lagrangian.
This space is actually \Kh as well, as can be seen by going to the frequently used parametrization
\bea
&& e^\gf=\2\left( S+\bS -2C\bC\right) \quad \,\qquad \gc = C+\bC \nn \\
&& \gs =\frac{\I}{2}\left(S-\bS + C^2 -\bC^2\right) \qquad \gvf=-\I\left(C-\bC\right) \label{eq:realb1} \pt
\eea
In these coordinates the \Kh potential is given by 
\equ{
\wK=-\ln\left(S+\bS -2 C\bC\right) \nn
}
and the line element on the target space by
\equ{
ds^2=e^{2\wK}\left(dS d\bS -2C dS d\bC -2\bC d\bS dC + 2(S+\bS)dC d\bC\right) \nn \km
}
the pullback of which, to $\cM^{1,3}$,  precisely gives the Lagrangian.  The $8$ isometries which leave the Lagrangian invariant were analyzed in \cite{deWit:1990na, deWit:1992wf}. The shift isometries of the R-R scalars and the axion mentioned above form a subgroup of the isometry group. This group is known as the Heisenberg group and acts on the fields as
\bea
&& S\to S+\I \ga + 2 \bge C + |\ge|^2 \nn \\[1mm]
&& C\to C+\ge \label{eq:heis1} \km
\eea
where the finite $\ga$ and $\ge$ are respectively real and complex. In the real basis \R{realb1} the Heisenberg group acts as
\equ{
\gf\to\gf \,  \quad \gc\to \gc+\gg \,\quad \gvf\to \gvf + \gb \,\quad \gs \to \gs-\ga-\gg \gvf \label{eq:heis2}  \km
}
$\gb=\I (\ge-\bge)$ and $\gg=(\ge+\bge)$.
The generators of the Heisenberg group obey the commutation relations
\equ{
[\gd_\ga, \gd_\gb]=[\gd_\ga, \gd_\gg]=0 \qquad [\gd_\gb, \gd_\gg]=\gd_\ga \nn \pt
}

Furthermore there is a $U(1)$ symmetry which acts as a rotation on $\gvf$ and $\gc$, together with a compensating transformation on $\gs$. Its finite transformation can be determined from the results in \cite{Becker:1995kb, Davidse:2003ww} and reads
\bea
&& \gvf \to \cos(\gd) \,\gvf + \sin(\gd) \,\gc \qquad \gc \to \cos (\gd) \,\gc -\sin (\gd)\, \gvf \km \nn \\ [1.5mm]
&& \gs \to \gs -\4 \sin (2 \gd) \left(\gc^2 -\gvf^2\right) + \sin^2 (\gd)\, \gc \gf \label{eq:u1tr} \pt
\eea
Four of the eight isometries have now been specified.\\
The isometries of the Heisenberg group  are expected to survive at the perturbative level. The reasons for expecting this are as follows. The Heisenberg group does not act on the dilaton, i.e., on $g_s$. This means that we can examine the Heisenberg group order by order in perturbation theory. Furthermore, the shifts in $\gc$ and $\gvf$ are preserved because they originated from the gauge transformation of the $3$-form $A_3$ in ten dimensions, see \R{3form1e}. The field strength of this $3$-form will involve its derivative, $dA_3$. The gauge symmetry is therefore given by $\gd A_3=d\gL_2$, where $\gL_2$ is a $2$-form. If one works out $dA_3$ using \R{3form1e} one finds (among others) the terms $d\gx^0 \ga_0 +d\tgx_0 \gb^0$ for which the gauge symmetry becomes
\equ{
\gx^0\to \gx^0 + \gg \qquad \tgx_0 \to \tgx_0+ \gb \nn \pt
}
We gave the constant shifts the same name as in \R{heis2}, to which these transformations correspond. We can think about the shifts in the axion in a similar way. \\
Finally there are the $4$ remaining isometries we do not give here. They involve non-trivial transformations on the dilaton and hence will change the string coupling constant. We will not consider these isometries nonperturbatively, in fact it is not even known what happens to them perturbatively. 
Although the isometries of the Heisenberg group (\ref{eq:heis1}, \ref{eq:heis2})  are unaffected by quantum corrections, which appear at one-loop only in the string frame see \cite{Antoniadis:2003sw, Strominger:1997eb, Antoniadis:1997eg, Anguelova:2004sj}, nonperturbative effects will quantize these isometries. This will be discussed in chapter \ref{ch3}.

\section{The double-tensor multiplet}
\label{secDTM1}
Instead of working with the UHM we will find it convenient to work with the so-called \emph{double-tensor multiplet}, or DTM for short.
As the name suggests, the DTM contains $2$ tensors and arises by dualizing a R-R scalar \mbox{($\gvf$ or $\gc$)} from the tensor multiplet to an additional tensor. We shall start by discussing the DTM in Lorentzian signature and the actual dualization process. In the next section we shall give its Euclidean formulation and the reason why we choose the Euclidean DTM over the Euclidean UHM.\\
The bosonic part of the DTM Lagrangian is given by
\equ{
e^{-1}\cL_{DT}=-R -\2 \pu{\gm}\gf\p{\gm}\gf -\4 F^{\gm\gn}F_{\gm\gn} -\2 e^{-\gf} \pu{\gm}\gc \p{\gm}\gc + \2 M^{IJ} H_I^\gm H_{\gm J} \label{eq:DTM1} \km
}
where we have dualized $\gvf$ into another $H$. Comparing with the UHM gives $\gf^I=(\gvf, \gs) \leftrightarrow H^I=(H^1, H^2)$. In effect we can go from the UHM to the DTM, the reason is that the UHM possesses two commuting isometries over which we can dualize, for instance the shift symmetries in $\gvf$ and $\gs$ corresponding to $\ga$ and $\gb$. The DTM also possesses such symmetries since $B_1$ and $B_2$ only appear through their derivatives. \\
The kinetic term (metric) for the $3$-form field strengths $H_I^\gm=\2 \gve^{\gm\gn\gr\gs}\p{\gn}B_{\gr\gs I}$ is given by
\equ{
M^{IJ}=e^\gf \left(%
\begin{array}{cc}
  1 & -\gc \\
  -\gc & e^\gf + \gc^2 \\
\end{array}%
\right) \label{eq:tensormetric1} \pt
}
The remaining two scalars $\gf$ and $\gc$ parametrize the coset $Sl(2,\mR)/O(2)$, see appendix \ref{appdtm1} for the various target manifolds. However, the two tensors break this $Sl(2,\mR)$ symmetry to a $2$-dimensional subgroup generated by rescalings of the tensors and by the remaining generator of the Heisenberg algebra \R{heis2} acting on $\gc$ and $B_1$ as
\equ{
\gc \to \gc + \gg \qquad B_1 \to B_1 + \gg B_2 \label{eq:tensorisom1} \km
}
$\gf$ and $B_2$ are invariant. Summarizing: in going from the UHM to the DTM two of the shift symmetries of the Heisenberg group, namely $\ga$ and $\gb$ (see \R{heis2}), have been used to dualize to two tensors, which now in turn have shift symmetries. The remaining shift symmetry $\gg$ now acts on $\gc$ and $B_1$ as in \R{tensorisom1}.\\
We will sometimes use the `Heisenberg invariant'
\equ{
\hH_1\equiv H_1 -\gc H_2 \label{eq:heisinv1} \km
}
which, as the name suggests, is invariant under \R{tensorisom1}.\\

Let us now finally be more specific about the process of dualization, for more details see \cite{Theis:2003jj}. In general one can dualize the scalars of a bosonic sigma model into tensors if they appear only through their derivatives. This means that the target manifold has a set of Abelian Killing vectors\footnote{Actually,  the Killing vectors have to leave the complex structures invariant and the isometries have to commute with supersymmetry. For a quaternionic manifold this is automatically the case. For more details see \cite{deWit:2001bk}.}, 
\equ{
\gd_\gTh \gf^{\hA} =\gTh^I k_I^\hA(\gf) \qquad [k_I, k_J]=0 \nn \pt
}
We can choose coordinates $\gf^\hA=(\gf^A, \gf^I)$, where $\gf^A=(\gf, \gc)$,  such that these transformations act as constant shifts on the $\gf^I$ and leave the $\gf^A$ invariant. When writing $\gf^\hA$  we mean to denote all the scalars in the UHM, i.e. $\gf^\hA=(\gf,\gc,\gvf,\gs)$.  $\gf^I$ are the scalars in the UHM which are dualized into the two tensors of the DTM and $\gf^A$ are the scalars in the DTM. As the notation suggests, tensors can be dualized as well if they appear solely through their field strengths. Furthermore they guarantee commuting isometries in the corresponding scalar sigma model, so if we go from the DTM to the UHM we have the two commuting shift isometries discussed earlier. \\
This procedure also works at the perturbative level, as mentioned before, since the Heisenberg group is preserved, \cite{Antoniadis:2003sw}. At the nonperturbative level things are slightly more subtle, we will discuss this. \\
To dualize from the UHM to the DTM one replaces the derivatives on $\gf^I$ by covariant derivatives. Add Lagrange multipliers $B_{\gm\gn I}$ with  $H^\gm_I=\frac{1}{2}\gve^{\gm\gn\gr\gs}\p{\gn}B_{\gr\gs I}$. Integrating out these multipliers sets the gauge fields to zero and gives the original action back.  Alternatively one can integrate out the gauge fields, giving the dual action. Explicitly: consider the bosonic part of the UHM, 
\equ{
e^{-1}\cL_{UH} =-\2 G_{\hA\hB} \pu{\gm}\gf^{\hA} \p{\gm}\gf^\hB -R -\4 F^{\gm\gn}F_{\gm\gn}  \nn \km
}
in which $G_{\hA\hB}$ is the scalar (quaternionic) metric one can read off from \R{UHM1}. Now we gauge two isometries by replacing $\p{\gm}\gf^I \to \p{\gm}\gf^I -A_{\gm}^I$ and by adding the Lagrange multiplier term $H_I^\gm A_\gm^I \,,\,I=1,2$. The Lagrangian density becomes
\bea
e^{-1}\cL &=&-\2 G_{AB}\p{\gm}\gf^A \pu{\gm}\gf^B -G_{AI} \p{\gm}\gf^A\left(\pu{\gm}\gf^I -A^{\gm I}\right) -R -\4 F^{\gm\gn}F_{\gm\gn} \nn \\[1mm]
&&\, -\2 G_{IJ}\left(\p{\gm}\gf^I -A_\gm^I\right) \left(\pu{\gm}\gf^J -A^{\gm J}\right) -H^\gm_I A_\gm^I \label{eq:temp2a} \km
\eea 
remember that $A=1,2$.
The next step is to integrate out the gauge field, the equation of motion for $A_\gm^I$ is
\equ{
A^I_\gm=A^I_A\p{\gm}\gf^A + \p{\gm}\gf^I -H_\gm^I \nn \km
}
\label{sec:page1} where $A^I_A\equiv M^{IJ} G_{JA}$. Inserting this back into \R{temp2a} gives
\equ{
e^{-1}\cL_{DT}=-\2 \cG_{AB}\pu{\gm} \gf^A \p{\gm} \gf^B + \2 M^{IJ} H_I^\gm H_{\gm J} -R-\4 F^{\gm\gn}F_{\gm\gn}  \label{eq:DTM2}
}
where $M^{IJ}$ has been defined in \R{tensormetric1}, the relation with the UHM metric $G_{\hA\hB}$ is $M^{IJ}\equiv (G_{IJ})^{-1}$. The metric for the $2$ remaining scalars is also a subset of $G_{\hA\hB}$, namely 
\equ{
\cG_{AB}=G_{AB} -G_{AI} M^{IJ} G_{JB} = \left(%
\begin{array}{cc}
  1 & 0 \\
  0 & e^{-\gf} \\
\end{array}%
\right)
\label{eq:GAB1} \pt
}
Together these components form the hypermultiplet metric, or the other way around: the hypermultiplet metric decomposes as
\equ{
\label{eq:dual-metric}
  G_{\!\hat{A}\hat{B}} = \lp \cG_{AB} + A_A^I M_{IJ} A_B^J & A_A^K
  M_{KJ} \\[2mm] M_{IK} A_B^K & M_{IJ} \rp \pt
}
We have obtained the DTM by dualizing two scalars from the UHM to tensors. Perturbatively, the DTM guarantees\footnote{Because the Heisenberg group is preserved in perturbation theory.} two commuting shift symmetries in the dual UHM description. Nonperturbatively, however, the duality also involves the constant modes of the dual scalars $\gvf$ and $\gs$ by means of theta-angle-like terms. These are surface terms which have to be added\footnote{The dualization process is defined up to such surface terms.} to the DTM Lagrangian and are non-vanishing in the presence of instantons and anti-instantons. In Euclidean space they can be written as integrals over $3$-spheres at infinity:
\equ{
S_{\textrm{surf}}^E=\I \left(\gvf \int_{S^3_\infty} H_1 + \gs\int_{S^3_\infty} H_2 \right)=\I \gvf Q_1 + \I \gs Q_2 \label{eq:boun1} \km
}
The charges $Q_1$ and $Q_2$ are related to the membrane and fivebrane instantons associated with the two tensors $H_1$ and $H_2$, as we will discuss in chapter \ref{ch2b} (see \R{instcharges1}). $\gvf$ and $\gs$ are now some parameters that play the role of coordinates on the moduli space in the dual UHM theory. 
The dualization back to the UHM is performed by promoting $\gs$ and $\gvf$ to fields that serve as Lagrange multipliers enforcing the Bianchi identities on the tensors. This gives back the fields $\gvf$ and $\gs$ in the UHM, in the DTM there are no $\gvf$ and $\gs$ fields because they have been dualized to two tensors.\\
Boundary terms such as these are also added in $3$-dimensional gauge theories in the Coulomb phase where the effective theory can be described in terms of a vector which can be dualized into a scalar (the dual photon) in a similar manner as described above, see \cite{Seiberg:1996nz, Dorey:1997ij, Dorey:1998kq} and in particular \cite{Polyakov:1976fu, Polyakov:1987ez}.  \\

\section{Wick rotations and duality}
\label{wickrotI}
Having constructed the DTM we still need to analytically continue it. We use the Wick rotation to analytically continue (the details are given in appendix \ref{appwick}).  The result is
\equ{
e^{-1}\cL^E_{DT}=\2 \pu{\gm}\gf\p{\gm}\gf + \2 e^{-\gf}\pu{\gm}\gc \p{\gm}\gc + \2 M^{IJ} H_I^\gm H_{\gm J} + R+\4 F^{\gm\gn}F_{\gm\gn} \label{eq:DTM3} \pt
}
In performing our Wick rotation we have assumed that we are dealing with a class of metrics for which we can perform this procedure. In general this is a subtle issue. However, in the next chapter we will limit ourselves to flat space, for which the rotation is well defined. We will do so because the instanton configurations we will discuss must have vanishing energy momentum tensor. This requirement will be satisfied by working in flat space. We see that in general the presence of the Einstein-Hilbert term causes our otherwise positive definite Euclidean action to be unbounded from below. Again, in flat space this is not an immediate problem since the Ricci scalar vanishes. However, gravitational fluctuations around the instanton configuration could ruin this. \\
From this Lagrangian two Bogomol'nyi equations can be derived, one corresponding to a NS $5$-brane and another to a membrane instanton, which will be the subject of the next chapter. However before continuing,  we will discuss the reason for working with the Euclidean DTM, as promised.\\

Apart from the Einstein-Hilbert term, the action corresponding to \R{DTM3} is positive definite and has real saddle points for the $3$-forms, which we will construct in the next chapter. In principle one could work with the UHM as well, but then one would have to deal with imaginary saddle points. To see this, let us reconsider the dualization process of a single (for simplicity) $3$-form $H$ to a scalar, the generalization is straightforward.\\
In Minkowski space the path integral involving a $2$-form is given by
\bea
&&\int [dH] e^{\I S_H} \prod_x \gd[dH(x)] \label{eq:D1}\\
&& = \int [dH] \int [da] \exp\left( \I S_H + \frac{\I}{3!} \int d^4 x\, a \,\gve ^{\gm\gn\gr\gs}\p{\gm}H_{\gn\gr\gs}\right) \nn \pt \eea
The Bianchi identity is enforced by including a delta function in the path integral, which is given in its functional representation in the second line. Note that $a$ is just a (real) dummy variable at this point. The action is given by
\equ{
S_H=\frac{1}{3!}\int d^4 x \sqrt{-g} \left(-\2 H_{\gm\gn\gr}H^{\gm\gn\gr}\right) \nn \pt
}
The functional integral over $H$ is a Gaussian integral and can be explicitly performed. To do so shift the integration variable:
\equ{
H_{\gm\gn\gr}=\frac{1}{\sqrt{-g}}\,\gve_{\gm\gn\gr\gs}\pu{\gs}a + h_{\gm\gn\gr} \nn \km
}
where the first part is the familiar dualization relation between a $3$-form and a pseudoscalar in $4$ dimensions. The exponent in \R{D1} becomes 
\equ{
-\frac{\I}{2}\int d^4 x \,\p{\gm}a\pu{\gm}a -\2\frac{\!\I}{3!}\int d^4 x \sqrt{-g}\,h_{\gm\gn\gr}h^{\gm\gn\gr} \nn \pt
}
The functional integral over the fluctuations $h_{\gm\gn\gr}$ can easily be performed since they appear undifferentiated, so we are left with the dualization relation
\equ{
H_{\gm\gn\gr}=\frac{1}{\sqrt{-g}} \, \gve_{\gm\gn\gr\gs}\pu{\gr}a \nn \pt
}
In Euclidean space, there is no factor of $\I$ in front of $S_H$, see \R{D1}.  This has the consequence that one obtains an extra factor of $\I$ in the dualization relation:
\equ{
H^E_{\gm\gn\gr}=\frac{\I}{\sqrt{g}} \, \gve_{\gm\gn\gr\gs}\pu{\gr}a \nn \pt
}
This factor of $\I$ is crucial in ensuring that a positive $3$-form action in Euclidean space dualizes correctly to a positive scalar action. \\
Hence we see that  performing a real saddle point calculation for the DTM in Euclidean space, corresponds to an imaginary saddle point calculation for the UHM, for more details see \cite{Burgess:1989da}.\\
Note that this factor of $\I$ one needs in the integral representation of the delta function in Euclidean space is exactly provided for by the way we define our Wick rotation. That is to say, after Wick rotating our Lagrange multiplier term precisely  has the right factor of $\I$, see \R{tempa21}, and we thus have a consistent framework: dualizing commutes with the analytic continuation.

\clearpage{\pagestyle{empty}\cleardoublepage}

\chapter{The instanton solutions}
\label{ch2b}
In the previous chapter we have shown how the tensor multiplet (and its equivalent formulation the DTM and UHM) arises in Calabi-Yau compactifications. We will proceed to demonstrate explicitly how NS $5$-branes and membranes appear in the DTM closely following \cite{Theis:2002er, Davidse:2003ww}.\\
We derive two Bogomol'nyi bounds for the DTM. Then we shall construct NS $5$-brane solutions corresponding to the first bound and membrane solutions corresponding to the second Bogomol'nyi bound.

\section{The Bogomol'nyi bound}
\label{sectionBb1}
It is convenient to use form notation in deriving the Bogomol'nyi bound. The Euclidean DTM Lagrangian \R{DTM3} is then written as
\equ{
\cL^E_{DT}=d^4 x \sqrt{g} R + \2 |d\gf|^2 + \2 e^{-\gf} |d\gc|^2 + \2 M^{IJ} \,^\star H_I\wedge H_J \label{eq:DTM5} \pt
}
Note that we have dropped the term $\4 F^{\gm\gn}F_{\gm\gn}$, since from now on we will choose a vanishing graviphoton and focus our attention exclusively on the scalar-tensor sector. 
Defining
\equ{
H\equiv \left(%
\begin{array}{c}
  H_1 \\
  H_2 \\
\end{array}%
\right) \qquad
E\equiv
\left(%
\begin{array}{c}
  d\gf \\
  e^{-\gf/2}d\gc \\
\end{array}%
\right)\qquad
N\equiv
e^{\gf/2}\left(%
\begin{array}{cc}
  0 & e^{\gf/2} \\
  1 & -\gc \\
\end{array}%
\right) \nn \km
}
such that $N^T N=M$ one can rewrite \R{DTM5} as
\equ{
\cL^E_{DT}=d^4 x \sqrt{g} R+\2 \,^\star \left(N ^\star H + OE\right)^T \wedge \left(N ^\star H+OE\right) + H^T\wedge N^TOE \nn\km
}
where $O$ is some scalar dependent orthogonal matrix. We explicitly include this matrix since $N$ and  $E$ are defined only up to $O(2)$ rotations. This formulation shows that the action is bounded from below\ by
\equ{
S^E \geq \int_\cM \left(d^4 x \sqrt{g}R+H^T\wedge N^T OE \right)\label{eq:bb1} \km
}
The second term in \R{bb1} is topological because it is independent of the spacetime metric. The action attains its lowest value for configurations satisfying the \emph{Bogomol'nyi bound} 
\equ{
^\star H=-N^{-1}O E \label{eq:bb2} \pt
}
Naturally this condition must imply the equations of motion, which will fix the matrix $O$. Furthermore, field configurations satisfying \R{bb2} have vanishing energy-momentum tensor which allows us to restrict ourselves to the case $g_{\gm\gn}=\gd_{\gm\gn}$.
The equations of motion for the tensors, i.e. $d(M ^\star H)=0$, are satisfied if
\equ{
d\left(N^T OE\right)=0 \nn \km
}
which also guarantees that the topological term in \R{bb1} is closed and can therefore be written as a total derivative. Consequently this term does not affect the equations of motion for the scalars, which are determined by demanding that the Bianchi identity holds ($dH=0$) which gives upon using \R{bb2} 
\equ{
d\left(N^{-1}O ^\star E\right)=0 \nn \pt
}
We have enough information to determine the possible $O$'s. The first solution is given by
\equ{
O_1=\pm\left(%
\begin{array}{cc}
  1 & 0 \\[1mm]
  0 & -1 \\
\end{array}%
\right)
\label{eq:Ofb} \pt
}
The second solution is given by
\equ{
O_2=\pm\frac{1}{|\gt'|}\left(%
\begin{array}{cc}
  \Re \gt' & -\Im \gt' \\[1mm]
  \Im \gt' & \Re \gt'\\
\end{array}%
\right) \label{eq:Omb} \km
}
with $\gt'\equiv (\gc-\gc_0) + 2\I e^{\gf/2}$ and $\gc_0$ a real integration constant.

\section{The NS $5$-brane}
\label{secns5b1}
Using \R{Ofb} the  Bogomol'nyi bound for the $5$-brane takes the form
\equ{
\left(%
\begin{array}{c}
  H_{\gm 1} \\[1mm]
  H_{\gm 2} \\
\end{array}%
\right)=
\pm \p{\gm} \left(%
\begin{array}{c}
  e^{-\gf} \gc \\[1mm]
  e^{-\gf} \\
\end{array}%
\right)
\label{eq:Bbound1}
}
and as this configuration defines a lower bound of the action,  it is the background we will be expanding around. The $+$ corresponds to instantons and the $-$ to anti-instantons. 
The second equation in \R{Bbound1} comes from the NS sector and specifies the NS $5$-brane, compare with \R{nsbrane2}. The first equation determines the R-R background in which the $5$-brane lives. \\
It is often useful to use the basis of \R{heisinv1} in which the Bogomol'nyi bound takes the form $\hH_{\gm 1}=\pm e^{-\gf}\p{\gm}\gc$. \\
Since the tensors have to obey the Bianchi identity, the scalars in \R{Bbound1} have to obey Laplace-like equations. One could either use source terms for these equations or excise points $\{x_i\}$ from the flat spacetime $\mR^4$, we will use the latter method. These excised points correspond to the locations of the instantons. We find multi-centered solutions of the form
\bea
&& e^{-\gf} =e^{-\gf_\infty} + \sum_i \frac{|Q_{2i}|}{4\gp^2 (x-x_i)^2} \label{eq:sol1} \\
&& e^{-\gf} \gc =e^{-\gf_\infty} \gc_\infty + \sum_i \frac{Q_{1i}}{4\gp^2 (x-x_i)^2} \label{eq:sol2}  \km
\eea
where $Q_{1i}\,,Q_{2i}\,, \gc_\infty$ and $\gf_\infty$ are independent integration constants. In equation \R{sol1} we write $|Q_{2i}|$ because the exponential function has to be positive everywhere in spacetime. In our conventions the string coupling constant is identified as $g_s\equiv e^{-\gf_\infty/2}$. As already anticipated, the two charges are defined by integrating the tensor field strengths $H_{\gm\gn\gr I}=-\gve_{\gm\gn\gr\gs}H^\gs_I$ over $3$-spheres at infinity, 
\equ{
Q_I=\int_{S^3_\infty} H_I \km \qquad I=1,2 \label{eq:instcharges1}
}
which are related to \R{sol1} and \R{sol2} via \R{Bbound1}. 
Explicitly calculating \R{instcharges1} gives
\equ{
Q_2=\mp \sum_i |Q_{2i}| \qquad Q_1=\mp \sum_i Q_{1i} \nn\km
}
which means that for instantons $Q_2$ is negative and for anti-instantons positive. $Q_1$ can be anything since the combination $e^{-\gf} \gc$ does not have to be positive. \\
The (anti-)instanton action \R{bb1} is found to be
\equ{
S_{\textrm{cl}}=\pm \int_{\p{}\cM}\left(\gc H_1 -\left(e^\gf + \2 \gc^2\right) H_2\right) \label{eq:instactioncl} \km
}
where the integral is over the boundaries of $\mR^4$:  $\p{}\cM=S^3_{\infty} \cup \sum_i S^3_i$, i.e. the sphere at infinity together with the  infinitesimal spheres around the excised points. 
This action is finite only if $\gc(x)$ is finite near the excised points:
\equ{
\gc_i\equiv \lim_{x\to x_i}\gc(x) =\frac{Q_{1i}}{|Q_{2i}|} \label{eq:38b} \km
}
which is finite if $Q_{2i}\neq 0$, for nonvanishing $Q_{1i}$. This implies that the integrated Heisenberg invariants \R{heisinv1} vanish, 
\equ{
\hQ_{1i}\equiv Q_{1i}-\gc_i |Q_{2i}|=0 \label{eq:heisinv3} \pt
}
The result for the action is 
\equ{
S_{\textrm{cl}}=\frac{|Q_2|}{g_s^2} + \2 \sum_i |Q_{2i}| \left(\gc_\infty -\gc_i\right)^2 \km \qquad \gc_\infty\equiv \lim_{x\to\infty} \gc(x)  \pt \label{eq:instaction2}
}
This action has an inversely quadratic dependence on $g_s$ which is precisely as expected for a $5$-brane wrapped around a CY $3$-fold, \cite{Becker:1995kb}.

If we consider a single-centered instanton around $x_0$, the action simplifies to 
\equ{
S_{cl}=\frac{|Q_2|}{g_s^2}\left(1+\2 g_s^2 (\gD \gc)^2\right) \label{eq:sc5b1} \km
}
where $\gD\gc\equiv \gc_{\infty}-\gc_0$.
 In contrast to the dilaton, $\gc$ remains finite and thus regular at the origin, so no source term can be associated to it. We can therefore regard this as a R-R background in which the instanton lives. Consequently, the `bare' instanton configuration is obtained by turning this background off, which lowers the value of the action. Turning it off means taking $\gc_\infty=\gc_0$ and using \R{sol2} one finds that this implies that $\gc(x)$ is constant everywhere: $\gc(x)=\gc_0$. \\
The actual instanton calculation performed in chapter \ref{ch3} will be for a single-centered instanton with action \R{sc5b1}. The multi-centered solutions\footnote{In chapter \ref{ch3} we  demonstrate that these solutions can be constrained further by requiring them to preserve half of the supersymmetries.} can then be obtained by making a multipole expansion around the single-centered one, the dominant term of which corresponds to the single-centered one.\\
To complete action \R{sc5b1} we have to combine it with the theta-angle terms of \R{boun1} which we can rewrite as
\equ{
S_{\textrm{surf}}=\I \gvf Q_1 + \I \gs Q_2 =\mp  \I \left(\gs + \gc_0 \gvf\right)|Q_2| + \I \gvf \hQ_1 \km \label{eq:boun2}
}
the total single-centered instanton action thus becomes
\equ{
S^\pm_{\textrm{inst}}=S_{\textrm{cl}} + S_{\textrm{surf}} \pt\label{eq:boun5}
}
The reason for rewriting \R{boun2} in this way is because we will associate $\hQ_1$ to the membrane charge in the following section. The surface term allows us to distinguish instantons from anti-instantons in the action, again with the $+$ denoting the former and $-$ the latter. 

\section{The membrane}
\label{mem1}
The second Bogomol'nyi bound is somewhat more complicated and becomes, using \R{Omb},
\equ{
\left(%
\begin{array}{c}
  H_{\gm 1} \\[1mm]
  H_{\gm 2} \\
\end{array}%
\right)
=\pm \frac{1}{|\gt'|}\left(%
\begin{array}{c}
  \gc(\gc-\gc_0)\p{\gm}e^{-\gf} + e^{-\gf}(\gc+\gc_0)\p{\gm}\gf + 2 e^\gf\p{\gm}e^{-\gf} \\[1mm]
  (\gc-\gc_0)\p{\gm} e^{-\gf} + 2 e^{-\gf}\p{\gm} \gc \\
\end{array}%
\right) 
\label{eq:bbound2} \pt
}
Remember that $\gt'\equiv (\gc-\gc_0) + 2\I e^{\gf/2}$ and $\gc_0$ is an arbitrary constant which will be identified below with the previously introduced $\gc_0$ for a single-centered instanton.  It is convenient to consider the Heisenberg combination
\equ{
H_{\gm 1}-\gc_0 H_{\gm 2} =\pm \p{\gm}h \qquad h\equiv e^{-\gf} |\gt'|\label{eq:II} \km
}
the Bianchi identities imply that $h$ must be harmonic and positive everywhere. We can write $\gc$ as follows:
\equ{
\gc-\gc_0=e^\gf\sqrt{h^2-4e^{-\gf}} \km \label{eq:Iib}
}
where we have taken the positive branch, the negative branch only differs by some unimportant minus signs in the following calculations. The Bogomol'nyi equation for $H_{\gm 2}$ now becomes
\equ{
H_{\gm 2}=\pm \frac{1}{\sqrt{h^2-4e^{-\gf}}}\left(2e^{-\gf}\p{\gm}h -h\p{\gm} e^{-\gf}\right) \label{eq:Iic}
}
and together with the Bianchi identities, the fact that  $h$ is harmonic gives
\equ{
\left(h^2 -4e^{-\gf}\right) \p{\gm}\pu{\gm}e^{-\gf}  + 2\p{\gm}e^{-\gf}\pu{\gm}e^{-\gf} -2h \p{\gm}h\pu{\gm}e^{-\gf} + 2 e^{-\gf}\p{\gm}h\pu{\gm}h=0 \label{eq:III} \pt
}
For simplicity we limit ourselves to spherically symmetric solutions\footnote{Some multi-centered solutions were constructed in \cite{matthijs}.} for $h$:
\equ{
h=e^{-\gf_\infty}|\gt_\infty'| + \frac{|\hQ_1|}{4\gp^2 (x-x_0)^2} \label{eq:h1}\km
}
which validates the assumption that the dilaton depends on the coordinates only through $h$.\\
Differentiating \R{III} allows us to solve for $\gf$: 
\equ{
e^{-\gf}=a h^2 + b h + c \nn \km
}
where $a,b,c$ are three integration constants. Combining this with \R{Iic} yields $c=-\gb^2$ where $\gb\equiv \pm Q_2/|\hQ_1|$, $b=-\gb \sqrt{1-\ga}$ and $\ga=4a$. Since $H_{\gm 2}$ is real and $e^{-\gf}$ must be positive this requires that $0\leq \ga \leq 1$ where $\ga=0$ must be treated separately. \\
Evaluating $\gD \gc$ using \R{Iib} fixes $\ga$ in terms of the charges and asymptotic values of the fields, 
\equ{
\ga=1-\frac{\left(\gD \gc -2\gb e^{\gf_\infty}\right)^2}{|\gt'_\infty|^2} \qquad |\gt'_\infty|^2=(\gD\gc)^2 + \frac{4}{g_s^2} \nn \pt
}
Furthermore, the solution for $\gc$ can be directly read off from \R{Iib}, using the information obtained above one can check that $\lim_{x \to x_0} \gc(x)$ is indeed $\gc_0$, as in the case of the $5$-brane instanton. \\
Note that, contrary to the $5$-brane system, $\gc$ does require a source at the excised point. Since $\ga$ has to lie in the interval $[0,1]$ we must have
\equ{
\frac{\gD\gc -|\gt'_\infty|}{2e^{\gf_\infty}} \leq \gb \leq \frac{\gD\gc +|\gt'_\infty|}{2e^{\gf_\infty}} \label{eq:V} \km
}
for fixed $g_s$ and (positive) $\gD \gc$. \\
All the integration constants of the solution have now been expressed in terms of the charges and the asymptotic values of the scalar fields. The formula for the action \R{bb1} gives us
\equ{
S=|\gt'_\infty| \left( |\hQ_1| + \2 \gD \gc Q_2\right) \nn \km
}
which is manifestly positive by virtue of \R{V}. The simplest form of the action is obtained by switching off the R-R background by taking $Q_2=\gD \gc=0$ and including the appropriate theta-angle term\footnote{Remember that in the DTM formulation the $\gf^s$ (i.e. $\gvf$ or $\gc$) are just parameters.}, which gives 
\equ{
S_{\textrm{inst}}=\frac{2|Q_1|}{g_s} + \I \gf^s Q_1^s \label{eq:VI} \km
}
where $\gf^s$ is either $\gvf$ or $\gc$ and $Q^s_1$ either $Q^\gvf_1$ or $Q^\gc_1$. Note the factor of $2$ in \R{VI}, which will be very nicely confirmed in chapter \ref{ch4}. As discussed in section \ref{secDTM1} the imaginary term is related to a surface term that arises in the dualization process. In the dual UHM formulation the parameter $\gvf$ (or $\gc$) in \R{VI} can be identified with the value of $\gvf$ (or $\gc$) at infinity. Its presence breaks the shift symmetry in $\gvf$ (or $\gc$) to a discrete subgroup.\\
The explanation of the choice of theta-angle terms in \R{VI} is as follows. Instead of dualizing from the tensor multiplet to the double tensor multiplet using the shift symmetry in $\gvf$, as we did in section \ref{secDTM1}, we could as well dualize using the shift symmetry in $\gc$. Or in the case of going from the UHM to the DTM, dualizing over $\gs$ and $\gc$ . \\
The point is that dualizing over $\gc$, i.e. using the shift generated by $\gg$ (see \R{heis2}), would have given a (formally) different $2$-form with associated field strength and consequently a different charge. This means that there are two different charges: $Q^\gvf_1$ and $Q^\gc_1$, the first charge corresponds to the shift $\gg$ and the second to $\gb$.  Hence they are associated to two different membranes (see also \cite{Becker:1999pb}).\\
This means that we can either dualize to a DTM in which the Bogomol'nyi bound leads to a membrane configuration with charge $Q^\gvf_1$, or to a configuration with charge $Q^\gc_1$. We cannot dualize over the two scalars simultaneously, since the shift symmetries in $\gc$ and $\gvf$ do \emph{not} commute. This fact will be re-derived  from a string theory point of view in chapter \ref{ch4} (it will have to do with the fact that the membrane can wrap either along the one or the other supersymmetric cycle).
The $5$-brane charge $Q_2$ corresponds  to the shift in $\gs$ by $\ga$.\\

This action also has exactly the right dependence on $g_s$, \cite{Becker:1995kb}, for a membrane instanton with charge $\hQ_1$, note that $\hQ_1=Q_1$ if $Q_2=0$.\\ 
One could add $5$-brane charge which raises the action until, for fixed $g_s$ and $\gD \gc$, one reaches $\ga=0$ in \R{V}. From that point on, the solution is no longer valid and one must restrict to the $5$-brane, without membranes. \\[3mm]

{\large \textbf{A short summary and outlook}}\\

We have constructed the $5$-brane and membrane solutions in $N=2$ supergravity in $4$ dimensions. These solutions can be identified with the $5$-brane and membrane respectively, since we know from string theory,  \cite{Witten:1995ex,Shenker:1990uf,Becker:1995kb}, what their actions should look like in four dimensions (in terms of $g_s$ anyway) and we find agreement. As explained above, we will perform an  instanton calculation for the $5$-brane in chapter \ref{ch3} and compute in this way the instanton corrections to the UHM. To do this it is necessary to calculate the (Euclidean) \susy transformations, Killing spinors and instanton measures, for this particular background. All of this will be  done in the subsequent chapter.\\

In principle one could try to do the same for the membrane solution, that is, perform an instanton calculation as for the $5$-brane. Instead we will use knowledge of the isometries of the the UHM to directly construct nonperturbative corrections to the UHM, without the usual instanton calculations. Comparison to string theory permits us to identify these corrections with loop expansions around membrane instantons. This will be the subject of chapter \ref{ch4}. 

\clearpage{\pagestyle{empty}\cleardoublepage} 
\chapter{The NS $5$-brane}
\label{ch3} 
In this chapter we  compute the instanton effects originating from the NS $5$-brane, using the DTM formulation.  The underlying idea is that, for small coupling constant $g_s$, the path integral is dominated by the configurations of lowest Euclidean action and one may proceed by expanding around these configurations. \\
Naturally, the simplest such configuration is the ordinary perturbative vacuum of the theory, which has $S_E=0$. However, as we have seen in chapter \ref{ch2b}, there are other minima of the action we have to expand about. These minima are the ones corresponding to the $5$-brane and membrane. For reasons that will become clear later on, we will not follow such a program for the \mem and we will restrict ourselves to the $5$-brane only.\\
We expand the action up to second order around the instanton configuration, which will be discussed further in section \ref{instmeasure}. Ideally one would like to compute the determinant of the resulting quadratic operator acting on the fluctuations, thus doing a  one-loop computation. However, we are dealing with a difficult non-linear sigma model coupled to supergravity and the resulting quadratic operator is rather complicated, therefore we will not compute its determinant. \\
The instanton solutions \R{sol1} and \R{sol2} break the translation invariance of the theory, since they are located at specific points $\{x_i\}$, these are the so-called \emph{collective coordinates}. Furthermore, the instanton solution partially breaks supersymmetry, which gives rise to fermionic collective coordinates. We will discuss the relation between broken symmetries and collective coordinates in section \ref{ubs}.\\
As we shall see in section \ref{instmeasure}, the collective coordinates are related to zero modes of the quadratic operator acting on the fluctuations. This means that one has to be careful in constructing the path integral measure. \\
In section \ref{instmeasure} we will construct the path integral measure suitable for a ($1$-loop) calculation in the presence of the NS $5$-brane instanton. We shall see that we have to trade the integration over a certain set of quantum fluctuations (zero modes of the quadratic operator) for integrations over the collective coordinates. Apart from the integration over the position (the bosonic collective coordinates) there are also the integrations over the fermionic collective coordinates, these are Berezin integrals. Such integrals are very restrictive because they will only be nonzero if we compute correlation functions which contain enough fermions to `soak up' the Berezin integrals.\\
This leads us to consider  specific correlators in section \ref{sect_correl}. By computing these correlators we can construct the effective action\footnote{Part of the effective action actually, namely the corrections to the kinetic terms of the scalars and tensors and some vertices. To compute corrections to the other terms in the action \R{sa1} one can apply supersymmetry.}. To be precise: the action of the DTM (with $N=2$ local supersymmetry) with $5$-brane instanton corrections.\\
In section \ref{sect_modul} we shall consider the consequences for the moduli space of the UHM, which is what we were after all along, and examine the breaking of certain isometries of the Heisenberg algebra to a discrete subgroup.\\
We have to keep in mind that we are approximating a NS \5b wrapped along the Calabi-Yau by this instanton. Stated differently, in string theory the \fb instanton is described by an embedding of the $6$-dimensional worldvolume into the $10$-dimensional space $\mR^4\times Y_3$ such that the worldvolume ends up entirely on the Calabi-Yau. The embedding maps are then thought of as the collective coordinates of the \fb and performing  a genuine path integral would involve doing an integration over these maps. This amounts to a path integral over the worldvolume theory  in the \sugra background $\mR^4\times Y_3$, as advocated in \cite{Becker:1995kb}, see also \cite{Harvey:1999as}.\\
Due to the complicated and somewhat mysterious nature of that worldvolume theory, this would be  difficult. We will not try to include such worldvolume effects and limit ourselves to an integration over the collective coordinates in their capacity as positions in Euclidean space. \\

First we will present some background material, in section \ref{sec1}, on the general $N=2$ supergravity theory coupled to tensors and scalars. This is the theory into which the DTM fits. In section \ref{dtm_susy} we will restrict ourselves to the case of only the DTM for which we will be doing the calculations of the rest of the chapter. Many technical details and calculations have been directed to the appendices, they will be referred to as needed. 

\section{Supersymmetry}
\label{sec1}
In chapter \ref{ch2} we have explained that we will be performing the NS $5$-brane instanton calculation in the Euclidean DTM. One can Wick rotate to Lorentzian signature and dualize back to the UHM, so that we do not need the (Euclidean) supersymmetric UHM. \\
The most general form of the  action for  scalars and tensors coupled to $N=2$ supergravity has been constructed in \cite{Theis:2003jj} for spacetimes with Lorentzian signature. In Euclidean space the action is given by
\begin{align} 
  & e^{-1} \cL  = \notag\\
  &\, \frac{1}{2\kappa^2}\, R + \frac{1}{4}\, \cF^{\mu\nu}
    \cF_{\mu\nu} + \frac{1}{2}\, \cG_{AB}\, \hat{D}^\mu \phi^A\,
    \hat{D}_\mu \phi^B + \frac{1}{2}\, M^{IJ} \cH^\mu_I\,
    \cH_{\mu J} \notag \\[2mm]
  &\, - \I A_A^I H^\mu_I\, \partial_\mu \phi^A \notag 
     + \I \gve^{\mu\nu\rho\gs} (\cD_\mu \psi_\nu^i \gs_\rho
    \bgps_{\gs i} + \psi_\gs^i \gs_\rho \cD_\mu \bgps_{\nu i}) \notag\\[1.6mm]
  &\, +
    \frac{\I}{2}\, h_{a\ba}\, (\gl^a \gs^\mu \cD_\mu \bgl^\ba -
    \cD_\mu \gl^a \gs^\mu \bgl^\ba) +\I \kappa M^{IJ} \cH^\mu_I\, (g_{Jia}\, \psi_\mu^i \gl^a +
    \text{c.c.})  
    \notag \\[1.8mm]
  &\, - \kappa\, \cG_{AB} (\hat{D}_\mu \phi^A + \partial_\mu \phi^A)\,
    (\gg^B_{ia}\, \gl^a \gs^{\mu\nu} \psi_\nu^i + \text{c.c.}) +
    \I \kappa M^{IJ} \cH^\mu_I\, (g_{Jia}\, \psi_\mu^i \gl^a +
    \text{c.c.}) \notag \\[2mm]
  &\, - \I \frac{\kappa}{2\sqrt{2}}\, (\tilde{\cF}^{\mu\nu} + \tilde{F}^{
    \mu\nu})\, (\psi_\mu^i \psi^{}_{\nu i} + \bgps_\mu^i \bgps^{}_{
    \nu i}) + \frac{\I\kappa}{2\sqrt{2}}\, \cF_{\mu\nu}\, (\cE_{ab}\,\gl^a
    \gs^{\mu\nu} \gl^b - \text{c.c.}) \notag \\[2mm]
  &\, + \I M^{IJ} k_{Ja\ba}\, \gl^a \gs^\mu \bgl^\ba \big[ \cH_{\mu
    I} + \I \kappa\, (g_{Iib}\, \psi_\mu^i \gl^b + \text{c.c.})
    \big] \notag \\[2.5mm]
  &\, + \kappa^2 M^{IJ} (g_{Iia}\, \psi_\mu^i \gl^a + \text{c.c.})\,
    (g_{Jjb}\, \gl^b \gs^{\mu\nu} \psi_\nu^j + \text{c.c.})
    \notag \\[1.2mm]
  &\, + \frac{\kappa^2}{8}\, (\cE_{ac} \cE_{bd}\, \gl^a \gl^b\ \gl^c
    \gl^d + \text{c.c.}) - \frac{1}{4}\, V_{ab\,\ba\bb}\, \gl^a
    \gl^b\, \bgl^\ba \bgl^\bb  \pt \label{eq:sa1}
\end{align} 
This action is based on  $n_T$ tensors $B_{\gm\gn I} \,,\, I=1, \ldots, n_T$ and $4n-n_T$ scalars $\gf^A$, $A=1, \ldots, 4n-n_T$, together with $2n$ $2$-component spinors $\gl^a$, $a=1,\ldots, 2n$, called the hyperinos. Furthermore there are the fields from the supergravity multiplet: the vielbeins $e^{\,\,m}_\gm$, the graviphoton $A_\gm$ and the gravitinos $\gps^i_\gm\,,\,i=1,2$.
We have performed the Wick rotation as discussed in appendix \ref{appwick}.\\
Our conventions are such that all the field dependent quantities, such as $M^{IJ}$, $\cG_{AB}$ and so on, are the same as in the Lorentzian case. Sign changes or different factors of $\I$ are never absorbed, but always written explicitly.\\
The condition that the \susy algebra closes and that the action is invariant, imposes constraints on (and relations between) the various quantities appearing in the action. They are  the same as in the Lorentzian case. We list a number of them in appendix \ref{appsusy1}, where we will  also specify the various covariant derivatives.  For more details see \cite{Theis:2003jj}. 

In this chapter we only need transformation rules and \susy relations for the instanton background. We will work up to linear order in the fermions. 
The supersymmetry transformations of the scalars are given by
\equ{
\gd_\ge \gf^A=\gg_{ia}^A \ge^i \gl^a + \bgg^{i\,\,\,A}_{\,\ba}\bge_i \bgl^\ba \label{eq:sctr1}  \pt
}
The transformation rules for the fermions (up to linear order in the fermions) are
\bea
\gd_\ge\gl^a&=&\left(\I\p{\gm}\gf^A W_A^{ai} + H_{\gm I}f^{Iai}\right)\gs^\gm \bge_i +\ldots \nn \\[1.5mm]
\gd_\ge\bgl^\ba&=&\left(\I\p{\gm}\gf^A \bW_{A i}^{\ba} + H_{\gm I}\bff^{Ia}_{\,\,\,\,i}\right)\bgs^\gm \ge^i +\ldots \label{eq:str2} \km
\eea
with scalar dependent functions $\gg$, $W$ and $f$. We have left out higher order terms in the fermions, denoted by the ellipses.\\
The transformation of the tensors is given by
\begin{equation} \label{eq:de_B_loc}
 \gd_\ge B_{\mu\nu I} = 2\I\, g_{Iia}\, \ge^i \gs_{\mu\nu} \gl^a
 - 4 \kappa^{-1} {\Omega_I}^i{}_j\, \ge^j \gs_{[\mu} \bgps_{\nu]i}
 + \text{c.c.} \pt
\end{equation}
The transformations of the supergravity multiplet are given by
 \begin{align}
&  \gd_\ge e_\mu{}^m  = \I \kappa\, (\ge^i \gs^m \bgps^{}_{\mu i} -
    \psi_\mu^i \gs^m \bge_i) \notag \\[2.5mm]
&  \gd_\ge A_\mu = \I\sqrt{2\,}\, (\ge_i \psi_\mu^i + \bge^i \bgps_{\mu
    i}) \notag \\[2mm]
&  \gd_\ge \psi_\mu^i  = \kappa^{-1}\, \cD_\mu \ge^i + \frac{1}{\sqrt{2\,}}\,
    \gve^{ij} F^+_{\mu\nu}\, \gs^\nu \bge_j - \I \kappa^{-1}
    H_{\mu I} \Gamma^{Ii}{}_j\, \ge^j + \dots \notag \\[0.5mm]
& \gd_\ge \bgps_{\mu i}  = \kappa^{-1}\, \cD_\mu \bge_i + \frac{1}
    {\sqrt{2\,}}\, \gve_{ij} F^-_{\mu\nu}\, \bgs^\nu \ge^j + \I \kappa^{-1}
    H_{\mu I} \Gamma^{Ij}{}_i\, \bge_j + \dots \km \nn
 \end{align}
where, in the last two lines, we have denoted the (anti-) selfdual
graviphoton field strengths by $F_{\mu\nu}^{\pm}=\half(F_{\mu\nu}\pm
\tilde{F}_{\mu\nu})$ and we have dropped fermion bilinears.

The quantities $g_{Iia}$, $\W{A}{ai}$, etc.\ appearing in the
above equations are functions of the scalar fields and satisfy the relations
\R{rel1}--\R{EWidentity}. Moreover, we have the relation
 \begin{equation} \label{eq:Gamma-Omega}
  \Gamma^{Ii}{}_j = M^{IJ} {\Omega_J}^i{}_j \km
 \end{equation}
between the coefficients which appear in the supersymmetry
transformations of the gra\-vi\-tinos and tensors, respectively.

Note that the fermions (and gravitini) are no longer related to each other by complex conjugation. If we write `c.c.' in the action or anywhere else, we mean the analytic continuation of the complex conjugated expressions in Lorentzian signature. We work in the so-called $1.5$ order formalism. This means that  the spin connection is a function of  other fields which is determined by its own (algebraic) field equation.

\section{The supersymmetric DTM}
\label{dtm_susy}
We will discuss the supersymmetry transformation rules of the DTM at linear order in the fermions. 
Action \R{sa1} is a general action for scalars and tensors coupled to $N=2$ supergravity. The action of the DTM is a specific example of  action \R{sa1}. We can obtain it by taking $n=1$ and $n_T=2$ so there are $2$ scalars and $2$ tensors, as discussed in section \ref{instsol}. Furthermore there are $2$ spinors $(\gl^1\,,\,\gl^2)$, $2$ gravitini $(\gps_\gm^1\,,\,\gps_\gm^2)$ and their conjugate  counterparts. Specifying, as in section \ref{instsol}, to
\equ{
M^{IJ}=e^\gf \left(%
\begin{array}{cc}
  1 & -\gc \\
  -\gc & e^\gf + \gc^2 \\
\end{array}%
\right)
\qquad 
\cG_{AB}= \left(%
\begin{array}{cc}
  1 & 0 \\
  0 & e^{-\gf} \\
\end{array}%
\right)
\qquad 
A^I_A=0
\label{eq:DTM_metrics}
}
we obtain the bosonic part of the DTM. From now on we will work in units of $\gk^2=\2$ and rescale the \susy parameters $\ge^i$ by a factor of $\sqrt{2}$ for convenience.\\
Note that we have set\footnote{$A^I_A\equiv M^{IJ}G_{JA}$ and arises in the dualization process from the UHM to the DTM, as  discussed in section \ref{secDTM1}.} $A^I_A$ equal to zero. Strictly speaking we could have allowed for a nonvanishing connection $A^I_A$ with trivial field strength  $F_{AB}^{\,\,\,\,I}=2\p{[A}A^I_{B]}=0$. Such connections are pure gauge and lead to total derivatives in the action, which could be dropped in perturbation theory. Nonperturbatively they can be nonvanishing and lead to imaginary theta-angle-like terms. We have discussed and included such terms separately in \R{boun1} and \R{boun2}, so it suffices to set $A^I_A=0$. In appendix \ref{appdtm1} we list the functions $g_{Iia}$, $\W{A}{ai}$, etc.\ for the DTM.\\
The linearized Euclidean \susy transformations of the fermions can be written as
\begin{alignat}{2}
  \gd_\ge \gl^a & = \I\sqrt{2\,}\, E_\mu^{ai}\, \gs^\mu \bge_i\ , &\qquad
    \gd_\ge \bgps_{\mu i} & = 2 \bar{D}_{\mu\,i}{}^j\, \bge_j +
    \gve_{ij} F^-_{\mu\nu}\, \bgs^\nu \ge^j \notag \\[2pt]
  \gd_\ge \bgl^\ba & = \I\sqrt{2\,}\, \bar{E}_{\mu\,i}^\ba\, \bgs^\mu \ge^i\
    , &\qquad \gd_\ge \psi_\mu^i & = 2 {D_\mu}^i{}_j\, \ge^j +
    \gve^{ij} F^+_{\mu\nu}\, \gs^\nu \bge_j\km \label{susyMN}
 \end{alignat}
where we have introduced
 \begin{alignat}{2} \label{defMN}
  & E_\mu^{ai}  = \partial_\mu \phi^A W_{\!A}^{ai} - \I H_{\mu I} f^{Iai}\ ,
    &\qquad \bar{D}_{\mu\,i}{}^j & = \gd^j_i \nabla_{\!\mu} - \partial_\mu
    \phi^A \Gamma_{\!A}{}^j{}_i + \I H_{\mu I} \Gamma^{Ij}{}_i
    \notag \\[1mm]
  & \bar{E}_{\mu\,i}^\ba  = \partial_\mu \phi^A \bar{W}_{\!Ai}^\ba - \I
    H_{\mu I} \bar{f}^{I\ba}{}_i\ , &\qquad {D_\mu}^i{}_j & =
    \gd^i_j \nabla_{\!\mu} + \partial_\mu \phi^A \Gamma_{\!A}{}^i{}_j -
    \I H_{\mu I} \Gamma^{Ii}{}_j\km
 \end{alignat}
with $\nabla_{\!\mu}$ the Lorentz-covariant derivative. The observation that $\bar{E}_\mu$ and $D_\mu$ are related to
their counterparts $E_\mu$ and $\bar{D}_\mu$ according
to\footnote{Note that in the second identity the covariant derivatives
in $D_\mu$ and $\bar{D}_\mu$ are in the same representation of Spin(4),
whereas in \eqref{susyMN} they are not.}
 \begin{equation} \label{eq:MNrel}
  \bar{E}_{\mu\,j}^\ba = - h^{\ba a} \cE_{ab}\, E_\mu^{bl}\, \gve_{lj}\
  ,\qquad {D_\mu}^i{}_j = \gve^{ik} \bar{D}_{\mu\,k}{}^l\, \gve_{lj} \km
 \end{equation}
 will prove  useful. \\
The first identity is due to the relation \R{EWidentity}, while
the second is a consequence of $SU(2)$-covariant constancy of
$\gve_{ij}$.

More explicitly, we have, at the linearized level,
 \begin{align}
  \gd_\ge \lp \gl^1 \\[2mm] \gl^2 \rp & = \I \lp e^{-\phi/2} \partial_\mu
    \chi - e^{\phi/2} \hat{H}_{\mu1} & \partial_\nu \phi + e^\phi
    H_{\nu2} \\[2mm] - \partial_\mu \phi + e^\phi H_{\mu2} & e^{-\phi/2}
    \partial_\nu \chi + e^{\phi/2} \hat{H}_{\nu1} \rp \lp \gs^\mu
    \bge_1 \\[2mm] \gs^\nu \bge_2 \rp \notag \\[3mm]
  \gd_\ge \lp \bgl^1 \\[2mm] \bgl^2 \rp & = \I \lp e^{-\phi/2} \partial_\mu
    \chi + e^{\phi/2} \hat{H}_{\mu1} & \partial_\nu \phi - e^\phi
    H_{\nu2} \\[2mm] - \partial_\mu \phi - e^\phi H_{\mu2} & e^{-\phi/2}
    \partial_\nu \chi - e^{\phi/2} \hat{H}_{\nu1} \rp \lp \bgs^\mu \ge^1
    \\[2mm] \bgs^\nu \ge^2 \rp \label{eq:ferm-transf}
 \end{align}
for the matter fermions, and
 \begin{align}
 & \gd_\ge \lp \psi_\mu^1 \\[2mm] \psi_\mu^2 \rp  = \lp 2 \nabla_{\!
    \mu} + \half e^\phi H_{\mu2} & - e^{-\phi/2} \partial_\mu \chi +
    e^{\phi/2} \hat{H}_{\mu1} \\[2mm] e^{-\phi/2} \partial_\mu \chi +
    e^{\phi/2} \hat{H}_{\mu1} & 2 \nabla_{\!\mu} - \half e^\phi
    H_{\mu2} \rp \lp \ge^1 \\[2mm] \ge^2 \rp + \dots \notag
    \\[3mm]
&  \gd_\ge \lp \bgps_{\mu1} \\[2mm] \bgps_{\mu2} \rp  = \lp 2
    \nabla_{\!\mu} - \half e^\phi H_{\mu2} & - e^{-\phi/2} \partial_\mu
    \chi - e^{\phi/2} \hat{H}_{\mu1} \\[2mm] e^{-\phi/2} \partial_\mu
    \chi - e^{\phi/2} \hat{H}_{\mu1} & 2 \nabla_{\!\mu} + \half
    e^\phi H_{\mu2} \rp \lp \bge_1 \\[2mm] \bge_2 \rp + \dots
    \label{eq:gravitino-trans}
 \end{align}
for the gravitinos, where we have omitted the graviphoton terms.\\
We end this section by giving the fermionic equations of motion, at the
linearized level. For the hyperinos we find
 \begin{align}
  & \I \gs^\mu \cD_\mu \bgl^\ba + H_{\mu I}\, \bar{\Gamma}^{I\ba}
    {}_{\bb}\, \gs^\mu \bgl^\bb + \frac{\I}{2}\, h^{\ba a}
    \cE_{ab}\, F^+_{\mu\nu}\, \gs^{\mu\nu} \lambda^b = - \frac{1}
    {\sqrt{2\,}}\, \gs^\nu \bar{E}^{\ba}_{\mu\,i} \bgs^\mu \psi_\nu^i
    \label{eq:hyperino-EOM} \\[2pt]
  & \I \bgs^\mu \cD_\mu \gl^a + H_{\mu I}\, \Gamma^{Ia}{}_{b}\, \bgs^\mu
    \gl^b - \frac{\I}{2}\, h^{\ba a} \bar{\cE}_{\ba\bb}\, F^-_{\mu
    \nu}\, \bgs^{\mu\nu} \bgl^\bb = -\frac{1}{\sqrt{2\,}}\, \bgs^\nu
    {E}^{ai}_{\mu} \gs^\mu \bgps_{\nu\,i}\pt
 \end{align}
For the definition of the covariant derivative $\cD_\gm$ see appendix \ref{appsusy1}.
What makes these different from the usual Dirac-like equation is the
presence of the mixing term with the (anti-) selfdual graviphoton field
strength and the inhomogeneous gravitino term originating from its
coupling to the rigid supersymmetry current of the double-tensor
multiplet. This will become important in the discussion of the fermionic
zero modes.
The gravitino field equations read
 \begin{align}
  & \I \gve^{\mu\nu\rho\gs} \gs_\rho \cD_\gs \bgps_{\nu\,i} - \gve^{\mu
    \nu\rho\gs} H_{\gs I} \Gamma^{Ij}{}_i\, \gs_\rho \bgps_{\nu j}
    - \I F^{\mu\nu-} \psi_{\nu\,i} = \frac{1}{2\sqrt{2\,}}\, h_{a\ba}
    \bar{E}^{\ba}_{\nu\,i}\, \gs^\nu \bgs^\mu \lambda^a
    \label{eq:gravitino-EOM} \\[2pt]
  & \I \gve^{\mu\nu\rho\gs} \bgs_\rho \cD_\gs \psi_\nu^i + \gve^{\mu\nu
    \rho\gs} H_{\gs I} \Gamma^{Ii}{}_j\, \gs_\rho \psi_\nu^j + \I
    F^{\mu\nu+} \bgps_\nu^i = - \frac{1}{2\sqrt{2\,}}\, h_{a\ba} E^{ai}_{
    \nu}\, \bgs^\nu \gs^\mu \bgl^\ba\ . \label{eq:gravitino-EOM2}
 \end{align}
Note that one can combine the first two terms on the left-hand side
into the operator ${D_\mu}^i{}_j$, as defined in \eqref{defMN}.

\section{(Un)broken supersymmetry}
\label{ubs}
Often a specific solution to the equations of motion will break most (or all) of the symmetries of the theory. The symmetries which are not broken by the solution are called \emph{unbroken} of \emph{residual} symmetries and form a symmetry group. The symmetries which are broken by the solution are called the \emph{broken} symmetries and can be used to generate new solutions.\\
For instance, the single-centered\footnote{The multicentered case is similar, but we will mainly use the single-centered one in this chapter.} instanton solution \R{ss1} is not translationally invariant because it is located at a specific point $\{x_0\}$. This is sometimes phrased by saying that the instanton `breaks translational invariance'. Because the underlying theory is translationally invariant, this will manifest itself in the degeneracy of the solutions related to each other by translations. Indeed, the action \R{sc5b1} does not depend on $\{x_0\}$. These solutions can be translated into each other by acting with the broken symmetry. This leads to the notion of \emph{collective coordinates}. These are the coordinates generated by the broken symmetries, in this case simply the position $\{x_0\}$.\\
Similarly, the solutions of a \sugra theory are in general not invariant under the \susy transformations that leave the theory invariant. The solutions that  preserve part of the supersymmetries are often called BPS\footnote{After Bogomol'nyi-Prasad-Sommerfield \cite{Prasad:1975kr, Bogomolny:1975de}. For more information on BPS solutions and supersymmetry see for example \cite{Weinberg:2000cr, Stelle:1998xg} and references therein.} solutions. \\
Schematically (local) supersymmetry transformations take the form
\bea
&& \gd_\ge B \sim \ge F \nn \\[1mm]
&& \gd_\ge F \sim \p{}\ge + B\ge \label{eq:vbs1} \km
\eea
where $B$ stands for bosons and $F$ for fermions. Typically one is interested in purely bosonic configurations. According to the general definition given above, the bosonic solutions will be supersymmetric if there is an $\ge(x)$ such that \R{vbs1} vanishes. The bosonic fields are trivially invariant (because the fermionic fields are zero) and one only needs to examine the equation
\equ{
\gd_\ge F \sim (\p{} + B)\ge=0 \pt \label{eq:ks1}
}
The commutator of $2$ such spinors will give an (infinitesimal) Killing vector which generates an isometry of the bosonic background. Equation \R{ks1} is often called the Killing spinor equation.\\

As we shall see in the next subsection, the instanton configuration preserves only part of the supersymmetries. The supersymmetric vacuum is the trivial one with all fields equal to zero. The broken supersymmetries will  generate new solutions characterized by their fermionic collective coordinates. Thus if we start with a purely bosonic configuration which is a solution to the equations of motion, acting with the broken supersymmetries generates fermionic fields. Naturally, this new configuration is also a solution to the equations of motion because the broken supersymmetries are symmetries of the theory.\\
There is a lot more to say about this subject, and we refer to the literature, e.g. \cite{Ortin:2004ms, Mohaupt:2000mj, Stelle:1998xg} and the references therein. In the following we present a detailed inspection of the (un)broken supersymmetries in the background of our instanton solution. The case of the anti-instanton is similar.

\subsection{Unbroken supersymmetry}
\label{unbroken}
Which supersymmetries leave the instanton solution (\ref{eq:sol1}, \ref{eq:sol2}, \ref{eq:Bbound1}) invariant, or equivalently, which supersymmetries are left unbroken by this solution? 
The background is determined by the instanton solution for the bosonic fields as in (\ref{eq:sol1}, \ref{eq:sol2}, \ref{eq:Bbound1}). The fermionic fields are all equal to zero.
The supersymmetry transformation rules of the bosonic fields contain the fermionic fields, which means that the bosons are always invariant under supersymmetry transformations. To find the unbroken supersymmetries we therefore only have to examine the conditions
\equ{
\gd_\ge \gl^a=\gd_\ge \bgl^\ba=\gd_\ge\gps_\gm^i=\gd_\ge\bgps_{\gm i}=0 \label{eq:cond3a}
}
where $\{a,\ba,i\}=1,2$. This will put certain constraints on the \susy parameters $\bge_i(x)$ and $\ge^i(x)$. We will focus on the $\bge_i(x)$ because the $\ge^i(x)$ can easily be obtained from those as we shall see. \\

It is convenient to consider $\gd_\ge\bgps_{\gm 2}$ first. Using \R{gravitino-trans} and \R{Bbound1} we obtain 
\equ{
0=\2 \gd_\ge\bgps_{\gm 2}=\left(\p{\gm}-\4 \p{\gm}\gf\right)\bge_2 \nn \km
}
defining $\bge_2\equiv e^{1/4\gf}\bget_2$ gives
\equ{
e^{\gf/4}\p{\gm}\bget_2 =0 \nn \km
}
which means that $\bget_2$ is a constant spinor. We can rewrite the $\gl^a$ variations \R{ferm-transf} as
\bea
&&\gd_\ge \gl^1=-2\I \gs^\gm \left(\p{\gm} -\2 \p{\gm}\gf -\4 e^\gf H_{\gm 2}\right)\bge_2 + \I \gs^\gm \gd_\ge \bgps_{\gm 2} \label{eq:ft1} \\
&& \gd_\ge \gl^2=+2\I \gs^\gm \left(\p{\gm} -\2 \p{\gm}\gf +\4 e^\gf H_{\gm 2}\right)\bge_1 - \I \gs^\gm \gd_\ge \bgps_{\gm 1} 
\label{eq:ft2} \pt
\eea
Upon using \R{Bbound1} these equations simplify. Equation \R{ft1} becomes
\equ{
\gd_\ge\gl^1=-2\I \gs^\gm\left(\p{\gm}-\4 \p{\gm}\gf\right)e^{\gf/4}\bget_2 =0 \nn\km
}
where we used that $\bge_2\equiv e^{1/4\gf}\bget_2$  and the fact that $\gd_{\bge_2}\bgps_{\gm 2}=0$. Because this equation is identically zero, it imposes no new constraints. Next on the list is $\gd_\ge \bgps_{\gm 1}$ which is, using \R{Bbound1} and the result for $\bge_2$
\equ{
0=\2\gd_\ge \bgps_{\gm 1}=\left(\p{\gm}+\4 \p{\gm}\gf\right) \bge_1 - e^{-\gf/4}\p{\gm}\gc \,\bget_2 =e^{-\gf/4}\left(\p{\gm}\bget_1 -\p{\gm}\gc\, \bget_2\right) \nn
}
where we have defined $\bge_1\equiv e^{-\gf/4}\bget_1$ in the last step, $\bget_1$ is some spinor. So if $\bget_1$ is related to $\bget_2$ through
\equ{
\p{\gm}\bget_1=\p{\gm}\gc \,\bget_2 \label{eq:srel1} \km
}
we have $\gd_\ge \bgps_{\gm 2}=\gd_\ge \bgps_{\gm 1}=\gd_\ge \gl^1=0$. This leaves only one equation, namely $\gd_\ge \gl^2=0$. Using \R{ft2} , \R{Bbound1} and the above results we find
\equ{
0=\gd_\ge \gl^2=\gs^\gm\left(\p{\gm} -\frac{3}{4} \p{\gm}\gf\right) e^{-\gf/4} \bget_1\km \nn
}
which means that $\p{\gm}\bget_1=\p{\gm}\gf \,\bget_1$ which is solved by $\bget_1=e^\gf \bc_1$ for some constant spinor $\bc_1$, giving $\bge_1=e^{3\gf/4}\bc_1$. Note that a more general solution for $\bget_1$ is
\equ{
\bget_1=e^\gf \bc_1 + e^\gf \p{\gn}h\,\bgs^\gn \gx \nn \km
}
with $h$ a harmonic function and $\gx$ a constant spinor. However we still have to deal with equation \R{srel1} and we did not find any non-trivial solutions for the $\p{\gn}h\bgs^\gn \gx$ part. Therefore we only consider $\bget_1=e^\gf \bc_1$. Relation \R{srel1} then becomes
\equ{
e^\gf\p{\gm}\gf\, \bc_1=\p{\gm}\gc \,\bget_2 \nn \km
}
which is only satisfied for 
\equ{
\gc=ae^\gf + b \label{eq:temp3a} \km
}
with some constants $a,b$ and $\bc_1 =a \bget_2$. 

In the calculations above we have not used single-centeredness of our instanton solutions, so $\gf$ and $\gc$ were still determined by \R{sol1} and \R{sol2}. Now we see that equation \R{temp3a} imposes a new constraint because we can only write $\gc$ in this particular form if, see \R{38b},  all $\gc_i$ are equal. Note that \R{temp3a} follows automatically, for some constants $a$ and $b$ that we will determine in a moment,  if one imposes spherical symmetry as in that case the harmonic function $e^{-\gf}\gc$ depends linearly on the harmonic function $e^{-\gf}$. This means that we are dealing with a single-centered instanton. Equations \R{sol1} and \R{sol2} then simplify:
\bea
&& e^{-\gf}=e^{-\gf_\infty} + \frac{|Q_2|}{4\gp^2 (x-x_0)^2} \label{eq:ss1} \\
&& e^{-\gf}\gc=e^{-\gf_\infty}\gc_\infty +\frac{Q_1}{4\gp^2 (x-x_0)^2}  \label{eq:ss2}
\eea
and the last equation can indeed be rewritten as
\equ{
\gc=e^{-\gf_\infty}\gD \gc e^\gf + \gc_0=g_s^2\gD\gc e^\gf +\gc_0 \nn \km
}
with $\gD\gc\equiv \gc_\infty-\gc_0$ and $\gc_0=\frac{Q_1}{|Q_2|}$, see \R{38b}. We can thus make the identifications $a=g_s^2 \gD \gc$ and $b=\gc_0$. The action for multi-centered instantons \R{instaction2} reduces to single-centered one \R{sc5b1}. \\
In the following we will always consider single-centered instantons, or equivalently, spherically symmetric ones.\\
The same analysis can be performed for anti-instantons, the total result is
\bea
&& \bge_1 =e^{\gf/2}(g_s^2\gD\gc)^{(1\pm 1)} e^{\pm\gf/4} \bget \label{eq:ks11}\\[2mm]
&& \bge_2=\pm e^{\gf/2}(g_s^2\gD\gc)^{(1\mp 1)} e^{\mp \gf/4} \bget \label{eq:ks2} \km
\eea
where the (lower) upper sign corresponds to (anti-)instantons. We have given $\bget_2$ the new name $\bget$. We can write this result concisely as $\bge_i(x)=u_i(x) \bget$, where the $u_i(x)$ can be read off from \R{ks11} and \R{ks2}.\\
The Killing spinors of opposite chirality are then given by $\ge^i(x)=\gve^{ij} u_j(x) \get$ with $\get$ another (unrelated) constant spinor. This immediately follows from \R{MNrel}: if $u_i$ are (spinless) zero modes of $E_\gm$ and $\bD_\gm$, then $\gve^{ij}u_j$ are zero modes of $\bE_\gm$ and $D_\gm$.
The conclusion of this analysis is that the NS $5$-brane instanton in flat space leaves one half of the supersymmetries unbroken. In other words, the instanton solution is a BPS configuration. \\
Although saying that the instanton preserves half of the supersymmetries is the standard way in which to phrase this result, it is slightly misleading. Namely, we go from $4$ arbitrary spinors $(\bge_i, \ge^i)$ to $2$ constant spinors $(\bget, \get)$ together with a very specific space dependence via  $e^\gf$. So we really go from $4$ times `infinitely many' supersymmetries, to $2$ very specific ones.\\
Note that the trivial solution with both $e^{-\gf}$ and $\gc$ constant, preserves all the supersymmetries. Thus the instanton configuration tends asymptotically to the maximally supersymmetric vacuum.\\

\subsubsection{The membrane}
\label{mem4a}
For the sake of completeness we briefly discuss the unbroken supersymmetries in the case of the membrane instanton configuration in a similar fashion, although we will not need them explicitly. As explained in section \ref{mem1}, the simplest membrane instanton solution is characterized by taking $Q_2=\gD \gc=0$ which means that $H_{\gm 2}=0$. The solution is then specified by
\equ{
H_{\gm 1}=\pm \p{\gm} h\nn\km
}
with $h$ given by \R{h1}.   To find the unbroken supersymmetries we again have to examine the conditions \R{cond3a}, but now in the membrane background. For convenience we define
\equ{
\gl^\pm\equiv\2 (\gl^1 \pm \gl^2) \qquad \gps_\gm^\pm\equiv\2 (\gps^1_\gm \pm \gps^2_\gm) \qquad \ge^\pm\equiv\2(\ge^1 \pm\ge^2) \nn
}
and similarly for barred quantities. The supersymmetry variations \R{ferm-transf} and \R{gravitino-trans} become
\bea
&& \gd_\ge \bgl^+=-2\I \p{\gm}\gf \,\bgs^\gm \ge^- \,\quad\qquad \gd_\ge \bgl^-=0 \nn \\
&& \gd \gps_\gm^+=2(\p{\gm} -\2 \p{\gm}\gf)\ge^+ \qquad \gd_\ge \gps_\gm^-=2(\p{\gm} +\2 \p{\gm}\gf)\ge^- \nn \pt
\eea
The unbroken supersymmetries are then given by
\equ{
\ge^+=e^{\gf/2}\get^+ \qquad \ge^-=0 \nn \km
}
with $\get^+$ a constant spinor. This means that the membrane instanton preserves half of the supersymmetries as well, it is also a BPS solution. The broken supersymmetries can be used to generate solutions for the fermions as will be shown in the case of the $5$-brane instanton in the next section. We will not demonstrate this explicitly for the membrane.

\subsection{Broken supersymmetries}
\label{brokensusy}
Now we will turn our attention to the broken symmetries. As we have explained, these will generate new solutions. Specifically, starting from a purely bosonic background (as determined by the instanton solution (\ref{eq:sol1}, \ref{eq:sol2}, \ref{eq:Bbound1})) the broken supersymmetries will generate solutions for the fermions. Alternatively, if we are going to generate new (classical) solutions for the fermions, we can also try to construct them by directly solving the equations of motion (linear in the fermions).\\
This will be the plan for this section: we will solve the equations of motion (up to linear order in the fermions) for the fermions  and then show  how to obtain these solutions by means of the broken supersymmetries. The  solutions therefore will depend on fermionic \emph{collective} coordinates. These will appear as integration constants if one uses the equations of motion and as supersymmetry parameters in the other approach. The importance of these solutions will become clear in the next section, where we will see that a more apt name for them is \emph{zero modes}.

\subsubsection{The hyperinos}
First we shall solve the equations of motion for the hyperinos, which are coupled to the gravitinos, (\ref{eq:hyperino-EOM}, \ref{eq:gravitino-EOM}). For the DTM we have (see appendix \ref{appdtm1}) $\G{A}{a}{b}=\gG^{1a}_{\,\,\,\,\,b}=0$ and  $\gG^{2a}_{\,\,\,\,\,b}$ is given by \R{Gamma2ab}. Using \R{Bbound1} we obtain\footnote{Unless stated otherwise we choose the instanton background, i.e., the plus sign. The calculation with the minus sign is of course similar.}
\begin{equation} \label{eq:5br-zeromode-eqn1}
  \I \gs^\mu \lp \partial_\mu \bgl^1 - \tfrac{3}{4} \partial_\mu \phi\, \bgl^1
  \\[2mm] \partial_\mu \bgl^2 + \tfrac{3}{4} \partial_\mu \phi\, \bgl^2 \rp = -
  \gs^\nu \bgs^\mu \lp e^{-\phi/2} \partial_\mu \chi\, \psi_\nu^1 + \partial_\mu
  \phi\, \psi_\nu^2 \\[2mm] 0 \rp 
 \end{equation}
and similarly for the $\gl^a$:
\begin{equation} \label{eq:5br-zeromode-eqn2}
  \I \bgs^\mu \lp \partial_\mu \gl^1 + \tfrac{3}{4} \partial_\mu \phi\, \gl^1
  \\[2mm] \partial_\mu \gl^2 - \tfrac{3}{4} \partial_\mu \phi\, \gl^2 \rp = \bgs^\nu
  \gs^\mu \lp 0 \\[2mm] \partial_\mu \phi\, \bgps_{\nu1} -e^{-\phi/2} \partial_\mu
  \chi\, \bgps_{\nu2} \rp \pt
 \end{equation}
It is convenient to start with $\bgl^2$ and $\gl^1$ because they decouple from the gravitinos. To this end, consider the more general operator
\equ{
\label{eq:Dk}
  \slsh{D}_k \equiv \sigma^\mu (\partial_\mu - k \partial_\mu \phi) = e^{k\phi}
  \p\!\!\!/\, e^{-k\phi}\ ,\qquad k \in \fieldR\pt
  }
The zero modes of $\slsh{D}_k$ are in one to one correspondence to those of $\sder$. But whereas the zero modes $\bgz$ of $\sder$ ($\sder \bgz=0$) are not normalizable\footnote{Normalizable zero modes $\cZ$ must satisfy $\int_0^\infty dr r^3 |\cZ|^2 < \infty$. This implies that $\cZ$ must go to zero as $r^{-5/2}$ or faster at infinity and it may not diverge faster that $r^{-3/2}$ at the origin.} the corresponding  modes of $\slsh{D}_k$ ($\gl=e^{k\gf}\bgz$) are only normalizable for appropriate values of $k$.\\
In flat Euclidean space the only solution for $\bgz$ is a constant spinor. However, when the origin is cut out, as it is in our case, there is a nontrivial solution: 
\equ{
\bgz(x)=2\I \p{\gm}h(x) \bgs^\gm \gx \nn \km
}
where $\gx$ is a constant spinor, $h$ is a harmonic function and we included the factor of $2\I$ for later convenience. This is the only solution as one can show by writing $\bgz$ as
\equ{
\bgz(x)=2\I f_\gm(x) \bgs^\gm \gx \nn
}
for an arbitrary real function $f_\gm$. The equation $\sder\bgz=0$ then imposes
\equ{
\p{[\gm}f_{\gn]}=\p{\gm}f^\gm=0 \pt\nn
}
The constant solution cannot lead to a normalizable solution and must therefore be discarded. The possible normalizable solutions to $\slsh{D}_k\bgl=0$ are given by
\equ{
\bgl=2\I e^{k\gf}\p{\gm}h \bgs^\gm \gx \nn \pt
}
As we are dealing with spherical solutions (\ref{eq:ss1}, \ref{eq:ss2}), the only harmonic function available is given by $e^{-\gf}$. Thus the normalizable zero modes to $\slsh{D}_k$ are given by
\equ{
\bgl=2\I e^{k\gf} \p{\gm}e^{-\gf}\bgs^\gm \gx \qquad k\geq \frac{3}{4} \nn \km
}
where $k\geq 3/4$ can be found by inspecting the asymptotic behaviour. 

Using these results we can now easily write down the solutions for the hyperinos:
\equ{
\bgl^2=0 \qquad \gl^1=0 \label{eq:hzm1} \pt
}
Because their equations of motion correspond to $k=-3/4$, which would lead to non-normalizable zero modes, they must be set to zero.\\
If there would be no gravitinos present in (\ref{eq:5br-zeromode-eqn1}, \ref{eq:5br-zeromode-eqn2}) the solutions for $\bgl^1$ and $\gl^2$ would be given by 
\equ{
\bgl^1=2\I e^{3\gf/4}\p{\gm} e^{-\gf} \bgs^\gm \gx^1 \qquad
\gl^2=2\I e^{3\gf/4}\p{\gm} e^{-\gf} \gs^\gm \bgx^2 \label{eq:hzm2}\km
}
with $\gx^1$ and $\bgx^2$ constant spinors.\\
As we will demonstrate now, there are in fact no normalizable solutions for the gravitinos.

\subsubsection{The gravitinos}
In the instanton background the equations of motion for the gravitinos (\ref{eq:gravitino-EOM}, \ref{eq:gravitino-EOM2}) become, using appendix \ref{appdtm1}, 
\equ{
\I \gve^{\mu\nu\rho\gs} \gs_\rho \lp \partial_\gs \bgps_{\nu1} + \tfrac{1}{4}
  \partial_\gs \phi\, \bgps_{\nu1} - e^{-\phi/2} \partial_\gs \chi\, \bgps_{\nu2}
  \\[2mm] \partial_\gs \bgps_{\nu2} - \tfrac{1}{4} \partial_\gs \phi\, \bgps_{\nu2}
  \rp = \frac{1}{2} \lp e^{-\phi/2} \partial_\nu \chi \\[2mm] \partial_\nu \phi \rp
  \gs^\nu \bgs^\mu \gl^1 \pt \label{eq:G1}
  }
We notice that only $\gl^1$ couples to the gravitino, but we just \R{hzm1} concluded that $\gl^1=0$. This means that the gravitinos decouple. If we define 
\equ{
\bgps_{\gm 2}\equiv e^{\gf/4} \bgz_{\gm 2} \nn \km
}
we can, using \R{B4}, rewrite the second equation in \R{G1} as
\equ{
\gs^\gm\left(\p{\gm}\bgz_{\gn 2} -\p{\gn}\bgz_{\gm 2}\right)=0 \label{eq:H1} \pt
}
There are many solutions to this equation, e.g. $\bgz_{\gm 2}=\p{\gm}\bgz_{2}$, but we still have to impose a (supersymmetry) gauge. We choose to work with the standard gauge
\equ{
\bgs^\gm\gps^i_\gm=\gs^\gm \bgps_{\gm i}=0 \km \label{eq:susygauge1}
}
in particular $\gs^\gm\bgz_{\gm 2}=0$. Equation \R{H1} simplifies to $\sder \bgz_{\gm 2}=0$ which has the solution
\equ{
\bgz_{\gm 2}=\p{\gm}\p{\gn}h \,\bgs^\gn \gx \km \label{eq:susyclass1}
}
with $\gx$ a constant spinor and $h$ a harmonic function. For a given solution one can still act with a supersymmetry transformation. These supersymmetry transformations can either transform the solution for the gravitino in such a way that it still respects the gauge \R{susygauge1}, or not. We must therefore check  if there are residual supersymmetry transformations that produce  new solutions for the gravitino that still obey the gauge condition \R{susygauge1}.\\
The supersymmetry transformations for the gravitinos \R{gravitino-trans} simplify to
\bea
&&\gd_\ge\bgps_{\gm 1}=2\left(\p{\gm} + \4 \p{\gm}\gf\right)\bge_1-2e^{-\gf/2}\p{\gm}\gc \,\bge_2 \nn \\[1mm]
&& \gd_\ge\bgps_{\gm 2}=2\left(\p{\gm}-\4 \p{\gm}\gf\right)\bge_2 \km \label{eq:grtr2}
\eea
where we again used the values for the instanton background and \mbox{appendix \ref{appdtm1}}. We can write $\gd_\ge \bgps_{\gm 2}$ as \equ{
\gd_\ge \bgps_{\gm 2}=2 \p{\gm}\left(e^{-\gf/4}\bge_2\right) \km \nn
}
a total derivative, which vanishes in \R{H1}. We should also impose the gauge condition, i.e.
\equ{
0=\gs^\gm\left(\gd_\ge\bgps_{\gm 2}\right) =2\sder \left(e^{-\gf/4}\bge_2\right) \label{eq:gauge1} \pt
}
The solutions are given by a constant or by the derivative of a harmonic function. Acting with such a residual supersymmetry transformation gives a solution which is contained within the class of solutions \R{susyclass1}. Therefore the only solution for $\bgps_{\gm 2}$ is given by
\equ{
\bgps_{\gm 2}=e^{\gf/4}\p{\gm}\p{\gn}h \,\bgs^\gn \gx \nn \km
}
where we can again choose $h=e^{-\gf}$ for spherically symmetric harmonic functions.\\
However, this solution is not normalizable because it diverges too fast at the origin. Consequently there is no normalizable solution for $\bgps_{\gm 2}$ and we must set it to zero in the hyperino equations of motion. \\
A similar analysis shows that there are no normalizable solutions for $\bgps_{\gm 1}$ and the unbarred gravitinos either, so we set them to zero in \R{5br-zeromode-eqn1} and \R{5br-zeromode-eqn2}. 
Summarizing, we have the following results
\bea
&&\bgps_{\gm i}=\gps^i_\gm =\gl^1=\bgl^2=0 \nn \\[2mm]
&&\bgl^1=2\I e^{3\gf/4}\p{\gm} e^{-\gf} \bgs^\gm \gx \quad \qquad
\gl^2=2\I e^{3\gf/4}\p{\gm} e^{-\gf} \gs^\gm \bgx \nn   \km
\eea
with two constant\footnote{Note that we use a slightly different notation now, compare with \R{hzm2}. The spinors $\gx$ and $\bgx$ are unrelated.} spinors $\gx$ and $\bgx$.

\subsubsection{Using the broken supersymmetries}
We have constructed all the solutions to the  equations of motion (linear in the fermions) with certain boundary conditions\footnote{Namely that they are normalizable, which imposes certain conditions on their asymptotic behaviour.}. We shall now show that we can generate these solutions with the broken supersymmetries.\\
As it turns out, the broken supersymmetries are contained in the residual supersymmetry transformations left after gauge fixing the gravitinos, i.e. those $\bge_1$ and $\bge_2$ satisfying $\gs^\gm \gd_{\ge}\bgps_{\gm 1}=\gs^\gm \gd_{\ge}\bgps_{\gm 2}=0$. The nontrivial solutions to this equation are given by 
\equ{
\bge_2=e^{\gf/4}\bget \label{eq:ngbr1}\km
}
with $\p{\gm}\bget=0$. This is precisely the Killing spinor from \R{ks2}, so it leaves the hyperinos invariant.  The broken supersymmetries for $\bge_1$ are determined by those transformations that leave $\bgps_{\gm 1}$ invariant and preserve the gauge condition \R{gauge1} but act nontrivially on the hyperinos. Using \R{grtr2} and \R{ngbr1} we find
\equ{
\bge_1=e^{-\gf/4}\left(\gc \bget + \bgx'\right) \nn\km
}
where $\bgx'$ is another constant spinor. If we insert $\bge_1$ and $\bge_2$ into the transformations of the $\gl^a$ we find that (after a redefinition $\bgx'=\bgx-\gc_0 \bget$) the spinor proportional to $\bget$ is precisely the Killing spinor \R{ks11}. The spinor\footnote{The supersymmetry parameters which generate new solutions for the hyperinos are summarized in \R{brsp1}.} proportional to $\bgx$ does generate a new solution for $\gl^2$:
\equ{
^{(1)}\gl^1=0 \quad \qquad ^{(1)}\gl^2=-2\I e^{-\gf/4}\p{\gm}\gf\, \gs^\gm \bgx \label{eq:kst1} \km
}
where we included a factor of $2\I$ for convenience. These are precisely the solutions obtained earlier, see (\ref{eq:hzm1}, \ref{eq:hzm2}). Equation \R{MNrel} relates the Killing spinors of opposite chirality\footnote{Remember that is really just a very concise way of doing the same calculation for the barred sector.}, which gives
\equ{
\ge^1=0 \qquad \quad \ge^2=-e^{-\gf/4}\gx \nn \pt
}
These generate
\equ{
^{(1)}\bgl^1=-2\I e^{-\gf/4}\p{\gm}\gf \,\bgs^\gm \gx \qquad \quad ^{(1)}\bgl^2=0 \label{eq:kst2}\km
}
also a perfect match with \R{hzm2}.\\
Note that these solutions depend on $x$ and the collective coordinate $\{x_0\}$. We have introduced some new notation: the superscript $^{(1)}$ left of the $\gl$'s indicates that this solution is obtained by applying a supersymmetry transformation once. The resulting function therefore contains one Grassmann collective coordinate (GCC). We omit the superscript $^{(0)}$.\\
Later on we will construct the scalar superpartner of the solutions for the hyperinos found above. The solution for the superpartner can be found by using \R{sctr1} and the solutions for the hyperinos and the supersymmetry parameter which generated them. These solutions for scalars will carry the label $^{(2)}$ because they contain two fermionic collective coordinates (one from the supersymmetry parameters and one from the hyperinos).\\
We can perform a  similar analysis for the anti-instanton in which case we find
\bea
&& ^{(1)}\gl^2=^{(1)}\bgl^1=0 \nn \\[1mm]
&& ^{(1)}\gl^1=-2\I e^{-\gf/4}\p{\gm}\gf\,\gs^\gm\tilde{\bgx} \qquad  ^{(1)}\bgl^2=+2\I e^{-\gf/4}\p{\gm}\gf\,\bgs^\gm\tilde{\gx} \nn \pt
\eea
For future convenience, we will introduce a very compact notation that links the hyperino labels $`1'$ and $`2'$ to the (anti-)instanton labels $(-) +$. \\
This is done in such a way that the hyperino labels $1$ and $2$ are denoted by upper and lower indices respectively. These indices are then further specified by indicating the background, i.e.  instanton or anti-instanton. In this notation the absence of fermionic zero modes is expressed by the equations $^{(1)}\gl^{\pm}=0$ where the upper index is associated to the first hyperino in the instanton $(+)$ background and the lower label to the second hyperino in the anti-instanton $(-)$ background. Similarly we have $^{(1)}\bgl^{\mp}=0$. For the broken supersymmetry parameters we have $\ge^\pm=\bge_\mp=0$ and for the $\gl$'s we can write
\equ{
^{(1)}\gl^{\mp}=-2\I e^{-\gf/4}\p{\gm}\gf \,\gs^\gm \bgx^\mp \qquad \quad
^{(1)}\bgl^{\pm}=\mp2\I e^{-\gf/4}\p{\gm}\gf \,\bgs^\gm \gx^\mp \label{eq:brfermions1}\pt
}
These are generated by 
\equ{
\ge^\mp =-e^{-\gf/4}\gx^\mp \qquad \quad \bge_\pm=\pm e^{-\gf/4}\bgx^\pm \label{eq:brsp1} \km
}
where the fermionic collective coordinates $\bgx^\mp$ are two independent constant spi\-nors which distinguish between instantons and anti-instantons, similarly for $\gx^\pm$.\\
Where it is clear from the context what is meant, we shall drop the $\pm$ indices on $\gx$ and $\bgx$ for clarity.\\
As a last remark in this section, note the difference with Yang-Mills theories where fermionic zero modes only appear in one chiral sector. In the  situation described above they are evenly distributed over the two (barred and unbarred) sectors.

\section{The instanton measure}
\label{instmeasure}
In the previous section we have seen how broken symmetries and collective coordinates are related. In this section we shall examine the consequences collective coordinates have for the path integral. \\
Consider a generic bosonic system without gauge invariance, with fields $\gF^M$ and Euclidean action $S[\gF]$. We can expand the fields around the instanton solution (which solves the equation of motion) as follows:
\equ{
\gF^M(x)=\gF^M_{cl}(x,C) + \gF_{qu}^M(x,C) \label{eq:field1} \pt
}
The collective coordinates are collectively denoted by $C^{i_0}$ and the fluctuations $\gF_{qu}^M$ depend on them as well. Note however that the field as a whole, i.e. $\gF^M(x)$ does not. 
For a general field $\gF^M(x)$, collective coordinates do not exist as such. These coordinates only become meaningful when the fields are expanded around a configuration which minimizes the action. Only then can some symmetries be broken, giving rise to collective coordinates. The collective coordinates in turn parametrize (by construction) only the classical solution and the fluctuations.
Expanding the action around the instanton background gives
\equ{
S=S_{cl} + \2\int d^4x\,\gF^M_{qu} M_{MN} \gF^N_{qu} + \cO(\gF^3) \label{eq:exp1} \km
}
where $M_{MN}$ results from taking the second variation of the action with respect to the  fields and is given by  $M_{MN}=\cG_{MP}\gD^P_{\,\,N}$. $\gD^P_{\,\,N}$ is a Hermitian operator with respect to the inner product for the fields defined by $\cG_{MN}$ and can explicitly be calculated by expanding in the fluctuations. Being Hermitian it possesses a basis of eigenfunctions $F_i^M$ in which we can expand the fluctuations: 
\bea
&&  \gD^M_{\,\,N} F^N_i=\ge_i F^M_i \label{eq:4a} \\[3mm]
&&  \gF^M_{qu}=\sum_i\gx_i F_i^M \label{eq:4b} \pt
\eea
There is a big caveat however, because the operator $\gD$ is guaranteed to have zero modes, which we can see by starting with the equations of motion $\frac{\gd S}{\gd \gF}\big|_{\gF_{cl}}=0$ and deriving with respect to the collective coordinates. 
\equ{
0=\frac{\gd}{\gd C^{i_0}} \frac{\gd S}{\gd \gF^N}\Big|_{\gF_{cl}}=\frac{\gd^2 S}{\gd \gF^M \gd \gF^N}\Big|_{\gF_{cl}} \pp{\gF^M_{cl}}{C^{i_0}} \nn \km
}
which means that $Z^N_{i_0}\equiv \pp{\gF^N_{cl}}{C^{i_0}}$ is a null vector\footnote{Admittedly, the question whether there are zero modes which cannot be obtained in this fashion remains an unsolved problem.} of $\gD$. \\
We see that if we have a solution to the equations of motion and take the derivative with respect to the collective coordinates, we obtain a zero mode. In the previous section we concluded that we can either generate fermionic solutions to the equations of motion by directly solving them or by applying the broken supersymmetries, naturally the results agreed. The collective coordinates were nothing but the fermionic parameters $\gx$ and $\bgx$ and for reasons which have just become clear we referred to those solutions as zero modes.\\
The following general treatment of zero modes  in the path integral and the relation with collective coordinates focuses on bosonic fields for simplicity. The fermionic case is similar.\\

Some of the modes in \R{4a} and \R{4b} will have zero eigenvalue $\ge$ but \emph{nonzero} fluctuation coefficient $\gx$. It is convenient to split \R{4b} into two sets\footnote{These sets are mutually orthogonal, in fact we can always choose an entirely orthogonal basis for the fluctuations, because $\gD$ is a Hermitian operator.} :
\equ{
\gF^M_{qu}=\sum_{\iz} \gx_\iz F_\iz^M + \sum_\iq \gx_\iq F_\iq^M \nn \km
}
such that the $\{\iz\}$ run over the zero mode fluctuations $F^M_\iz$ which have $\ge_\iz=0$ but $\gx_\iz\neq 0$ and the $\{\iq\}$ run over the non-zero mode fluctuations $F^M_\iq$. We see that we can identify $Z^M_\iz=F^M_\iz$. \\
We can define an inner product for the fluctuations as follows:
\equ{
U_{ij}\equiv \int d^4 x F^M_i \cG_{MN}F_j^N \nn \pt
}
The action \R{exp1} then becomes\footnote{The part involving the zero mode fluctuations is zero, because $\ge_\jz=0$. Thus these fluctuations are also `zero modes' in the sense that they represent physical fluctuations in field space which do not change the value of the action.}
\equ{
S=S_{cl} + \2 \sum_{i,j} \gx_j \gx_i \ge_j U_{ij}
=S_{cl} + \2 \left(\sum_{\iz, \jz} \gx_\jz \gx_\iz \ge_\jz U_{\iz \jz} + \sum_{\iq, \jq}\gx_\jq \gx_\iq \ge_\jq  U_{\iq\jq}\right) \nn .
}
The path integral measure is defined as 
\equ{
\int [\d\gF^M]\equiv \int \sqrt{\det U_0} \prod_\iz \frac{\d\gx_\iz}{\sqrt{2\gp}} \big[\d\tgF^{A}\big] \label{eq:pim1} \pt
}
We have separated the zero and non-zero fluctuations. The measure of the latter is indicated by $\big[\d\tgF^{A}\big]$, corresponding to an integral over the fluctuation coefficients $\gx_\iq$. Including the zero modes in this measure would lead to a determinant of $M$ which would equal zero, thus invalidating this approach. Instead, the non-zero modes now lead to an \emph{amputated} determinant $\det{\!}'\gD$. The $U_0$ is the inner product over the zero modes only, it actually defines a metric on the moduli space of collective coordinates. \\
The integral over the zero modes must be treated separately, by converting it into an integral over the collective coordinates by a Faddeev-Popovish trick, \cite{Gervais:1974dc, Tomboulis:1975gf}. This means that we insert the number $1$,  suitably represented, into the path integral measure \R{pim1}. A suitable representation for $1$ is
\equ{
1=\int \prod_\iz \d C^\iz \left| \det \pp{f_\iz}{C^\jz}\right| \prod_\jz \gd \left[f_\jz (C)\right] \label{eq:one1} \km
}
which holds for any set of (invertible) functions $f_\iz$. We identify the $C$ with the collective coordinates (as in \R{field1}) and we let the labels run over the zero mode set. 
An apt choice for the functions $f_\iz$ is 
\equ{
f_\iz \equiv -\int d^4x \,\gF^M_{qu} \cG_{MN} F_\iz^N =- \sum_\jz \gx_\jz U_{\jz\iz} \pt\nn
}
Taking the derivative yields
\equ{
\pp{f_\iz(C)}{C^\kz} =-\int d^4 x \left\{\pp{\gF^M_{qu}}{C^\kz} \cG_{MN}F^N_\iz + \gF^M_{qu} \p{C^\kz}\left(\cG_{MN}F^N_\iz\right)\right\} \nn 
}
and because $\gF^M$ does not depend on the collective coordinates $C$, see \R{field1}, we can replace $\gF^M_{qu}$ by $-\gF^M_{cl}$ in the first term, which gives
\bea
\pp{f_\iz(C)}{C^\kz} &=&-\int d^4 x \left\{-F^M_\kz \cG_{MN}F^N_\iz+\gF^M_{qu} \p{C^\kz}\left(\cG_{MN}F^N_\iz\right)\right\} \nn \\
&=&U_{\kz\iz}-\int d^4 x \gF^M_{qu} \p{C^\kz}\left(\cG_{MN}F^N_\iz\right) \nn\pt
\eea
Equation \R{one1} can thus be expressed as
\equ{
1=\!\int \!\prod_\iz \d C^\iz \left|\det \left\{ U_{\kz\iz}-\int \! d^4 x \,\gF^M_{qu} \p{C^\kz}\left(\cG_{MN}F^N_\iz\right)\right\}\right| \prod_\jz \gd\left[\sum_\iz \gx_\iz U_{\iz\jz}\right] \nn 
}
which has to be inserted into \R{pim1}. The delta function over the $\gx_\iz$'s forces them to be zero, because $U_{\iz\jz}$ is invertible and thus has no zero eigenvectors. This means that $\gF^M_{qu}$ in the second term in the determinant only contains the genuine quantum fluctuations which gives this term an extra factor of $\hbar$ (had we kept $\hbar$ in the game) meaning that it contributes at $2$-loop instead of $1$-loop. We will therefore drop this term, see also \cite{Amati:1978wu, Yaffe:1978wx, Morris:1984zi}.\\
The total measure\footnote{Note that his measure is invariant under general coordinate transformations on the moduli space.} becomes
\equ{
\int  [\d\gF^M] e^{-S}=\int \prod_\iz \frac{dC^\iz}{\sqrt{2\gp}} \sqrt{\det U_0} \, e^{-S_{cl}}\left(\det{\!}'\gD\right)^{-\2} \nn \pt
}
This is the general procedure to obtain the one-loop measure in the presence of instantons, for reviews of instanton calculations in supersymmetric theories, see \cite{Dorey:2002ik, Belitsky:2000ws}.\\

\subsection{The bosonic measure}
The general treatment above can be directly applied to the NS $5$-brane instanton solution. Let us start with the bosonic collective coordinates. For the single centered instanton there are $4$ collective coordinates, namely the location of the instanton $x_0^\gm$ in $\mR^4$.  This means that the moduli space is four dimensional and has a $4$-dimensional metric which we shall denote by $(U_0)_{\gm\gn}$. This metric receives contributions from both the scalars and the tensors. In the above treatment we considered a generic bosonic system with fields $\gF^M$. For the instanton solution we must consider the fields $(\gf^A, B_{\gm\gn I})$, so we can identify
\equ{
\gF^M=\{\gf^A, B_{\gm\gn I}\} \qquad \textrm{and}\qquad \cG_{MN}=\left(%
\begin{array}{cc}
  \cG_{AB}(\gf_{cl}) & 0 \\[1mm]
  0 & M^{IJ}(\gf_{cl}) \\
\end{array}%
\right)
\nn \pt
}
The Hermitian operator reads
\equ{
\gD^M_{\,\,N}=\left(%
\begin{array}{cc}
  -\gd^A_{\,\,B}\p{}^2 + \ldots & \ldots \\[1mm]
  \ldots & -\gd_I^{\,\,J} \p{}^2+\ldots\\
\end{array}%
\right)
\km
}
the ellipses stand for operators at most linear in derivatives. In the scalar sector we have a block diagonal metric $\cG_{AB}$, see \R{GAB1}, hence we can compute the contributions coming from $\gf$ and $\gc$ separately. For the case of the dilaton zero mode, i.e. $ \pp{\gf}{x_0^\gm}$,  we have to compute
\equ{
U^{(\gf)}_{\gm\gn}=\int \!d^4 x \pp{\gf}{x_0^\gm}\pp{\gf}{x_0^\gn} =\int \!d^4 x \frac{x_\gm x_\gn}{r^2} e^{-2\gf}\left(\p{r}e^{-\gf}\right)^2 =\4 \gd_{\gm\gn} \int \!d^4 x e^{2\gf} \left(\p{r}e^{-\gf}\right)^2 \label{eq:gf1} 
}
where we used $\pp{\gf}{x_0^\gm}=e^\gf\p{\gm}e^{-\gf}$ and the spherical symmetry of the single-centered instanton. To compute the last expression in \R{gf1} we use the slightly more general integral
\equ{
I_p\equiv \int d^4 x h^{-p}(\p{r}h)^2 \qquad h\equiv h_\infty + \frac{Q}{4\gp^2 r^2} \nn
}
for some power $p>1$. Using $\frac{dr}{r^3}=-2\gp^2 \frac{dh}{Q}$ this integral becomes
\equ{
I_p=\left(\frac{Q}{2\gp^2}\right)^2 \textrm{Vol}(S^3)\int_0^\infty dr r^{-3} h^{-p}=Q \int_{h_\infty}^\infty dh h^{-p} =\frac{Q}{p-1}h_\infty^{1-p} \nn \km
}
which diverges for $p\leq 1$. Applying this to \R{gf1} yields the result
\equ{
U^{(\gf)}_{\gm\gn}=\frac{|Q_2|}{4g_s^2} \gd_{\gm\gn} \nn \pt
}
Similarly, we have for the $\gc$ zero mode, 
\equ{
\pp{\gc}{x_0^\gm}=g_s^2 \gD\gc e^{2\gf}\p{\gm}e^{-\gf} \nn\km
}
the metric
\equ{
U^{(\gc)}_{\gm\gn}=\int d^4 \, x e^{-\gf} \pp{\gc}{x_0^\gm}\pp{\gc}{x_0^\gn}=\frac{|Q_2|}{8} (\gD\gc)^2 \gd_{\gm\gn} \nn \km
}
remember that $\cG_{\gc\gc}=e^{-\gf}$. \\
The tensors are slightly more subtle. Tensors have gauge symmetries which have to be gauge-fixed, we choose the background gauge condition
\equ{
\pu{\gm}\left(M^{IJ}B^{qu}_{\gm\gn J}\right)=0 \label{eq:bgg1} \pt
}
The instanton configurations are solutions to the classical, gauge invariant equations of motion and taking derivatives with respect to $x_0^\gm$ do not yield zero modes of the \emph{gauge-fixed} operator $\gD^M_{\,\,N}$ in general. Therefore we consider the following zero modes
\equ{
Z_{\gm\gn I\gr}\equiv \pp{B_{\gm\gn  I}}{x_0^\gr} - 2\p{[\gm}\gL_{\gn]I\gr} \nn \km
}
where we added the second part to make sure the $Z$ obey the background gauge condition \R{bgg1}. If we then choose $\gL_{\gn I \gr}=B_{\gn \gr I}$ we obtain 
\equ{
Z_{\gm\gn I\gr}=-H_{\gm\gn\gr I}=\gve_{\gm\gn\gr\gs}H_I^\gs \nn \km
}
which manifestly satisfies \R{bgg1} because of the classical tensor field equations, see section \ref{sectionBb1}. The tensorial part of the metric consequently becomes
\equ{
U^{(B)}_{\gm\gn}=\2 \int d^4 x\, M^{IJ}Z^{\gr\gs}_{\,\,\,I\gm}Z_{\gr\gs J\gn}=\int d^4 x \,M^{IJ}\left(\gd_{\gm\gn}H^\gr_I H_{\gr J} -H_{\gm I}H_{\gn J}\right) \nn \km
}
spherical symmetry and the Bogomol'nyi equation \R{Bbound1} give
\equ{
U^{(B)}_{\gm\gn}=\frac{3}{4} \gd_{\gm\gn} \int d^4 x \left(e^\gf \hH_1^\gr \hH_{\gr 1} + e^{2\gf} H_2^\gr H_{\gr 2}\right) =3\left(U^{(\gc)}_{\gm\gn} +U^{(\gf)}_{\gm\gn}\right) \nn \km
}
where $\hH^\gm_I$ was defined in \R{heisinv1}. 
The total bosonic metric becomes
\equ{
(U_0)_{\gm\gn}=U^{(\gf)}_{\gm\gn}+U^{(\gc)}_{\gm\gn}+U^{(B)}_{\gm\gn}=S_{cl}\gd_{\gm\gn} \nn \km
}
with $S_{cl}$ given in \R{sc5b1}. This compact result  is similar to the case of Yang-Mills instantons.\\
With these results we can finally write down the bosonic part of the single-centered (anti-)instanton measure:
\equ{
\int \frac{d^4 x_0}{(2\gp)^2} (\det U_0)^{\2} e^{-S^\pm_{\inst}}\left(\det{\!}' \gD\right)^{-\2} =\int \frac{d^4 x_0}{(2\gp)^2} S^2_{cl} e^{-S^\pm_{\inst}} \left(\det{\!}' \gD\right)^{-\2} \label{eq:bos-meas} \pt
}

\subsection{The fermionic measure}
In section \ref{brokensusy} we have shown that there are (nonzero)  solutions for some of the hyperinos and none for the gravitinos. This means that the equations of motion (linear in the fermions) are given by (\ref{eq:5br-zeromode-eqn1}, \ref{eq:5br-zeromode-eqn2}) with zeroes on the right hand sides. We can reproduce these equations of motion by starting from the action
\equ{
S=\int d^4 x \,\I \gl^a\left(\slsh{D}_{3/4}\right)_{a\bb} \bgl^\bb \nn \km
}
where 
\equ{
\left(\slsh{D}_{3/4}\right)_{a\bb}\equiv \left(%
\begin{array}{cc}
  \slsh{D}_{3/4} & 0 \\[2mm]
  0 & \slsh{D}_{-3/4} \\
\end{array}%
\right) \nn 
}
and $\slsh{D}_{\pm3/4}$ has been defined in \R{Dk}.\\
When expanding the fermionic fields as in \R{field1}, the quadratic part of this action (which includes the zero modes) becomes
\equ{
S_2=\int d^4 x \,\I \gl^a_{qu}\left(\slsh{D}_{3/4}\right)_{a\bb} \bgl_{qu}^\bb \nn \pt
}
We have to proceed along the same lines as for the bosons. Contrary to $\gD^M_{\,\,N}$ however, $\slsh{D}_k$ is not an Hermitian operator, but it satisfies\footnote{This is different from one's usual twisted Dirac operator, because we twist with an anti-Hermitian connection. As a result, it is not clear how to calculate the index of such an operator.}
\equ{
\left(\slsh{D}_k\right)^\dagger=\slsh{\bar D}_{-k} \nn \pt
}
On the other hand, the operators
\equ{
M_k\equiv \slsh{D}_{\!-k}\,\slsh{\bar
D}_k \qquad \quad {\bar M}_k\equiv\slsh{\bar D}_{\!-k}\,\slsh{D}_k \nn 
}
are Hermitian. Furthermore, the spectrum of nonzero modes of $M_{-3/4}$ and $\bM_{3/4}$ is identical, similarly for $M_{3/4}$ and $\bM_{-3/4}$. This can be seen as follows, let $F^1_i$ and $F^2_i$ denote a basis of eigenfunctions of $M_{-3/4}$ and $M_{3/4}$ respectively. $\bF^1_i$ and $\bF^2_i$ denote a basis of eigenfunctions of $\bM_{3/4}$ and $\bM_{-3/4}$. \\
The eigenfunctions of $M_{-3/4}$ and ${\bar M}_{3/4}$
are then related, with the same eigenvalue $\gve_i^1=\bar\gve_i^1\neq 0$,
by 
\equ{
{\bar F}^1_i=(\gve_i^1)^{-1/2}\slsh{\bar D}_{\!-3/4}F^1_i \qquad F^1_i=(\gve_i^1)^{-1/2}\slsh{D}_{3/4}{\bar F}^1_i \nn \pt
}
Similarly, the spectrum of nonzero modes of $M_{3/4}$ and ${\bar
M}_{-3/4}$ is identical and the relation between the eigenfunctions is
given by 
\equ{
{\bar F}^2_i=(\gve_i^2)^{-1/2}\slsh{\bar D}_{3/4}F^2_i \qquad F^2_i=(\gve_i^2)^{-1/2}\slsh{D}_{\!-3/4}{\bar F}^2_i \nn \pt
}
Here we assumed  for simplicity that the eigenvalues are positive. Bearing in mind that both $M_{3/4}$
and ${\bar M}_{3/4}$ have zero modes, together with the fact that the
fermion zero modes are in $\gl^2$ and $\bgl^1$, we can expand the
fermions in a basis of eigenfunctions (suppressing spinor indices),
 \begin{equation}
  \gl^a_\mathrm{qu} = \sum_i\, \xi^a_i F_i^a\ ,\qquad
  \bgl^\ba_\mathrm{qu} = \sum_i\, \bar{\xi}_i^a \bar{F}^\ba_i\ ,
 \end{equation}
with $\xi_i^a$ and $\bar{\xi}^a_i$ anticommuting (there is no sum over
$a$). Substituting this into the action and using the relation between the
different eigenfunctions as discussed above we get
 \begin{equation}
  S_2 = \I \sum_{a,i,j} \xi^a_i\, U_{ij}^{aa}\, (\gve_j^a)^{1/2}\,
  \bar{\xi}^a_j\ ,\qquad U_{ij}^{ab} \equiv \int\! \d^4x\ F_i^aF_j^b\ .
 \end{equation}
We then define the fermionic part of the path-integral measure as
(up to a sign from the ordering of the differentials)
 \begin{equation}
  [\d\gl]\, [\d\bgl] \equiv \prod_a \prod_i\, \d\xi_i^a\, \d
  \bar{\xi}_i^a\, (\det U^{aa})^{-1}\ ,
 \end{equation}
such that the fermion integral gives the Pfaffians of $\slsh{\bar
D}_{3/4}$ and $\slsh{D}_{3/4}$ in the nonzero mode sector. In the zero
mode sector, we are left over with an integral over the four GCCs (Grassmann collective coordinates). These 
are combined into two spinors, multiplied by the inverses\footnote{Note that to ensure invariance under reparametrizations of the GCCs we have to use the inverse determinant on the moduli space of GCCs, instead of the square root, as in the case of bosonic collective coordinates \R{pim1}.} of the norms
of the zero modes. The zero mode eigenfunctions have the form
$Z^2_{\alpha\beta'}=\p{}^{(1)}\!\gl^2_\alpha/\partial\bgx^{\beta'}$ given in
\R{kst1}, so that we find for their inner product
 \begin{align}
  U^{22}_{\alpha'\beta'} & = \int\! \d^4x\, {Z^{2\ga}}_{\alpha'}\,
    Z^2_{\ga\beta'} = -4 \int\! \d^4x\, e^{-\phi/2} \partial_\mu \phi\,
    \partial_\nu \phi\, (\gve\,\bgs^\mu \gs^\nu)_{\alpha'\beta'} \notag \\[1mm]
  & = 4\, \gve_{\alpha'\beta'} \int\! \d^4x\, e^{3\phi/2} (\partial_r
    e^{-\phi})^2 = \frac{8\, |Q_2|}{g_s}\, \gve_{\alpha'\beta'}\ .
 \end{align}

The fermionic measure on the moduli space of collective coordinates
then is
 \begin{equation} \label{eq:ferm-meas}
  \int\! \d^2\xi\, \d^2\bar{\xi}\ \Big(\frac{g_s}{8\,|Q_2|}\Big)^2\
  \big( \det{\!}' M_{3/4}\ \det{\!}' \bar{M}_{3/4} \big)^{1/2} \pt 
 \end{equation}
The convention is that $\d^2\xi\equiv\d\xi_1\,\d\xi_2$. \\
The complete measure is then given by combining \R{ferm-meas} with
\R{bos-meas} (which contains the determinant due to the fluctuations in the hypermultiplet) this gives 
\begin{equation} \label{eq:meas}
  \int \frac{d^4x_0}{(2\pi)^2}\ \int\! \d^2\gx\, \d^2\bgx\ \Big(
  \frac{g_s\, S_\cl}{8\,|Q_2|} \Big)^2\, K_\text{1-loop}^\pm\,
  e^{-S^\pm_\inst} \pt
 \end{equation}
 
Although the classical values in the instanton background of the gravitational and vector multiplets are trivial (flat metric and vanishing vector multiplets) their quantum effects cannot be ignored. This is a complicated calculation which is beyond the scope of our computations. Moreover, these loop effects would have to be computed in the full ten-dimensional string theory. We will be pragmatic and denote with $K_\text{1-loop}^\pm$ the
ratio of all fermionic and bosonic determinants in the
one-(anti-)instanton background.

\section{Correlation functions}
\label{sect_correl}
In this section instanton effects to the effective action will be computed by calculating correlators that receive instanton corrections. The (single-centered anti-)instanton measure \R{meas} is used to compute the instanton contributions to the correlators. We again notice that it contains an integral over the four GCCs. Hence a generic correlation function $\langle A \rangle$ will only be nonzero if the `$A$' is able to saturate the fermionic measure. Clearly there will be at least a nonzero four-point fermion correlation function, which will correct the four point vertex. Diagrammatically, such a four-point vertex   consists of four fermion zero modes connected to an instanton at position $x_0$ which is integrated over. If one computes this diagram, one can read off the four-index tensor that determines the four-fermi terms in the effective action, i.e., one can see how $V_{ab\ba\bb}$ would get corrected, see \R{sa1}.\\
This would be difficult to do in practice due to the fact that we are working in the $1.5$ order formalism as explained in section \ref{sec1}. Additional four-fermi terms are hidden in the spacetime curvature scalar $R(\go)$ as a function of the spin connection. Moreover, the four-fermi correlator would merely give information about the target-space curvature-like terms rather than the fundamental objects $M^{IJ}$, $\cG_{AB}$ and $A^I_A$. The latter are much more interesting because they determine the metric of the hypermultiplet, see \R{dual-metric}.\\
We can  obtain instanton corrections to these fundamental quantities by studying the GCC dependence of the scalars and tensors. For instance, the correlator \equ{
\langle \gf^A(x) \gf^B(y)\rangle \nn
}
will give one-loop corrections  to $\cG_{AB}$ in the presence of an instanton. But what to insert for $\gf^A(x)$?\\
Obviously this $\gf^A(x)$ must be something with two GCCs. We must again use the broken supersymmetries which generate fluctuations that  are related by supersymmetry to the purely bosonic instantons and are genuine zero modes which leave the action unchanged as we saw in section \ref{brokensusy}. We observed that applying them once (at linear order) generated the fermionic zero modes, see \R{brfermions1} and  \R{brsp1}. To these solutions for the hyperinos correspond bosonic superpartners. These are obtained by acting twice with the broken supersymmetries on the scalars. Put differently, substitute the values for the hyperinos and the broken supersymmetry parameters (\R{brfermions1} and  \R{brsp1}) into \R{sctr1}. This induces a quadratic GCC dependence in the scalars, i.e., the scalar superpartner contains two GCCs. Similarly, by using \R{de_B_loc} the tensor superpartners can be constructed.
The relevant correlators to study will thus be $2$-point functions of scalars and tensors. \\

At second order in the GCCs the scalars are given by (see \R{sctr1})
\begin{equation}
  {}^{(2)}\!\phi^A = \frac{1}{2}\, \gd_\ge^2 \phi^A|_\cl = \frac{1}{\sqrt{2\,}}
  \big( \gg^A_{ia}(\phi_\cl)\, \ge^i\, {}^{(1)}\!\gl^a + \bar{\gg}^i
  {}^A_\ba(\phi_\cl)\, \bge_i\, {}^{(1)}\!\bgl^\ba \big) \km \nn
 \end{equation}
where we write $\frac{1}{2}\, \gd_\ge^2 \phi^A|_\cl$ because one has to exponentiate the infinitesimal (broken) supersymmetry transformations\footnote{Remember that we have rescaled the supersymmetry parameters by a factor of $\sqrt{2}$.}. \\
In section \ref{brokensusy} we have seen that $^{(1)}\gl^\pm=^{(1)}\bgl^\mp=0$ and for the broken supersymmetry parameters $\ge^\pm=\bge_\mp=0$. This means that only terms proportional to $\gg_{\mp\mp}^A$ and $\bgg^{\pm A}_{\,\,\pm}$ contribute, they are both zero for the dilaton. Only $\gc$ gets corrected (at this order)
 \begin{equation}\label{eq:2chi} 
  {}^{(2)}\!\chi = 2\I\, \partial_\mu \phi\, \gx \gs^\mu \bgx\ ,\qquad
  {}^{(2)}\!\phi = 0 \pt
 \end{equation}
Due to our conventions for the fermionic zero modes chosen in
\R{brfermions1}, this expression for $\chi$ is the same in the instanton and anti-instanton background.
Analogously, the second order corrections of the tensors follow from \R{de_B_loc}. The instanton and anti-instanton cases yield, up to a sign, the same answer,
 \begin{equation} \label{eq:2B1}
  {}^{(2)}\!B_{\mu\nu 1} = \mp 2\I\, \gve_{\mu\nu\rho\gs}\, \partial^\rho
  e^{-\phi}\, \xi \gs^\gs \bgx\ ,\qquad {}^{(2)}\!B_{\mu\nu 2} = 0 \km
 \end{equation}
notice again that only the R-R sector is turned on.
One can easily check that the Bogomol'nyi  equation \R{Bbound1} still holds at this order in the GCCs:
\begin{equation}
  {}^{(2)}\!H_{\mu1} = \pm \partial_\mu \big( e^{-\phi}\, {}^{(2)}\!\chi
  \big) \km
 \end{equation}
the second equation in \R{Bbound1} is trivially satisfied.
It may surprise  those familiar with instanton calculus,
that the equations of motion are satisfied without any fermion-bilinear
source term. One would expect such a source term to be present, since
\R{2chi} and \R{2B1} are obtained by acting with those broken
supersymmetries that also generate the fermionic zero modes. This is
typically what happens with the Yukawa terms in $N=2$ or $N=4$ SYM
theory in flat space; in that theory the adjoint scalar field is found by
solving the inhomogeneous Laplace equation with a fermion-bilinear
source term. The fermionic zero modes in the presence of a YM instanton
then determine the profile and GCC dependence of the adjoint scalar
field. Some references where this is discussed in more detail are given
in \cite{Dorey:1996hu, Dorey:2002ik, Dorey:1999pd}.

In the case at hand, the fermion bilinear source term actually vanishes
when the zero modes are plugged in. To see this, let us  consider
the tensors, for which the full equations of motion read
 \begin{align}
  e^{-1}\, \frac{\gd S}{\gd B_{\mu\nu I}} = \gve^{\mu\nu\rho\gs} \partial_\rho
    \big[ & M^{IJ} \cH_{\gs J} - \I A_A^I \partial_\gs \phi^A +
    \frac{\I}{\sqrt{2\,}} M^{IJ} (g_{Jia} \psi_\gs^i \gl^a + \text{c.c.})
    \notag \\*
  & + \I M^{IJ} k_{Ja\ba}\, \gl^a \gs_\gs \bgl^\ba\, \big]\ .
 \end{align}
The fermionic zero modes we have found above do not enter these
equations directly, because ${}^{(1)}\psi_\mu^i={}^{(1)}\bgps_{\mu
i}=0$, and the two matrices $M^{IJ} k_{Ja\ba}$ are diagonal (actually
zero for $I=1$) but for $a=\ba$ either ${}^{(1)}\!\gl^a$ or ${}^{(1)}
\!\bgl^\ba$ vanishes. Hence, up to second order in the GCC, only the
bosonic fields contribute. This is consistent with the fact that the
BPS condition still holds at this order. A similar analysis can be
done for the equations of motion for the scalars.

\subsection{$2$-point functions}
We have seen above that the objects of prime interest are bilinears in the R-R fields $^{(2)}\gc$ and $^{(2)}H_1^\gm$.  In this section we will compute correlators of these R-R combinations. In the next section we will compute combinations of one R-R field with the zero modes $^{(1)}\gl^\mp$ and $^{(1)}\bgl^\pm$.\\
The total measure for the single-centered (anti-) instanton has been found in the previous section, see \R{meas}. Using this measure, one computes a correlator by integrating the fields over the collective coordinates. However, computing the correlator of two $\chi$'s for instance (as given in \R{2chi}) is not straightforward due to the integration over $x_0$. For simplicity we therefore take 
a large distance limit. We then express the fields (for convenience)  in terms of propagators, which will enable us to read off the effective vertices from the correlation functions by stripping off the external legs. For the bosons we find
\begin{align}
 & {}^{(2)}\!\chi(x)   = - 2\I\, |Q_2|\, g_s^{-2}\, \xi \gs^\mu \bgx\,
    \partial_\mu G(x,x_0)\, \big( 1 + \dots \big) \notag \\[1.4mm]
 & {}^{(2)}\!H^\mu_1(x)   = \mp 2\I\, |Q_2|\, \xi \gs^\nu \bgx\, \big(
    \partial^\mu \partial_\nu - \gd^\mu_\nu \partial^2 \big) G(x,x_0)\km \label{eq:exact1}
 \end{align}
where $G(x,x_0)=1/4\pi^2(x-x_0)^2$ is the massless scalar propagator.

In the first equation in \R{exact1} we only keep the
leading term in the large distance expansion valid when $(x-x_0)^2\mg
|Q_2|/4\pi^2 g_s^2$. In this limit the dilaton is effectively given by
$e^{-\phi}\approx e^{-\phi_\infty}=g_s^2$ and similarly $\chi\approx
\chi_\infty$.  
So the fields are replaced by their asymptotic values
and these will be used to describe the asymptotic geometry of the moduli
space in a next section. 
The second equation in \R{exact1} is exact: for correlators involving ${}^{(2)}\!H^\mu_1$ it is not necessary to make a large distance approximation.
For correlators involving the fermions it is also necessary to make a large distance approximation. In this approximation the fermion zero modes are given by
\begin{align} \label{eq:ferm_prop}
  {}^{(1)}\gl^\mp_\alpha(x) & = - 2\, |Q_2|\, g_s^{-3/2} S_{\alpha
    \beta'} (x,x_0)\, \bgx^{\beta'} \big( 1 + \dots \big) \notag
    \\[1.4mm]
  {}^{(1)} \bgl^\pm_{\beta'}(x) & = \pm 2\, |Q_2|\, g_s^{-3/2}\,
    \xi^\alpha S_{\alpha\beta'}(x,x_0)\, \big( 1 + \dots \big) \km
 \end{align}
where the ellipses again indicate terms of higher order in the large distance expansion and $S(x,x_0)=-\I{\p\!\!\!/}\,G(x,x_0)$ is the $\gl\bgl$ propagator.\\
Let us begin with the purely bosonic correlators. With the GCC measure
$\d\mu_\xi\equiv\d^2\xi\,\d^2\bgx$ $\big(g_s/8|Q_2|\big)^2$ from
\R{ferm-meas} and the Fierz identity $\xi\gs_\mu\bgx\,\xi\gs_\nu\bgx
=-\2\gd_{\mu\nu}\,\xi\xi\,\bgx\bgx$, we find in the large distance
limit
 \begin{equation}
  \int\! \d\mu_\xi\, {}^{(2)}\chi(x)\, {}^{(2)}\chi(y) = \frac{1}{8
  g_s^2}\, \partial_0^\mu\, G(x,x_0)\, \partial^0_\mu\, G(y,x_0)\km \nn
 \end{equation}
where we have replaced $\p{\gm}$ with $-\partial^0_\mu\equiv -\p{}/\partial x^\mu_0$, denoting the derivative with respect
to the bosonic collective coordinates. Using ${}^{(2)}\phi=0$, we then
obtain for the leading semiclassical contribution to the correlation
function of two scalars in the one-(anti-)instanton background
 \begin{align}
  \langle \phi^A(x)\, \phi^B(y) \rangle_{\textrm{{\scriptsize inst}}} & = g_s^{-2}\, \gd^A_\chi
    \gd^B_\chi\! \int\! d^4x_0\ Y_\pm\, \partial_0^\mu\, G(x,x_0)\,
    \partial^0_\mu\, G(y,x_0) \notag \\[1mm]
  & = g_s^{-2}\, Y_\pm\, \gd^A_\chi \gd^B_\chi\, G(x,y)\pt
    \label{eq:cor_chi-chi}
 \end{align}
Here we denote (remember that the difference between $S_\cl$ and
$S^\pm_\inst$ is given by the surface terms (\ref{eq:boun5}, \ref{eq:boun2}))
 \begin{equation} \label{eq:Y-inst}
  Y_\pm \equiv \frac{1}{32\pi^2}\, S_\cl^2\, e^{-S^\pm_\inst}
  K_\text{1-loop}^\pm\ ,
 \end{equation}
which is small for small string coupling constant $g_s$. Since
translation invariance implies that neither $S_\cl$ nor
$K_\text{1-loop}^\pm$ depends on the collective coordinates $x_0$, we
were allowed to integrate by parts and use $\partial_0^2\,G(x,x_0)=-\gd(x-
x_0)$. There is no boundary term because the domain of integration
covers all of $\mR^4$ with no points excised (the instanton can be located at any point) and the integrand vanishes at
infinity.

The result \R{cor_chi-chi} is to be compared with the
propagator derived from an effective action with instanton and
anti-instanton corrected metric $\cG_{AB}^\eff=\cG_{AB}+\cG_{AB}^\inst$,
with $\cG_{AB}$ as in \R{DTM_metrics}. Similarly we write for the
inverse $\cG^{AB}_\eff=\cG^{AB}+\cG^{AB}_\inst$, with $\cG_{AC}\cG^{CB}
=\delta_A^B$. At leading order in $Y_\pm$, we find 
 \begin{equation} \label{eq:inv-cG_eff}
  \cG^{AB}_\inst = \lp 0 & 0 \\[2mm] 0 & g_s^{-2} (Y_+ + Y_-) \rp\pt
 \end{equation}
Note that since $Y_-=(Y_+)^\star$, instanton and anti-instanton
contributions combine into a real correction\footnote{We are assuming
here that $K_-=(K_+)^\star$. Presumably, the one-loop determinants $K_\pm$
only differ by a phase coming from the fermionic determinants. If this
phase can be absorbed in the corresponding surface terms
\R{boun2}, the instanton and anti-instanton determinants are
real and equal. \label{ftnote10}}. This result receives of course
corrections from perturbation theory and from terms that become
important beyond the large distance approximation in which $e^{-\phi}
\approx g_s^2$. Such terms play a role when inverting the result of
\R{inv-cG_eff} to obtain the effective metric $\cG_{AB}^\eff$. They
correspond to higher order powers in $Y_\pm$ and interfere with
multi-centered (anti-)instanton effects. Dropping all these subleading
terms we find
 \begin{equation} \label{eq:cG_eff}
  \cG_{AB}^\eff =  \lp 1 & 0 \\[2mm] 0 & e^{-\phi} - g_s^2 (Y_+ + Y_-)
  \rp  
  \km
 \end{equation}
for more details see appendix \ref{propcor}. \\

Having computed the correlator of two R-R scalars in the (anti-)instanton background, we now turn to the correlator of two R-R tensors using \R{exact1}. This will give instanton corrections to the tensor metric $M^{IJ}$. The integration over the GCCs gives 
 \begin{equation}
  \int\! \d\mu_\xi\, {}^{(2)}H_{\mu1}(x)\, {}^{(2)}H_{\nu1}(y) =
  \frac{g_s^2}{8}\, G_{\mu\rho}(x,x_0)\, G^\rho_{\,\nu}(y,x_0)\ ,
 \end{equation}
where $G_{\mu\nu}(x,x_0)=\big(\partial_\mu\partial_\nu-\gd_{\mu\nu}\partial^2\big)G(x,
x_0)$ is the gauge-invariant propagator of the dual tensor field strengths. We then have to integrate over $x_0$, use ${}^{(2)}H_{\mu2}=0$ and the convolution property\footnote{This convolution is defined as \equ{G_{\gm\gr} \star G^\gr_{\,\gn}=\int d^4z\, G_{\gm\gr}(x,z) G^\gr_{\,\gn}(y,z)=G_{\gm\gn}(x,y) \nn\pt}}  $G_{\mu\rho}\star
G_{\,\nu}^\rho=G_{\mu\nu}$, it follows that
 \begin{equation}
  \langle H_{\mu I}(x)\, H_{\nu J}(y) \rangle_{\textrm{{\scriptsize inst}}} = g_s^2\, Y_\pm\,
  \gd_I^1\, \gd_J^1\, G_{\mu\nu}(x,y) \pt \label{eq:MIJ1}
 \end{equation}
{}From the right-hand side we read off the (anti-) instanton correction
to the inverse metric $M_{IJ}$, which multiplies the tensor propagators.
We find for the sum
 \begin{equation} \label{eq:M_eff}
  M_{IJ}^\inst = g_s^2\, (Y_+ + Y_-)\, \gd_I^1\, \gd_J^1\ ,
 \end{equation}
In the large distance approximation (for which $\chi\approx
\chi_\infty$) we then obtain
 \begin{equation}
  M^{IJ}_\eff = M^{IJ}- g_s^{-2}\, (Y_+ + Y_-) \lp 1 & -\chi_\infty
  \\[2mm] -\chi_\infty & \chi^2_\infty \rp\km \label{eq:tisc1}
 \end{equation}
with $M^{IJ}$ as in \R{DTM_metrics}. This seems to suggest that both
R-R and NS-NS sectors get corrections in front of the tensor kinetic
terms. However, when expressed in terms of $\hat{H}_1=H_1-\chi H_2$,
the tensor kinetic terms in the effective action simplify to
 \begin{equation}
  e^{-1} \cL_\eff = \frac{1}{2}\, \big( e^\phi - g_s^{-2} 
  (Y_+ + Y_-) \big)\, \hat{H}^\mu_1 \hat{H}_{\mu1} + \frac{1}{2}\,
  e^{2\phi} H^\mu_2 H_{\mu2} + \dots\ ,
 \end{equation}
In this basis, which is the one to distinguish between fivebrane and
membrane instantons (see the discussion in section \ref{mem1}),
the NS-NS sector does not receive any instanton corrections.\\

There is also the combination of the tensor and scalar (see \R{exact1}) we can consider. 
The GCC integration over this mixed bosonic combination yields
 \begin{equation}
  \int\! \d\mu_\xi\, {}^{(2)}H_{\mu1}(x)\, {}^{(2)}\chi(y) = \mp
  \frac{1}{8}\, G_{\mu\nu}(x,x_0)\, \partial_0^\nu\, G(y,x_0) \km
 \end{equation}
which vanishes when integrated over $x_0$ thanks to the Bianchi
identity $\partial^\mu G_{\mu\nu}=0$. We conclude that
 \begin{equation}
  \langle H_{\mu I}(x)\, \phi^A(y) \rangle = 0\ .
 \end{equation}
This was to be expected, since for constant coefficients $A_A^I$ the
vertex 
\equ{
-\I A_A^I H^\mu_I\partial_\mu\phi^A \nn
}
is a total derivative and therefore does not contribute to the propagator. However, we will argue later 
that instantons \emph{must} induce such a vertex with field-dependent
coefficients. To determine this vertex explicitly we would have to
go beyond the leading and large distance expansion, see appendix \ref{inst-vielb}.

The calculation of the bosonic 2-point functions is sufficient to
determine the instanton-corrected moduli space metric, which we present
in the next section. The full geometry, or the full effective action, does not follow from the metric alone, but also from various
connections that appear e.g.\ in the supersymmetry transformation rules.
 As we now show, these connections can
be read off from the three-point functions.

\subsection{3-point functions} 
First we compute $\Gamma^{Ia}{}_b$ in the instanton background. It appears in the effective action \R{sa1}
through the relation \R{k-Gamma} and measures the strength of the
coupling between the tensors and the fermions. We therefore compute
 \bea
  &&\langle \gl_\alpha^a(x)\, \bgl_{\beta'}^\bb(y)\, H_{\mu I}(z) \rangle \nn \\[2mm]
  &&\,\,= -\I \frac{|Q_2|}{g_s}\, Y_\pm \gd^a_\mp\, \gd^\bb_\pm\, \gd_I^1\!
  \int\! \d^4x_0\, \big[ S(x,x_0) \bgs^\nu S(y,x_0) \big]_{\alpha
  \beta'}\, G_{\mu\nu}(z,x_0)\pt \nn 
 \eea
These two correlators induce an effective vertex $-h_{a\ba}
(\Gamma^I_\inst)^a{}_b\,\gl^b\gs^\mu\bgl^\ba H_{\mu I}$ with
coefficients
 \begin{equation} \label{eq:Gamma_inst}
  (\Gamma_\inst^I)^a{}_b = -\I \frac{|Q_2|}{g_s}\, M^{I1}_\infty\,
  (Y_+\, \gd^a_2 h_{b1} + Y_-\, \gd^a_1 h_{b2}) = -\I \frac{|Q_2|}
  {g_s}\, M^{I1}_\infty \lp 0 & Y_- \\[2mm] Y_+ & 0 \rp\ .
 \end{equation}
Here we have used that $h_{a\bb}$ is not corrected at leading
order\footnote{The 2-point function of two fermion insertions vanishes
in the semiclassical limit.}. We also used the notation that $M^{I
J}_\infty$ stands for $M^{IJ}$ with the fields replaced by their
asymptotic values at infinity.

The last two correlators contribute to the connection $\G{A}{a}{b}$,
which appears in the covariant derivative on the fermions
\R{covderlam}. This connection was zero on tree-level, see
\R{DTM-connections}, but it receives instanton corrections as
follows from
\bea
&&\langle \gl_\alpha^a(x)\, \bgl_{\beta'}^\bb(y)\, \phi^A(z) \rangle \nn \\[2mm]
  &&\,\,= \mp \I \frac{|Q_2|}{g_s^3}\, Y_\pm \gd^a_\mp\, \gd^\bb_\pm\,
  \gd^A_\chi\! \int\! \d^4x_0\, \big[ S(x,x_0) \bgs^\mu S(y,x_0)
  \big]_{\alpha\beta'}\, \partial_\mu\, G(z,x_0) \pt \nn
\eea
This corresponds to an effective vertex $-\I h_{a\ba}(\Gamma^\inst_{\!A}
)^a{}_b\,\gl^b\gs^\mu\bgl^\ba\,\partial_\mu\phi^A$ with
 \begin{equation}
  (\Gamma^\inst_{\!A})^a{}_b = \frac{|Q_2|}{g_s^3}\, \cG_{A
  \chi}^\infty \lp 0 & -Y_- \\[2mm] Y_+ & 0 \rp \pt\nn
 \end{equation}

The above connections induce instanton corrections to the curvature
tensors that appear in the four-fermi couplings of the effective action
\R{sa1}. Indeed, the fact that these curvatures receive instanton
corrections also follows from the computation of 4-point functions of
fermionic insertions. These results should be consistent  with
the instanton corrections to the curvatures as determined by the connections. For reasons explained in
the beginning of this section, checking this consistency may be a
complicated task.

Note however, that there is another four-fermi term in
\R{sa1} proportional to the product of two antisymmetric
tensors $\cE_{ab}$. It is easy to see that this tensor cannot receive
instanton corrections since it multiplies only $\gl^a$ in the action,
not $\bgl^\ba$. Due to the even distribution of fermionic zero modes
among $\gl^a$ and $\bgl^\ba$ there are thus no non-vanishing correlation
functions that could induce an effective vertex involving $\cE_{ab}$. A
similar argument shows that the connections $\Gamma^{Ii}{}_j$ do not get
corrected: they occur in the action only in combination with gravitinos.
 For example in the vertex $2\Gamma^{Ii}{}_j H^{\mu\nu\rho}_I\psi^j_\mu
\gs_\nu^{}\bgps_{\rho i}$, hidden in the square of the supercovariant
field strengths of the tensors \R{sucoH}, which have no zero modes to
lowest order in the GCC\@. Correlation functions of fields corresponding
to vertices involving $\Gamma^{Ii}{}_j$ then do not saturate the GCC
integrals and vanish. If we were to continue the procedure of sweeping
out solutions by applying successive broken supersymmetry
transformations to the fields, the gravitinos could obtain a GCC
dependence at third order, but then the number of GCCs in the
correlators of interest exceeds the number of degrees of freedom and
they therefore vanish as well. However, due to
\R{gamma-omega1} the coefficients ${\Omega_I}^i{}_j$ do get
corrected:
 \begin{equation}
  (\Omega_I^\eff)^i{}_j = M_{IJ}^\eff\, \Gamma^{Ji}{}_j =
  {\Omega_I}^i{}_j + g_s^2 (Y_+ + Y_-) \gd_I^1\, \Gamma^{1i}_{\infty
  j}\ .
 \end{equation} 
These quantities appear in the supersymmetry transformations of the 
tensors \R{de_B_loc}.

\section{The universal hypermultiplet moduli space} 
\label{sect_modul}

In order to determine the instanton corrections to the universal
hypermultiplet, we first Wick-rotate back to Lorentzian signature and then
dualize the tensors $H_I$ into two pseudoscalars $\phi^I=(\varphi,\gs)$,
using the same notation as in section \ref{instsol}, i.e.\ $\varphi$ is a R-R
field and $\gs$ the NS axion. If we combine the latter and $\phi^A=(\phi
,\chi)$ into a four-component field $\phi^{\hat{A}}=(\phi^A, \phi^I)$,
then in this basis the universal hypermultiplet metric reads, see \R{dual-metric}
 \begin{equation}\label{eq:dual-metric2}
  G_{\!\hat{A}\hat{B}} = \lp \cG_{AB} + A_A^I M_{IJ} A_B^J & A_A^K
  M_{KJ} \\[3mm] M_{IK} A_B^K & M_{IJ} \rp\pt
 \end{equation}
Using \R{cG_eff}, \R{M_eff} and $A_A^I=0$, we find for the
asymptotic effective Lagrangian
 \begin{align}
&  e^{-1} \cL_\mathrm{UH} = - \frac{1}{2}\, (\partial_\mu \phi)^2 - \frac{1}   {2}\, e^{-\phi} (1 - g_s^2 e^\phi Y)\, (\partial_\mu \chi)^2 \nn \\[1mm]
  &- \frac{1}{2}\, e^{-\phi} (1 + g_s^2 e^\phi Y)\, (\partial_\mu \varphi)^2
     - \frac{1}{2}\, e^{-2\phi} (\partial_\mu \gs + \chi \partial_\mu
    \varphi)^2 + \dots\km \label{eq:lag1}
 \end{align}
where the ellipses stands for subleading terms and $Y=Y_++Y_-$ is the
sum of the instanton and anti-instanton contributions, as introduced in
\R{Y-inst}. It can be written as
 \begin{align} \label{eq:total-Y}
  Y & = \frac{1}{32\pi^2}\, S_\cl^2\, e^{-S_\cl} \big( e^{\I{\hat\gs}
    |Q_2|} K_\text{1-loop}^+ + e^{-\I{\hat\gs |Q_2|}} K_\text{1-loop}^-
    \big)\notag \\[1.5mm]
  & = \frac{1}{16\pi^2}\, S_\cl^2\, e^{-S_\cl} K_\text{1-loop}
    \cos(\hat\gs Q_2)\km
 \end{align}
where we introduced $\hat\gs\equiv\gs+\chi_0\varphi$ such that $Y$
is periodic in $\hat\gs$. The second equality in \R{total-Y} holds
only under the reality assumption made in footnote \ref{ftnote10}. 
Furthermore note that only the R-R sector receives corrections from the
NS5-brane instanton.

\subsection{The metric and isometries} 
\label{vze}
In this section we present the instanton corrected line element of the quaternionic target space of the UHM. 
First we  write down the general form of the line element, which is given by
(using $A_A^I=0$)
 \begin{equation*}
  d s_\mathrm{UH}^2 = G_{\!\hat{A}\hat{B}}\, d\phi^{\hat{A}} \otimes
  d\phi^{\hat{B}} = \cG_{AB}\, d\phi^A \otimes d\phi^B + M_{IJ}\,
  d\phi^I \otimes d \phi^J\pt \nn
 \end{equation*}
We remind the reader that the classical metric is given by the line
element
 \begin{equation}
  d s_\mathrm{UH}^2 = d\phi^2 + e^{-\phi} d\chi^2 + e^{-\phi}
  d\varphi^2 + e^{-2\phi} (d\sigma + \chi d\varphi)^2 \km \label{eq:hf4t1}
 \end{equation}
and describes the homogeneous quaternionic space $SU(1,2)/
U(2)$, see \R{uhmgeo1}. As discussed in section \ref{instsol}, the isometry group
$SU(1,2)$ can be split into three categories. First, there is a
Heisenberg subgroup of shift isometries,
 \begin{equation} \label{eq:Heis-alg2}
  \phi \rightarrow \phi \ ,\qquad \chi \rightarrow \chi + \gamma\ ,
  \qquad \varphi \rightarrow \varphi + \beta\ ,\qquad \sigma
  \rightarrow \sigma - \alpha - \gamma\, \varphi\km
 \end{equation}
where $\alpha$, $\beta$, $\gamma$ are real (finite) parameters. This
Heisenberg group is preserved in perturbation theory \cite{Strominger:1997eb, Antoniadis:1997eg}. We have
not discussed these perturbative corrections, which only appear at one-loop in the
string frame, here. They are discussed in \cite{Antoniadis:2003sw, Anguelova:2004sj} and should be
added to our final result for the metric.\\
Second, there is a U(1) symmetry \R{u1tr} that acts as a rotation on $\varphi$
and $\chi$, accompanied by a compensating transformation on $\sigma$,
\bea
&& \gvf \to \cos(\gd) \,\gvf + \sin(\gd) \,\gc \qquad \gc \to \cos (\gd) \,\gc -\sin (\gd)\, \gvf \km \nn \\ [1.5mm]
&& \gs \to \gs -\4 \sin (2 \gd) \left(\gc^2 -\gvf^2\right) + \sin^2 (\gd)\, \gc \gvf \label{eq:delta-isom} \pt
\eea
We now present the instanton corrected moduli space metric. As shown
above, instanton effects are proportional to $Y$, given by
\R{total-Y}, and depend on the instanton charge $Q_2$ and the R-R
background specified by $\chi_0$.  Moreover, also the asymptotic values
of the fields, $g_s$ and $\chi_\infty$, appear. They are treated as
coordinates in the asymptotic regime of the moduli space, i.e., where
$\chi =\chi_\infty$ and $e^{-\phi}=g_s^2$. For fixed values of
$\chi_0$ and $Q_2$, the moduli space metric is given by
 \begin{equation} \label{eq:nonpert-metric}
  d s_\mathrm{UH}^2 = d\phi^2 + e^{-\phi} (1-Y) d{\chi}^2 +
  e^{-\phi} (1+Y) d\varphi^2 + e^{-2\phi} (d\gs + \chi
  d\varphi)^2\km
 \end{equation}
up to subleading terms. This metric therefore satisfies the constraints
from quaternionic geometry only up to leading order\footnote{As we have checked explicitly.}. It remains to be seen to what extent the
quaternionic structure can fix these subleading corrections. The result written in \R{nonpert-metric} depends on $Q_2$
and on the chosen R-R background. To obtain the full moduli space metric,
one must sum over all instanton numbers $Q_2$. It would be very
interesting to do this sum explicitly and to see of which function we
have the asymptotic limit. Unfortunately, for that we need more
knowledge of the one-loop determinants and the subleading corrections, which is not available at present.

We can also deduce the leading-order instanton corrections to the
vielbeins and other geometric quantities. These can be computed from the
vielbeins that determine the double-tensor multiplet geometry, which we
give in appendix~\ref{inst-vielb}.\\

What happens to the isometries \R{Heis-alg2} and \R{delta-isom}? For the Heisenberg group, this
amounts to investigating which isometries are broken by the quantity
$Y$, as the other terms are invariant. First we focus on the
$\gamma$-shift in $\chi$. For a given, fixed R-R background $\chi_0$, the
$\gamma$-shift is broken completely. This is because $Y$ is proportional
to $S_\cl$, which contains $\Delta\chi=\chi_\infty-\chi_0$, see
\R{sc5b1}. However, this symmetry can be restored if we
simultaneously change the background as $\chi_0\rightarrow\chi_0+
\gamma$. Since $\chi_0$ is subject to a quantization condition (see
section \ref{secns5b1}), this induces a quantization condition on
the possible values for $\gamma$. This means that the $\gamma$-shift is
broken to a discrete subgroup.\\
With this in mind, we find that under the action of a generic element in
the Heisenberg group the metric is invariant only if the following
quantization condition is satisfied:
 \begin{equation} \label{eq:quant-cond}
  \alpha - (\chi_0 + \gamma) \beta = \frac{2\pi n}{|Q_2|}\km
 \end{equation}
with $n$ an integer. As explained before, the $\gamma$-dependence is
not relevant here since we could shift the R-R background again. As for the
other two isometries generated by $\alpha$ and $\beta$, only a linear
combination is preserved. Stated differently, the $\beta$-isometry is
preserved as a continuous isometry if we accompany it by a compensating
$\alpha$-shift, where $\alpha$ is determined from \R{quant-cond}.\\
If we solely perform an $\alpha$-transformation, only a discrete
$\fieldZ_{|Q_2|}$ subgroup survives as a symmetry. In fact, since the
full metric includes a sum over $Q_2$, only shifts with $\alpha=
2\pi n$ are unbroken. In conclusion, for the Heisenberg group one
isometry remains continuous and two are broken to discrete subgroups.
This is precisely in line with the proposal made in \cite{Anguelova:2004sj}.\\
The remaining isometry we discuss is \R{delta-isom}. Since the last
term in \R{nonpert-metric} is invariant by itself, we should only
look at the R-R sector. Due to the fact that $Y$ is independent of
$\varphi$, but depends on $\chi^2$, this continuous rotation symmetry
is broken. In fact, the terms proportional to $Y$ break this isometry
down to the identity $\delta=0$ and the discrete transformation with
$\delta=\pi$,
 \begin{equation}
  \chi \rightarrow -\chi\ ,\qquad \varphi \rightarrow -\varphi\ ,
  \qquad \gs \rightarrow \gs\pt
 \end{equation}
This conclusion is different from \cite{Becker:1999pb}, where also $\delta=\pi/2$
was claimed to survive as an isometry. 
It is not excluded though, that the full answer may have more symmetries. This full answer should contain (all the) subleading corrections and perhaps membranes.
 At the end of appendix \ref{inst-vielb} we in fact argue
that the connection $A_A^I$ becomes nonzero at subleading order, away
from the asymptotic region. After dualization, this induces new terms
in the metric, as follows from \R{dual-metric2}, so one has to
reanalyze the breaking of isometries. Clearly, this is an interesting
point that deserves further study.\\
Finally note the existence of another discrete isometry which
changes the sign in $\chi$ (or $\varphi$) together with a sign flip
in $\gs$. This is because the (leading) instanton plus anti-instanton
corrections are even in $\chi$ and $\gs$. This discrete isometry is
however not part of (a discrete subgroup of) $SU(1,2)$.

\section{Short summary and outlook}
Having reached our goal, \R{nonpert-metric}, we need to reflect on its meaning. In the range of approximations we are working in,  using  supergravity and the large distance approximation for the fields (see \R{exact1} and \R{ferm_prop}), this is a solid result. Note that one should actually combine our result with the one-loop correction found in \cite{Antoniadis:2003sw}.  We shall elaborate on this one-loop correction in the next chapter.\\
To make further progress, one should really perform this calculation in string theory, which then includes the worldvolume theory of the $5$-brane, and calculate the one-loop determinants. Another generalization is the case of more hypermultiplets, which corresponds to more general Calabi-Yau manifolds with $h^{(1,2)}\neq 0$. To advance in the UHM case, one should try to take a closer look at the constraints coming from the quaternionic geometry. We saw in section \ref{hypergeo} that the definition of a four (real) dimensional quaternionic manifold is that it is Einstein and has (anti-)selfdual Weyl curvature. It is easy to check that the classical metric, \R{hf4t1} obeys these conditions. If we include the $5$-brane contributions as in \R{nonpert-metric}, these conditions are still obeyed up to leading order in $Y$. In fact, this already imposes some constraints on $Y$. One could try to systematically analyze these conditions and find the corrections to the metric which obey the constraints. This is exactly what we will do in the next chapter where we will construct the possible deformations of the metric allowed by the quaternionic constraints.

 \clearpage{\pagestyle{empty}\cleardoublepage}

\chapter{The membrane}
\label{ch4} 
In the previous chapters we discussed the classical geometry of the UHM metric and the corrections due to NS $5$-brane instantons. In this chapter we compute the corrections arising from membranes wrapped along $3$-cycles of the internal rigid CY $3$-fold.\\
This will be done in a different way from the semiclassical instanton calculation of chapter \ref{ch3}. The reason is that the most general $4$-dimensional quaternionic metric with (at least) one isometry is known and can be used to describe the membrane corrected metric. As it turns out, this metric is governed by the \emph{Toda equation}, a complicated nonlinear differential equation in three variables. We will construct solutions to this equation and demonstrate that they correspond to  membranes by comparing to results from string theory. \\
Having obtained these membrane corrections (to the effective action) we can proceed by gauging the isometry, which corresponds to a certain \emph{flux compactification}. This gauging will induce a scalar potential  in the low energy effective action. It turns out that the membrane corrections lead to a (meta-stable) de Sitter vacuum.

\section{Toda equation and the UHM}
\label{todasec1}
Let us recall some results from  the previous chapters. We saw in equation \R{VI} that the simplest membrane action is given by
\equ{
S_{\inst}=\frac{2|Q_1|}{g_s} + \I \gf^s Q_1^s \label{eq:act1} \km
} 
with $\gf^s Q_1^s$ given either by $\gvf Q_1^\gvf$ or $\gc Q_1^\gc$. This means that there are two distinct membrane instantons depending on whether one dualizes over $\gvf$ or $\gc$ (see also sections \ref{mem1} and \ref{secDTM1}). We shall confront this with string theory. We note again the factor of $2$ appearing in front of $|Q_1|$, we will come back to this in section \ref{stcal1}.\\
The presence of the theta angles in \R{act1} breaks the shift symmetry in $\gvf$ (or $\gc$) to a discrete subgroup. Furthermore, we have seen that the theta angle term of the NS $5$-brane is proportional to the axion $\gs$, see \R{boun2}, which means that the continuous shift symmetry in $\gs$ remains exact as long as we do not switch on $5$-branes. In other words, in the absence of fivebrane instantons, the quantum corrected UHM moduli space will be a quaternionic manifold with a shift isometry in $\gs$. Such manifolds have been classified  in terms of a single function.\\
In \cite{Przanowski:1991ru} Przanowski derived the general form of $4$-dimensional quaternionic manifolds with at least one isometry, this was later re-derived by Tod \cite{Tod:1995vm}. The Przanowski-Tod (PT) metric reads, in adapted coordinates,
\equ{
ds^2=\frac{1}{r^2}\left[f dr^2 + f e^h(du^2 + dv^2) + \frac{1}{f}(dt+\gTh)^2\right] \label{eq:tod1} \km
}
where the isometry acts as a constant shift on $t$. The function $h=h(r,u,v)$ is determined by the $3$-dimensional continuous \emph{Toda equation}
\equ{
(\p{u}^2 +\p{v}^2)h + \p{r}^2 e^h=0 \label{eq:tod2} \pt
}
The function $f=f(r,u,v)$ is given in terms of $h$ as follows
\equ{
f=-\frac{3}{2\gL}(2-r\p{r}h) \label{eq:tod3} \pt
}
The $1$-form $\gTh=\gTh(r,u,v)=\gTh_r dr+ \gTh_u du + \gTh_v dv$ is a solution to the equation
\equ{
d\gTh=(\p{u} f dv-\p{v}f du)\wedge dr + \p{r}(f e^h) du\wedge dv \label{eq:tod4} \pt
}
Manifolds with such a metric are Einstein with anti-selfdual Weyl tensor and thus\footnote{Conversely one can check that starting with \R{tod1} and demanding that the metric is Einstein and has an anti-selfdual Weyl tensor forces $h, f$ and $\gTh$ to obey (\ref{eq:tod2}, \ref{eq:tod3}, \ref{eq:tod4}) respectively.} quaternionic (see section \ref{hypergeo}). The constant $\gL$ in \R{tod3} is the `cosmological' constant of the target-space, $R_{AB}=\gL G_{AB}$.\\

To apply the above results to our situation we have to recast the line element \R{hf4t1} into the PT form \R{tod1}, i.e., we have to  identify  the PT coordinates with the UHM moduli. 
If we consider the Heisenberg algebra of isometries \R{Heis-alg2}, we see that we can either identify $t$ with $\gs$ or with $\gvf$. In the first case the shift is generated by $\ga$ and in the second case by $\gb$. This leads to two possible representations of the PT metric that describe the same moduli space. We call these bases the membrane and the fivebrane bases respectively. In the membrane basis, which is the relevant basis, we identify the coordinate $t$ with $\gs$ such that the $\ga$-shift symmetry is manifest. The reason is the absence of $5$-brane instantons, which ensures a continuous $\ga$-shift symmetry. The coordinates can then be chosen as
\equ{
t=\gs \qquad r=e^\gf \qquad u=\gc \qquad v=\gvf \label{eq:fmap1} \pt
}
In this basis the classical moduli space metric of the UHM \R{hf4t1}\footnote{This metric has $\gL=-3/2$.} corresponds to the solution $e^h=r$, which gives $f=1$ and $\gTh=u dv$. Note that $\gTh$ is only defined up to an exact form.\\
Previously (see section \ref{vze} for instance) we have mentioned that there are perturbative ($1$-loop only) corrections to the UHM \cite{Strominger:1997eb, Antoniadis:2003sw}. We can easily incorporate those in the PT framework by using the following observation. The Toda equation is insensitive to  constant shifts in $r$ (or in $u$ and $v$ for that matter) which means that $h(r+c,u,v)$ is also a solution to the Toda equation, $c\in \mR$. Applying this to the classical solution $r=e^h$, gives
\equ{
e^h=r+c \qquad f=\frac{r+2c}{r+c} \qquad \gTh=u dv \label{eq:pertu1} \pt
}
This exactly reproduces the $1$-loop corrected metric (in the string frame) of \cite{Antoniadis:2003sw} provided we identify
\equ{
c=-\frac{4\gz(2)\gc(Y_3)}{(2\gp)^3}=\frac{h^{(1,2)}-h^{(1,1)}}{6\gp} \nn \km
}
where $\gc(Y_3)$ is the Euler number of the CY $3$-fold on which the type IIA string theory has been compactified (see \R{euler1}). We consider only rigid Calabi-Yau's so $h^{(1,2)}=0$, which gives the  bound
\equ{
c<0 \label{eq:c2} \pt
}
The PT coordinate $r$ is related to $\gr$ in \cite{Antoniadis:2003sw} through $r=\gr^2-c=e^\gf$, the relation between the fields and PT coordinates receives no (perturbative) quantum corrections.\\
Note that for $c<0$ the metric becomes negative-definite because $f$ becomes negative for $r<2|c|$. As a result, we have to restrict ourselves to the open interval $2|c|<r<\infty$. 

\subsubsection{Relation with other work}
The $1$-loop correction to the classical UHM  was derived in \cite{Antoniadis:2003sw} by searching for deformations that respect both the Heisenberg group and the quaternionic structure of the UHM. This can be done systematically because the most general form of a $4$-dimensional quaternionic manifold with (at least) two commuting  isometries has been constructed in \cite{Calderbank:2001uz}. This metric is expressed in terms of a single function which is determined by a linear differential equation. This is different from the PT metric that is governed by a  nonlinear differential equation \R{tod2}.\\
We should also mention work  of Ketov \cite{Ketov:2001gq, Ketov:2001ky, Ketov:2002vr} in which membrane (and five-brane) solutions are discussed. An important difference with our work is that Ketov works on the basis that the membrane instantons preserve two isometries,  which allows him to use \cite{Calderbank:2001uz}. However, we have seen explicitly that membrane instantons preserve only one continuous isometry. That is, we can either use the shift in $\gvf$ or $\gc$ to dualize to the DTM for which we can construct membrane instanton solutions, with action \R{act1}. In section \ref{stcal1} we shall see that this corresponds to either wrapping the membrane over the `$\cA$-cycle' or `$\cB$-cycle'. In the end one should sum them both up to obtain the effect of the two membrane\footnote{So if one considers only one membrane instanton there are two isometries left, although they are not the same as the one Ketov uses.} (instantons), to correlation functions for instance. In that situation both isometries are broken to a discrete subgroup and only the shift in $\gs$ is left, in the absence of $5$-brane instantons at least.

\subsubsection{The $5$-brane}
By now the reader might wonder why we have not used the PT metric to calculate $5$-brane corrections to the UHM moduli space. This would  require that we identify $t$ with $\gvf$ (or $\gc$). However, then the mapping of the other PT coordinates to the classical (plus one-loop) UHM metric cannot be consistently realized to describe the fivebrane.\\
Current investigations indicate that this is because of the fact that four-dimensional quaternionic metrics with (at least) one isometry fall into two classes, see \cite{Przanowski:1991ru}. The one class constitutes the PT framework as described above and the other class involves a different differential equation, instead of the Toda equation. This class should probably be used to describe the fivebrane.

\section{Solutions to the Toda equation}

We would like to find a solution that mimics an (infinite) series of exponential corrections describing membrane instantons. We have seen that the real part of the action \R{act1} is inversely proportional to $g_s$. In PT coordinates this real part is given by $2\sqrt{r}$. Therefore we make a fairly general ansatz of the form
\equ{
e^h=r+\sum_{n\geq 1}\sum_m f_{n,m}(u,v)r^{-m/2 +\ga}e^{-2n \sqrt{r}} \label{eq:ansatz1} \km
}
where we expand in the ($1/g_s$ part of the) instanton action: $(e^{-S_\inst})^n$. A slightly more general ansatz is given by replacing the exponent with $e^{-\gd n \sqrt{r}}$ where $\gd$ is a constant. However, it turns out that this ansatz only satisfies the Toda equation (with certain functions $f$ that we will construct shortly) if $\gd=2$. 
There exist interesting solutions to the Toda equation, see \cite{Calderbank:1999ad, Bakas:1996gf}, but with the mapping of PT coordinates to the fields that we use \R{fmap1}, these do not describe physically interesting solutions. In particular, they do not describe a membrane instanton expansion. 
As observed earlier, we can always shift $r$ with a constant to produce a new solution. This means that we can first find a solution for $e^h(r,u,v)$ and then take $r\to r+c$ in this solution, thus incorporating the one-loop correction in one go.
The ansatz, apart from the classical piece, consists of a double sum. There is the sum over $n$ which gives an expansion in $e^{-2/g_s}$, meaning that $n$ is the instanton number. For each $n$ there is an expansion in $m$. This is a perturbative (loop) expansion in a given  instanton sector, since it goes as $r^{-m/2}=(g_s)^m$ and the sum $m$ runs over integers. To account for the possibility that $m$ does not\footnote{As is the case in \cite{Ooguri:1996me} for instance.} run over integers, we have added a parameter $\ga \in [0,\2)$. We will later show that the Toda equation is satisfied for  $\ga=0$ only, see appendix \ref{toda1}. 
At each instanton level $n$ there is, by definition, a lowest  value $m_n$ that defines the leading term in the expansion
\equ{
f_{n,m}(u,v)=0 \qquad \textrm{for}\quad m<m_n\nn\pt
}
Because we have spelled out the $r$ dependence explicitly, the Toda equation gives a series of differential equations for the $f_{m,n}(u,v)$. These equations can be solved iteratively, order by order in $n$ and $m$ to any desired order. To examine the Toda equation \R{tod4} in detail we first write it in the equivalent form
\begin{equation} \label{eq:Toda2}
  e^h \big( \partial_u^2 + \partial_v^2 + e^h \partial_r^2 \big) e^h - (\partial_u e^h)^2
  - (\partial_v e^h)^2 = 0\pt
 \end{equation}
Using  ansatz \R{ansatz1} equation \R{Toda2} can be written as
\begin{align}\label{eq:4.5}
  0 = \sum_{n,m} &\ r^{-m/2}\, e^{-2n\sqrt{r}}\, \Big\{ (\Delta +
    n^2)\, f_{n,m+2} + n\, a_{m+2}\, f_{n,m+1} + b_{m+2}\, f_{n,m}
    \notag \\[-2pt]
  & + \sum_{n',m'} e^{-2n'\!\sqrt{r}}\, \big[ 2n\, a_{m'+1}\, f_{n',
    m-m'-1} + 2 b_{m'+2}\, f_{n',m-m'-2} \notag \\[-6pt]
  & \mspace{112mu} + f_{n',m-m'}\, (\Delta + 2n^2) - \nabla f_{n',m-m'}
    \cdot \nabla \big] f_{n,m'} \notag \\[4pt]
  & + \sum_{n',m'} \sum_{n'',m''} e^{-2(n'+n'')\sqrt{r}}\, f_{n,m'}
    f_{n',m''}\, \big[ n^2 f_{n'',m-m'-m''-2} \notag \\[-8pt]
  & \mspace{94mu} +  n\, a_{m'+1}\, f_{n'',m-m'-m''-3} + b_{m'+2}\,
    f_{n'',m-m'-m''-4} \big] \Big\} \km
 \end{align}
where $\nabla\equiv (\partial_u,\partial_v)$, $\Delta\equiv \nabla^2$ and
 \begin{equation}\label{eq:4.6}
  a_m \equiv \half\, (2m-1)\ ,\qquad b_m \equiv \quart\, m(m-2)\pt
 \end{equation}
Equation \R{4.5} is organized in terms of single-, double- and triple sums. We can write a generic exponential in \R{4.5} as  $e^{-2N\sqrt{r}}$.
In the $(N=1)$-instanton sector only the single-sum terms contribute. The double- and triple-sums have to be taken into account beginning with the $(N=2)$- and $(N=3)$-instanton sectors, respectively. We will save some details for appendix \ref{toda1} where we prove that $m_n\geq-2$ for all $n$ and $\ga=0$.\\

\subsection{The one-instanton sector}
\label{1insts}
We shall start with the one-instanton sector, $N=1$, in which case equation \R{4.5} reduces to
\begin{equation} \label{eq:f1m}
  (\Delta + 1)\, f_{1,m} + a_m\, f_{1,m-1} + b_m\, f_{1,m-2} = 0\pt
 \end{equation}
If for simplicity we consider  first a one-dimensional truncation, we can prove that the general solution is given by
\begin{equation} \label{eq:solf1m}
  f_{1,m}(x) = \Re \sum_{s\geq 0}\, \frac{1}{s!\,(-2)^s}\, k_{1,m}(s)\,
  G_s(x) \km
 \end{equation}
 where $x\in \{u,v\}$. The coefficients are defined recursively as
 \begin{equation}\label{eq:recurse-rel}
  k_{1,m}(s+1) = a_m k_{1,m-1}(s) + b_m k_{1,m-2}(s) \pt
 \end{equation}
The $G_s(x)$ are complex functions related to spherical Bessel functions of the third kind, see appendix \ref{toda1}. The coefficients $k_{1,m}(s)$ have as their lowest values in $s$
\equ{
k_{1,m}(0)=A_{1,m} \nn \km
}
which are the complex integration constants originating from the homogeneous part of \R{f1m}. By definition we have $f_{1,m}=0$ for $m<m_1$, which implies that $k_{1,m}(s)=0$ and in particular $A_{1,m}=0$ for $m<m_1$. The recursion relation \R{recurse-rel} furthermore implies $k_{1,m}(s>m-m_1)=0$, which means that the highest monomial in $x$ contained in $f_{1,m}(x)$ is of order $m-m_1$. This is nicely illustrated by the first two solutions
\begin{align}
 & f_{1,m_1}(x)  = \Re \big\{ A_{1,m_1} e^{\I x} \big\}\, , \notag
    \\[2pt]
 & f_{1,m_1+1}(x)  = \Re \big\{ A_{1,m_1+1} e^{\I x} + \half a_{m_1
    +1} A_{1,m_1}\, \I x\, e^{\I x} \big\}\pt \label{eq:solf1m0}
 \end{align}
To solve \R{f1m} for a $(u,v)$-dependent solution one can proceed by separation of variables, which yields a basis of solutions. The most general solution is then obtained by superposition and can be written as
\bea
  && f_{1,m}(u,v) = \int\! d\lambda\, \Re \sum_{s\geq 0}\, \frac{1}{s!\,
    (-2\, \omega^2)^s}\, k_{1,m}(s, u; \lambda)\, G_s(\omega v)\ ,
    \notag \\[1.5mm]
  &&k_{1,m}(s+1, u; \lambda) = a_m k_{1,m-1}(s, u; \lambda) + b_m k_{1,
    m-2}(s, u; \lambda)\km\label{eq:solf2m}
 \eea 
where
 \begin{equation}
  k_{1,m}(0, u ; \lambda) = B_{1,m}(\lambda)\,   A_{1,m}
  (\lambda) e^{\I\lambda u} \pt \nn 
 \end{equation} 
The basis is parametrized by a continuous real parameter $\gl$ which is integrated over in \R{solf2m} and $\go\equiv \sqrt{1-\gl^2}$. $A_{1,m}(\gl)$ and $B_{1,m}(\gl)$ are arbitrary complex integration constants which determine the `frequency spectrum' of the solution. 

The $(u,v)$-dependent solution \R{solf2m} is too general for our physical problem, the reason being that the general form with its integration over $\gl$ has products of exponents in $\I u$ and $\I v$. These correspond in the supergravity description to the theta-angle-like terms. However as we were reminded of by \R{act1}, we can have either the $\gc$ or the $\gvf$ theta-angle and corresponding charges switched on. One either has the `$u$-instanton' ($\gc$) or the `$v$-instanton' ($\gvf$), not both simultaneously. Consequently we must restrict \R{solf2m} to reflect this property. \\
This can be done by letting the coefficient functions peak around $\gl=0$ and $\gl=1$. If we set
\begin{equation}
  A_{1,m}(\lambda) = A_{1,m}\ ,\qquad B_{1,m}(\lambda) = \delta
  (\lambda)\km \nn
 \end{equation}
we obtain \R{solf1m} as a function\footnote{We will refer to this configuration as the `v-instanton'. The v-anti-instanton has opposite signs in the exponentials: $e^{-\I v}$ instead of $e^{\I v}$. Similarly for the `u-instanton'.} of $v$: $f_{1,m}(v)$ with $k_{1,m}(0)=A_{1,m}$.
Similarly one can produce a solution independent of $v$ ($x=u$) by taking
\equ{
A_{1,m}(\gl)=\gd(\gl-1) \qquad B_{1,m}(\gl)=B_{1,m} \nn \km
}
which produces $f_{1,m}(u)$ with $k_{1,m}(0)=B_{1,m}$ as integration constants. Naturally, the sum of these two solutions is a solution to \R{f1m} as well. The one-instanton sector $e^h$ can therefore completely be expressed in terms of the one-dimensional solutions as
 \bea
  &&\exp [h(r,u,v)] = \exp [h_\text{pert}(r)] + \exp [h_\text{1-inst}(r,u)
  ] + \exp [h_\text{1-inst}(r,v)] + \ldots \nn \\ [2.9mm]
 && \exp [h_\text{pert}(r)] = r + c \nn \\[2mm]
 && \exp [h_\text{1-inst}(r,u)]  = e^{-2 \sqrt{r+c}}\, \sum_{m\geq m_1}
    f_{1,m}(u)\, (r+c)^{-m/2} \label{eq:9.1}
\eea
and similarly for $h_\text{1-inst}(r,v)$. We see that the $u$- and the $v$-instanton contribute separately to $e^h$ and thus to the line element \R{tod1}. The ellipses denote the higher $(n>1)$ instanton contributions. We have applied the shift in $r$: $r\to r+c$ to incorporate the one-loop result. 
For later reference, we also give the leading order expression for
$e^h$ in the regime $r\mg 1$ (small string coupling). To leading order
in the semi-classical approximation, the instanton solution \R{9.1}
reads
 \begin{equation} \label{eq:eh-exp}
  e^h = r + c + \frac{1}{2}\, r^{-m_1/2}\, \big( A_{1, m_1}\, e^{\I v}
  + A_{1, m_1}^\star\, e^{-\I v} + B_{1, m_1}\, e^{-\I u} + B_{1, m_1}^\star\,
  e^{\I u} \big)\, e^{-2\sqrt{r}} + \ldots 
 \end{equation}
Note that we need to include both instantons and anti-instantons to
obtain a real solution. 
To find the leading-order instanton corrected hypermultiplet metric, we
first compute the leading corrections to $f$ defined in \R{tod3}:
 \bea 
  && f = \frac{r + 2c}{r+c} + \label{eq:finst} \\[1.5mm]
  && \frac{1}{2}\, r^{-(m_1 + 1)/2}\, \big( A_{1,
  m_1}\, e^{\I v} + A_{1,m_1}^\star\, e^{-\I v} + B_{1,m_1} \, e^{-\I u}
  + B_{1,m_1}^\star\, e^{\I u} \big)\, e^{-2\sqrt{r}} + \ldots  \nn
  \eea
Substituting this result into \R{tod4} yields the leading
corrections to the $\Theta$ $1$-form. Setting
 \begin{equation}
  \Theta = u\, \d v\ + \Theta_\text{inst}\km
 \end{equation}
these are given by
\begin{equation}\label{eq:Tinst}
 \Theta_\text{inst} = r^{-m_1/2}\, e^{-2 \sqrt{r}}\, \Im \{ A_1\,
  e^{\I v}\, \d u + B_1\, e^{-\I u}\, \d v \}\ + \ldots\pt
\end{equation}
The leading order corrections to the hypermultiplet scalar metric are
then obtained by substituting these expressions into the PT metric
\R{tod1}.\\

\subsection{The two-instanton sector}
We now briefly discuss the $N=2$ sector. The Toda equation requires at
this level
 \begin{align}\label{eq:N=2sector}
  0 & = (\Delta + 4)\, f_{2,m} + 2a_m\, f_{2,m-1} + b_m\, f_{2,m-2}
    \notag \\[2pt]
  & \tab + \sum_{m'} \big[ f_{1,m-m'-2} + a_{m'+1}\, f_{1,m-m'-3}
    + b_{m'+2}\, f_{1,m-m'-4} \notag \\[-6pt]
  & \mspace{76mu} - \nabla f_{1,m-m'-2} \cdot \nabla \big] f_{1,m'}\ ,
 \end{align}
where we have used \R{f1m} for $\Delta f_{1,m'}$ in the double sum.
We have not derived the general solution to these equations in closed
form. The one-dimensional truncation is straightforward to
solve order by order in $m$. At lowest order\footnote{In appendix
\ref{toda1} we show that $-2\leq m_n\leq m_{n'}$ for $n\geq n'$.}
$m_2$ we have
 \begin{equation} \label{eq:2m0eq}
  (\Delta + 4)\, f_{2,m_2} + \gd_{m_2,-2}\, \big[ (f_{1,m_2})^2 -
  (\nabla f_{1,m_2})^2 \big] = 0\ .
 \end{equation}
Note that the inhomogeneous term is present only for the lowest possible
value $m_2=-2$. The one-dimensional truncation yields the equation
 \begin{equation}
  (\partial_x^2 + 4)\, f_{2,m_2}(x) + \gd_{m_2,-2}\, \Re \big\{ A_{1,m_2}^2
  e^{2\I x} \big\} = 0\ ,
 \end{equation}
where we have inserted the solution \R{solf1m0} for $f_{1,m_1}(x)$.
The general solution then reads
 \begin{align}
  f_{2,m_2}(x) & = \Re \big\{ A_{2,m_2} e^{2\I x} + \tfrac{1}{4}
    \gd_{m_2,-2}\, A_{1,m_2}^2\, \I x\, e^{2\I x} \big\} \notag
    \\[1mm]
  & = \Re \big\{ A_{2,m_2} G_0(2x) - \tfrac{1}{8} \gd_{m_2,-2}\, A_{1,
    m_2}^2 G_1(2x) \big\}\ ,
 \end{align}
$A_{2,m_2}$ being a further complex integration constant. 

The solution for $m>m_2$ can now be constructed by solving the
appropriate equation arising from \R{N=2sector}. Based on
\R{solf2m} we can also construct the general $(u,v)$-dependent
solution for $f_{2,m_2}(u,v)$. The idea is to decompose the products of
$\cos(\lambda_1 u)\,\cos(\lambda_2 u)$, etc., appearing in the
inhomogeneous part of \R{2m0eq} into a sum of $\cos$ and $\sin$
terms using product formulae for two trigonometric functions. We can
then construct the full inhomogeneous solution by superposing the
inhomogeneous solutions for every term in the sum. We refrain from
giving the result,  since it is complicated and not particularly
illuminating.

We conclude this subsection by giving an argument that the iterative
solution devised above indeed gives rise to a consistent solution of the
Toda equation. The general equations which determine a new $f_{n,m}(u,
v)$ are two-dimensional Laplace equations to the eigenvalue $n^2$
coupled to an inhomogeneous term, which is completely determined by the
$f_{n,m}(u,v)$'s obtained in the previous steps of the iteration
procedure. These equations are readily solved, e.g., by applying a
Fourier transformation. It then turns out that the iteration procedure
is organized in such a way that any level in the perturbative expansion
\R{4.5} determines one `new' $f_{n,m}(u,v)$, i.e., there are no
further constraints on the $f_{n,m}(u,v)$ determined in the previous
steps. This establishes that the perturbative approach indeed extends to
a consistent solution of the Toda equation \R{tod2}.

\subsection{The isometries}
\label{sect4.3}
Based on the Toda solution \R{9.1} we now discuss the breaking of
the Heisenberg algebra \R{Heis-alg2} in the presence of membrane
instantons to a discrete subgroup. We start with the shift symmetry in the axion $\sigma
\rightarrow\sigma-\alpha$. By identifying $t=\sigma$, this shift 
corresponds to the isometry of the PT metric, so that it cannot
be broken by the instanton corrections.

Analyzing the $\beta$ and $\gamma$-shifts is more involved. Under the
identification \R{fmap1} the $\beta$-shift acts as $v\rightarrow
v+\beta$. Taking the leading order one-instanton solution
\R{eh-exp}-\R{Tinst}, we find that $e^h$ as well as the
resulting functions $f$ and $\Theta$ appearing in the metric depend on
$v$ through $e^{\pm\I v}$ or $dv$ only. These theta-angle-like terms
break the $\beta$-shift to the discrete symmetry group\footnote{This agrees with earlier observations made in
\cite{Becker:1999pb}.}
$\mZ$. Going beyond the leading instanton corrections by taking
into account higher loop corrections around the single instanton will
 generically break the $\beta$-shift completely. This is due to the
appearance of polynomials in $v$ multiplying the factors $e^{\pm\I v}$.
We point out that by setting the integration constants
multiplying the terms odd in $v$ to zero, there is still an unbroken
$\mZ_2$ symmetry. This symmetry is defined by $v\rightarrow-v$, $t\rightarrow-t$, which amounts to 
interchanging $v$-instantons and anti-instantons.

To deduce the fate of the $\gamma$-shift, $u\rightarrow u+\gamma$,
$t\rightarrow t-\gamma v$, we first observe that $t\rightarrow t-\gamma
v$ implies that the combination $dt+udv$ is invariant.
Applying the same logic as for the $\beta$-shift above, we then find
that the one-loop corrections of a single $u$-instanton break the
$\gamma$-shift to the discrete symmetry $\mZ$, which will be
generically broken by higher order terms appearing in the loop
expansion. Similar to the $\beta$-shift, however, we can arrange the
constants of integration appearing in the solution in such a way that
there is also a $\mZ_2$ symmetry. We expect that these two
$\mZ_2$ symmetries could play a prominent role when determining
(some of) the coefficients appearing in the solution \R{9.1} from
string theory.

\section{String theory}
So far we have calculated membrane effects from a supergravity point of view. We already noted that this construction by itself was too general and needed to be tailored to the physical problem. Therefore we would like to compare these results with a microscopic string derivation of membrane effects. 
In \cite{Becker:1995kb} Becker, Becker and Strominger did precisely that, i.e., they considered the compactification of M-theory on a CY $3$-fold (and a circle) with membranes wrapped along $3$-cycles. These membranes give rise to certain corrections to $4$-fermion correlation functions. This is not surprising: we already saw in section \ref{unbroken} on page \pageref{mem4a} that the membrane preserves one half  of the supersymmetries. Conversely, it breaks one half of the supersymmetries. The resulting four fermionic zero modes lead to nonvanishing $4$-fermion correlators. \\
We will review the analysis of \cite{Becker:1995kb} and explicitly evaluate their result for the $4$-fermion correlation functions in the case of a rigid Calabi-Yau manifold. Then we shall compute the same object, but starting with the results obtained in the previous sections. We will find that these results agree beautifully. \\

\subsection{The string calculation}
\label{stcal1}
In section \ref{11d} we gave the $11$-dimensional supergravity action. Let us write it down again in Euclidean signature with the conventions of \cite{Becker:1995kb}
\equ{
S_{11}=\frac{1}{2\gp^2}\int d^{11}x \sqrt{g} \big[ -R + \frac{1}{48} (dA_3)^2\big] + \frac{\!\I}{12\gp^2}\int A_3\wedge dA_3 \wedge dA_3 \label{eq:11dsugra} \km
}
where $A_3$ is the $3$-form potential, as in \R{11dsugra1}, note that we leave out the hat.  We work in units in which the $11$-dimensional Planck length  equals one.\\
The membrane solutions can effectively be described (for scales large compared to the thickness of the membrane) by its worldvolume action. This action is completely fixed by supersymmetry \cite{Bergshoeff:1987cm}\footnote{See also \cite{Bergshoeff:1987qx, Duff:1990xz, Hughes:1986fa}.} to be
\bea
S_3&=&\int d^3 \gs \sqrt{h}\big[ \2 h^{\ga\gb} \p{\ga}X^M \p{\gb}X^N g_{MN} \nn \\[1mm]
&& -\frac{\I}{2} \bgTh \gG^\ga \nabla_\ga \gTh + \frac{\I}{3!}\ge^{\ga\gb\gg}A_{MNP}\p{\ga}X^M \p{\gb}X^N \p{\gg}X^P + \ldots\big] \nn \km
\eea
where $\gs^\ga$ with $\ga,\gb, \gg=1,2,3$ are the worldvolume coordinates and $h_{\ga\gb}$ is the (auxiliary) worldvolume metric with Euclidean signature. The $X^M(\gs)$ with $M,N=1,\ldots,11$ describe the membrane configuration: the maps embedding the membrane into the $11$-dimensional space. $\gTh$ is an $11$-dimensional Dirac spinor. Only the leading terms in powers of the fermions are given, ellipses  denote higher order terms. \\
The global fermionic symmetries of $S_3$ act on the membrane fields as
\equ{
\gd_\ge \gTh=\ge \qquad \gd_\ge X^M=\I \bge \gG^M \gTh \nn \km
}
where $\ge$ is a constant spinor in $11$ dimensions. The worldvolume theory is also invariant under the so-called $\gk$ symmetries, local fermionic transformations
\equ{
\gd_\gk \gTh=2 P_+ \gk(\gs) \qquad \gd_\gk X^M=2\I \bgTh \gG^M P_+ \gk(\gs) \label{eq:kappa1} \km
}
with $\gk$ a spinor in $11$ dimensions. The projection operators 
\equ{
P_\pm=\2\left(1\pm \frac{\I}{3!}\ge^{\ga\gb\gg}\p{\ga}X^M\p{\gb}X^N\p{\gg}X^P\gG_{MNP}\right) \nn \km
}
obey (see \cite{Becker:1995kb})
\equ{
P_{\pm}^2=P_\pm \qquad P_+ P_-=0 \qquad P_+ + P_-=1 \label{eq:kappa2} \pt
}
A general bosonic membrane configuration $X(\gs)$ breaks all the global supersymmetries generated by $\ge$. Some unbroken supersymmetries remain only if they are compensated by a  $\gk$-transformation such that
\equ{
\gd_\ge \gTh + \gd_\gk \gTh=0 \nn \pt
}
If we apply $P_-$ and use \R{kappa1} we find that the \susy parameter which leaves the configuration invariant must obey
\equ{
P_-\ge=0 \label{eq:kappa3} \pt
}
Such spinors are left invariant under $P_+$, that is, $P_+\ge=\ge$ as can be seen using \R{kappa2}. On the other hand, if we have a spinor $\ge_0$ which obeys $P_+\ge_0=0$, \R{kappa2} gives $P_-\ge_0=\ge_0$. So this spinor breaks \susy and will therefore generate zero modes, compare with the discussion in section \ref{ubs}. \\

Condition \R{kappa3} must be examined in the case when we compactify on a CY $3$-fold $Y_3$ while the membrane is wrapped along its $3$-cycles. To this end introduce complex coordinates $X^m, X^\bn$ $m,\bn=1,2,3$ for the Calabi-Yau manifold. Furthermore, let $\ge_+=(\ge_-)^\star$ be two covariantly constant $6$-dimensional spinors 
with opposite chirality, $\ge_+$ has by definition positive chirality. Their normalization can be chosen such that
\equ{
\gg_{mnp}\ge_+=e^{-\cK}\gO_{mnp}\ge_- \qquad \gg_{\bm np}\ge_+=2\I K_{\bm [n}\gg_{p]} \ge_+ \qquad \gg_{\bm}\ge_+=0 \label{eq:5H} \km
}
with $K$ the \Kh form, see section \ref{kahlergeo}. We have introduced 
\equ{
\cK\equiv\2 \left(\cK_V-\cK_H\right) \label{eq:Kpot1} \km
}
where $\cK_H$ is the \Kh potential on the moduli space of complex structures\footnote{This potential was already introduced in equation \R{kahmod1}.} and $\cK_V$ the potential on the \Kh moduli space
\equ{
\cK_H=-\ln\left(\I\int_{Y_3} \gO\wedge \bgO\right) \qquad \cK_V=-\ln\left(\frac{4}{3}\int_{Y_3} K\wedge K\wedge K\right) \label{eq:Kpot2} \pt
}
If we split up the spinor $\ge$ in a $6$- and a $5$-dimensional\footnote{Later we will go to four spacetime dimensions, for now it does not matter.} part as $\ge_\gth\otimes \gl$ with $\gl$ the $5$-dimensional spinor and 
\equ{
\ge_\gth\equiv e^{\I\gth}\ge_+ + e^{-\I \gth}\ge_- \nn \km
}
$P_-\ge=0$ implies $P_-\ge_\gth=0$. We can examine this condition using \R{kappa2} and \R{5H} which give two conditions: 
\bea
&& \p{[\ga}X^m \p{\gb]} X^\bn K_{m\bn}=0 \label{eq:condi1} \\[2mm]
&& \p{\ga}X^m\p{\gb}X^m \p{\gs}X^p \gO_{mnp}=e^{\I \gvf}e^\cK \gve_{\ga\gb\gg} \label{eq:condi2} \km
\eea
with $\gvf\equiv 2\gth + \gp/2$. Thus the pullback of the \Kh form onto the membrane must vanish and the membrane volume element has to be proportional to the pullback of $\gO$ onto the membrane. If these two conditions are met, the map $X(\gs)$ is called a \emph{supersymmetric cycle} because then $P_-\ge=0$ is satisfied and supersymmetry is (partially) preserved.\\

We will consider  the case where the membrane is wrapped around such a supersymmetric cycle, denoted by $\cC_3$, such that we have four broken and four unbroken supersymmetries.\\
As in the previous chapter, the broken supersymmetries will give rise to corrections to the low energy effective action. We saw that the broken \susy parameters are those $\ge_0$ such that $P_+\ge_0=0$. With these spinors we can generate zero modes for the fermions in the hypermultiplets\footnote{The hypermultiplet sector is (almost) identical in four and five spacetime dimensions, see page \pageref{spinors}.} of the low energy effective theory. In \cite{Becker:1995kb} they are denoted by $\gc^I$; these are symplectic-Majorana spinors (see page \pageref{spinors}). The correlator
\equ{
\langle(\bgc_I\gc_J)(\bgc_K \gc_L)\rangle_\inst \label{eq:corf4a} \km
}
gives rise to membrane instanton corrections to a symmetric tensor $\cR_{IJKL}$ with indices $I,\ldots, L=1,\ldots,2n$, with $n=h^{(1,2)}+1$.  By constructing vertex operators for the spinors $\gc^I$ the correlator can be evaluated giving the correction
\equ{
\gD_{\cC_3}\cR_{IJKL}=N e^{-S_\inst}\int_{\cC_3}d_I\int_{\cC_3}d_J\int_{\cC_3}d_K\int_{\cC_3}d_L \label{eq:corrs1} \km
}
where $N$ is some normalization factor which includes an unknown determinant which may very well depend on $g_s$. The instanton action $S_\inst$ is given by
\equ{
S_\inst =e^{-\cK}\left|\int_{\cC_3}\gO\right| + \I \int_{\cC_3}A_3 \label{eq:535b} \km
}
where $A_3$ is the $3$-form potential from $11$-dimensional supergravity, see \R{11dsugra}. The $d_I$ form a real symplectic basis of $H^3(Y_3,\mZ)$ which obeys
\equ{
\int d^I\wedge d^J=\gve^{IJ} \nn\km
}
with $\gve^{IJ}$ the invariant antisymmetric tensor\footnote{Remember that we can use this tensor to raise and lower indices of the fermions, in this case of the $\gc_I$.} of $Sp(2h^{(1,2)}+2)$ and $I,J=1,\ldots,2h^{(1,2)}+2$. Note that this is  the same as \R{basissym1}, although for a different basis.
Equation \R{corrs1} is the object we have to evaluate explicitly. Introduce a basis of real $3$-cycles $(\cA^a, \cB_a)$ of $H_3(Y_3,\mZ)$ satisfying
\bea
&&\int_{\cA^a}\ga_b=\int_{Y_3}\ga_b\wedge \gb^a=\gd^a_b \nn \\[1mm]
&&\int_{\cB_b}\gb^a=\int_{Y_3} \gb^a\wedge \ga_b=-\gd^a_b \nn\\[1mm]
&&\int_{\cA^a}\gb^b=\int_{\cB_a}\ga_b=0 \label{eq:c5b} \pt
\eea
We have used the basis of harmonic $3$-forms $(\ga_a, \gb^b)$ with $a,b=0,\ldots,h^{(1,2)}$ as in \R{basissym1}. Note that to avoid confusion with the notation of \cite{Becker:1995kb} we have used the indices $a,b$. \\
For rigid Calabi-Yau manifolds $h^{(1,2)}=0$ and because we have only the one index $a=0$, we might as well omit it. As in section \ref{CY} we then proceed to define the periods of the holomorphic $3$-form $\gO$ as
\equ{
Z=\int_\cA\gO \qquad \cG=\int_\cB\gO \nn
}
and thus $\gO=Z\ga-\cG(Z)\gb$. The $Z$ are projective coordinates and we have only one $Z$, so we can take it to be equal to $1$. We choose the normalization such that\footnote{This choice of $\cG$ also ensures that $\cK_H$ is real.} $\cG=-\I Z^2$, $\gO$ then becomes 
\equ{
\gO=\ga+\I\gb \label{eq:tc5c} \pt
}
Together with the conventions in \R{c5b} this results in the volume of $Y_3$ being normalized to $1$. In general we have, see \cite{Bodner:1990zm}, for a CY $3$-fold
\bea
\int\gO\wedge\bgO &=&\left(\frac{1}{3!}\right)^2\int d^6\gx \,\gO_{ijk}\bgO_{\bl\bm\bn}\gve^{ijk}\gve^{\bl\bm\bn}\nn\\
&=&-\I \left|\!\left|\gO\right|\!\right|^2 \int d \textrm{Vol}(Y_3) \nn \\
&=&-\I \left|\!\left|\gO\right|\!\right|^2 \textrm{Vol}(Y_3)  \label{eq:volf2} \km
\eea 
with $\gx$ some complex coordinates on $Y_3$. This means that the volume form equals
\equ{
d \textrm{Vol}(Y_3)=\frac{\gO\wedge\bgO}{-\I  \left|\!\left|\gO\right|\!\right|^2 }=\ga\wedge\gb \label{eq:volf1} \km
}
where we have used \R{tc5c} and $\left|\!\left|\gO\right|\!\right|^2\equiv \frac{1}{3!}\gO_{ijk}\bgO^{ijk}=2$ . If we now use \R{c5b} we see that Vol$(Y_3)=\int\ga\wedge\gb=1$.\\ 
Our next job is to find the supersymmetric cycle $\cC_3$ for the rigid Calabi-Yau manifold. We shall demonstrate that the real $3$-cycles $\cA$ and $\cB$ are themselves supersymmetric cycles. \\
The first condition \R{condi1} states that the pullback of the \Kh form of $Y_3$ onto the membrane, that is onto $\cA$ or $\cB$, has to vanish.\\
To prove this, let us generalize the argument given in section $2.2$ of \cite{Becker:1995kb}. Suppose $Y_3$ has an antiholomorphic involution: an isometry $\gth: Y_3 \to Y_3$ such that $\gth^2=\Id$ and $\gth^\star J=-J$ with $J$ the complex structure. The effect on the \Kh form will then be $\gth^\star K=-K$, see \R{khform1}. The fixed points Fix$(\gth)$ of $\gth$ constitute a special Lagrangian submanifold\footnote{For more information on special Lagrangian manifolds and calibrated geometry see \cite{Joyce:2001xt, Gauntlett:2003di}. In \cite{Harvey:1982xk} it was shown that the two conditions (\ref{eq:condi1}, \ref{eq:condi2}) are actually equivalent.} and the pullback of the \Kh form  behaves as
\equ{
K\big|_{\textrm{Fix}(\gth)} =\gth^\star K\big|_{\textrm{Fix}(\gth)}=-K\big|_{\textrm{Fix}(\gth)} \nn \km
}
which holds if $K\big|_{\textrm{Fix}(\gth)}=0$. Note that this (discrete) isometry is nothing but complex conjugation: interchanging barred and unbarred coordinates
\equ{
\gth: X^m\to X^\bm \nn \pt
}
We see from \R{compls} that (the action of) $J$ indeed reverses sign under this isometry. The fixed points of $\gth$ are then the real submanifolds, specifically the real $3$-cycles $\cA$ and $\cB$. \\

The second condition \R{condi2}, which states that the pullback of $\gO$ must be proportional to the volume form of $\cC_3$, is also satisfied for $\cA$ and $\cB$. From equation \R{tc5c} we see that the pullback of $\gO$ onto cycle $\cA$ or $\cB$ is either (proportional to) $\ga$ or $\gb$ respectively. Equation \R{volf1} states that the volume form on $Y_3$ is given by $\ga\wedge \gb$. The Riemann bilinear identity
\equ{
\int_{Y_3}\gPs\wedge\gX=\sum_a \left(\int_{\cA^a}\gPs\int_{\cB_a}\gX -\int_{\cB_a}\gPs\int_{\cA^a}\gX\right) \nn \km
}
for arbitrary $3$-forms $\gPs$ and $\gX$, can be applied to the volume integral giving
\equ{
\textrm{Vol}(Y_3)=\int_{Y_3}\ga\wedge\gb=\int_{\cA}\ga\int_\cB\gb \nn\pt
}
Thus $\ga$ and $\gb$ correspond to the induced volume forms on $\cA$ and $\cB$ respectively. We can therefore conclude that $\cA$ and $\cB$ are the supersymmetric cycles.\\
Suppose we take a linear combination of the cycles $\cA$ and $\cB$: 
\equ{
\cC_3=m\cA+n\cB \label{eq:susyc1} \km
}
with $m$ and $n$ integers, then this is not a supersymmetric cycle. The reasoning is as follows. We concluded that the (induced) volume forms on $\cA$ and $\cB$ are $\ga$ and $\gb$ respectively. The volume of $m$ times $\cA$ then becomes
\equ{
\int_{m\cA}\ga=m\int_\cA \ga=m \nn \km
}
using \R{c5b}, similarly for $n\cB$. Together they determine the volume of the $\cC_3$ cycle to be $m+n$. On the other hand, integrating (the pullback of) $\gO$ gives
\equ{
\int_{\cC_3}\gO=m-\I n \label{eq:OonC} 
}
which is not proportional to $m+n$. The two are only proportional if either $m$ or $n$ is zero.
Consequently, a membrane wrapping $\cA$ and $\cB$ simultaneously is not a supersymmetric configuration and will not contribute to $\cR_{IJKL}$. The membrane needs to be wrapped either on $\cA$ or on $\cB$ to be supersymmetric.\\

Having identified the supersymmetric cycles, we can evaluate \R{535b} and \R{corrs1}.
The integral over $\gO$ in $S_\inst$ \R{535b} is easily evaluated using \R{OonC}
\equ{
\left|\int_{\cC_3}\gO\right|=\sqrt{m^2+n^2} \label{eq:go1} \pt
}
To do the integral of $A_3$ over $\cC_3$ in \R{535b} we must expand the $3$-form potential $A_3$ as
\equ{
A_3=C_3 + v\ga +u\gb \km\label{eq:spacetimec1}
}
where $C_3$ is the spacetime $3$-form potential dual to $e_0$. $v$ and $u$ are functions of the four dimensional spacetime coordinates, we left out the vector multiplet terms. This is the same expansion as in \R{3form1e} (see also the discussion on page \pageref{hereI}). Evaluating this $A_3$ on the general cycle $\cC_3$, see \R{susyc1}, results in
\equ{
\I\int_{\cC_3}A_3=\I mv-\I nu \label{eq:temp5ac} \pt
}
In calculating $\cK$, see \R{Kpot1}, we must be a bit careful. The $\cK_H$ part is easy, using \R{volf1} and \R{volf2} we obtain
\equ{
\I\int_{Y_3}\gO\wedge\bgO=2 \label{eq:eqi}\pt
}
To compute $\cK_V$ we use a different (but equivalent) formula for the volume of a Calabi-Yau manifold:
\equ{
\textrm{Vol}(Y_3)=\frac{1}{3!}\int_{Y_3} K\wedge K\wedge K \nn\pt
}
This formula is not directly applicable because we have to compactify\footnote{We could have done this in an earlier stage, but it does not matter for the decomposition \R{spacetimec1} since that is with respect to the Calabi-Yau. This is why we left out the hat on $A_3$ from the beginning, compare also with \R{hgh1}.} one dimension, since we are still in $5$ dimensions and we would like to compute the corrections to the geometry of the UHM in $4$ spacetime dimensions. In the string frame this is achieved (compare with \R{stf1}) by decomposing the $10$-dimensional metric as
\equ{
ds^2_{11}=e^{-2\gf/3}(dx_{11}+A_m dx^m) + e^{\gf/3}ds^2_{10} \label{eq:compa1} \km
}
where the $10$-dimensional part incorporates $Y_3$. For the \Kh form it has the consequence that it picks up a dilaton dependence:
\equ{
K=e^{\gf/3}K_{\textrm{s.f.}} \nn \km
}
where $K_{\textrm{s.f.}}$ is the \Kh form in the string frame. This means that the we can write
\equ{
\frac{4}{3}\int_{Y_3}K\wedge K\wedge K=8\textrm{Vol}(Y_3)e^\gf=8e^\gf \nn\km
}
as measured in the string frame. Using this result and \R{eqi} we calculate
\equ{
e^{-\cK}=2e^{\gf/2} \nn \pt
}
The total result for $S_\inst$ using the PT variable $r$ can thus be written as
\equ{
e^{-S_\inst}=e^{-2\sqrt{m^2+n^2}\sqrt{r}}e^{-\I mv+\I nu} \label{eq:stinsta1} \pt
}
Observe that we really can interpret the $m$ and $n$ as instanton charges. Keep in mind though that for a supersymmetric configuration only one instanton can be switched on at a time, as we already anticipated in section \ref{mem1}. 
Comparing the result for $S_\inst$ in \R{stinsta1}, which originated from \R{535b}, with the supergravity result \R{act1}, we find agreement. In \R{stinsta1} we should consider only $m$ or $n$ unequal to zero. Similarly, in \R{act1} we have either the membrane charge due to dualizing with respect to $\gvf$ or to $\gc$. We see that in the string picture the charges ($n$ and $m$) are quantized, i.e., they are integers. Furthermore, not only is the $g_s$($=\sqrt{r}$) behaviour correct, but the numerical factor of $2$ matches as well.\\
Lastly there is the tensorial structure of \R{corrs1}. The supersymmetric cycle is either $\cA$ or $\cB$ which means that the $d^I$ ($I=1,2$) should all be equal, $\cA$ giving corrections to $\cR_{1111}$ and $\cB$ to $\cR_{2222}$. \\
Summarizing, we can write for the corrections to $\cR$ due to the two supersymmetric cycles:
\bea
&& \gD_\cA \cR_{1111}=N e^{-2|m| \sqrt{r}}e^{-\I mv} \nn \\[1.5mm]
&& \gD_\cB \cR_{2222}=N e^{-2|n| \sqrt{r}}e^{+\I nu} \label{eq:Rres1} \km
\eea
the other combinations are zero.
In the next subsection we will make the comparison between the string theory result and the results in the PT framework more precise.

\subsubsection{A note on the spinors}
\label{spinors}
In \R{corf4a} the correlator is written down which corrects the tensor $\cR$ as in \R{corrs1}. This expression is valid for five spacetime dimensions, with Lorentzian signature. We have evaluated \R{corrs1} giving the result \R{Rres1}. In the process we went down to four spacetime dimensions, by compactifying on a circle \R{compa1}. In five spacetime dimensions the $2n$ spinors $\gc^I$ are \emph{symplectic-Majorana} spinors. The point is that in four space and one time dimension, one cannot impose a Majorana condition. But if one has an even number of Dirac spinors $\gc^I$, one can impose a reality condition:
\equ{
\gc^{I\star}=\gO_{IJ}\cB \gc^J \label{eq:temp5a} \km
}
with $I=1,\ldots,2n$. The matrix $\gO$ is real and satisfies
\equ{
\gO_{IJ}\,\gO_{JK}=-\gd_{IK} \nn \pt
}
Equation \R{temp5a} is consistent if $\cB^\star \cB=-1$. The $\cB$ matrix has to do with the fact that if the set $\{\gG^a\}$ represents the Clifford algebra, in a certain number of dimensions with a certain signature, then $\{\pm\gG^{a\star}\}$ does so as well. The $\cB$ matrices then provide a similarity transformation relating the $\gG^a$ and $\gG^{a\star}$
\equ{
\cB_{\pm}\gG^a\cB_\pm^{-1}=\pm \gG^{a\star} \nn \pt
}
The existence of the $\cB_\pm$ depends on the number of dimensions and the signature. In five spacetime dimensions they do exist and obey $\cB^\star_\pm \cB_\pm=-1$, so we can either use $\cB_-$ or $\cB_+$ in \R{temp5a}, which is why we denoted it by $\cB$.
The $\cB_\pm$ matrices are related to the charge- and the Dirac conjugation matrix. For a nice account see \cite{Ortin:2004ms}.\\
If one then goes to four spacetime dimensions, one can impose the normal Majorana condition, which reads in this notation
\equ{
\gc^{I\star}=\cB\gc^I \nn \pt
}
The reduction of the spinors can be obtained by the following identification
\equ{
\bgl^\ba \sim (1+\gg_5)\gc^I \qquad \gl^a\sim (1-\gg_5)(\gO\gc)^I \nn \km
}
where $a,\ba=1,\ldots,2n$ as in chapter \ref{ch3}. With this notation we mean to indicate that a single $\gc^I$ gives two spinors: $\gl^a$ and $\bgl^\ba$. The projection operators $(1\pm\gg_5)$ thus give us chiral spinors. Because under complex conjugation the indices are interchanged we indeed have a system of $2n$ Majorana spinors. For more information on the relation between the geometry of hypermultiplets in five and four spacetime dimensions and the reduction, see \cite{Rosseel:2004fa}.

\subsection{Comparing to the PT metric}
We explicitly constructed the corrections to the completely symmetric tensor (see \R{Rres1}) and we must compare this with the instanton corrections to a completely symmetric tensor in the four-dimensional $N=2$ supergravity theory.
This tensor has been constructed in \cite{Bergshoeff:2004nf, Bergshoeff:2002qk}. The symmetric tensor $\cW_{abcd}$ (where the indices are $Sp(2n,\mR)$ indices, with $n=1$ in our case) can be obtained from the curvature decomposition 
\begin{equation} \label{eq:r.1}
  R_{\hA\hB\hC\hD} = \nu (R^{SU(2)})_{\hA\hB\hC\hD} + \frac{1}{2}\, {L_{\hD\hC}}^{a
  b}\, \cW_{abcd}\, {L_{\hA\hB}}^{cd}\pt
\end{equation}
In our conventions, see appendix \ref{geopt}, we have $\gn=-1/2$. 
$R_{\hA\hB\hC\hD}$ is the curvature tensor of the quaternionic manifold, so $\hA,\ldots,\hD=1,\ldots,4n$ and the $SU(2)$ part of the tensor is given by
\begin{equation}
  (R^{SU(2)})_{\hA\hB\hC\hD} = \frac{1}{2}\, g_{\hD [\hA}\, g_{\hB ]\hC} + \frac{1}{2}\,
  K^r_{\hA\hB}\, K^r_{\hD\hC} - \frac{1}{2}\, K^r_{\hD[\hA}\,
  K_{\hB ]\hC}^r\km \nn
 \end{equation}
with 
 \begin{equation}
  {L_{\hA\hB a}}^b = V_{\hA ia}\, \bar{V}^{ib}_\hB\pt
 \end{equation}
We can solve \R{r.1} for $\cW_{abcd}$ by using the inverse relation for ${L_{\hA\hB}}^{ab}$:
 \begin{equation}
  - \frac{1}{2}\, V^{i\hB}_c\, V^{\hA}_{id}\, {L_{\hA\hB}}^{a
  b} = \gd^a_c \, \delta^b_d\pt \nn
 \end{equation}
The resulting expression for $\cW_{abcd}$ then reads:
 \begin{equation}
  \cW_{abcd} = \frac{1}{2}\, \gve^{ij}
  \gve^{kl}\, V_{id}^\hA\, V_{jc}^\hB\, V_{kb}^\hD\, V_{l
  a}^\hC \big( R_{\hA\hB\hC\hD} - \nu (R^{SU(2)})_{\hA\hB\hC\hD} \big)\pt
 \end{equation}
Starting from the general PT metric \R{tod1} we can compute the rather complicated curvature tensor and by using our expressions for the vielbeins and complex structures, see appendix \ref{geopt}, we can compute $\cW_{abcd}$. The result is 
\begin{align} \label{r.2}
  \cW_{1111} & = 4 r^2 f^{-3} e^{-h} \big[ f (\partial^2_{\bz} f - \partial_{\bz}
    h\, \partial_{\bz} f) - 3 (\partial_{\bz} f)^2 \big] \notag \\[2pt]
  \cW_{2111} & = r^2 f^{-3} e^{-h/2} \big[ 2 f\, \partial_r \partial_{\bz} f - 3
    (f\, \partial_r h + 2 \partial_r f)\, \partial_{\bz} f + f^2 \partial_r \partial_{\bz} h \big]
    \notag \\[2pt]
  \cW_{2211} & = - r^2 f^{-3} \big[ f \big( r \partial^3_r h - (\partial_r h)^2
    \big) - 4 e^{-h}\, \partial_z f\, \partial_{\bz} f + 2 (\partial_r f)^2 \big]
    \notag \\[2pt]
  \cW_{2221} & = - r^2 f^{-3} e^{-h/2} \big[ 2 f\, \partial_r \partial_z f - 3
    (f\, \partial_r h + 2 \partial_r f)\, \partial_z f + f^2 \partial_r \partial_z h \big] \notag
    \\[2pt]
  \cW_{2222} & = 4 r^2 f^{-3} e^{-h} \big[ f (\partial^2_z f - \partial_z h\, \partial_z
    f) - 3 (\partial_z f)^2 \big]\pt
 \end{align}
We have introduced the complex variable $z\equiv u+\I v$ to keep the expressions compact\footnote{The fact that we have such a compact expression is absolutely delightful given the \emph{very} unwieldy result for the curvature tensor, which we will not give.}. Note that $\cW_{1111}=\cW_{2222}^\star$, $\cW_{2111}=-\cW_{1222}^\star$ and $\cW_{2211}$ is real.\\
Given this general form of $\cW_{abcd}$ we compute its instanton contributions. We find that at the perturbative level (the classical plus $1$-loop correction but no instantons) specified  by \R{pertu1}, the only nonvanishing component is given by
\begin{equation}
  \cW_{2211} = - \frac{r^3}{(r+2c)^3}\pt \nn
 \end{equation} 
As for the leading order instanton corrections: the $1$-instanton sector up to $1$-loop, specified by \R{eh-exp} and \R{finst}, gives the following contributions to $\cW_{abcd}$
\begin{align} \label{eq:r.3}
  \Delta \cW_{1111} & = \hN\, \big[ A_1\, e^{\I v} + A_1^\star\, e^{-\I v}
    - B_1\, e^{-\I u} - B_1^\star\, e^{\I u} \big] \notag \\[2pt]
  \Delta \cW_{1112} & =  \hN\, \big[ A_1\, e^{\I v} - A_1^\star\, e^{-\I v}
    + \I (B_1\, e^{-\I u} - B_1^\star\, e^{\I u}) \big] \notag \\[2pt]
  \Delta \cW_{1122} & =  \hN\, \big[ A_1\, e^{\I v} + A_1^\star\, e^{-\I v}
    + B_1\, e^{-\I u} + B_1^\star\, e^{\I u} \big] \notag \\[2pt]
  \Delta \cW_{1222} & =\hN\, \big[ A_1\, e^{\I v} - A_1^\star\, e^{-\I v}
    - \I (B_1\, e^{-\I u} - B_1^\star\, e^{\I u}) \big] \notag \\[2pt]
  \Delta \cW_{2222} & = \hN\, \big[ A_1\, e^{\I v} + A_1^\star\, e^{-\I v}
    - B_1\, e^{-\I u} - B_1^\star\, e^{\I u} \big]\pt
 \end{align}
We have defined $\hN\equiv (r+c)^{(1-m_1)/2} e^{-2\sqrt{r+c}}$. Comparing the $r$-dependence of $\hN$ with the $r$-dependence of $N$ in \R{Rres1} then fixes the value of $m_1$. In particular, if $N$ is  $r$-independent $m_1=1$, in order to match $\hN$ to $N$.\\
Let us repeat \R{Rres1} for the case of the the single instanton:
\equ{
 \gD_{\cA}\cR_{1111}=N e^{-2\sqrt{r}}e^{-\I v} 
\qquad \gD_{\cB}\cR_{2222}=N e^{-2\sqrt{r}}e^{+\I u} \label{eq:Rres1b} \pt
}
This indicates that the integration coefficient $A_1$ must be associated to the $\cA$-cycle and $B_1$ to the $\cB$-cycle. \\
Yet \R{r.3} seems very different from the string result \R{Rres1b}.
However, the latter result expresses something different: it is the result from \emph{one} membrane wrapping \emph{one} supersymmetric cycle, either $\cA$ or $\cB$. In contrast,  \R{r.3} includes both types of instantons and anti-instantons. Moreover, the fermion frame used by \cite{Becker:1995kb} is different from ours. So result \R{Rres1b} is in a different frame from ours. Still, these frames can differ at most by a local $SU(2)$ ($Sp(2,\mR)$) rotation. 
The rotation group has to be compatible with the
reality condition imposed on the pair of symplectic Majorana spinors
coupling to $\cW_{abcd}$ and preserve fermion bilinears. These conditions then
lead to the fact that the most general transformation is given by $SU(2)$, this is discussed in \cite{Cortes:2003zd}.
We can parametrize such a $SU(2)$ rotation by 
\begin{equation}
  U = \lp e^{\I\xi} \cos\eta & e^{\I\rho} \sin\eta \\[1mm]
  -e^{-\I\rho}  \sin\eta & e^{-\I\xi} \cos\eta \rp\km \nn
 \end{equation}
where the parameters $\eta$, $\xi$, $\rho$ can depend on the scalars, although we will find that this will not be necessary. 
Consider the contribution to $\gD\cW$ arising only from the $\cB$-instanton. This requires setting $A_1=0$. Upon performing a global $SU(2)$ rotation of the fermion frame with parameters $\get=\gp/4$ and $\gx=\gr+\gp/2$ we obtain
\equ{
\gD\tilde{\cW}_{1111}=-4 \hN B^\star_1 e^{\I u} \qquad \gD \tilde{\cW}_{2222}=-4 \hN B_1 e^{-\I u} \km \label{eq:trW1}
}
with the other components vanishing identically. Note that the remaining free parameter $\gr$ in the transformation only induces a phase on the components of $\gD \tilde{\cW}_{abcd}$. We can set it to zero for convenience. This reproduces\footnote{Take $u\to -u$, which corresponds to an orientation reversal of the cycle, see \R{temp5ac}.} \R{Rres1b}, which is due to one instanton without anti-instantons. The correction \R{trW1} incorporates both the instanton and anti-instanton, which necessarily appear in different sectors of the theory\footnote{Just as in the case of the NS $5$-brane for which the zero modes in the fermions due to the $5$-brane instanton were in a different sector than those due to the $5$-brane anti-instanton see \R{brfermions1}.}.\\
Likewise, we can consider the contribution of the $\cA$-instanton by setting $B_1=0$. In this case the transformation $\get=\gp/4$ and $\gx=\gr=0$ leads to the correction
\equ{
\gD \tilde{\cW}_{1111}=4\hN A_1 e^{\I v} \qquad \gD \tilde{\cW}_{2222}=4 \hN A^\star_1 e^{-\I v} \label{eq:trW2} \km
}
with the other components vanishing identically. This is again of the form predicted by string theory \R{Rres1b}, although in a different fermionic frame from \R{trW1}. Summing these two corrections involves rotating some of the contributions into the proper fermionic frame. Hence, our result precisely agrees with the one obtained in \cite{Becker:1995kb}.\\
In order to sum the two contributions, we go to the $\cB$-instanton frame, i.e., the frame in which we obtained \R{trW1}, but now we also include $A_1$. The corrections to the $\cW$-tensor are then given by
\bea
&& \gD \tilde{\cW}_{1111}=-\hN\left(A_1 e^{\I v} + A^\star_1 e^{-\I v} + 4 B^\star_1 e^{\I u}\right) \nn \\[1mm]
&& \gD \tilde{\cW}_{1112}=\hN\left(A_1 e^{\I v} - A^\star_1 e^{-\I v}\right)  \nn \\[1mm]
&& \gD \tilde{\cW}_{1122}=-\hN\left(A_1 e^{\I v} + A^\star_1 e^{-\I v}\right) \nn \\[1mm]
&& \gD \tilde{\cW}_{1222}=\hN\left(A_1 e^{\I v} - A^\star_1 e^{-\I v}\right) \nn \\[1mm]
&& \gD \tilde{\cW}_{2222}=-\hN\left(A_1 e^{\I v} + A^\star_1 e^{-\I v} + 4 B^\star_1 e^{\I u}\right) \label{eq:trW4}  \pt
\eea
This result implies that the four fermionic zero modes $\gps_1, \gps_2$ arising from a membrane wrapping the $\cA$ and the $\cB$ cycle, respectively, are not orthogonal. 
If in the $\cB$-frame we denote the two zero modes giving rise to the $e^{\I u}$ corrections by $\gps_1$, then the corrections proportional to $e^{\I v}$ arise from the zero modes $\gps_1-\gps_2$. This zero mode configuration produces all the signs appearing in \R{trW4}, since the $\cA$-anti-instanton has its zero modes in $\gps_1+\gps_2$.

\section{Compactification with fluxes}
\label{fluxes}
In chapter \ref{ch2} we sketched the compactification from $10$ to $4$ dimensions, \emph{without} fluxes. However, in a compactification  also more complicated field strengths can be considered. This introduces more parameters and is referred to as compactification \emph{with} fluxes.\\
Ideally one would like to compactify while turning on all possible fluxes, because this gives the most general effective action in four dimensions.  It is a rather extended subject and we will refer to the literature for more information, specifically \cite{Louis:2002ny, Kachru:2004jr, Grana:2005jc} and references therein.\\
We will illustrate the mechanism by considering again the $2$-form potential from section \ref{fourd}. In \R{Bexp1}  the $10$-dimensional $2$-form $\hB_2$ was expanded as $\hB_2=B_2 + b^i \go_i$, where the  $\go_i$ are $(1,1)$-forms and the $b^i$ scalar fields in four dimensions. The corresponding field strength  is defined as
\equ{
H_3\equiv d B_2 + db^i \go_i \nn \pt
}
Remember that $d\go_i=0$, so this quantity obeys the Bianchi identity $dH_3=0$. More generally we can include fluxes in the definition of the field strength:
\equ{
H_3=dB_2 + db^i \go_i + H_3^{\mathrm{flux}} \label{eq:flux1} \km
}
where $H_3^{\mathrm{{flux}}}$ is a $3$-form on the Calabi-Yau $3$-fold
\equ{
H_3^{\mathrm{flux}} \equiv p^A \ga_A + q_A \gb^A \nn 
}
and the $(\ga_A, \gb^A)$ again form a real basis of $H^3(Y_3)$. This gives rise to extra flux parameters $p^A$ and $q_A$ which have to be constant. They are obtained by integrating $H_3^{\mathrm{flux}}$ over the $3$-cycles of the Calabi-Yau.  In the $4$-dimensional effective action these parameters gauge certain fields, namely the scalars $\gx^a$ and $\tgx_a$ from the hypermultiplets \R{hyperm1} with $a=1,\ldots,h^{(1,2)}$. Concretely, the action will now contain terms
\equ{
D_\gm \gx^a\equiv \p{\gm}\gx^a -p^a A_\gm \qquad D_\gm \tgx^a\equiv \p{\gm}\tgx^a -q^a A_\gm \km \nn
}
where $A_\gm$ is the graviphoton, similarly for the axion (see \cite{Louis:2002ny}). In addition there will be a potential containing these parameters. The parameters gauge certain isometries of the scalar manifold \R{scm1}. The scalar manifold\footnote{We will ignore the vector multiplets, and thus $\cM_V$, for the moment.} possesses a number of isometries. For instance, we have seen in section \ref{instsol} that the quaternionic target space of the classical (and perturbative) UHM has a number of shift symmetries acting on the scalars, see \R{heis1}. These are collectively denoted by
\equ{
\gd \gf^\hA=\gL^I k^\hA_I(\gf) \nn \km
}
where the $\gf^\hA$ are the $4$ scalars in the UHM\footnote{The generalization to more hypermultiplets is straightforward.}. $k^\hA_I$ are the Killing vectors and $\gL^I$ their parameters, the $I$ denotes the various isometries. For example, the action of the $\ga$ shift on $\gs$ in \R{heis1} can be written as
\equ{
\gd \gs=\gL^I k_I^\gs =\gL^\ga k_\ga^\gs \nn \km
}
with $\gL^\ga=-\ga$ and  $k^\gs_\ga=1$. \\
To gauge such isometries one introduces gauge covariant derivatives
\equ{
D_\gm \gf^\hA \equiv \p{\gm}\gf^\hA-k^\hA_I(\gf) A^I_\gm \label{eq:gaugecov1} \km
}
where $A^I_\gm$ are the vectors from the vector multiplets plus the graviphoton. In addition one has to gauge the various connections, which give rise to gauged curvatures. Furthermore, a potential has to be added to the action in order to preserve supersymmetry.\\
The part of the potential which involves quaternionic quantities is of the form
\equ{
V=e^{\wK} X^I\bX^J\left(2\gk^{-2} G_{\hA\hB} k_I^\hA k_J^\hB -3 P_I^r P_J^r\right) \label{eq:potI} \pt
}
The \Kh potential $\wK$ is given by \R{kahlerpot1} and the $X^I$  ($I=0,\ldots,n_v$ with $n_v$ the number of vector multiplets) are holomorphic functions of the scalars\footnote{These scalars are defined as $t^i\equiv b^i+\I v^i$, see the footnote on page \pageref{spk2}.} $t^i$ in the vector multiplets, precisely the sections from \R{Osection1}. 
The Killing vectors $k^\hA_I(\gf)$ satisfy the Killing equations which in $N=2$ supergravity can be solved in terms of a triplet of Killing prepotentials $P^r_I$, $r=1,2,3$, also called `moment maps'. See appendix \ref{geopt} for more information on the potential and moment maps.\\

By sketching  the effect of turning on a flux in $H_3$, \R{flux1},  we have introduced two different topics: compactification with fluxes and gauged supergravity. The reason for introducing the latter is that  the effect including fluxes has on the $4$-dimensional effective theory, can be reproduced by gauging the $4$-dimensional \sugra theory one obtains by compactification  without fluxes. This means that the powerful machinery of gauged supergravity  can be used to study the effect of fluxes in the internal Calabi-Yau manifold on the effective theory. For more details on $N=2$ gauged supergravity we refer to \cite{D'Auria:1990fj, Andrianopoli:1996cm, Ceresole:2000jd, deWit:2001bk} and references therein.\\
There are a few points that must be made. In gauged supergravity, the hypermultiplet scalars can acquire charges under vectors, as in \R{gaugecov1}. This means, contrary to the ungauged case, that the hypermultiplets and vector multiplets no longer decouple. However, the structure of the Lagrangian is encoded in the same data as before the gauging with the addition of the Killing vectors encoding the isometries of the scalar manifold. In spite of the presence of a potential, no mixing of the metrics is allowed. Therefore the metric of the vector multiplets still receives no string loop corrections. \\
Furthermore, the presence of fluxes in general distorts geometry of the Calabi-Yau causing it to change into something else. In such cases the appropriate internal manifold is a generalized Calabi-Yau manifold, see \cite{Behrndt:2005vi, Louis:2005te}.\\

It turns out that the case we are considering corresponds to a very simple type of flux compactification. Remember that we are in a situation in which membranes break all the continuous isometries of the UHM save one. The $4$-dimensional quaternionic scalar manifold of the UHM only has the shift in the axion $\gs$ left, or in PT coordinates $t$. As demonstrated in \cite{Louis:2002ny}, the presence of a nontrivial spacetime part in the $3$-form $A_3$ gauges this shift. In other words, the presence\footnote{Note that in \cite{Bodner:1990zm} the spacetime filling $C_3$ was set to zero.} of $C_3$   gauges the $\ga$-shift in $t$, see \R{fmap1} and \R{spacetimec1}.\\
The reason is that the field strength $d C_3$ of a $3$-form in four spacetime dimensions is dual to a $0$-form field strength and carries no local degrees of freedom. One can therefore eliminate it from the action by dualizing it to a constant. We will denote this constant by $e_0$ (see also \cite{Kachru:2004jr}). In the $4$-dimensional action the derivative of the axion will then become a covariant derivative:
\equ{
\p{\gm}\gs\to D_\gm \gs\equiv \p{\gm}\gs + 2e_0 A_\gm \nn \km
}
where $A_\gm$ is again the graviphoton. Comparing to \R{gaugecov1} we see that $k^\hA=-2 e_0 \gd^{\hA\gs}$, where we have left out the index $I$ because we are considering just  one isometry. The precise form of this Killing vector is not important, as we shall see, and is taken equal to $e_0 \gd^{\hA\gs}$.  Since we will only be gauging this one isometry, we only need one vector field and thus can suffice with the graviphoton. No vector multiplets need to be used. 
In the $10$-dimensional supergravity theory we have the field strength corresponding to $A_3$: $F_4=dA_3$, see \R{3form1e}. This is (Hodge) dual to a $6$-form field strength $F_6=\!\! \,^\star F_4$. Consequently, the nontrivial spacetime part $dC_3$ is equivalent to turning on $F_6$ flux in the Calabi-Yau manifold. Conversely one could say that turning on $F_6$ flux can be dealt with after reducing on the Calabi-Yau by considering a nontrivial spacetime part $C_3$. This means that the $F_6$ flux does not backreact on the internal Calabi-Yau  geometry. \\
In string theory the inclusion of fluxes can lead to tadpoles, but it was shown in \cite{Kachru:2004jr} that only by turning on $F_0$ and $H_3$ flux simultaneously in a type IIA compactification tadpoles can arise, so that is not an issue here.

The potential \R{potI} simplifies considerably in our case. As we are considering the UHM with only one isometry, we have only nontrivial components of $P^r_I$ for $I=0$. Furthermore, there are no vector multiplets present and the potential simplifies to
\equ{
V=4G_{\hA\hB} k^\hA k^\hB -3P^r P^r \nn \km
}
as in appendix \ref{geopt}. The labels $\hA,\hB$ run over the PT coordinates. Substituting the results for the the moment maps \R{moment2} and the Killing vector \R{Kv2} we obtain 
\equ{
V=\frac{1}{r^2}\left(\frac{4}{f} -3\right)e_0^2 \label{eq:potA} \pt
}
We see that the constant $e_0$ which multiplies the Killing vector  appears as an overall factor, hence we will set it to one.

\section{The potential}
We are now in a position to study the effect of membranes on the $4$-dimensional theory in a very concrete manner: through the potential. This potential will consist of three parts:
\equ{
V=V_{\class} +V_\text{1-loop} + V_\textrm{inst} \label{eq:Vtot} \pt
}
Using \R{pertu1} we calculate
\equ{
V_\class=\frac{1}{r^2}  \qquad  V_\textrm{1-loop}=-\frac{4c}{r^2(r+2c)} \nn \km
}
we will discuss $V_\textrm{inst}$ shortly, first we shall discuss $V_\class$ and $V_\text{1-loop}$.

\subsubsection{The perturbative potential}
\begin{figure}[t]
  \renewcommand{\baselinestretch}{1}
  \epsfxsize=0.48\textwidth
  \begin{center}
  \leavevmode
  \epsffile{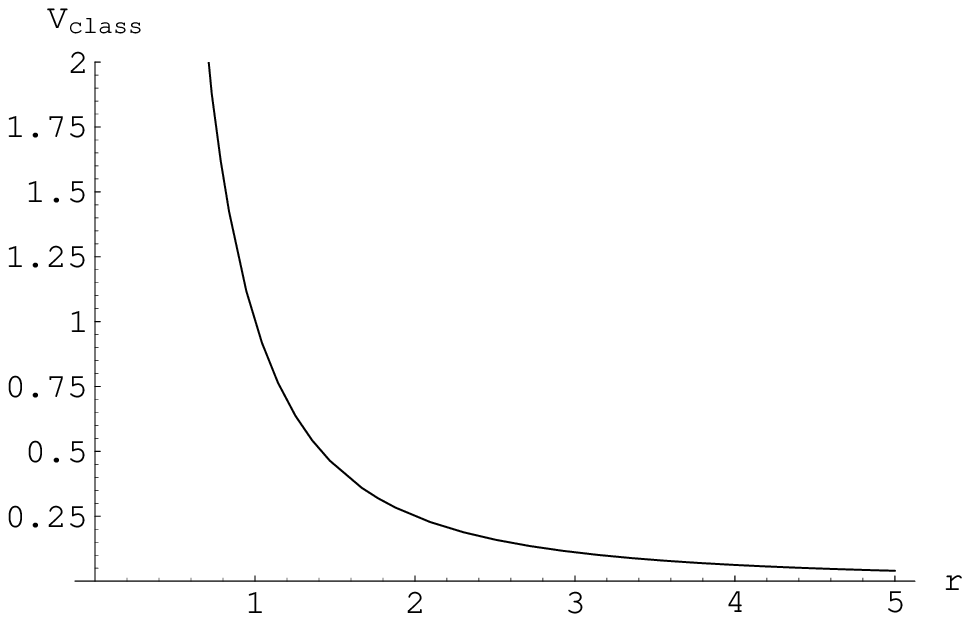} \,  
  \epsfxsize=0.48\textwidth
  \epsffile{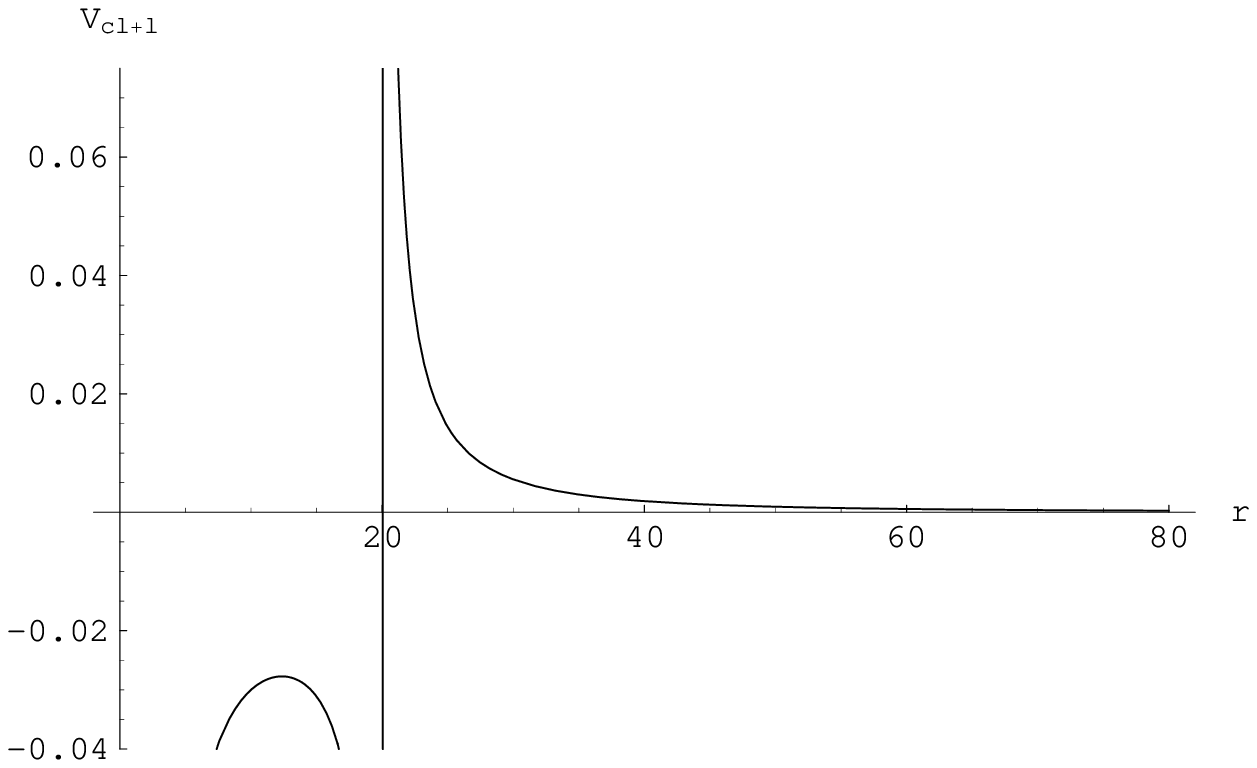}
  \end{center}
  \parbox[c]{\textwidth}{\caption{\label{eins}{\footnotesize The scalar potential $V_\class$ (left) and
$V_\class + V_\textrm{1-loop}$ (right) for $c=-10$. Including the $1$-loop correction
with $c<0$ leads to a pole at $r=-2c$.}}} 
\end{figure}

$V_\class$ is a positive monotonically decreasing function with a typical `runaway' behaviour in $r$. For $r\to 0$ or $g_s \to \infty$ (remember that $g_s =1/\sqrt{r}$) it diverges.  $V_\class$ is displayed in figure \ref{eins}, it has no vacua except the trivial one at $r=\infty$.
\\
If we add the $1$-loop contribution the behaviour changes drastically. In figure \ref{eins} $V_\class + V_\textrm{1-loop}$ is displayed for the `generic' value $c=-10$. For different values of $c$ the location of the pole shifts but the (qualitative) properties of the potential remain the same. We found in \R{c2} that $c$ is always smaller than zero. Since the region $0<r<|2c|$ has to be discarded, see the discussion below \R{c2},  we also have to discard this region of the potential. For $r>|2c|$ the behaviour of the potential is similar to the behaviour of $V_\class$. \\
These two pieces of the potential are independent of $(u,v,t)$ which means that these scalars parametrize flat directions. The absence of $u$ and $v$ is just a consequence of their absence in \R{pertu1}. The fact that the potential does not depend on $t$ is a different matter. Because we have gauged the isometry in $t$, gauge invariance implies that the potential cannot depend on $t$. 
Consequently we only have to stabilize $u$ and $v$ in the potential to lift the flat directions, since $t$ is stabilized no matter what. We now demonstrate that we can stabilize $u$ and $v$ by taking into account the membrane contributions.

\subsubsection{The potential including membrane effects}
$V_\textrm{inst}$ is the part containing the contributions from the infinite instanton expansion. If we limit ourselves to small string coupling, that is $r\mg 1$, a good approximation is given by (\ref{eq:eh-exp}, \ref{eq:finst}).
We will consider the leading contributions coming from the one-instanton sector. Furthermore, we will consider the effect of the $v$-instanton sector while switching off the $u$-instanton sector. We can do this because both sectors enter separately in exactly the same manner in the potential, as can be seen from (\ref{eq:9.1}, \ref{eq:eh-exp}, \ref{eq:finst}). Therefore we can first study the effect of one sector and then draw similar conclusions for the other sector. \\
To leading order the instanton part of the potential becomes
\equ{
 \label{eq:8.5}
  V_\textrm{1-inst} = -4\, r^{-(m_1+5)/2}\, \big( \hat{A}_{1,m_1} \cos(v)
  - \tilde{A}_{1,m_1} \sin(v) \big)\, e^{-2 \sqrt{r}}\km
}
where we have used \R{finst}.  We have defined $A_{1,m_1}=\hat{A}_{1,m_1}+\I\tilde{A}_{1,m_1}$ and have set $B_{1,m_1}=0$. $V_\inst$ has no poles except at $r=0$. In fact, this 
is an artefact  of the expansion in \R{8.5}. If we would not have expanded the
denominator which contained $(r+2c)$, $V_\textrm{1-inst}$ also would develop a
singularity at $r=-2c$, which even dominates over the one in
$V_\textrm{1-loop}$, and the potential is no longer bounded from below.
Resolving this singularity presumably requires resumming the entire
instanton expansion to obtain expressions which are valid at small
values of $r\le -2c$. It is unclear how to resum the expansion and we will continue to work with the expanded
expressions \R{8.5}. Note that resolving singularities
by nonperturbative effects has been shown to work in the context of
the Coulomb branch of three-dimensional gauge theories with eight
supercharges \cite{Seiberg:1996nz, Dorey:1997ij, Dorey:1998kq, Chalmers:1996xh}. In these cases the moduli space is
hyperk\"ahler instead of quaternionic.\\
Furthermore, since $V_\textrm{inst}$ is exponentially suppressed for large $r$, the large $r$ behaviour of $V$ is dominated by $V_\class + V_\textrm{1-loop}$. This means that for $r\to\infty$ $V$ still approaches $0$ from above. The total potential is bounded from below and diverges to $+\infty$ for $r\downarrow |2c|$. Due to the transcendental nature of the potential we have to analyze the vacuum structure numerically. Furthermore, since we do not know anything about the $g_s$ dependence of the determinant in the background of instantons, contained in the factor $N$ in \R{corrs1}, we will choose $m_1=-2$, see \R{r.3}. This is of little importance since the qualitative behaviour of the potential is not very sensitive to this value. We can obtain similar results for different values of $m_1$.\\

\begin{figure}[th]
\renewcommand{\baselinestretch}{1}
\epsfxsize=0.6\textwidth
\begin{center}
\leavevmode
\epsffile{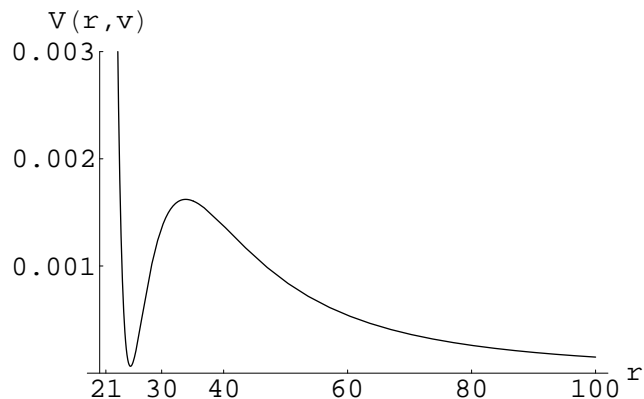} \,  
\end{center}
\parbox[c]{\textwidth}{\caption{\label{zwei}{\footnotesize A view of $V(r,v)$ in the $r$-direction
along $v=0$.}}}
\end{figure}

We can simplify this potential by imposing the symmetry which interchanges $v$-instantons with $v$-anti-instantons: $v\to -v$, $t\to -t$, see section \ref{sect4.3}. This means that we have to take $\tA_{1,m_1}=0$. Furthermore, we can take $\hA_{1,m_1}$ positive because negative values merely correspond to shifting $v\to v+\gp$. The local minimum in the $v$-direction is then located at $v=0$. Knowing the minimum in the $v$-direction, we can seek out the minimum in the $r$-direction. This depends on the value of the one remaining parameter $\hA_{1,m_1}$. For a given value of $c$ we can distinguish three different cases characterized by two values for $\hA_{1,m_1}$: $A_\textrm{min}$ and $A_\textrm{max}$. These have to be determined numerically for each value of $c$. If $\hA_{1,m_1} < A_\textrm{min}$ there is no local minimum in the $r$-direction except for the one at $r=\infty$. The (qualitative) picture is just that of $V_\class + V_\textrm{1-loop}$ in figure \ref{eins}. For $\hA_{1,m_1}>A_\textrm{max}$ we do find a minimum in the $r$-direction. The value of the potential at this point, let us call it $r_\textrm{AdS}$, is negative\footnote{In the next section we will explain the names for  these minima.}. The most interesting situation occurs when $A_\textrm{min} < \hA_{1,m_1} < A_\textrm{max}$. In such a situation the minimum of the potential, occurring at $r_\textrm{dS}$,  is positive, which is  significant as we will learn in the next section. As in the previous case this stabilizes both the $v$ and the $r$ modulus. We are thus able to stabilize all the moduli of the UHM. This situation is depicted in figure \ref{zwei} for the value $c=-10$ and $\hA_{1,m_1}=9867$. 
Note that the only difference with the case $\hA_{1,m_1}>A_\textrm{max}$ is that the minimum is now positive. 
The effect of increasing $\hA_{1,m_1}$ is to move the location of the minimum to the left: closer to the singularity. This means that the value for which the string coupling is smallest is obtained by taking $\hA_{1,m_1}=A_\textrm{min}$. In table \ref{t1} we give the relevant values for the parameters for the cases $c=-6/\gp \approx -1.9$ and $c=-10$. 
\begin{table}
  \centering
\begin{tabular}{|c|c|c|}
 \hline
   & $c = -1.9$ & $c = -10$ \\ \hline
  $A_\textrm{min}$ & $53.0$ & $8180$ \\ 
  $A_\textrm{max}$ & $60.8$ & $9900$ \\
  $r_\mathrm{dS}$ & $7.4$ & $27.5$\\ \hline
\end{tabular}
  \caption{{\footnotesize Illustrative values for
$A_\textrm{min}$, $A_\textrm{max}$, and $r_\textrm{dS}$ for the
$\wZ$-manifold $h^{(1,2)}=0$, $h^{(1,1)}=36$, $c\approx-1.9$) and
a fictional rigid CY$_3$ where $c=-10$, corresponding to $h^{(1,1)}=
\cO(100)$. For $A_\textrm{min}<\hat{A}_{1,m_1}<A_\textrm{max}$ the potential \R{Vtot}
has a meta-stable dS vacuum at which all hypermultiplet moduli are
stabilized.}}\label{t1}
\end{table}
The first case corresponds to the so-called $\wZ$-manifold (see for instance \cite{Candelas:1993nd}). The case $c=-10$ corresponds to a fictitious rigid CY $3$-fold with $h^{(1,1)}\approx \cO(100)$.\\
We see from table \ref{t1} that decreasing $|c|$ also decreases the values for $A_\textrm{min}$ and $A_\textrm{max}$ while their relative difference stays about the same. Decreasing $|c|$ also moves $r_\textrm{dS}$ closer to the singularity. This is a generic feature and does not depend on the particular values of $c$ in the table. \\

\begin{figure}[th]
\renewcommand{\baselinestretch}{1}
\epsfxsize=1.0\textwidth
\begin{center}
\leavevmode
\epsffile{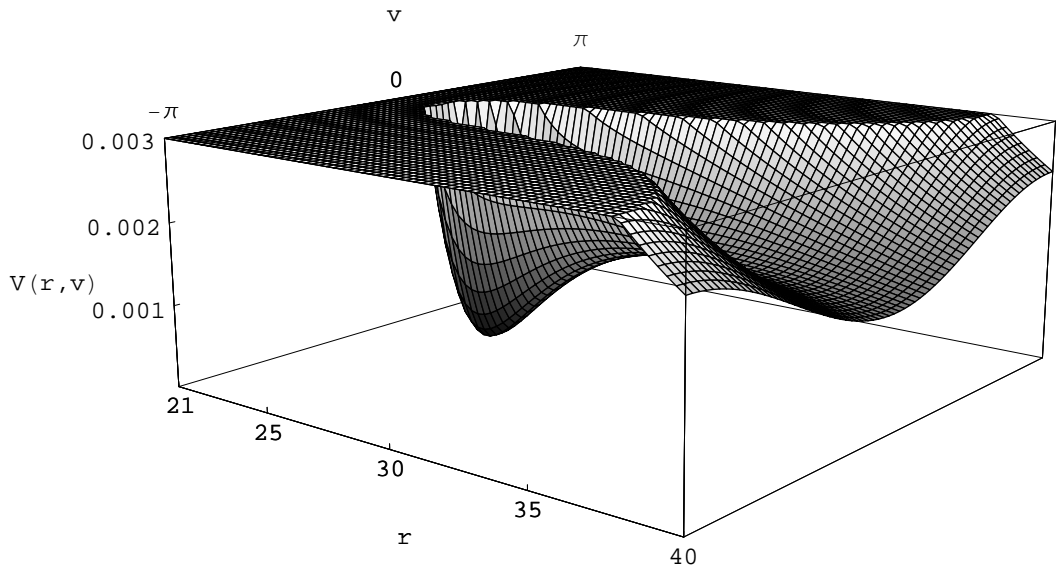} \,  
\end{center}
\parbox[c]{\textwidth}{\caption{\label{drei} A detailed view of the meta-stable dS minimum of
the potential $V(r,v)$ in the $r$ and $v$ direction.}}
\end{figure}

To check that the minimum in the $r$-direction is also a minimum in the $v$-direction, we plot the total potential \R{Vtot} in the neighbourhood of the minimum, see figure \ref{drei}. To obtain this graph we used the values $c=-10$ and $\hA_{1,m_1}=9867$, as before.
If we would extend the graph in the $v$-direction we would see that it is periodical due to the cosine in \R{8.5}. This periodicity can be lifted by including higher order terms in $r$. One can check that by taking such higher order terms into account one can still construct a minimum with positive value. The values of the other minima (the `copies') are generically lowered to negative values.

\section{De Sitter space}
\label{Sitter}
De Sitter spacetime (dS) is a maximally symmetric solution to Einstein's equations with a positive cosmological constant $\gL$ (see \cite{Spradlin:2001pw, Hawking} for more details). We can describe $d$-dimensional de Sitter space as a hyperboloid embedded in flat $(d+1)$-dimensional Minkowski space $\cM^{d,1}$. If $\cM^{d,1}$ has coordinates $X^A, A=0, \ldots, d$ and metric $\get_{AB}=\textrm{diag}(-1,1,\ldots,1)$, the hypersurface 
\equ{
\get_{AB}X^AX^B=R^2 \nn 
}
defines a $d$-dimensional dS$_d$, with `radius' $R$. For the metric on dS$_d$ we can use the induced metric from the embedding space $\cM^{d,1}$. One can show that dS$_d$ is an  Einstein space with positive scalar curvature and an Einstein tensor satisfying 
\equ{
G_{\gm\gn} + \gL g_{\gm\gn}=0 \nn \km
}
where 
\equ{
\gL=\frac{(d-2)(d-1)}{2R^2} \nn 
}
is the cosmological constant. \\

We have been considering an $N=2$ supergravity theory in four spacetime dimensions. We have discussed that  gauging an isometry gives rise to a potential (the relevant terms are given in \R{app9t1a} and \R{app9t1b}). If we take the variational derivative with respect to $g_{\gm\gn}$ of the (hypermultiplet) action we find
\equ{
0=\frac{1}{\sqrt{g}}\frac{\gd S}{\gd g_{\gm\gn}}=R^{\gm\gn}-\2 g^{\gm\gn} R + T^{\gm\gn} \nn \km
}
where $T^{\gm\gn}$ is the variational derivative of the part of the action containing the fields and the potential. This means that when considering the scalar fields at the minimum of the potential, $T^{\gm\gn}$ contributes to an effective cosmological constant term:
\equ{
g^{\gm\gn} V_{\textrm{min}}(r,u,v) \nn \pt
}
$V_{\textrm{min}}(r,u,v)$ is the constant value of the potential at its minimum (we denoted the fields as in the previous section).
When this value is larger than zero (if this is the only contribution to the vacuum energy) it can be associated to a de Sitter spacetime\footnote{Alternatively, if the minimum of the potential is less than zero it corresponds to an Anti-de Sitter space, hence the name $r_{\textrm{AdS}}$.}.\\
Supergravity cannot be realized in a de Sitter space. This means that the $N=2$ supergravity theory has no supersymmetry for a de Sitter  vacuum (see e.g. \cite{Pilch:1984aw}).
We will not discuss the breaking of supersymmetry in the case of an AdS or Minkowski\footnote{This corresponds to a zero cosmological constant, so a vanishing value of the potential at its minimum.} vacuum. There is much more to say about de Sitter vacua in supergravity and supersymmetry breaking, see for instance \cite{Ferrara:1995gu, Fre:1996js, Fre':2003gd, Fre:2002pd}. \\
The reason that de Sitter space is relevant, is that astronomical observations indicate that our own universe is in a de Sitter phase, see \cite{Trodden:2004st, Peebles:2002gy} and references therein for more information. Therefore it is exciting that we can reproduce a de Sitter vacuum from our supergravity setup. Note that one really should compute the coefficient $\hA_{1,m_1}$ in string theory because in our supergravity approach it is still a free parameter. This means that only if the result is such that $A_\textrm{min} < \hA_{1,m_1}<A_\textrm{max}$ we can conclude  that we can produce a de Sitter vacuum from string theory.\\
Naturally we are not the first to have searched for a de Sitter vacuum in string theory (see \cite{Fre:2002pd, Fre':2003gd}). There is, for instance, the work of `KKLT' \cite{Kachru:2003aw}, to which we will compare our results.\\
The work of KKLT consists of three parts. Firstly, all moduli, apart from the \Kh moduli, are fixed by the inclusion of fluxes. Secondly, the \Kh moduli (the volume modulus to be precise) are stabilized by nonperturbative instanton effects, but at an AdS vacuum. Thirdly, a positive energy contribution in the form of anti D$3$-branes\footnote{They work in type IIB string theory.} is added to lift the minimum of the potential to a positive value, i.e., a dS vacuum.\\
In our approach the inclusion of a single R-R $3$-form spacefilling flux provides a potential. Classically this is of the runaway type in the dilaton and both R-R scalars $\gc$ and $\gvf$ are flat directions. Only when we include the membrane corrections (plus the one-loop corrections) does this change, leading to a (meta-)stable vacuum. Depending on $\hA_{1,m_1}$ this is an AdS, Minkowski  or dS vacuum. \\
Note that, in contrast to KKLT, we do not have to include additional positive energy contributions by hand. This contribution is already provided for by the background flux at the classical level. Furthermore, by making a suitable choice for the numerical parameters corresponding to the one-loop determinant around a one-instanton background, the moduli can be stabilized in a meta-stable de Sitter vacuum at small string coupling constant $g_s\ll 1$. This is consistent with the fact that we only took into account the terms dominating for small $g_s$ (\ref{eq:finst}). \\

We have examined a simple model in which we have one hypermultiplet and no vector multiplets. As this hypermultiplet is present in any $N=2$ supergravity theory in four dimensions obtained by compactifying type IIA supergravity on a Calabi-Yau manifold, the results are expected to have  validity for more general Calabi-Yau compactifications as well (that is $h^{(1,2)}\neq 0$). 
A next step would then be to include vector multiplets and hypermultiplets and consider more general fluxes,  e.g. $2$- and $4$-form fluxes related to the even homology cycles of the Calabi-Yau manifold, thus stabilizing the K\"ahler moduli\footnote{Remember that these moduli reside in the vector multiplets.}. 
One could also try to include the nonperturbative effects coming from the $5$-brane, as calculated in the previous chapter, into the potential.

\clearpage{\pagestyle{empty}\cleardoublepage}

{
\renewcommand{\chaptermark}[1]{\markboth{\thechapter\ #1}{}}
\fancyhead{}
\fancyhead[LE, RO]{\thepage}
\fancyhead[CO]{\slshape\leftmark}
\fancyhead[CE]{\slshape\leftmark}
\appendix
\chapter{Symplectic groups and quaternions}
\label{quatapp}
The quaternions are the associative algebra $\mH=\langle 1, i_1, i_2, i_3\rangle \cong \mR^4$ with multiplication $i_1 i_2=-i_2 i_1=i_3, i_2 i_3=-i_3 i_2 =i_1, i_3 i_1=-i_1 i_3=i_2$ and $i_1^2=i_2^2=i_3^2=-1$. 
If we have a quaternion $x=x_0+\sum_{j=1}^3 x_j i_j$ then its conjugate is given by $\bx=x_0 -\sum_{j=1}^3 x_j i_j$. 
The \emph{symplectic} group, $Sp(n)$ is the subgroup of $Gl(n, \mH)$, the invertible $n\times n$ quaternionic matrices, which preserves the standard Hermitian form on $\mH^n$: 
\equ{
\langle x,y\rangle =\sum_{i=1}^n \bx_i y_i \nn \pt
}
This means that if $A\in Sp(n)$, $\bA^T A=A\bA^T=\Id$, it is  the quaternionic unitary group $U(n,\mH)$, sometimes called the \emph{hyperunitary} group. $Sp(n)$ is a real Lie group of dimension $n(2n+1)$, compact and (simply) connected. The Lie algebra is given by the $n\times n$ quaternionic matrices $B$ that satisfy
\equ{
B+\bB^T=0 \nn \pt
}
A different, but closely related, type of symplectic group is $Sp(2n,F)$ the group of degree $2n$ over a field $F$, in other words, the group of $2n \times 2n$ symplectic matrices with entries in $F$ and with group operation that of matrix multiplication. 
If $F$ is the field of real/complex numbers the Lie group $Sp(2n,F)$ has real/complex dimension  $n(2n+1)$. \\
Since all symplectic matrices have unit determinant, the symplectic group is a subgroup of the special linear group $Sl(2n,F)$, in fact for $n=1$ this means that $Sp(2,F)=Sl(2,F)$. If $D\in Sp(2n,F)$ then
\equ{
D^T \cC D= \cC  \qquad \textrm{with} \qquad \cC=
\left(%
\begin{array}{cc}
  0 & \Id \\
  -\Id & 0 \\
\end{array}%
\right)\pt
\label{eq:sym1}
}
The Lie algebra of $Sp(2n,F)$ is given by the set of $2n \times 2n$ matrices $E$ (over $F$) that satisfy
\equ{
\cC E + E^T \cC=0 \nn \pt
}
The symplectic group $Sp(2n,F)$ can also be defined as the set of linear transformations of a $2n$-dimensional vector space (over $F$) that preserve a nondegenerate antisymmetric bilinear form. This precisely leads to the property \R{sym1}.\\
\emph{Unitary-symplectic} groups are the intersection of unitary groups and symplectic groups, both acting in the same vector space. That is, the elements of the unitary-symplectic groups are elements in both unitary and symplectic groups: 
\equ{
U\!Sp(n,n)=U(n,n;\mC) \cap Sp(2n;\mC) \nn \pt
}
Explicitly we can construct such groups by considering \R{sym1} with $D$ a $2n\times 2n$ complex matrix which has to satisfy the additional condition
\equ{
D^\dagger \cH D=\cH \qquad \textrm{with} \qquad \cH=\left(%
\begin{array}{cc}
  \Id_{n\times n} & 0 \\
  0 & -\Id_{n\times n}\\
\end{array}%
\right) \label{eq:iso3} \pt
}
The general block form of this $D$ is
\equ{
D=\left(%
\begin{array}{cc}
  T & V^\star \\
  V & T^\star \\
\end{array}%
\right) \label{eq:block1}
}
making \R{iso3} equivalent to
\bea
&& T^\dagger T -V^\dagger V =\Id \nn \\
&& T^\dagger V^\star -V^\dagger T^\star =0 \nn \pt
\eea
These groups are isomorphic to symplectic groups in the hyperunitary sense:
\equ{
U\!Sp(2n) \cong Sp(n) \nn \pt
}
Finally there is the following isomorphism
\equ{
Sp(2n;\mR) \cong U\!Sp(2n) \label{eq:iso1} \km
}
from which we see that $Sp(2n;\mR) \cong Sp(n)$.\\

A quaternionic manifold is Riemannian but not necessarily complex, which means that the holonomy group $Sp(n)\times Sp(1)$ is contained in $O(4n)$ as a subgroup. On the other hand, hyperk\"ahler manifolds are in fact K\"ahler (with respect to all three complex structures, see also \R{Kcond1}) and must therefore have their holonomy contained in $U(2n)$.  In terms of holonomy groups, the difference between the two types of manifolds is the factor of $Sp(1)$. This arises from the fact that \R{quatal1} is defined up to local $SO(3)\cong Sp(1)$ rotations. This feature is absent in the hyperk\"ahler case since the complex structures are covariantly constant and globally defined. 
For more information on the relation between holonomy groups, symplectic groups and quaternions, see \cite{Aspinwall:2000fd, Fre:1995dw, Besse, CY, Gilmore}. 

\clearpage{\pagestyle{empty}\cleardoublepage} 
\chapter{The Wick rotation}
\label{appwick} 
In this appendix we  clarify the Wick rotation introduced in section \ref{instsol}, furthermore the calculations in chapter \ref{ch3} will be performed in Euclidean space and require the use of appropriate fermions.\\
First let us Wick rotate the DTM Lagrangian \R{DTM2}. In Minkowskian space the action appears in the path integral as $e^{\I S}$ with $\get_{\gm\gn}=$diag$(-+++)$. We rotate $t\to \I t\equiv x_4$ which implies $\pp{}{t}=\I \pp{}{x_4}$ and the integration measure becomes $d^4x=-\I d^4x_E$. This means that a massless scalar appears in the path integral as
\equ{
\exp\{\I \int d^4x \left(-\2 \p{\gm}\gf \pu{\gm}\gf\right)\} \to \exp\{-\int d^4x_E \2 \p{\gm}^E \gf \pu{\gm}_E \gf\} \equiv\exp\{-S_E\} \nn
}
where in the second half the indices run over $1$ to $4$. 
We will always extract an overall minus sign out of the action giving a damped exponential, since the action has become positive definite. \\
Having seen how scalars transform we turn to the double tensor term:
\equ{
\cL_{dt}=\2 M^{IJ}H_I^\gm H_{\gm J}\km \qquad H^\gm_I=\2 \gve^{\gm\gn\gr\gs}\p{\gn}B_{\gr\gs I}=\frac{1}{3!} \gve^{\gm\gn\gr\gs}H_{\gn\gr\gs}\nn
}
Using $\gve^{\gm\gn\gr\gs}\gve_{\gm\gl\gt\gd}=-3! \gd^{[\gn}_{[\gl}\gd^\gr_\gt \gd^{\gs]}_{\gd]}$, the conventions $\gve^{0123}=-\gve_{0123}=1$ and $\gve^{1234}=1$, we can rewrite this (up to unimportant $I,J$ quantities) as
\equ{
H_{\gn\gr\gs}H^{\gn\gr\gs}=3H_{0ij}H^{0ij} + H_{ijk} H^{ijk} \nn \km
}
where $i,j=1,2,3$.  Now we Wick rotate according to
\equ{
B_{0i}=\I B_{4i} \quad \to \quad H_{0ij}=\I H_{4ij} \nn 
}
and since $H_{0ij}H^{0ij}=H_{4ij}H^{4ij}$ the double tensor term becomes
\equ{
\cL_{dt}^E=\2 H^E_{\gm I} H^E_{\gm J} M^{IJ} \nn \pt
}
Then there is the scalar-tensor term
\bea
\cL_{st}&=&-A^I_A H^\gm_I \p{\gm} \gf^A \nn \\
&\sim&-\2 \left(\gve^{0\gn\gr\gs}H_{\gn\gr\gs} \p{0}\gf + 3\gve^{i0\gr\gs}H_{0\gr\gs}\p{i}\gf\right) \nn \pt
\eea
Using $\gve^{0\gn\gr\gs}=-\gve^{4\gn\gr\gs}$ and $\gve^{i0\gr\gs}H_{0\gr\gs}=-\I\gve^{i4\gr\gs}H_{4\gr\gs}$ we obtain
\equ{
\cL^E_{st}=-\I A^I_A H_I^{\gm E} \p{\gm} \gf^A \label{eq:tempa21} \pt
}
When using these expressions we will leave out the superscript $E$ and understand that $\gm=1,\ldots,4$. \\

Lastly there is the matter of the spinors. The spinors we work with are Lorentzian $2$-component Weyl spinors, say $\gl_\ga, \bgl^{\dga}$, related to each other by complex conjugation. However, in Euclidean space the Lorentz group factorizes as $Spin(4)\cong SU(2)\times SU(2)$ and to each $SU(2)$ belongs a $2$-component spinor, $\gl_\ga$ and $\bgl^{\ga'}$ which are not related to each other by complex conjugation, they constitute inequivalent representations of the Lorentz group.  The same remarks hold for the \susy parameters $\ge^i$ and $\bge^i$.
The sigma matrices $\gs^\gm$ and $\bgs^\gm$ have lower and upper indices respectively, i.e., $(\gs^\gm)_{\ga\ga'}$ and $(\bgs^\gm)^{\ga'\ga}$ where we instead of dotted indices use indices with a $'$ to denote the inequivalent representations. In Euclidean space these matrices take the form 
\equ{
\gs^\gm=(\vec \gs, -\I) \qquad \bgs^\gm=(-\vec \gs, -\I) \nn
} 
where $\vec \gs$ are the Pauli matrices, note that this is consistent with $\gs^0=-\I \gs^4$. We have the properties (slightly different from Lorentzian signature) 
\equ{
\gs^\gm \bgs^\gn =-g^{\gm\gn} + 2 \gs^{\gm\gn} \qquad \2 \gve^{\gm\gn\gr\gs}\gs_{\gr\gs}=\gs^{\gm\gn} \nn \km
}
where $\gs^{\gm\gn}\equiv \2 \gs^{[\gm}\bgs^{\gn]}$ and
\equ{
\gs^\gm\bgs^\gn\gs^\gr =g^{\gm\gr}\gs^\gn -g^{\gn\gr} \gs^\gm -g^{\gm \gn}\gs^\gr + \gve^{\gm\gn\gr\gs}\gs_\gs \nn \pt
}
Other often used properties of the sigma matrices are
\bea
&& (\gs)_{\ga\gb'}^\gm (\gs_\gm)_{\gg \ga'}=-2\gve_{\ga\gg}\gve_{\gb' \ga'} \nn \\
&& (\bgs^\gm)^{\gb' \ga}(\gs_\gm)_{\gg\ga'}=-2\gd^\ga_\gg \gd^{\gb'}_{\ga'} \nn \\
&& (\gs^{\gm\gn}\gve)_{\ga\gb} (\gs_\gn)_{\gg\ga'}=-\gve_{\gg(\gb}\gs^\gm_{\ga)\ga'} \label{eq:B4} \pt 
\eea
Finally, the gravitini and the graviphoton are rotated as follows
\equ{
\gps_0^i=\I\gps^i_4 \qquad \bgps_{0i}=\I  \bgps_{4i} \qquad A_0=\I A_4 \nn \pt
}
For more information on the conventions we used, see Wess and Bagger \cite{Wess:1992cp}.

\clearpage{\pagestyle{empty}\cleardoublepage}

\chapter{Scalar-tensors and supersymmetry}
\label{appsusy1} 
In this appendix we will list some of the relations (as determined by supersymmetry) between the various quantities which appear in the action. We will only give the ones needed for our calculations, consult \cite{Theis:2003jj} for more information. For more details  on the construction of such non-linear sigma models see also \cite{deWit:1998zg, deWit:1999fp, Bagger:1983tt}. We will also expand a little on the geometry of the hypermultiplets as determined by supersymmetry. \\
To start with some algebraic relations
\bea
&&  \gg_{ia}^A\, \W{A}{bj} + g_{Iia}\, f^{Ibj} = \gd_i^j\, \gd_a^b
    \notag \\[2mm]
&&  \gg_{ia}^A\, \bbW{Aj}{\ba} + g_{Iia}\, \bar{f}^{I\ba}{}_j + (i
    \leftrightarrow j) = 0\km \label{eq:rel1}
 \eea
for contractions over $A$ and $I$, and
 \begin{equation}
  \begin{pmatrix} \gg_{ia}^A\, \W{B}{aj} & \gg_{ia}^A\, f^{Jaj} \\[2pt]
  g_{Iia}\, \W{B}{aj} & g_{Iia}\, f^{Jaj} \end{pmatrix} + \text{c.c.}
  (i \leftrightarrow j) = \gd_i^j \begin{pmatrix} \gd_B^A & 0 \\[2pt]
  0 & \gd_I^J \end{pmatrix} \pt \label{rel2}
 \end{equation}
Furthermore, we have
\begin{equation} \label{eq:Gga_Mg}
  \cG_{AB}\, \gg_{ia}^B = h_{a\bb}\, \bbW{Ai}{\bb}\ ,\qquad M^{IJ}
  g_{Jia} = h_{a\ba} \bar{f}^{I\ba}{}_i\pt
 \end{equation}
These relations imply among others that
 \bea
&&  \cG_{AB} = h_{a\bb}\, \W{A}{ai}\, \bbW{Bi}{\bb}\ ,\qquad
    M^{IJ} = h_{a\bb}\, f^{Iai} \bar{f}^{J\bb}{}_i \notag \\[2mm]
&&  \gd_i^j\, h_{a\bb} = \cG_{AB}\, \gg^A_{ia}\, \bar{\gg}^j{}^B_\bb
    + M^{IJ} g_{Iia}\, \bar{g}^j_{J\bb} \pt \label{eq:GMh}
 \eea
Barred and unbarred objects are related by complex conjugation which raises or lowers the $SU(2)$ index $i$, e.g. $\bW^\bb_{Bi}=(W^{bi}_B)^\star$.
An interesting relation is the following:
\begin{equation} \label{eq:F}
  F_{AB}{}^I = 2\I M^{IJ} k_{Ja\ba} \bbW{Ai}{\ba}\, \W{B}{ai} - 2
  h_{a\ba} \bbW{Ai}{\ba}\, \Gamma^{Ii}{}_j \W{B}{aj}\ ,
 \end{equation}
where the field strength is defined as $F_{AB}{}^I=2\partial^{}_{[A}
A_{B]}^I$. This field strength also measures the nonvanishing of the
covariant derivatives of $\gamma^A_{ia}$ and $\W{A}{ai}$, that is
\equ{
\cD_A \W{B}{ai} =-\2 F_{AB}{}^I \bar{g}^i_{I\ba} h^{\ba a} \label{eq:tor1} \km
}
where the covariant derivative $\cD_A$ contains connections $\G{A}{a}{b}$ and $\gG_{AB}^{\,\,\,\,C}$ and similarly for $\gamma^A_{ia}$.   The right-hand side of \R{tor1} is antisymmetric in $A, B$ and can be interpreted as the torsion of the target space connection. Interestingly enough $F_{AB}{}^I=0$ classically, but in the presence of instantons it is nonzero, as we will see in appendix \ref{inst-vielb}.

The higher
order fermion terms in the supersymmetry transformation rules
\R{str2} contain tensors $\Gamma^{Ia}{}_b$ that satisfy
 \begin{equation} \label{eq:k-Gamma}
  M^{IJ} k_{Ja\ba} = \I h_{b\ba} \Gamma^{Ib}{}_a \pt
 \end{equation}

Finally, one can define a covariantly constant tensor
\begin{equation} \label{eq:def-E-tensor}
  \cE_{ab} = \half \gve^{ji}\, (\cG_{AB}\, \gg^A_{ia}\, \gg^B_{jb}\
  + M^{IJ} g_{Iia}\, g_{Jjb})\ ,
 \end{equation}
which satisfies
 \begin{equation} \label{eq:EWidentity}
  \cE_{ab} \lp \W{A}{bi} \\[2pt] f^{Ibi} \rp = \gve^{ij} h_{a\ba}
  \lp \bbW{Aj}{\ba} \\[2pt] \bar{f}^{I\ba}{}_j \rp \pt
 \end{equation}
This tensor appears explicitly in the four-fermi terms in the
supergravity action \R{sa1}. The covariant derivatives of the fermions are given by
\bea
&& \cD_\gm \gl^a=\nabla_\gm \gl^a + \p{\gm}\gf^A \G{A}{a}{b} \gl^b \nn \\
&& \cD_\gm \gps_\gn^i=\nabla_\gm \gps_\gn^i  + \p{\gm}\gf^A \G{A}{i}{j} \gps_\gn^j \label{eq:covder1} \km
\eea
where $\nabla_\gm$ is the Lorentz covariant derivative.
The connection coefficients $\gG$ specified to the case of the DTM are given in appendix \ref{appdtm1}.\\
In the action appear certain `supercovariant' quantities such as $\cH^\gm_I$
\equ{
\cH^\gm_I=\2 \gve^{\gm\gn\gr\gs}\left[\p{\gn}B_{\gr\gs I}+4\OO{I}{i}{j} \gps^j_\gn\gs_\gr\bgps_{\gs i}-2\I \gk\left(g_{Iia}\gps^i_\gn \gs_{\gr\gs}\gl^a -\textrm{c.c}\right)\right] \label{eq:sucoH} \pt
}
The difference with the familiar $H^I_\gm$ is precisely given by terms quadratic in fermions which can be dropped, as we are working at linear order in the fermions. The general supersymmetry rules for the hyperinos contain these $\cH^\gm_I$ and simplify to \R{str2} when working at linear order in the fermions. Similarly for the supercovariant derivative of the scalars in \R{sa1}
\equ{
\hD_\gm \gf^A=\p{\gm}\gf^A-\gk\left(\gg_{ia}^A\gps^i_\gm \gl^a + \bgg^{iA}_\ba \bgps_{\gm i} \bgl^\ba\right) \nn \km
}
where again the second term can be dropped. In the same fashion, the supercovariant field strength $\cF_{\gm\gn}$ reduces to $F_{\gm\gn}$.\\

\subsubsection{Scalars versus tensors}

 Let us for a moment remind ourselves where the scalar-tensor system comes from. It can be obtained by starting with the $n$ hypermultiplets coupled to $N=2$ supergravity. This theory contains $4n$ scalars together with fermionic fields. In section \ref{instsol} we sketched how to dualize the UHM with its $4$ scalars to the DTM with its $2$ scalars and $2$ tensors. The same procedure can be followed  if one starts with $4n$ scalars and dualizes $n_T$ of them into tensors, thus ending up with the scalar-tensor system containing $n_T$ tensors $B_{\gm\gn I}$, $I=1, \ldots, n_T$ and $4n-n_T$ scalars $\gf^A$, $A=1, \ldots, 4n-n_T$. Naturally this has an effect on the fermionic sector  of the hypermultiplet theory. For instance, in \R{covder1} we gave the covariant derivatives on the fermions with certain connections. These connections have their origin in the theory with $n$ hypermultiplets where the covariant derivatives are given by
\equ{
D_\gm \gl^a=\nabla_\gm \gl^a + \p{\gm}\gf^\hA \OO{\hA}{a}{b} \gl^b \qquad D_\gm \gps^i_\gn=\nabla_\gm \gps^i_\gn + \p{\gm}\gf^\hA \OO{\hA}{i}{j} \gps^j_\gn \label{eq:covderlam} \km
}
with $\hA=1,\ldots, 4n$. These are the Lorentz and $SU(2)\otimes Sp(2n, \mR)$-covariant derivatives. Specifically, $\OO{\hA}{a}{b}$ is the $Sp(2n,\mR)$ connection and $\OO{\hA}{i}{j}$ the $SU(2)$ connection. The connections in the scalar-tensor theory \R{covder1} are related to these connections as follows
\bea
&& \G{A}{i}{j}=\OO{A}{i}{j} -A^I_A \OO{I}{i}{j} \nn \\
&& \G{A}{a}{b}=\OO{A}{a}{b} -A^I_A \OO{I}{a}{b} \nn \\
&&  \gG^{Ii}_{\,\,\,\,j}=M^{IJ} \OO{J}{i}{j} \label{eq:gamma-omega1}\km
\eea
where as in section \ref{instsol},
\equ{
A^I_A\equiv M^{IJ} G_{JA} \qquad M^{IJ}\equiv (G_{IJ})^{-1} \nn \km
}
with $G_{\hA\hB}$ the metric on the quaternionic space. We note that the geometric structure of the target space  of the scalar-tensor theory is not as clear as the geometry of the hypermultiplet theory. \\

\subsubsection{Hypergeometry}

In chapter \ref{ch1} the hypergeometry was described. As indicated there, the geometry for the target space of the hypermultiplets is determined by supersymmetry. \\
Let us therefore consider $n$ hypermultiplets without tensors ($n_T=0$) in the case of Lorentzian signature, we will follow \cite{DeJaegher:1997ka} but use the slightly different notation of \cite{Theis:2003jj}. The supersymmetry transformation of the (real) $4n$ scalars is
\equ{
\gd_\ge \gf^\hA= \gg^\hA_{ia}\ge^i \gl^a + \bgg^{i\hA}_\ba \bge_i \bgl^\ba  \km \label{eq:gft1}
}
with $2n$ positive-chirality spinors $\gl^a$ and $2n$ negative-chirality spinors $\bgl^\ba$, they are related by $\bgl^\ba=(\gl^a)^\star$ together there are then $2n$ Majorana\footnote{Remember that  a Majorana spinor in the Weyl basis is written as $\gPs_M=\left(%
\begin{array}{c}
  \gc_\ga \\
  \bgc^{\dga}\\
\end{array}%
\right)
$.} spinors. The two supersymmetry parameters $\ge^i$ obey the relation $\bge_i=(\ge^i)^\star$, with $i=1,2$. \\
The transformation of the scalars is thus manifestly real, because also for the inverse vielbein $\gg$ the indices are barred and raised, respectively, under complex conjugation, i.e. $(\bgg^{i\hA}_\ba)^\star=\gg_{ia}^\hA$. The $\gd_\ge \gf^\hA$ therefore take values in a real $4n$-dimensional tangent bundle $T$.\\
The supersymmetry algebra is invariant under the automorphism group $SU(2)_R \times U(1)_R$ which rotates the supersymmetry generators into each other. This group is often referred to as the $R$-symmetry group, hence the label $R$. This means that the supersymmetry parameters $\ge^i$ take their value in the $SU(2)_R$ group\footnote{The supersymmetry variations are consistent with a $U(1)$ chiral invariance under which the scalars remain invariant, whereas the fermion fields and the supersymmetry transformation parameters transform, this is the $U(1)_R$ part of the $R$-symmetry group.}. As it concerns local transformations it is really an $SU(2)$-bundle.  So the $\ge^i$ take their values in the $SU(2)$-bundle. Note that for rigid supersymmetry the $\ge^i$ are constant and are therefore global sections of the $SU(2)$-bundle, which is equivalent to the triviality of this bundle.\\
In addition to the $\gg$'s there are related objects $V$ which appear in the supersymmetry variation of the fermions
\equ{
\gd_\ge\gl^a=\I \hD_\gm \gf^\hA V_\hA^{ai} \gs^\gm \bge_i -\gd_\ge \gf^\hA \OO{\hA}{a}{b} \gl^b \nn \pt
}
Supersymmetry imposes a number of relations and constraints on these objects, among which
\equ{
\gg_{ib}^\hA V_\hA^{aj}=\gd^j_i \gd^a_b \qquad \gg_{ia}^\hA V_\hB^{ai}=\gd^\hA_\hB \nn \km
}
implying that $\gg$ and $V$ are each others inverse. Furthermore one has
\equ{
G_{\hA\hB}=h_{a\bb} V^{ai}_\hA \bV^\bb_{\hB i} \qquad h_{a\bb}=\2 G_{\hA\hB}\gg_{ia}^\hA \bgg_\bb^{i\hB} \label{eq:app4t1} \km
}
where $h_{a\bb}$ is the Hermitian metric for the fermions, determining the kinetic terms as in \R{sa1}. As before $G_{\hA\hB}$ determines the kinetic terms for the hypermultiplet scalars.
This means that if we interpret $G_{\hA\hB}$ as the (real) metric on the target space, we can view the $V^{ai}_\hA$ as the vielbeins on it. In addition one can construct a covariantly constant antisymmetric tensor
\equ{
\gve_{ab}=\2 \gve^{ji} G_{\hA\hB} \gg_{ia}^\hA \gg_{jb}^\hB \nn \km
}
which satisfies
\equ{
\gve_{ij}\gve_{ab}V^{jb}_\hA=G_{\hA\hB} \gg^\hB_{ia}=h_{a\bb} \bV^\bb_{\hA i} \nn \pt
}
Because $\bV_{\hA i}^\bb=(V^{bi}_\hA)^\star$ the complex conjugated vielbein $V^\star$ is related to $V$, i.e., 
\equ{
\bV^\bc_{\hA i}=h^{\bc a}\gve_{ab} V_{\hA i}^b \nn \pt
}
In other words, we can use $h^{\ba a}$ and $\gve_{ab}$ to change barred- into unbarred indices. Consequently the vielbeins are pseudoreal and the $a$ indices are $Sp(2n,\mR)$ indices\footnote{A pseudoreal representation is a group representation that is equivalent to  its complex conjugate, but that is not a real representation.  One can show that the definition of `symplectic' and `pseudoreal' are equivalent, see for a very compact review for instance \cite{Okubo:1989vn}.}. Observe that we now have an explicit representation for the $\gve_{ab}$, see also \R{metrict1}, with the right properties.\\
The vielbein $V^{ai}_\hA$ maps the tangent bundle $T$ with indices $\hA$ to a product of two symplectic vector bundles with indices $a$ and $i$. Consequently, the geometric quantities such as the connection and curvature decompose into an $SU(2)$ and an $Sp(2n,\mR)$ part. By considering the antisymmetrized action of two covariant derivatives on the covariantly constant (inverse) vielbein one can compute \cite{Bagger:1983tt} that the Riemann curvature tensor of $T$, $R_{\hA\hB\hC\hD}$, decomposes into the $SU(2)$ and $Sp(2n,\mR)$ curvatures, $R_{\hA\hB ab}$ and $R_{\hA\hB ij}$ respectively.
The holonomy is then contained in $SU(2)\times Sp(2n,\mR)$. The tangent bundle $T$ is thus effectively decomposed as the product of two vector bundles $T=SU(2)\otimes Sp(2n,\mR)$ as in section \ref{hypergeo}. Note that the tensor product of two pseudoreal representations is a real representation, as it should be because $T$ is a real $4n$-dimensional manifold. We can also use the vielbeins to construct complex structures which satisfy the quaternionic algebra \R{quatal1}, another way of demonstrating the fact that we are dealing with a quaternionic target space. We will explicitly construct them in \R{quat2f3}.\\

Consider again rigid supersymmetry. In this case the \susy parameters are constant and the $SU(2)$-bundle is trivial: a zero connection form and $R_{\hA\hB ij}$ equal\footnote{In section \ref{hypergeo} the $SU(2)$ curvature is denoted by $\gO^r$.} to zero. This means that the holonomy is contained solely in $Sp(2n,\mR)$, by definition a hyperk\"ahler manifold. On the other hand, it can be shown \cite{Bagger:1983tt} that if the $\ge^i$ are nonconstant the $SU(2)$-curvature is nonzero. Summarizing: the above relations which determine the quaternionic geometry follow entirely from supersymmetry. \\

In the above we have tacitly  introduced the so-called \emph{chiral notation}, which amounts to keeping track of spinor chiralities through writing the $SU(2)_R$ index as an upper or lower index, \cite{deWit:1978sh}. Thus, upper and lower $SU(2)_R$ indices are correlated with a fixed spinor chirality and with either the fundamental or antifundamental representation. Depending on the spinor, an upper index might be associated with left or with right chirality. As a consequence $SU(2)_R$ indices are raised and lowered by complex conjugation, or by using $\gve_{ij}$. 

\subsubsection{Notation}
We have introduced several objects using a different notation. If we leaf back to section \ref{hypergeo} where the geometry of hypermultiplets was discussed, we observe the introduction of a set of vielbeins  $V_i^{\,a}\equiv V_{i\,\,A}^{\,a} dq^A$. These vielbeins characterized the quaternionic geometry with inverse vielbeins given by $V^{i\,\,B}_{\,a}$. In this appendix a different notation for the vielbeins is used, because the conventions of \cite{Theis:2003jj} are followed (as in chapter \ref{ch3}). However, in section \ref{hypergeo} the slightly different conventions of \cite{Davidse:2005ef} (the subject of chapter \ref{ch4}) are followed, see also appendix \ref{geopt}. These two notations are easily related though. Comparing \R{app4t1} to \R{metrict1} and \R{metrict2} one observes that 
\equ{
V^{\,a}_{i\,\,A}\leftrightarrow V^{ai}_\hA \qquad V^{i\,\,A}_{\, a}\leftrightarrow \gg_{ia}^\hA \nn \km
}
for the vielbeins and their inverses. Note that the metric $h_{a\bb}$  is called $G_{a\bb}$ in section \ref{hypergeo}.

\clearpage{\pagestyle{empty}\cleardoublepage}

\chapter{The double tensor multiplet}
\label{appdtm1}
The DTM provides a solution to the
constraints in the previous appendix. In the following we list the
coefficient functions appearing in its classical action and
transformation laws which satisfy the relations given above. They
receive quantum corrections from instantons, some determined in this
thesis, and 1-loop effects \cite{Antoniadis:2003sw}. For the scalar zweibeins we have
 \begin{equation}
  \gamma^\phi_{ia} = (W_\phi^{ai})^\dag = \frac{1}{\sqrt{2\,}} \begin{pmatrix}
  0 & -1 \\[2pt] 1 & 0 \end{pmatrix}\ ,\qquad \gamma^\chi_{ia} = e^\phi
  (W_\chi^{ai})^\dag = \frac{1}{\sqrt{2\,}}\, e^{\phi/2} \begin{pmatrix} 1 & 0
  \\[2pt] 0 & 1 \end{pmatrix} \km  \nn
 \end{equation}
while the tensor zweibeins are given by, for $I=1,2$,
 \begin{equation}
  g_{1\,ia} = -\frac{\I}{\sqrt{2\,}}\, e^{-\phi} \begin{pmatrix} -e^{\phi/2} &
  \chi \\[2pt] \chi & e^{\phi/2} \end{pmatrix}\ ,\qquad g_{2\,ia} = -
  \frac{\I}{\sqrt{2\,}}\, e^{-\phi} \begin{pmatrix} 0 & 1 \\[2pt] 1 & 0
  \end{pmatrix}\km \nn
 \end{equation}
and
 \begin{equation}
  f^{1\,ai} = \frac{\I}{\sqrt{2\,}}\, e^{\phi/2} \begin{pmatrix} -1 & 0 \\[2pt]
  0 & 1 \end{pmatrix}\ ,\qquad f^{2\,ai} = \frac{\I}{\sqrt{2\,}}\, e^{\phi/2}
  \begin{pmatrix} \chi & e^{\phi/2} \\[2pt] e^{\phi/2} & -\chi
  \end{pmatrix}\pt \nn
 \end{equation}
One may check that these quantities satisfy the relations
\R{rel1}--\R{EWidentity}, with
 \begin{equation}
  h_{a\ba}=\begin{pmatrix} 1 & 0 \\[2pt] 0 & 1 \end{pmatrix}\ ,\qquad
  \cE_{ab}=\begin{pmatrix} 0 & -1 \\[2pt] 1 & 0 \end{pmatrix}\km \nn
 \end{equation}
where we also have taken $\gve^{12}=1$.

The target space connections for the double-tensor multiplet are
particularly simple:
 \begin{equation}\label{eq:DTM-connections}
  \G{A}{a}{b} = 0\ ,\qquad \G{\phi}{i}{j} = 0\ ,\qquad \G{\chi}{i}{j}
  = \frac{1}{2}\, e^{-\phi/2} \lp 0 & -1 \\[2pt] 1 & 0 \rp\pt
 \end{equation}
Since $F_{AB}{}^I=0$, the scalar zweibeins $\W{A}{ai}$, $\gg^A_{ia}$
are covariantly constant with respect to these connections. The tensor
$k_{Ia\ba}$ can be determined from \R{k-Gamma}, with
 \begin{equation} \label{eq:Gamma2ab}
  \Gamma^{1\,a}{}_b = 0\ ,\qquad \Gamma^{2\,a}{}_b = -\frac{3\I}{4}\,
  e^\phi \lp 1 & 0 \\[2pt] 0 & -1 \rp\pt
 \end{equation}
Other quantities are the gravitino coefficients in the supersymmetry 
transformations of the tensors
 \begin{equation}
  {\Omega_1}^i{}_j = \frac{\I}{4}\, e^{-\phi} \lp \chi & 2\,
  e^{\phi/2} \\[2pt] 2\, e^{\phi/2} & -\chi \rp\ ,\qquad
  {\Omega_2}^i{}_j = \frac{\I}{4}\, e^{-\phi} \lp 1 & 0 \\[2pt] 0 &
  -1 \rp  \nn 
 \end{equation}
and, by using \R{gamma-omega1}, the coefficients in the
transformations of the gravitinos
 \begin{equation}
  \Gamma^{1\,i}{}_j = \frac{\I}{2}\, e^{\phi/2} \begin{pmatrix} 0 & 1
  \\[2pt] 1 & 0 \end{pmatrix}\ ,\qquad \Gamma^{2\,i}{}_j = - \frac{\I}
  {4}\, e^{\phi/2} \begin{pmatrix} -e^{\phi/2} & 2 \chi \\[2pt] 2 \chi
  & e^{\phi/2} \end{pmatrix}\pt \nn
 \end{equation}
Just as in the universal hypermultiplet, the four-$\gl$ terms come with
field-in\-de\-pen\-dent coefficients,
 \begin{equation} \label{cR_DTM}
  \frac{1}{4}\, V_{ab\,\ba\bb}\, \gl^a \gl^b\, \bgl^\ba \bgl^\bb = -
  \frac{3}{8}\, \big( \gl^1 \gl^1\, \bgl^1 \bgl^1 - 2\, \gl^1 \gl^2\,
  \bgl^1 \bgl^2 + \gl^2 \gl^2\, \bgl^2 \bgl^2 \big)\pt
 \end{equation} 
 
\subsubsection{Dualization and target spaces} 
As we have mentioned several times, the geometry of the UHM is a quaternionic one, in Minkowski space at least. When dualizing one or two scalars  this must change, since the target manifold is parametrized by the scalars, see also \cite{Theis:2002er}. For the Lorentzian theory we have the following dualization chain, i.e. UHM $\to$ TM $\to$ DTM:
\equ{
\frac{SU(1,2)}{U(2)} \to \frac{SO(1,3)}{SO(3)} \cong \frac{Sl(2, \mC)}{SU(2)} \to \frac{Sl(2,\mR)}{O(2)} \nn \pt
}
The scalar manifold for the tensor multiplet corresponds to Euclidean $AdS_3$. After Wick rotating we have a very different chain:
\equ{
\frac{Sl(3,\mR)}{Sl(2,\mR) \times SO(1,1)} \to \frac{SO(2,2)}{SO(2,1)} \cong Sl(2,\mR) \to \frac{Sl(2,\mR)}{O(2)} \nn \km
}
where the tensor multiplet now corresponds to $AdS_3$. \\
The geometry of these constraints must be consistent with supersymmetry. These constraints are different for Lorentzian and Euclidean signatures\footnote{For Euclidean $N=2$ supergravity in four dimensions see \cite{Theis:2001ef}.}. In general the precise constraints on the geometry of Euclidean hypermultiplets scalars is not known. 

\clearpage{\pagestyle{empty}\cleardoublepage} 
\chapter{Corrections to the propagators}
\label{propcor}
In this section we compute \R{inv-cG_eff}. Knowing that the (anti-)instanton will give corrections to $\cG_{AB}$ we can include its general form and derive the corresponding propagator. Including such a term gives for the kinetic terms of the two scalars in the DTM
\equ{
\left(\cG_{AB}+\cG_{AB}^{\inst}\right)\p{\gm}\gf^A\pu{\gm}\gf^B \label{eq:prop0}\km
}
which has to be used to derive the instanton corrected propagator. Write this term as
\equ{
\gf^A\left[\cG^\infty_{AB} + \cG^{\inst}_{AB}\right] \Box \gf^B \label{eq:prop2} \km
}
which means that the propagator is given by 
\equ{
\left(\cG_\infty^{AB} +\cG_{\inst}^{AB}\right)G(x, y) \label{eq:prop3}
}
and comparing with $\langle \phi^A(x)\, \phi^B(y) \rangle_{\textrm{{inst}}}$ in \R{cor_chi-chi} tells us that
\equ{
 \cG^{AB}_\inst = \lp 0 & 0 \\[2pt] 0 & g_s^{-2} (Y_+ + Y_-) \rp \label{eq:prop4}\km
}
precisely \R{inv-cG_eff}. Note that in \R{prop2} we write $\cG^\infty_{AB}$, because we have to take the asymptotic (constant) values of the field dependent $\cG_{AB}$. Otherwise we would end up with all kinds of interaction terms, we are dealing with a non-linear sigma model after all. The computation for the instanton corrections to the metric of the tensors \R{tisc1} is similar.

\clearpage{\pagestyle{empty}\cleardoublepage}

\chapter{Instanton corrections to the vielbeins} 
\label{inst-vielb}

Although we are mostly interested in instanton corrections to the scalar
and tensor metrics, it is also worthwhile to compute the corrected
zweibeins $\W{A}{ai}$ etc. Once these are known, one can compute the
vierbeins on the quaternionic side by using the results of \cite{Theis:2003jj}.
Obviously, they are determined only up to $SU(2)$ rotations. We use
this fact to choose the components as simple as possible.

We begin with determining $\W{A}{ai}$ from the first relation in
\R{GMh},
\equ{
 \cG_{AB}=\Tr(W_{\!A}^\dag\,h^T\,W_{\!B}^{})\nn \pt
} Since as
noted above both $\cE_{ab}$ and $h_{a\ba}$ are given by their classical
expressions, \R{EWidentity} implies that $W_{\!A}$ (and $f^I$)
are of the form
 \begin{equation*}
  \W{A}{ai} = \lp u_A & v_A \\[2pt] - \bar{v}_A & \bar{u}_A \rp\pt\nn
 \end{equation*}
We solve $\cG_{\phi\phi}^\inst=\cG_{\phi\chi}^\inst=0$ by setting
$W_{\!\phi}^\inst=0$. Furthermore, we can choose $u_\chi$ real and
$v_\chi=0$ and $\cG_{\chi\chi}^\inst=-g_s^2 Y$ then implies
 \begin{equation} 
  W_{\!\chi}^{\eff\,ai} = \frac{1}{2\sqrt{2\,}}\, \big( 2 e^{-\phi/2} - g_s
  Y \big) \lp 1 & 0 \\[2pt] 0 & 1 \rp\pt\nn
 \end{equation}
The $\gg^A$ now follow from \R{Gga_Mg}, $\gg^A=\cG^{AB}W_{\!B}^\dag
\,h^T$,
 \begin{equation}
  \gg^\phi_{\eff\,ia} = \gg^\phi_{ia}\ ,\qquad \gg^\chi_{\eff\,ia}
  = \frac{1}{2\sqrt{2\,}}\, \big( 2 e^{\phi/2} + g_s^{-1} Y \big) \lp 1 & 0
  \\[2pt] 0 & 1 \rp\pt\nn
 \end{equation}

Determining the $g_I$ is similar, from 
\equ{
M_{IJ}=\Tr(g_I^\dag\,g_J^{}\,h^{-1T})\nn
}
 and $M^\inst_{12}=
M^\inst_{22}=0$ we conclude that $g^\inst_2=0$. For $g_1$ we take
 \begin{equation}
  g_{1ia}^\eff = g_{1ia} + \frac{\I}{2\sqrt{2\,}}\, g_s Y \lp 1 & 0 \\[2pt] 0
  & -1 \rp\pt \nn
 \end{equation}
The coefficients $f^I$  are given by the relation $f^I=M^{IJ}
g_J^\dag\,h^{-1T}$:
 \begin{equation}
  f^{1ai}_\eff = f^{1ai} + \frac{\I}{2\sqrt{2\,}}\, g_s^{-1} Y \lp 1 & 0
  \\[2pt] 0 & -1 \rp \quad f^{2\,ai}_\eff = f^{2\,ai} - \frac{\I}
  {2\sqrt{2\,}}\, \chi_\infty\, g_s^{-1} Y \lp 1 & 0 \\[2pt] 0 & -1 \rp . \nn
 \end{equation}

In section \ref{sect_modul} we have observed that the NS sector in the
effective action is not affected by instantons. With the above vielbeins
we can check whether this  holds for the supersymmetry
transformations as well. Indeed, $g_2^\inst=0$ and $\Omega_2^\inst=0$
imply that the transformation \R{de_B_loc} of the NS 2-form
$B_{\mu\nu2}$, dual to the axion $\gs$, is not corrected, while
$\gg^\phi_\inst=0$ leaves the dilaton transformation unchanged.\\
As explained above, we have access only to the asymptotic region of
moduli space, where instanton corrections are constant. One question we
can answer nevertheless is whether the metric components $A_A^I$, which
vanish classically and which receive no corrections at
leading order, receive subleading field-dependent corrections.
Although we are not able to determine them explicitly, we now show their
existence. Toward this end, we compute the leading instanton correction
to the curvature $F_{AB}{}^I=2\partial^{}_{[A}A_{B]}^I$ by using the above
results in identity \R{F},
 \begin{equation}
  F_{AB}{}^I = - 2 \Tr\big( W_{\!A}^\dag\, h^T\, \Gamma^I W_{\!B}^{}
  \big) - 2 \Tr\big( W_{\!A}^\dag\, h^T\, W_{\!B}^{} \Gamma^{It}
  \big) \pt \nn
 \end{equation}
This relation must also hold in the effective theory if supersymmetry is to be 
preserved. The first trace on the right contains the connection
$\Gamma^{Ia}{}_b$, the second trace the connection $\Gamma^{Ii}{}_j$.
As argued above, the latter is not corrected by instantons. It is then
readily verified that the second trace vanishes identically as a result
of $\Gamma^{Ii}{}_j$ being symmetric. On the other hand, the first trace
does receive a contribution from the instanton-corrected connection
\R{Gamma_inst}. We find
 \begin{equation}\label{eq:field-strength}
  F_{\chi\phi}^{\eff\,1} = \I \frac{|Q_2|}{g_s^2}\, (Y_+ - Y_-)\ ,
  \qquad F_{\chi\phi}^{\eff\,2} = - \chi_\infty F_{\chi\phi}^{\eff\,
  1}\km
 \end{equation}
the other components vanish because of antisymmetry. A nonvanishing field
strength implies a nonvanishing connection, hence we conclude that
$A_A^I$ does get corrected after all. This is not in contradiction to
what we found in section \ref{sect_correl}, where $A_A^I=0$ to leading
order. The way out is that to determine the connection from the field
strength, one has to integrate over the moduli space. For that, one
needs information about the subleading corrections that allow us to go
beyond the asymptotic region such that we can integrate and
differentiate. The derivative of this connection, taken at its
asymptotic value, should then coincide with the field strength in
\R{field-strength}. 

\clearpage{\pagestyle{empty}\cleardoublepage} 
\chapter{The Toda equation}
\label{toda1} 
This appendix collects several technical details about the solution of
the Toda equation constructed in section \ref{todasec1}. In the first part we prove that 
$m_n\geq -2$. The proof that $\alpha=0$ is given the second part. In the third part the derivation of the
one-instanton solution is given.

\subsubsection{The lower bound on $m_n$} 

In this subsection we establish $m_n\geq -2$. Our starting point is the
ansatz \R{ansatz1}, which we substitute into the Toda equation
\R{Toda2}. This results in the following power series
expansion\footnote{Here we have not performed the splitting into
instanton sectors yet.}
 \begin{align} \label{eq:C.10}
  0 = & \sum_{n,m} r^{-m/2+\alpha+1}\, e^{-2n\sqrt{r}} \big[ (\Delta +
    n^2) f_{n,m} + (n\, a_{m+1}\, r^{-1/2} + b_{m+2}\, r^{-1})
    f_{n,m} \big] \notag \\[2mm]
  & + \sum_{n,m} \sum_{n',m'} r^{-(m+m')/2 + 2\alpha}\, e^{-2(n+n')
    \sqrt{r}}\, \big[ f_{n',m'} (\Delta + 2n^2) f_{n,m}  \notag
    \\[0.5mm]
  & \mspace{100mu} - \nabla f_{n,m} \cdot \nabla f_{n',m'} + 2 (a_{m+1}
    \, r^{-1/2} + b_{m+2}\, r^{-1}) f_{n,m}\, f_{n',m'} \big] \notag
    \\[7mm]
  & + \sum_{n,m} \sum_{n',m'} \sum_{n'',m''} r^{-(m+m'+m'')/2 + 3\alpha
    -1}\, e^{-2 (n+n'+n'') \sqrt{r}} f_{n,m}\, f_{n',m'}\, f_{n'',
    m''} \notag \\[0mm]
  & \mspace{100mu} \times \big[ n^2 + n\, a_{m+1}\, r^{-1/2} + b_{m+2}\,
    r^{-1} \big]\ ,
 \end{align}
where we have extended the definitions for $a_m$, $b_m$ given in
\R{4.6} to non-zero $\alpha$:
 \begin{equation} \label{eq:C.2}
  a_m = \half\, (2m - 4\alpha - 1)\ ,\qquad b_m = \4\, (m - 2\alpha)
  (m - 2\alpha - 2)\ .
 \end{equation}

In order to obtain a bound on $m_n$ (for which the $f_{n,m_n}\neq 0$),
we extract the leading order contributions in the $r$-expansion arising
from the single, double and triple sum in \R{C.10}. Starting at
$n=1$ and working iteratively towards higher values $n=2,3,\ldots$,
we find that at a fixed value of $n$ these contributions are
proportional to
 \begin{align} \label{eq:C.11}
  &\text{single sum}\  \propto\ r^{-m_n/2 + \alpha + 1} \notag \\[2pt]
  &\text{double sum}\  \propto\ r^{-m_n + 2 \alpha} \notag \\[2pt]
  &\text{triple sum}\  \propto\ r^{-3 m_n /2 + 3 \alpha - 1}\ .
 \end{align}
Investigating the $m_n$-dependence of these relations, we find that for
$m_n\le -3$ the leading order term in $r$ arises from the triple sum,
which decouples from all the other terms in \R{C.10}.\\

We now assume that for a fixed value $n$ there exsists an $f_{n,m_n}\neq
0$ for $m_n\le -3$. Extracting the equation leading in $r$ from
\R{C.10}, we find that
 \begin{equation} \label{eq:C.12}
  n^2\, f^3_{n,m_n} = 0\ ,\qquad m_n \le - 3\ ,
 \end{equation}
which has $f_{n,m_n}=0$ as its only solution. Hence, we establish the
lower bound
 \begin{equation} \label{eq:C.13}
  m_n \geq -2
 \end{equation}
for all values of $n$ or, equivalently, all instanton
sectors\footnote{Notice that this argument is not quite sufficient to
also fix $\alpha=0$, as for $\alpha=1/4$ the single and triple sums do
not decouple, which has been crucial in establishing \R{C.12}.}.

\subsubsection{Fixing the parameter $\alpha$} 

When making the ansatz \R{ansatz1} in order to describe membrane
instanton corrections to the universal hypermultiplet, we included the
parameter $\alpha\in[0,1/2)$ to allow for the possibility that the
leading term in the instanton solution occurs with a fractional power of
$g_s$. Based on the plausible assumption that the perturbation series
around the instanton gives rise to a power series in $g_s$ (and not
fractional powers thereof) we now give a proof that a consistent
solution of the Toda equation requires $\alpha=0$.

Splitting \R{C.10} into instanton sectors gives us the following
analogue of \R{4.5}
 \begin{align}\label{eq:C.1}
  0 = \sum_{n,m} &\ r^{-m/2+\alpha}\, e^{-2n\sqrt{r}}\, \Big\{ (\Delta
    + n^2)\, f_{n,m+2} + n\, a_{m+2}\, f_{n,m+1} + b_{m+2}\, f_{n,m}
    \notag \\[-2pt]
  & + \sum_{n',m'} r^{\alpha}\, e^{-2n'\!\sqrt{r}}\, \big[\, 2n\,
    a_{m'+1}\, f_{n',m-m'-1} + 2 b_{m'+2}\, f_{n',m-m'-2} \notag
    \\[1mm]
  & \mspace{134mu} + f_{n',m-m'}\, (\Delta + 2n^2) - \nabla f_{n',m-m'}
    \cdot \nabla \big] f_{n,m'} \notag \\[4mm]
  & + \sum_{n',m'} \sum_{n'',m''} r^{2\alpha}\, e^{-2(n'+n'')
    \sqrt{r}}\, f_{n,m'} f_{n',m''}\, \big[ n^2 f_{n'',m-m'-m''-2}
    \notag \\[-1.5mm]
  & \mspace{94mu} +  n\, a_{m'+1}\, f_{n'',m-m'-m''-3} + b_{m'+2}\,
    f_{n'',m-m'-m''-4} \big] \Big\} \, .
 \end{align}
Based on this equation we can now make several observations. First, we
find that the $N=1$ sector of \R{C.1} still gives rise to
\R{f1m}, with the coefficients $a_m$, $b_m$ now replaced by
\R{C.2}. To lowest order, $m=m_1$, this is just the equation
 \begin{equation} \label{eq:C.3}
  (\Delta + 1) f_{1,m_1}(u,v) = 0\ .
 \end{equation}
Second, we observe that the equation describing the $N=2$ sector is
modified to
 \bea
  0&=&(\Delta + 4)\, f_{2,m} + 2a_m\, f_{2,m-1} + b_m\, f_{2,m-2}   + \sum_{m'} r^{\alpha} \big[ f_{1,m-m'-2}  \nn\\[1mm]
  && + a_{m'+1}\,
    f_{1,m-m'-3} + b_{m'+2}\, f_{1,m-m'-4} - \nabla f_{1,m-m'-2}
    \cdot \nabla \big] f_{1,m'} \pt\nn
\eea
Note that for $\alpha=0$, the sum appearing in the second line is just
an inhomogeneous term to the equations determining $f_{2,m}$. For $\alpha
\neq 0$, however, the sum decouples due to the different powers in $r$.
Therefore, in the case $\alpha\neq 0$, the sum gives rise to an
additional constraint equation, which is absent for $\alpha=0$. Since
the sum contains the $f_{1,m}$ only, this additional relation imposes a
restriction on the $N=1$ instanton solution. Upon using \R{C.3},
this additional constraint reads at the lowest level
 \begin{equation} \label{eq:C.5}
  f_{1,m_1}^2 - (\nabla f_{1,m_1})^2 = 0\ .
 \end{equation}
For $\alpha\neq 0$ a non-trivial 1-instanton solution has to satisfy
both \R{C.3} and \R{C.5}, so that for establishing $\alpha=0$
it suffices to show that these equations have no common non-trivial
solution.

Suppose that $f_{1,m_1}\neq 0$, which by definition of $f_{1,m_1}$ has
to hold. We then multiply \R{C.3} with $f_{1,m_1}$, giving
 \begin{equation*}
  0 = f_{1,m_1}\, \Delta f_{1,m_1} + f_{1,m_1}^2 = f_{1,m_1}\, \Delta
  f_{1,m_1} + (\nabla f_{1,m_1})^2 = \half \Delta f^2_{1,m_1}\km \nn
 \end{equation*}
where we have used \R{C.5} in the first step. In terms of complex
coordinates $z=u+\I v$ it is $\Delta=4\partial_z\partial_{\bar{z}}$, and the general
solution reads
 \begin{equation*}
  f_{1,m_1}^2(z,\bar{z}) = g(z) + \bar{g}(\bar{z})\ . \nn
 \end{equation*}
Substituting this back into \R{C.3}, we find
 \begin{equation*}
  0 = (\Delta + 1) f_{1,m_1} = f_{1,m_1}^{-3} \big[\! - \partial_z
  g(z)\, \partial_{\bar{z}} \bar{g}(\bar{z}) + (g(z) + \bar{g}(\bar{z}))^2
  \big]\ ,\nn
 \end{equation*}
which is equivalent to
 \begin{equation*}
  \partial_z g(z)\, \partial_{\bar{z}} \bar{g}(\bar{z}) = g(z)^2 + 2 g(z) \bar{g}
  (\bar{z}) + \bar{g}(\bar{z})^2\ .\nn
 \end{equation*}
Since the right-hand side of this expression contains terms which are
(anti-) holomorphic, whereas the left-hand side does not, we find that
the only solution is given by $g(z)=\I c$ with $c\in\fieldR$ constant.
Thus $f_{1,m_1}=0$, which contradicts our assumption and shows that the
ansatz \R{ansatz1} does \emph{not} give rise to a one-instanton
sector if $\alpha\neq 0$. Conversely, a non-trivial one-instanton sector
exists for $\alpha=0$ only, which then fixes $\alpha=0$.

\subsubsection{The one-instanton solution} 

The general one-dimensional solution in the one-instanton sector was
given in \R{solf1m}. The functions $G_s(x)$ introduced there are
defined by
 \begin{equation}
  G_s(x) = x^{s+1} h_{s-1}(x)\ ,\nn
 \end{equation}
where $h_s(x)=j_s(x)+\I y_s(x)$ are the spherical Bessel functions of
the third kind. For $s\geq 0$ the $G_s(x)$ have no poles, they read
 \begin{equation}
  G_0(x) = e^{\I x}\ , \qquad G_{s>0}(x) = 2^{-s}\, e^{\I x}\,
  \sum_{k=1}^{s} \frac{(2s-k-1)!}{(s-k)!\, (k-1)!}\, (-2\I x)^k\ .\nn
 \end{equation}
Using the properties
 \begin{equation}
  x^2 h_s'' + 2x\, h_s' + \big[ x^2 - s(s+1) \big] h_s = 0\ ,\qquad
  h_s' + \frac{s+1}{x}\, h_s = h_{s-1}\ ,\nn
 \end{equation}
we easily verify the relation
 \begin{equation}
  (\partial_x^2 + 1)\, G_s(x) = 2s\, G_{s-1}(x)\ .\nn
 \end{equation}
The proof of \R{solf1m} is now simple:
 \begin{align}
  (\partial_x^2 + 1) f_{1,m}(x) & = \Re \sum_{s\geq 0}\, \frac{1}{s!\,
    (-2)^s}\, k_{1,m}(s)\, (\partial_x^2 + 1) G_s(x) \notag \\
  & = - \Re \sum_{s\geq 1}\, \frac{1}{(s-1)!\,(-2)^{s-1}}\, k_{1,m}
    (s)\, G_{s-1}(x) \notag \\
  & = - \Re \sum_{s\geq 0}\, \frac{1}{s!\,(-2)^s}\, k_{1,m}(s+1)\,
    G_s(x) \notag \\
  & = - \Re \sum_{s\geq 0}\, \frac{1}{s!\,(-2)^s}\, \big[ a_m
    k_{1,m-1}(s) + b_m k_{1,m-2}(s) \big]\, G_s(x) \notag \\ 
  & = -\, a_m f_{1,m-1}(x) - b_m f_{1,m-2}(x)\ . \label{eq:proof1}
 \end{align}
For the general $(u,v)$-dependent solution given in \R{solf2m}, the
proof is almost identical.

\clearpage{\pagestyle{empty}\cleardoublepage} 
\chapter{The moment maps}
\label{geopt} 
In this appendix we will elaborate on the quaternionic geometry introduced in section \ref{hypergeo}. Since we will apply the quaternionic geometry to supergravity we will call  the coordinates $\gf^\hA$, $\hA=1,\ldots, 4n$ with $n$ the number of hypermultiplets. For example, the quaternionic $1$-forms are now expressed as $V_i^{\,a}\equiv V_{i\,\,\hA}^{\,a} d\gf^\hA$. We can rewrite the hyperk\"ahler $2$-forms of equation \R{quat2form1} as
\equ{
K^r=\frac{\I}{2} G_{\ba b} V_i^b\wedge \bar{V}^{j \ba} (\tau^r)^i{}_j\km \label{eq:quat2f3}
}
with $\gt^r$ the Pauli matrices, $r=1,2,3$, $i=1,2$ and $a,\ba=1,\ldots,2n$.  These $2$-forms satisfy the quaternionic algebra \R{quatal1}. As in \R{SU2curv1} we have the $SU(2)$ connection $1$-form $\go^r=\go^r_\hA d\gf^\hA$ with $SU(2)$ curvature\footnote{The convention for the $SU(2)$ connection and curvature is chosen to be the same as e.g.\ in \cite{deWit:2001bk}. With respect to \cite{Bergshoeff:2004nf, Bergshoeff:2002qk}, our $SU(2)$ connection is chosen (minus) twice
the one in \cite{Bergshoeff:2004nf, Bergshoeff:2002qk}, and therefore also the $SU(2)$ curvature is
(minus) twice as large.}
\equ{
\gO^r\equiv d\go^r - \2 \gve^{rst}\go^s \wedge \go^t \label{eq:SU2curv2} \pt
}
In the conventions of \cite{Bergshoeff:2004nf, Bergshoeff:2002qk} we write the Einstein property as
\equ{
R_{\hA\hB}=\frac{1}{4n} G_{\hA\hB} R  \nn \km
}
with $R$ the constant Ricci scalar. The relation between the $SU(2)$ curvature $2$-forms and the hyperk\"ahler $2$-forms \R{hyperkrel1} reads
\equ{
\gO^r=\gn K^r \qquad \textrm{with}\qquad \gn\equiv \frac{1}{4n(n+2)} R \label{eq:curv7} \pt
}
The $SU(2)$ curvature obeys
\equ{
d\gO^r=\gve^{rst}\go^s\wedge \gO^t \km \label{eq:Ocurv2}
}
compare with \R{condqh1} and \R{hyperkrel1}. 
For the potential we need the moment maps. They are defined from\footnote{Our definition of the moment map is the same as
in \cite{deWit:2001bk}. This normalization is different from \cite{Bergshoeff:2004nf, Bergshoeff:2002qk},
and our moment maps are (minus) two times the ones defined in
\cite{Bergshoeff:2004nf, Bergshoeff:2002qk}.}
\equ{
K^r_{\hA\hB} k_I^\hB=D_\hA P^r_I=\p{\hA}P^r_I-\gve^{rst}\go_\hA^s P^t_I \km \nn
}
where $I$ labels the various isometries and $D_\hA$ is the $SU(2)$ covariant derivative, as in \R{condqh1}. One can solve this relation for the moment maps which yields \cite{deWit:2001bk}
\equ{
P^r_I=-\frac{1}{2n\gn} K^r_{\hA\hB}D^\hA k^\hB_I \label{eq:moment1} \pt
}

In supergravity, the value of $\nu$ is fixed in terms of the
gravitational coupling constant. If we normalize the kinetic terms of
the graviton and scalars in the supergravity action as
 \begin{equation} \label{eq:norm-action}
  e^{-1} \cL_\text{kin} = - \frac{1}{2\kappa^2}\, R(e) - \frac{1}{2}\,
  G_{\hA\hB} \partial_\mu \phi^\hA\, \partial^\mu \phi^\hB\ ,
 \end{equation}
then local supersymmetry fixes $\nu=-\kappa^2$. This is in accordance
with \cite{Bergshoeff:2004nf, Bergshoeff:2002qk} and with \cite{Bagger:1983tt} after a rescaling of the metric
$G_{\hA\hB}$ with a factor 1/2. For the universal hypermultiplet, we will
work with conventions in which $\nu=-1/2$, so we set $\kappa^2=1/2$
below. To compare with \cite{deWit:2001bk}, we have to multiply the Lagrangian
\R{norm-action} by 2 and then set $\kappa^2=2$.\\

We now include the scalar potential that arises after gauging a single
isometry, so we can leave out the subscript $I$. The isometry can then be gauged by the graviphoton and in the
absence of any further vector multiplets, the relevant terms in the
Lagrangian are
 \begin{equation}
  e^{-1} \cL = - \frac{1}{2\kappa^2}\, R - \frac{1}{2}\, G_{\hA\hB} D_\mu
  \phi^\hA\, D^\mu \phi^\hB -V \label{eq:app9t1a} \km
 \end{equation}
the potential $V$ is given by
\equ{
V= \big( 2 \kappa^{-2} G_{\hA\hB}\, k^\hA k^\hB - 3\,
  \vec{P}\! \cdot\! \vec{P}\, \big) \label{eq:app9t1b} \pt
}
$D_\mu$ is the covariant derivative with respect to the gauged
isometry that corresponds to the Killing vector $k^\hA$. The factor of
$\kappa$ has to appear on dimensional grounds. For
$\kappa^2=2$ this agrees precisely with the result in \cite{deWit:2001bk}, we will set $\kappa^2=1/2$.

Our conventions are chosen such that they apply to the
universal hypermultiplet metric and the conventions used in \cite{Davidse:2004gg, Theis:2003jj}.
At the classical level we have
 \begin{equation}
  d s^2 = G_{\hA\hB}\, d\phi^\hA \otimes d\phi^\hB = d\phi^2 + e^{-\phi}
  (d\chi^2 + d\varphi^2) + e^{-2\phi} (d\sigma + \chi d\varphi
  )^2\pt \nn
 \end{equation}
For the corresponding Ricci tensor we find
 \begin{equation}
  R_{\hA\hB}= - \frac{3}{2}\, G_{\hA\hB}\pt\nn
 \end{equation}
The Ricci scalar is then $R=-6$ and therefore we have $\nu=-1/2$. This
implies that in these conventions we should set $\kappa^2=1/2$, which
is equivalent to a cosmological constant $\Lambda=-3/2$, see \R{tod3} on the 
quaternionic manifold. Note that the constant $\gn$ is identical to the $\gl$ introduced in \R{hyperkrel1}.

\subsubsection{Quaternionic geometry of the PT metric} 
\label{appA.2}
We present the quaternionic geometry of the UHM in the PT framework. The quaternionic properties of the PT metric can be demonstrated by
constructing the corresponding quaternionic 1-form vielbeins  which we parameterize as
 \begin{equation}
  V_i^a = \lp \bar{a} & - \bar{b} \\ b & a \rp\pt \nn
 \end{equation} 
Substituting this ansatz into \R{quatmet1} we obtain
 \begin{equation} \label{eq:metric1}
  \d s^2 = a \otimes \bar{a} + b \otimes \bar{b} + \text{c.c.}\pt 
 \end{equation}
Comparing this expression with the PT metric \R{tod1} leads to
 \begin{equation}
  a = \frac{1}{\sqrt{2}\, r}\, \big( f^{1/2}\, \d r + \I f^{-1/2}\, (\d t +
  \Theta) \big)\ ,\qquad b = \frac{1}{\sqrt{2}\, r}\, (f e^h)^{1/2}\,
  \big( \d u + \I\, \d v \big)\pt \nn
 \end{equation}  
The computation of the quaternionic 2-forms \R{quat2f3} then yields
 \begin{equation}
  K^1 = -\I (a \wedge b - \bar{a} \wedge \bar{b})\ ,\quad
  K^2 = a \wedge b + \bar{a} \wedge \bar{b}\ ,\quad
  K^3 = -\I (a \wedge \bar{a} + b \wedge \bar{b})\pt \nn
 \end{equation}
These satisfy the quaternionic algebra \R{quatal1}.
Using \R{curv7} and \R{Ocurv2}, we then determine the $SU(2)$
connection for the PT metric,
 \begin{gather}
  \go^1 = \frac{1}{r}\, e^{h/2} \d v\ ,\qquad \go^2
    = \frac{1}{r}\, e^{h/2} \d u\ , \notag \\
  \go^3 = -\frac{1}{2r}\, (\d t + \Theta) - \frac{1}{2}\,
    (\partial_v h\, \d u - \partial_u h\, \d v)\pt \nn
 \end{gather}

The PT metric has a shift symmetry in $t$. In coordinates $(r,u,v,t)$
the corresponding Killing vector is given by
 \begin{equation}
  k^A = (\, 0\, ,\, 0\, ,\, 0\, ,\, e_0\,)^\mathrm{T}\pt \label{eq:Kv2}
 \end{equation} 
The moment maps of this shift symmetry can be computed from \R{moment1}.
The result is \emph{independent} of the functions $f$, $h$, and $\Theta$
and reads
 \begin{equation}
  P^1 = 0\ ,\qquad P^2 = 0\ ,\qquad P^3 = \frac{e_0}{r}\pt \label{eq:moment2}
 \end{equation}

\clearpage{\pagestyle{empty}\cleardoublepage} 
}
\backmatter
\renewcommand{\chaptermark}[1]{\markboth{ #1}{}}
\fancyhead{}
\fancyhead[LE, RO]{\thepage}
\fancyhead[CO]{\slshape\nouppercase{\leftmark}}
\fancyhead[CE]{\slshape\nouppercase{\leftmark}}

\makeatletter
\def\thickhrulefill{\leavevmode \leaders \hrule height 1ex \hfill \kern \z@}
\def\@makechapterhead#1{%
  \vspace*{10\p@}%
  {\parindent \z@ \centering \reset@font
        \Huge \bfseries #1\par\nobreak
        \par
        \vspace*{10\p@}%
    \vskip 60\p@
  }}
\def\@makeschapterhead#1{%
  \vspace*{10\p@}%
  {\parindent \z@ \centering \reset@font
        \Huge \bfseries #1\par\nobreak
        \par
        \vspace*{10\p@}%
    \vskip 60\p@
  }}

\addcontentsline{toc}{chapter}{Bibliography}
\bibliographystyle{eigenutcaps}
\fontsize{11}{13pt}\selectfont
\bibliography{bib1}

\end{document}